\documentclass[12pt,twoside]{book} 

\usepackage{cranfieldthesis}

\usepackage{bm}
\usepackage{graphicx}
\usepackage{epstopdf}
\usepackage{amsmath}
\usepackage{amsfonts}
\usepackage[version=4]{mhchem}
\usepackage{siunitx}
\usepackage{longtable,tabularx}
\usepackage{booktabs}
\usepackage{multirow}
\usepackage{algorithm2e}
\usepackage{caption,subcaption}
\usepackage{mathtools}
\usepackage{mathrsfs} 
\usepackage{hyphenat}
\usepackage[shortcuts]{extdash}
\usepackage[sort&compress, numbers]{natbib}

\usepackage{titlesec}
\allowdisplaybreaks

\DeclareMathAlphabet{\mathpzc}{OT1}{pzc}{m}{it}

\DeclarePairedDelimiter\abs{\lvert}{\rvert}%
\DeclarePairedDelimiter\norm{\lVert}{\rVert}%
\makeatletter
\let\oldabs\abs
\def\abs{\@ifstar{\oldabs}{\oldabs*}}

\let\oldnorm\norm
\def\norm{\@ifstar{\oldnorm}{\oldnorm*}}
\makeatother

\let\savebigtimes\bigtimes
\let\bigtimes\relax
\let\saveamalg\amalg
\let\amalg\relax
\let\savedegree\degree
\let\degree\relax

\usepackage{mathabx}
\let\bigtimes\savebigtimes
\let\amalg\saveamalg
\let\degree\savedegree

\usepackage[usenames, dvipsnames]{color}
\definecolor{External}{RGB}{0,0,0}
\definecolor{Internal}{RGB}{0,0,0}
\definecolor{Both}{RGB}{0,0,0}
\definecolor{rita}{RGB}{0,0,0}
\definecolor{reviewer1}{RGB}{0,0,0}
\definecolor{reviewer2}{RGB}{0,0,0}
\definecolor{reviewer3}{RGB}{0,0,0}

\title{Multi-Fidelity Modelling of Low-Energy Trajectories for Space Mission Design}
\author{Rita Neves}
\date{March 2019}
\school{Aerospace, Transport and Manufacturing}
\course{Space Engineering}
\degree{PhD}
\academicyear{2016--2019}

\supervisortext{Supervisor: Dr. Joan-Pau S\'anchez \\ Associate Supervisor}  
\supervisor{Dr. \textcolor{Both}{Stephen} Hobbs}
\copyrightyear{2019}
\fulfilment{} 

\usepackage{fancyhdr}

\begin{document}


\pagestyle{plain} 

\frontmatter

\maketitle

\begin{abstract}

The proposal of increasingly complex and innovative space endeavours poses growing demands for mission designers. In order to meet the established requirements and constraints while maintaining a low fuel cost, the use of low-energy trajectories is particularly interesting. These paths in space allow spacecraft to change orbits and move with little to no fuel, but they are computed using motion models of a higher fidelity than the commonly used two-body problem. For this purpose, perturbation methods that explore the third-body effect are especially attractive, since they can accurately convey the system dynamics of a three-body configuration with a lower computational cost, by employing mapping techniques or exploring analytical approximations. 

The focus of this work is to broaden the knowledge of low-energy trajectories by developing new mathematical tools to assist in mission design applications. In particular, novel models of motion based on the third-body effect are conceived and classified by the forces they account for (conservative or non-conservative). The necessary numerical tools to complement the trajectory design are developed: this includes differential correction methods and targeting schemes, which take advantage of the Jacobian matrices derived from the presented models to generate full low-thrust control laws.

One application of this analysis focuses on the trajectory design for missions to near-Earth asteroids. Two different projects are explored: one is based on the preliminary design of separate rendezvous and capture missions to the invariant manifolds of libration point $L_2$. This is achieved by studying two specific, recently discovered bodies and determining dates, fuel cost and final control history for each trajectory. The other covers a larger study on asteroid capture missions, where several asteroids are regarded as potential targets. The candidates are considered using a multi-fidelity design framework. Its purpose is to filter through the trajectory options using models of motion of increasing accuracy, so that a final refined, low-thrust solution is obtained. The trajectory design hinges on harnessing Earth's gravity by exploiting encounters outside its sphere of influence, the named \textit{Earth-resonant encounters}.

An additional application explored in this investigation is the search and computation of periodic orbits for different planetary systems, following the current interest for missions involving distant retrograde and prograde orbits.

\textcolor{External}{In summary, this thesis presents four novel methods to model the third-body perturbation, distinct in their suitability for applications from real-time computations to long-term orbital predictions. These, together with the additionally developed tools for trajectory design, are applied in two asteroid mission cases. The developed Earth-resonant encounters allow for a very large increase in retrievable mass with respect to the state-of-the-art, namely for the cases of six near-Earth asteroids presented.}

\end{abstract}

\sstableofcontents

\sslistoffigures

\sslistoftables

\begin{listofabbreviations}
    \abbrev{2BP}{Two-Body Problem}
    \abbrev{ARRM}{Asteroid Redirect Robotic Mission}
    \abbrev{AU}{Astronomical Unit}
    \abbrev{CR3BP}{Circular-Restricted Three-Body Problem}
	\abbrev{CR4BP}{Circular-Restricted Four-Body Problem}    
    \abbrev{DPO}{Distant Prograde Orbit}
	\abbrev{DRO}{Distant Retrograde Orbit}	
	\abbrev{E-KM}{Euler-Keplerian Map}
	\abbrev{ESA}{European Space Agency}
	\abbrev{GVE}{Gauss' Variational Equations}
	\abbrev{GVE-3B}{Gauss' Variational Equations for the Third-Body Effect}
	\abbrev{JPL}{Jet Propulsion Laboratory}
	\abbrev{K3BP}{Keplerian Third-Body Potential}
    \abbrev{KM}{Keplerian Map}
    \abbrev{LCLM}{Low-Cost Likelihood Map}
    \abbrev{LPE}{Lagrange Planetary Equations}
    \abbrev{LPO}{Libration Point Orbit}
    \abbrev{LVLH}{Local Vertical, Local Horizontal}
    \abbrev{MPC}{Minor Planet Centre}
    \abbrev{NASA}{National Aeronautics and Space Administration}
    \abbrev{NEA}{Near-Earth Asteroid}
    \abbrev{NLP}{Non-Linear Programming}
    \abbrev{P3DRO}{Period-3 DRO}
    \abbrev{PAP-KM}{Periapsis-Apoapsis-Periapsis Keplerian Map}
    \abbrev{QPDRO}{Quasi-Periodic DRO}
    \abbrev{SOA}{State-of-the-Art} 
	\abbrev{STM}{State Transition Matrix}   
    \abbrev{VOP}{Variation of Parameters} 
    \abbrev{TOF}{Time of Flight} 
\end{listofabbreviations}

\chapter*{List of Symbols}
\addcontentsline{toc}{chapter}{List of Symbols}

\abbrev{$a$}{Semi-major axis}
\abbrev{$b$}{Semi-minor axis}
\abbrev{$C$}{Jacobi constant}
\abbrev{$E$}{Eccentric anomaly}
\abbrev{$e$}{Eccentricity}
\abbrev{$FM$}{Figure of merit}
\abbrev{$G$}{Gravitational constant}
\abbrev{$g_0$}{Gravitational acceleration}
\abbrev{$I_{SP}$}{Specific impulse}
\abbrev{$i$}{Inclination}
\abbrev{$J$}{Cost function}
\abbrev{$\text{\textbf{J}}$}{Jacobian matrix}
\abbrev{$L$}{Angular momentum}
\abbrev{$l$}{True longitude}
\abbrev{$\bm{l}$}{Eigenvector}
\abbrev{$M$}{Mean anomaly}
\abbrev{$m$}{Mass}
\abbrev{$n$}{Mean motion}
\abbrev{$O_{I xyz}$}{\textcolor{Both}{Inertial reference frame}}
\abbrev{$O_{R xyz}$}{\textcolor{Both}{Synodic reference frame}}
\abbrev{$O_{\Earth xyz}$}{\textcolor{Both}{Earth-pointing reference frame}}
\abbrev{$O_{r \theta h}$}{\textcolor{Both}{LVLH frame}}
\abbrev{$O_{eph}$}{\textcolor{Both}{Orbital plane frame}}
\abbrev{$o$}{\textcolor{Both}{Generalized momenta}}
\abbrev{$P$}{\textcolor{Both}{Number of periods}}
\abbrev{$p$}{\textcolor{Both}{Semi-latus rectum}}
\abbrev{$\mathpzc{R}$}{Disturbing function}
\abbrev{$\bm{r}$}{Position vector}
\abbrev{$\bm{s}$}{State vector}
\abbrev{$T$}{Orbital period}
\abbrev{$\mathpzc{T}$}{Kinetic energy}
\abbrev{$\mathpzc{U}$}{Potential energy}
\abbrev{$\bm{u}$}{Control vector}
\abbrev{$\bm{v}$}{Velocity vector}
\abbrev{$v_e$}{Exhaust velocity}
\abbrev{$_{\Earth}$}{Earth subscript}
\abbrev{$_{\Sun}$}{Sun subscript}
\begin{center}
	$\ast$
\end{center}
\abbrev{$\alpha$}{Phasing of the massless particle with the disturbing body}
\abbrev{$\Delta v$}{Velocity change}
\abbrev{$\theta$}{Angle formed by the line connecting primaries with the massless particle}
\abbrev{$\bm{\kappa}$}{Vector of orbital elements}
\abbrev{$\lambda$}{Eigenvalue}
\abbrev{$\mu$}{Gravitational parameter}
\abbrev{$\nu$}{True anomaly}
\abbrev{$\upsilon$}{Argument of latitude}
\abbrev{$\bm{\Phi}$}{State transition matrix}
\abbrev{$\Omega$}{Longitude of the ascending node}
\abbrev{$\omega$}{Argument of the periapsis}
\begin{center}
	$\ast$~$\ast$~$\ast$
\end{center}
\abbrev{$\text{\textbf{A}}$}{A matrix}
\abbrev{$\mathpzc{K}$}{A Keplerian element}
\abbrev{$\bm{z}$}{A vector}
\abbrev{$\norm{\bm{z}}$}{Euclidean norm of a vector}
\abbrev{$\bm{Z}$}{A free variables vector}

\chapter{Acknowledgements}

When I started this PhD journey, I had no idea what to expect of the three years ahead. I know now that it would have been impossible to get to the end without the help of all the people I'm naming here. I hope I have thanked all of you properly, and these words are just one more way to do so.

First of all, to my supervisor, Pau: my PhD experience wouldn't have been as good without your guidance. Thank you for always having time (and patience) to discuss things with me, for giving me freedom to pursue my own side projects, for challenging me with new opportunities, for editing so much of my written work and, mainly, for always encouraging me to do better.

To the friends who shared the PhD experience with me: Tiago, Eddie, Rob, Melissa, thank you for the coffee breaks, the times to complain and the unspoken understandings. To Cat, who has been supporting me since we were 6 and always found time for a chat and a musical in London. To my friends that encouraged me from afar, wherever you were: Ru, Spitafires, Francisco, In\^es, Sofia.

\vspace{5mm}

Para a minha fam\'ilia, por me obrigarem a explicar a minha pesquisa em almo\c{c}os de domingo e pela preocupa\c{c}\~ao que sempre tiveram comigo. M\~ae e Pai, obrigada por me aturarem, por virem visitar-me onde quer que esteja e por sempre me terem dado as condi\c{c}\~oes, os conselhos e a confian\c{c}a para arriscar. Isabel, por estarmos ligadas. N\~ao teria conseguido isto sem voc\^es.

Duarte, obrigada por me levares a passear quando est\'a bom tempo, por estares comigo todos os dias e saberes sempre o que preciso.
\mainmatter

\chapter{Introduction}
\label{chap:intro}
In the past few years, the space industry has been contemplating innovative missions that are yet to be achieved: the colonisation of Mars, space manufacturing and mining, on-orbit servicing and asteroid capture are examples amongst many.  Most of the necessary technologies for these missions have already been conceptualised; however, the global budget for space exploration is finite. Thus, in order to increase the feasibility of each concept, the associated mission cost must be the lowest possible. 

One of the main ways to minimise the overall cost of a space mission is to make sure the spacecraft trajectory design requires the least possible amount of fuel, while complying with the desired requirements: more fuel means more carried mass, and the latter is a valuable commodity in space. Thus, computing optimal trajectories is essential to obtain a feasible, affordable mission.

Designing a spacecraft's trajectory is, however, no simple task. On one hand, computing motion in space requires modelling the spacecraft's interaction with the environment (i.e. the Solar System) as accurately as possible. If the influence of all planets and celestial bodies is considered at once, the problem becomes intractable: thus, some simplifications must be adopted. Nevertheless, the dynamical behaviour of the system must remain analogous, since the challenges posed by the aforementioned prospective missions require a very accurate modelling of the sensitivities in multi-body dynamics. 

On the other hand, although increasing the level of detail will yield a higher accuracy, models of motion get more complex and computationally costly with the number of bodies considered. Since finding the optimal trajectory requires the calculation of a very large number of possibilities, the model should be as computationally inexpensive as possible for the search not to become infeasible.

In summary, in order to successfully tackle the challenges proposed by new innovative missions, it is very advantageous to employ low-computational cost methods that are still accurate enough to replicate the sensitivities of multi-body dynamics. This is particularly necessary when modelling low-energy trajectories: paths in the Solar System that cost little to no fuel to traverse. 

Finally, these low-cost, accurate models will be tested to tackle the design of some of the aforementioned innovative missions. The particular application scenario of this thesis is the case of asteroid capture and rendezvous trajectories, for which low-energy transfers are obtained and presented.

\section{Literature Review}

This section presents the background and related research for the main threads composing this dissertation: \textcolor{Both}{space trajectory design, low-energy trajectories}, models of motion for the three-body problem and asteroid missions. Additional references are presented in the following chapters, as necessary.

\subsection{\textcolor{Both}{Space Trajectory Design}}
\label{sec:opt_traj_design}

\begin{figure}[h!]
	\centering
	\includegraphics[width=\linewidth]{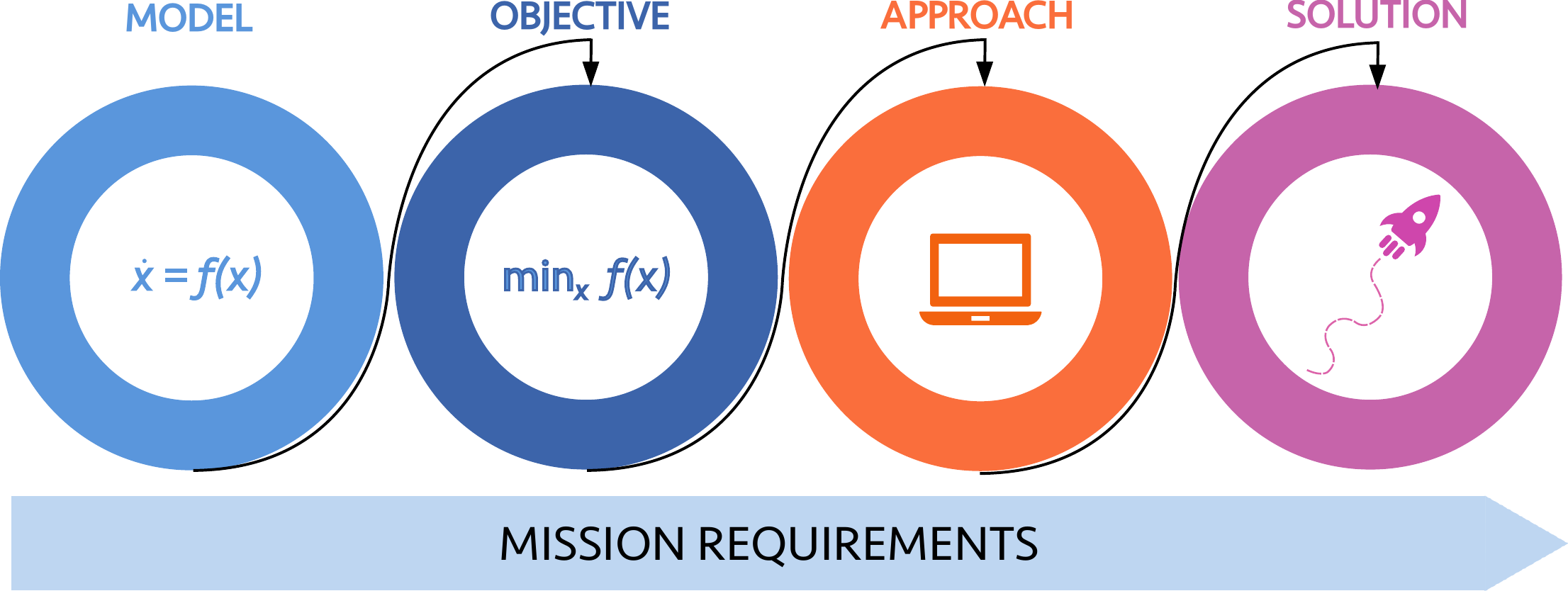}
	\caption{\label{fig:lin_process}Space mission design steps as a linear process}
\end{figure}

Designing a trajectory for a spacecraft to traverse is one of the first activities in the long life cycle of a space mission. The proposal of increasingly elaborate concepts, from asteroid capture to interplanetary cubesats, imposes several complex requirements and constraints to the trajectory design. Shirazi et. al. \cite{ltreview} define the latter as a procedure consisting in four steps: defining the motion model, establishing the objective function, deciding on the computational approach and, finally, solving the problem. This linear process can be visualised in Figure \ref{fig:lin_process}.

The first step of this process involves understanding the system dynamics by deriving the equations of motion for the spacecraft. The second step requires making decisions on the main drivers of the mission, by defining cost functions. Amongst other examples, these can be related to the amount of time the mission will take, the carried mass or fuel budgets, depending on the trade-offs established by the mission requirements.

The third and fourth steps have to do with the optimisation methods used to obtain a solution. Given the information about the dynamical system and constraints, the trajectory design problem is often formulated in literature as a common optimisation problem \cite{izzo_global}. The trajectory optimisation or \textit{optimal control problem} is composed of a collection of a given number of phases; for each of them, the system dynamics are described by a set of variables \cite{betts}:
\begin{align}
\bm{z} = \begin{Bmatrix}
\bm{s}(t)\\ \bm{u}(t)\\
\end{Bmatrix}
\end{align}
in which $\bm{s}(t)$ is the state vector and $\bm{u}(t)$ is the control vector. The dynamics of the system are described by the differential equations of motion, model-dependant:
\begin{align}\label{eq:dynconst}
\dot{\bm{s}} = f[\bm{s}(t), \bm{u}(t), t]
\end{align}

Furthermore, the solution must satisfy initial and final time conditions, algebraic constraints and bounds on the state and control variables. Having this figured out, the optimal control problem becomes the matter of determining the set of dynamical variables and times that minimise \textcolor{Internal}{a} \textit{cost function} $J$ for each phase:
\begin{align}
J = \bm{\Psi}[\bm{s}(t_0), t_0, t_1]
\end{align}	
The previously mentioned phases for the optimal control problem have to do with a possible partitioning of the time domain, which can be delimited by the bodies visited, the number of flybys or any other factor. Consequently, the dynamics cannot change within a phase, but may do so from one to another. As such, event constraints must be imposed between phases, so they can be linked in order to achieve a fully connected trajectory.

The characteristic that makes the trajectory design problem unique in its kind is the complex nature of the system dynamics, which limits the straightforward use of numerical methods to obtain an optimal solution, and allows the presence of a large number of locally optimal solutions \cite{izzo_global}. This definition of dynamical models for Eq. \eqref{eq:dynconst} depends on the propulsion system used in the spacecraft mission. There are two main types of propulsion models used when designing a space trajectory: impulsive and low-thrust systems.

In the case of impulsive trajectory design, the thrust phases are considered to be very short compared to the transfer time; thus, they can be approximated by singular events that change the spacecraft's velocity instantaneously, while its position remains fixed. The low-thrust optimal control problem is considerably different, since the thrust magnitude and direction have continuous time histories that must be determined \cite{conway}.

\subsubsection{Propulsion System Models}

An impulsive propulsion system is modelled as a series of trajectory arcs connected by instantaneous changes in the spacecraft's velocity, commonly known as $\Delta v$'s. In this way, the system's \textcolor{Internal}{control} vector is assumed to be zero ($\bm{u}(t) = 0$) and the manoeuvres are considered as sudden velocity increments ($\Delta v$) with zero burn times ($\Delta t = 0$). 

Under an impulsive thrust assumption, the $\Delta v$ quantity is proportional to the fuel consumption; the relationship between this velocity change and the fuel mass can be observed in \textcolor{External}{the Tsiolkovsky rocket equation:}
\begin{align}
m_f = m_0 \exp{\frac{-\Delta v}{g_0 I_{SP}}}\label{eq:tsiolkovsky}
\end{align}
\textcolor{Internal}{in which $m_0$ and $m_f$} are respectively the initial and final mass, $g_0$ is the standard gravitational acceleration and $I_{SP}$ is the specific impulse, characteristic of the engine.

This model is relatively simple to use in the simulation of trajectories with large accelerations and a rapid spacecraft response to commanded manoeuvres, which is typical of a chemical engine. These have a fixed amount of energy per unit mass, provided by the fuel carried from departure, which limits the available exhaust velocity ($v_e$). However, the rate at which energy can be supplied to the propellant is independent of its mass, so very high powers and thrust levels can be achieved\footnote{http://sci.esa.int/smart-1/34201-electric-spacecraft-propulsion/?fbodylongid=1535, Accessed 2019-01-10}.

The low-thrust model \textcolor{Internal}{can be} employed when the spacecraft's propulsion system is electric. Several types of electrical spacecraft engines exist and they can be divided into three categories, depending on the method used to accelerate the propellant: electrothermal, electrostatic, and electromagnetic \cite{2010low}. Since they work with this added energy source, their exhaust velocity $v_e$ (proportional to the specific impulse $I_{SP}$) can be much larger than that available to a chemical propulsion system. For instance, the highest specific impulse for a chemical propellant ever tested was 542 seconds \cite{combustion}, while electrical propulsion yields a much higher specific impulse (thousands of seconds). As such, low-thrust systems enable missions that would not be possible with the impulsive alternative. 

Most space missions carried out throughout history employ chemical propulsion, as opposed to electric propulsion \cite{sims}. Prior to the 1990's, interplanetary solar electric propulsion systems were not available. Since then, these have been utilized on a range of missions, from NASA's Deep Space 1 and Dawn missions to JAXA's Hayabusa \cite{hayabusa}. 

In contrast to the impulsive system design, the thrust vector in a low-thrust model is assumed to change with time. The added dimensionality to the equations makes the trajectory design much more complex \cite{ltreview}, requiring the solution to an optimal control problem in which the goal is to find the best control law for the mission design. Plus, the low-thrust $\Delta v$ differs from the impulsive one due to gravity losses, since the velocity change is no longer instantaneous \cite{fortescue}.

\subsubsection{Solving the Optimal Control Problem}

Following the classification made by Betts \cite{betts}, methods to solve the optimal control problem can generally be classified as either \textit{direct} or \textit{indirect} methods, although not every technique falls neatly into these categories.

Indirect methods are characterized by analytically solving the optimality conditions, stated in terms of the adjoint differential equations and associated boundary conditions, using the calculus of variations. This requires including in the problem the co-state variables (or adjoint variables or Lagrange multipliers), which are equal in number to the state variables, and their governing equations. As such, the size of the dynamical system doubles, making it even more complex to solve.

Direct methods do not require an analytic expression for the necessary conditions: instead, the dynamic variables are adjusted to directly optimize the objective function \cite{hargraves}. In order to do so, the variables are parametrised and discretised, so that the dynamical equations are integrated stepwise. This representation transforms the optimal control problem into a non-linear programming problem (NLP). 

In the direct method, since the problem is discretised, the cost function can be observed at each iteration and the search direction modified accordingly, to ensure that it is always decreasing. As such, the region of convergence for such a method may be considerably bigger than the one for an indirect method, the latter requiring a better initial guess.

Depending on the size of the trajectory optimisation problem, an initial guess should be provided in any case. Throughout this project, low-thrust trajectories are designed taking into consideration first-guess solutions based on impulsive trajectories. Conway \cite{conway} lists the characteristics to strive for in a 'reasonable' first guess. The latter has to satisfy: the dynamical equations of motion, any specified initial and terminal constraints and the upper and lower boundary conditions given to the NLP problem solver. 

Different approaches can be effectively used in order to obtain such an initial guess. Again, Conway \cite{conway} divides them into three categories:
\begin{itemize}
	\item Known Optimal Control Strategies
	\item Shape-Based Methods
	\item Evolutionary Methods 
\end{itemize}

Examples of the second and third items can respectively be found in \cite{dachwald} and \cite{petropoulos}. The trajectory design implemented in this project in Chapter \ref{chap:ast1} employs the first approach, adapting the particular work of Sims and Flanagan \cite{sims}.

\subsubsection{Orbital Manoeuvres}

An orbital manoeuvre is the use of propulsion systems to change the orbit of a spacecraft. The flight segments when the propulsion system is not actively being used are referred to as \textit{coasting}, while a sequence of manoeuvres that allows for the change from an initial to a target orbit is called an \textit{orbital transfer}. This section presents the mission design manoeuvres that will be used in the application scenarios of this thesis.

\paragraph{Lambert's Problem}\hspace{-4mm}: the design of an orbital transfer requires the determination of the specific orbit that goes through two points in space, at a certain time. The task of finding this orbit, knowing only the position and time vectors associated to these locations in space, is called the \textit{Lambert's Problem}. 

\begin{figure}[h]
	\centering
	\includegraphics[width=0.5\linewidth]{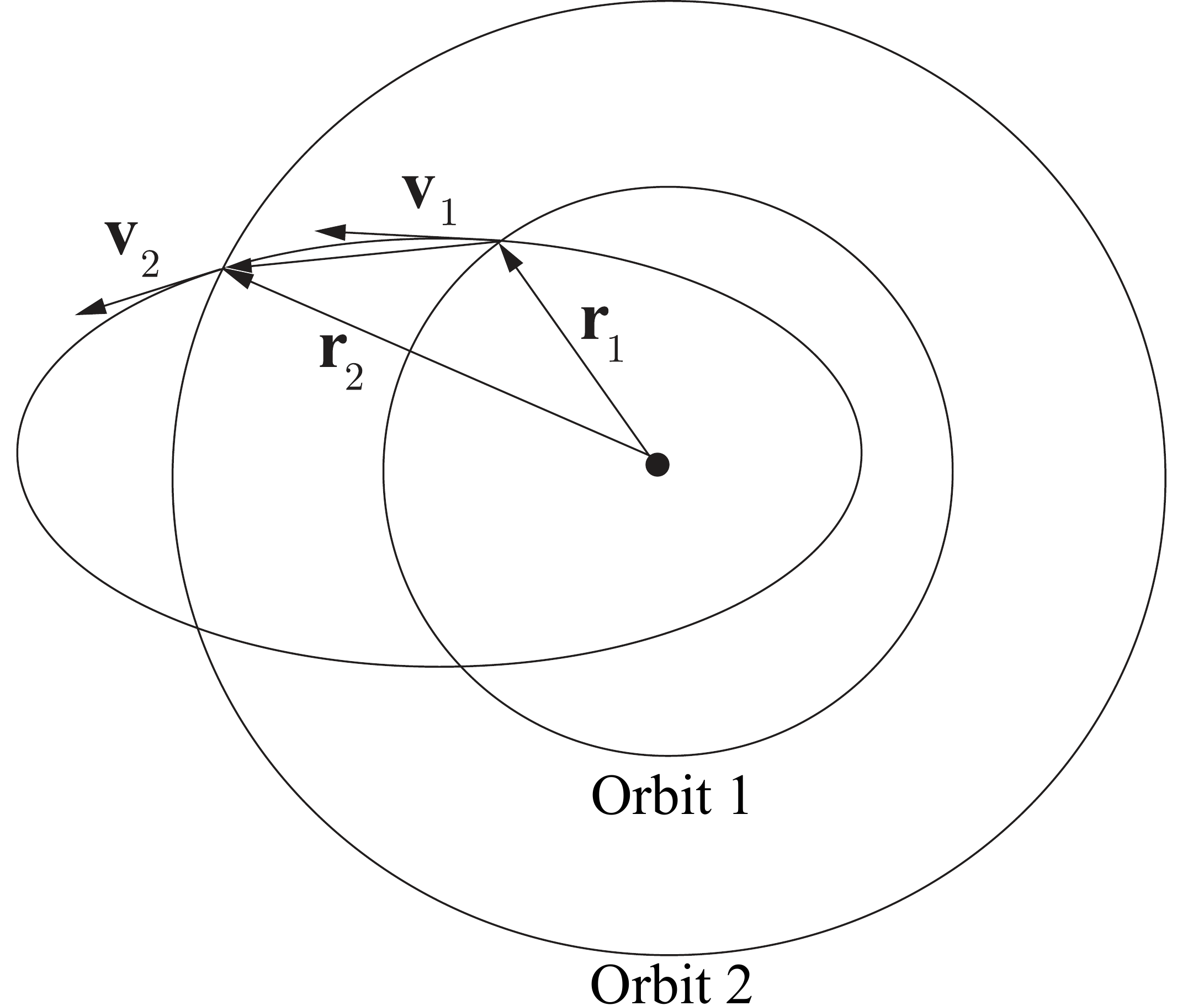}
	\caption{\label{fig:lambert}Example of a Lambert arc geometry connecting Orbit 1 to Orbit 2}
\end{figure}

A solution to the Lambert's problem is called a \textit{Lambert arc}. Thus, to compute a transfer between two orbits, the Lambert arc that connects their ephemerides at given times has to be computed, generally using the two-body problem (2BP) model to achieve this. Thus, it is possible to find the velocities for the spacecraft at the beginning and end of the trajectory ($\bm{v}_1$ and $\bm{v}_2$, which are then employed to compute the manoeuvre's $\Delta v$. The schematic of a Lambert arc can be found in Figure \ref{fig:lambert}).  

Thus, the design of an orbital trajectory from point A to point B requires the definition of many variables: initial and final ephemerides of the points and the parameters of the transfer orbit itself. This is done in such a way as to optimise a performance parameter, e.g. the minimization of the total $\Delta v$ of the manoeuvre or the maximization of the spacecraft's mass. Thus, the problem to compute an orbit becomes unequivocally an optimisation one, with the aforementioned variables acting as design parameters for a minimisation of the objective function.

Many ways to solve the Lambert's problem can be found in literature \cite{fundamentals, vallado, izzo_lambert, an_lambert}. This particular research project uses the one formulated in Battin \cite{battin}. The Lambert arc can then be posteriorly corrected to any other model of motion by using numerical targeting methods.

\begin{figure}[h]
	\centering
	\includegraphics[width=0.5\linewidth]{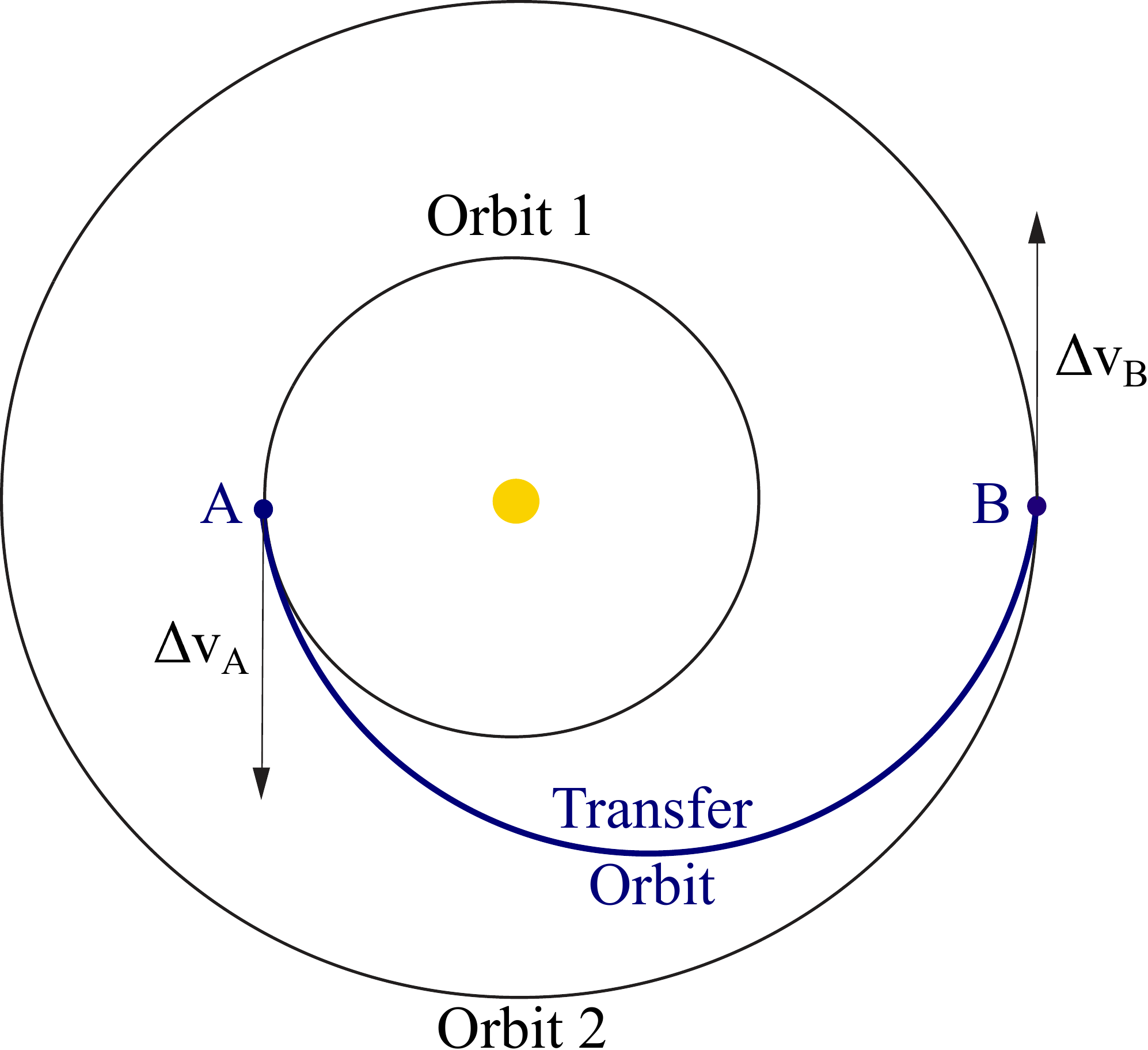}
	\caption{\label{fig:hohmann}Hohmann transfer geometry}
\end{figure}

\paragraph{Hohmann Transfer}\hspace{-4mm}: a Hohmann transfer is the particular solution of a Lambert arc that requires the least possible amount of $\Delta v$ for transfers between circular orbits. It is a bi-impulsive trajectory that takes a spacecraft from one circular orbit to another, developed by the German scientist Walter Hohmann in 1925. The design of a Hohmann transfer can be found in Figure \ref{fig:hohmann}.

A Hohmann transfer requires for the starting and destination points to be at particular locations in their orbits relative to each other. Space missions using a Hohmann transfer must wait for this required alignment to occur; furthermore, this is generally considered to be a very long transfer, which makes it unsuitable for use over very long distances.

\subsection{\textcolor{Both}{Low-Energy Trajectories}}

Many innovative mission concepts have been proposed since the beginning of the space race. Some of them have not yet been accomplished: asteroid retrieval missions and exploration with interplanetary cubesats are such examples \cite{cubesat, pablo}. Their ultimate realisation relies on many factors, one of them being the need for a very efficient trajectory design that yields the minimum fuel cost possible.  In order to achieve these requirements, one of the tools that astrodynamicists have at their disposal is the use of low-energy trajectories. 

Like the name suggests, these are paths in the Solar System that allow spacecraft to change orbits using little to no fuel, making them a useful resource in mission design and, in particular, interplanetary space travel \cite{belbruno}. As such, low-energy trajectories are intimately related with transport phenomena in the solar system \cite{koon_heteroclinic, transport1, transport2}. The computation of low-energy trajectories is only possible by studying the simultaneous interaction of at least three bodies, which can only be done with the development of increasingly complex models of motion. Vallado \cite{vallado} also adds that the use of techniques such as gravity assists or any other harnessing of gravitational energy are necessary to find this type of trajectory. Koon et. al. \cite{ballistic} stress out the importance of invariant manifolds and periodic orbits as building blocks for low-energy transfers, since they represent natural motions that take no fuel to traverse. Belbruno et. al. \cite{belbruno_hiten} focus instead on the concept of weak stability boundaries: transition regions, in the position-velocity space between gravitational fields of the bodies, where the dynamic effects on the spacecraft tend to balance \cite{belbrunoWSB}. 

The first successful use of low-energy trajectories in a mission was recorded in 1991: after the realisation that there was not enough fuel to perform the intended Hohmann manoeuvre, the Hiten probe was transferred to lunar orbit via a low-energy trajectory \cite{koon2008}, effectively rescuing the mission. Since then, several spacecraft missions have taken advantage of low-energy trajectories, mostly near the Earth or Moon: ISEE-3, WMAP, SOHO, Genesis, WIND, GRAIL, and several others. 

\subsection{\textcolor{Both}{The Three-Body Problem}}

As stated in Section \ref{sec:opt_traj_design}, the computation of a spacecraft trajectory requires both understanding the physical principles behind planetary motion and modelling the chaotic interactions between the surrounding celestial bodies. 

The motion of celestial objects is subject to their mutual gravitational attractions. The study of these forces is crucial to understand and predict the movements of moons, stars and to guide spacecraft to distant planets. This problem of determining the motions of many bodies interacting with each other is named the \textit{N-body problem} and is famous among astronomers and mathematicians for having no general analytical solutions for ${N > 2}$ \cite{vallado}. Nevertheless, specific solutions and simplifications have been eagerly sought after and occasionally discovered. Kepler was the first to achieve a concrete set of principles ruling space motion, in 1609. Kepler's equation describes the motion of a body orbiting another of much larger mass, which is commonly known as the \textit{restricted two-body problem}. In 1687, Newton formulated the equation that relates gravitational acceleration with the distance between any two planetary bodies: the \textit{law of universal gravitation}.

Following the same logic, an object moving under the influence of two other celestial bodies represents the \textit{three-body problem}. Leonhard Euler and Joseph-Louis Lagrange found all the known analytical solutions to an important subclass of the three-body problem, known as \textit{central configurations} (i.e. the gravitational acceleration vector produced on each mass by all the others points towards the common centre of mass and is proportional to the distance to it). However, work by Heinrich Bruns and Henri Poincar\'e in the late 1880s showed that a general arrangement of three or more bodies admits no analytical solution. A simplification on this question is the circular restricted three-body problem (CR3BP) \cite{battin}, which assumes that the studied body's mass is negligible when compared to the other two (known as the \textit{primaries}), that move in circular orbits. The solution to the CR3BP is computed using numerical methods (e.g Runge-Kutta algorithms). 

A different way of modelling the object's movement in a three-body system is by employing perturbation methods \cite{introcelmec}. These are used to compute the effect of an additional object besides the main attractive body, well outside the former's sphere of influence \cite{vallado}. In this way, instead of numerically integrating the orbits directly, only deviations from a two-body solution are studied. In certain cases, the trajectory obeys Kepler's equations outside of certain regions of movement, which are considered as being perturbed by the additional body. As such, this sectioning provides the clear advantage of being less computationally expensive than the CR3BP.

When employing a perturbation method, the additional planetary effect can be represented by a disturbing function that is then integrated in the equations of motion of the model. This is traditionally done by expanding the disturbing function in a power series. Some examples include using the ratio of the semi-major axes of the massless particle and disturbing body \cite{kozai, kaula}, the eccentricity and inclination \cite{laskar} or the system's gravitational parameter \cite{scheeres_multiple}.

\subsubsection{\textcolor{Both}{The Keplerian Map Method}}

The Keplerian map (KM) is an example of a perturbation method that uses the system's gravitational parameter for the expansion of the disturbing function. It is employed to compute the change in orbital elements due to the \textit{third-body effect}---the added perturbation of a planetary body besides the central one---throughout one period of the motion. 

Its development started with the works of Petrosky and Broucke \cite{petrosky} and of Chirikov and Vecheslavov \cite{chirikov}, to be used as a tool for description of long-term chaotic orbital behaviour of comets in nearly parabolic motion, in the neighbourhood of Jupiter. Later on, the same idea was exploited by different authors in astronomy and in atomic physics \cite{shevchenko2011kepler}. In the astrodynamics framework, the concept was then continued by Ross and Scheeres \cite{scheeres_multiple} to study distant flybys in the planar CR3BP. Later on, it was expanded by Alessi and S\'anchez \cite{alessi_semi} for three-dimensional applications.

\textcolor{Both}{This later iteration of the KM is computed as a semi-analytical method that employs a Picard's first iteration \cite{picard} on the Lagrange planetary equations (LPE), with the integrals being solved numerically over one orbital period. Its disturbing function is generated using the Keplerian third-body potential (K3BP), derived from the Hamiltonian of the CR3BP. The time taken to compute the integrals of this formulation was shown to always be less or comparable to the one needed to propagate the CR3BP \cite{alessi_semi}. However, the reference frame used for the KM computation is non-autonomous---as such, the time is a function of the true anomaly of the secondary body. As it is formulated by Alessi and S\'anchez \cite{alessi_semi}, this dependency poses a singularity in the KM equations and limits the range of its numerical integration intervals.}

\subsubsection{Periodic Orbits and Dynamical Structures}

The study of periodic orbits \textcolor{External}{in a three-body system} started as early as the 19\textsuperscript{th} century. After the investigation into the three-body problem by Euler and Lagrange, Darwin \cite{darwin_periodic} predicted the existence, in this model, of an infinity of periodic and quasi-periodic orbits. Later on, Str\"omgren and Makutuma \cite{stromgren} classified these into distinct orbital families, with different energy levels and shapes. By the second half of the 20\textsuperscript{th} century, H\'enon \cite{henon_periodic, henon_nonperiodic} further developed these orbital families into a consolidated notation, using the letters \textit{a}, \textit{c}, \textit{f} and \textit{g}. The first two include orbits around the $L_1$ and $L_2$ points--- the \textit{libration point orbits} (LPOs). Family \textit{g} corresponds to the group of prograde orbits around the secondary, \textit{distant prograde orbits} (DPOs), while \textit{f} contains the retrograde ones, commonly known as \textit{distant retrograde orbits} (DROs). H\'enon \cite{henon_nonperiodic} also classified \textit{quasi-periodic DRO} (QPDROs), from which natural examples can be found in the Solar System (as in the case of asteroid 2016 HO3 \cite{2016ho3}). However, these tend to be unstable and degrade into other types of motion throughout the years, making them not as interesting for mission design.

In the past few years, an effort to find and document DROs and DPOs in the Solar System has been carried out, together with the computation of linking transfers between them \cite{doedel, database, capdevila_phd}. Particularly, DROs are currently in high demand for mission design: due to their stability characteristics and relative location to the celestial bodies, they are desirable endgame orbits for missions such as asteroid capture or Mars transfers \cite{ARRM, dro_mars, perozzi2017distant}. Furthermore, they can be found in different models of motion. Examples include H\'enon \cite{henon_periodic}, which uses Hill's restricted form of the three-body problem in two dimensions; Zagouras \cite{zagouras_3d}, computing DROs in the CR3BP; and Scheeres \cite{scheeres4bp}, studying periodic orbits in the Hill four-body problem.

Considering that there is no closed-form equation that can describe a periodic orbit in a high-fidelity model, their generation is a purely numerical task. The latter is generally divided in three phases: first, a search for the points in space which most likely will be part of a periodic orbit. This is typically done using very extensive grid search mechanisms \cite{markellos_1974, tsirogiannis_grid} which, for fine grids, take a long time. Second, the differential correction of the orbit, so that the initial and final states match \cite{dro_algorithm}. Third, the continuation into families of orbits of similar characteristics to the original one.

\subsection{Asteroid Retrieval Missions}

Asteroids have been at the forefront of space exploration for many years, with the proposal and completion of missions such as JAXA's Hayabusa \cite{hayabusa}, NASA's Dawn \cite{dawn} and ESA's Rosetta \cite{rosetta}. There are many reasons for the current interest in these bodies: from the fact that their study may answer questions about the formation and evolution of the Solar System \cite{rosetta2}, to their profusion of potentially valuable resources and useful materials for space manufacturing \cite{RN10, elvis}. Furthermore, although main-belt asteroids are known since the early 19\textsuperscript{th} century, near-Earth asteroids (NEAs) have only been discovered by its end \cite{eros}. NEAs are now considered the easiest celestial bodies to reach from the Earth and, additionally, they may represent a potential impact threat to our planet \cite{RN1}, making asteroid retrieval studies desirable from a planetary defence standpoint.

The concept of asteroid retrieval missions envisages a spacecraft that rendezvous with an asteroid, lassos it and hauls it back to the Earth's neighbourhood, where it can be more easily accessed. The mission has clear synergies with all three of the above aspects of asteroid missions: science, planetary defence and resource utilization. Although this idea has been discussed since the 1960s \cite{C}, no mission to do so has yet been carried out. However, in 2013, NASA started the Asteroid Initiative, which included the asteroid redirect robotic mission (ARRM). This mission was initially planned to employ solar electric propulsion to haul an entire NEA, with an estimated mass around 1300 tonnes, to a DRO of the Earth-Moon system \cite{ARRM}. Currently, these activities have been put on hold and replaced instead with a mission to re-direct an asteroid using a kinetic impactor \cite{aida2}.

The trajectory design involved in prospective asteroid retrieval missions has been conceptualised by several authors throughout the years, with different destination orbits and target asteroids. S\'anchez et al. \cite{frontiers} survey the ideas put forward so far and summarise some convergence points between concepts: 

\begin{itemize}
	\item The mission should employ electric propulsion: even small NEAs, with diameters of about 10 meters, will likely weigh thousands of tonnes. As such, in the near to mid-future, systems with high exhaust velocity are the best equipped to provide the propulsive needs to move this amount of mass. Brophy et al. \cite{brophy2} employ this concept by studying the use of a solar electric propulsion system to move a theoretical small boulder, while Hasnain et al. \cite{N} investigate the required thrust to move asteroids into Earth-bound orbits.
	\item The target NEA should have a similar orbit to the one of the Earth, i.e. it should be energetically close to this planet. This yields a smaller retrieval cost than for asteroids that do not fit this requirement, as detailed in several asteroid capture cost lists \cite{G, I, pau_eros, pau_retrieval}.
\end{itemize}

\textcolor{External}{S\'anchez et al. \cite{frontiers} also highlight that the trajectory design for asteroid retrieval missions can be divided in two distinct phases: the Earth delivery trajectory and the endgame orbit. The former includes both the path from the Earth's departure to the asteroid rendezvous (outbound leg) and the trajectory to bring the body into the planet's neighbourhood (inbound leg). These segments are not considered in this work: instead, the focus is given to the trajectory up to the endgame orbit, i.e. the particular target where the captured asteroid is to be placed.}

\textcolor{External}{While many publications consider purely the energy requirements for permanent Earth capture \cite{A, I, N, RN18, R, RN16}, several works have studied many varied endgame orbits to place the asteroid in. By exploiting the dynamical richness of the Sun-Earth-Moon system, a target orbit that appears quite often is the DRO. Lunar DROs are the endgame orbits considered for the original ARRM concept, proposed in 2011 \cite{ARRM}. In Landau et al. \cite{O}, a round-trip trajectory to asteroid 2008 HU4 is shown to be able to retrieve up to 1,300 tons of material to a Lunar DRO.} 

\textcolor{External}{Although these periodic orbits present the benefit of being theoretically stable, the inherent instability of LPOs leads to the existence of hyperbolic invariant manifold structures connected to them. These can be used for the efficient targeting of retrieval transfers, presenting a possible benefit to the trajectory design. Considering these orbits as endgame, Garc\'ia Y\'arnoz et al. \cite{pau_eros} present a systematic approach to design impulsive capture transfers in the CR3BP into stable invariant manifolds of Sun-Earth LPOs. The classification of Easily Retrievable Objects is then given to all asteroids that can be captured with a total $\Delta v$ manoeuvre requiring less than 500 m$\cdot$s$^{-1}$, and an initial list of 12 such objects is provided. Consequently, Mingotti et al. \cite{mingotti2014combined} employ a CR3BP framework to solve the optimal control for the capture of the 12 EROs identified by the former publication, with either DROs or LPOs as target.}

\textcolor{External}{The gravitational perturbations of other celestial objects can also be exploited to achieve optimal capture trajectories. One of these bodies is the Moon, as studied by Gong and Li \cite{R} who, in a planar restricted three-body framework, characterize the orbital conditions that lead to a capture after a Moon fly-by. Tan et al. \cite{RN24} and Mingotti et al. \cite{mingotti2014combined} focus on captures in the Earth-Moon system. The former, by targeting Earth-Moon LPOs; the latter, considering both Lunar DROs and DPOs.}

\textcolor{External}{The Earth's influence in the capture of asteroids is also carefully considered by several authors. Bao et al. \cite{RN18} study Earth-Earth leveraging transfers, as well as Moon fly-bys, to facilitate the capture of asteroids by reducing the final retrieval $\Delta v$. The approach is relatively high-energy, in the sense that Lambert arcs and the patched conic approximation are considered. In contrast, Neves and Sánchez \cite{neves2018} present a methodology that exploits the Earth's third-body effect by designing Earth encounters occurring outside the planet's sphere of influence, using the KM method to model the gravitational perturbation.}

\section{Objectives}

The focus of this work is to expand the knowledge of low-energy trajectories by developing new mathematical tools to assist on mission design applications. \textcolor{External}{These tools are used to explore interesting periodic orbits and low-energy trajectories for innovative mission concepts, particularly asteroid retrieval and rendezvous.} 

The aforementioned mathematical tools will require dynamical models of motion of higher complexity than the 2BP, in order to capture the sensitivities of the gravitational interactions in the Solar System. A particular focus will be given to the KM model for the Sun-Earth system, in which the spacecraft navigates in a close encounter with the secondary.

The main objectives can be further expanded into tasks developed throughout the course of this work. These are as follows:

\begin{enumerate}
	\item Implement well-established, adequate models of motion for trajectory design in the Sun-Earth system: 2BP and the CR3BP.
	\item Analyse the Keplerian third-body potential (K3BP) to model conservative forces:
	\begin{itemize}
		\item Correct flaws \textcolor{External}{related to singularities} and study \textcolor{External}{possible improvements in computational speed and accuracy} to the existing KM approximation;
		\item \textcolor{External}{Develop novel formulations of this perturbation model for state estimation to encompass different types of solutions.}
	\end{itemize}
	\item Analyse the K3BP for modelling non-conservative forces:
	\begin{itemize}
		\item Study its advantages for controlled trajectory design;
		\item Create a framework for low-thrust motion optimization;
		\item Exploit its usage to design low-thrust trajectories for asteroid missions.
	\end{itemize}
	\item Apply the K3BP for the design of asteroid capture missions:
	\begin{itemize}
		\item Create a layered framework for multi-fidelity mission design, from a lower to a higher accuracy model of motion (i.e. a refinement process for the trajectory);
		\item Optimise the capture trajectory by exploiting Earth's perturbation.
	\end{itemize}
	\item Apply the K3BP in the modelling of periodic orbits:
	\begin{itemize}
	\item Study how it compares to the CR3BP in generating DROs and DPOs.
	\end{itemize}
\end{enumerate}

\section{Contributions}

This thesis focuses on the development of low-computational cost, accurate models of motion for multi-body trajectories and on their applications concerning asteroid mission design and periodic orbits. Therefore, the contributions of Chapters \ref{chap:k3bp} to \ref{chap:dro} are here acknowledged.

\textbf{Development and expansion of different perturbation models using the K3BP:} this function, used also to derive the well-known KM, is employed in the development of four novel methods. These formulations are categorized based on the forces at play and the way the equations are computed. Three of them are employed, in distinct approaches, to model the third-body perturbation for \textit{conservative forces}. The first is the analytical solution based on a Taylor expansion on the eccentricity, which is valid only for semi-major axis propagation in almost circular orbits. The second is the periapsis-apoapsis-periapsis-Keplerian map (PAP-KM), a method that solves the singularities presented by the original KM \cite{alessi_semi}, improving on its performance. The third is the Euler-Keplerian map (E-KM), which uses an Euler integrator to obtain the orbital behaviour at each time step. A very important characteristic of the formulations is that they can be hybridized and combined together, depending on the application scenario. For example, the analytical prediction of the semi-major axis evolution can be combined with the E-KM method for the evolution of the remaining orbital elements of motion, which can then be transformed into a predictor of the position and velocity of a spacecraft. This is potentially useful for real-time computations and application in GNC algorithms.

The fourth method was developed to model the third-body perturbation for \textit{non-conser\-va\-tive forces}, i.e. adding a thrusting acceleration. The resulting framework uses Gauss' variational equations (GVE) to express the orbital evolution in Keplerian elements, leading to an easy definition of bounds and boundary conditions that facilitate the convergence of the optimal control problem for a low-thrust mission.

\textbf{Creation of novel mission design tools:} \textcolor{External}{differential correctors that employ the models of motion developed throughout this research were formulated. These include: i) a single shooting method that changes an initial velocity to match a final position using the Keplerian third-body Jacobian; ii) a multiple shooting method, using the same Jacobian, that considers a state vector of position, velocity and acceleration---by establishing boundaries on initial and final positions and velocities, this fixed-time shooting method adjusts the control acceleration vector on each segment of the trajectory, making it a useful tool in the design of continuous thrust transfers;} iii) a single shooting scheme that corrects the final mean anomaly of the motion, by modifying the starting semi-major axis. The latter was developed to complement the research on Earth-resonant trajectories---\textcolor{Internal}{the term \textit{resonant} is here used to account for two consecutive encounters of the body with the Earth---}in which it was found that the angular phasing of the body with the Earth is directly correlated with the impact that the planet's perturbation has in modifying the body's orbital elements. Furthermore, by employing these techniques, a multi-fidelity design framework was developed. This managed to filter out \textcolor{External}{sub-optimal} preliminary trajectories using the KM, and then refine a small number of promising ones with a higher-fidelity method (in this work, the CR3BP), ending up with a robust trajectory design. In order to achieve a fast low-thrust transfer solution, a Sims-Flanagan inspired approach was devised and implemented. The resulting trajectory can be easily inserted into an optimal control solver for a quick and accurate solution.

\textbf{Exploitation of the third-body effect for asteroid capture:} the perturbation of the Earth, even when outside its sphere of influence, can be extremely significant for mission design in nearly resonant regimes of motion with the planet, i.e. when the analysed body moves in almost co-orbital fashion with the secondary. \textcolor{External}{Using a low-fidelity, perturbative model of motion, this disturbing acceleration} was used to optimise the trajectory design for asteroid retrieval, resulting in the so-called Earth-resonant capture trajectories. Using this technique, a list that highlights the fuel cost savings for several NEAs is presented.

\textbf{Search for periodic orbits around the secondary body:} a method was developed to showcase the probability of an initial state, in Keplerian elements, to be a periodic orbit. This was achieved using the PAP-KM model which, as previously stated, is valid only for the region outside the disturbing body's sphere of influence. As such, the study targets mainly orbits around the secondary (DROs and DPOs). The resulting grid search is presented as a low-cost likelihood map (LCLM) of orbital states.

\subsection{Publications}

The work developed throughout this research resulted in the \textcolor{External}{listed journal and conference publications}:

\textbf{Neves, R.}, S\'anchez, J. P. (2018). Multifidelity Design of Low-Thrust Resonant Captures for Near-Earth Asteroids. Journal of Guidance, Control and Dynamics, Vol. 42, No. 2, pp. 335-346.

S\'anchez, J. P., \textbf{Neves, R.}, Urrutxua, H. (2018). Trajectory Design for Asteroid Retrieval Missions: A Short Review. Frontiers in Applied Mathematics and Statistics, Stat. 4:44.

\textcolor{External}{\textbf{Neves, R.}, S\'anchez, J. P. (2018). Optimization of Asteroid Capture Missions Using Earth Resonant Encounters. In  Stardust Final Conference (pp. 3-16). Springer.}

\textcolor{External}{\textbf{Neves, R.}, S\'anchez, J. P.. Asteroid Capture Missions for Unattainable Targets Using Earth-Resonant Encounters. 68$^{th}$ International Astronautical Congress, 2017, Adelaide, Australia.}

\textbf{Neves, R.}. Asteroid Capture using Earth-Resonant Encounters: The Case of Asteroid 2011MD. Move an Asteroid Competition\footnote{Winner of the 2017 edition}, Space Generation Advisory Council.

\textcolor{External}{\textbf{Neves, R.}, S\'anchez, J. P., Colombo, C., Alessi, E.M.. Analytical and Semi-Analytical Approaches to the Third-Body Perturbation in Nearly Co-Orbital Regimes. 69$^{th}$ International Astronautical Congress, 2018, Bremen, Germany.}

\textcolor{External}{\textbf{Neves, R.}, S\'anchez, J. P. (2018). Gauss' Variational Equations for Low-Thrust Optimal Control Problems in Low-Energy Regimes. 69$^{th}$ International Astronautical Congress, 2018, Bremen, Germany.}

\textcolor{External}{Cano, J., Cunill, J., Diaz, A. J., Golemis, A., Gupta, S., Innes, D., Maiden, D., March, K., Rael, H., Shawe, J., Sierra, V., Torrents, A., Rossi, E. Z., Machuca, P., \textbf{Neves, R.}, S\'anchez, J. P.. ARTEMIS: A complete mission architecture to bridge the gap between humanity and near-Earth asteroids. 69$^{th}$ International Astronautical Congress, 2018, Bremen, Germany.}

\section{Structure of the Thesis}

This thesis is organised as follows:

\begin{itemize}
	\item Chapter 2: The well-known dynamical models for the two-body problem and the three-body problem are characterised. Specifically, the equations of motion of the 2BP and the CR3BP are presented, together with the integral of motion known as the Jacobi constant. Particular solutions, in the form of zero-velocity curves, equilibrium points and invariant manifolds are explored. Finally, the class of perturbation methods is presented, together with the concept of flow maps for the computation of orbital motion. 
	\item Chapter 3: The third-body effect is detailed and its importance in mission design is highlighted. The K3BP is presented in order to generate a disturbing function that describes this effect, when paired with classical perturbation techniques. Based on this, four novel models of motion are developed, categorised by the type of forces taken into consideration: conservative or non-conservative.
	\item \textcolor{Both}{Chapter 4: The first trajectory design application is introduced. Two distinct asteroid missions are designed: one for rendezvous and another for capture of two separate NEAs. The mission design showcases the usage of the K3BP formulation for non-conservative accelerations to obtain fully developed low-thrust motions. Further testing on the accuracy of this model is undertaken.}
	\item Chapter 5: The second trajectory design application is introduced. A multi-fidelity design of asteroid retrieval trajectories, including multiple Earth encounters, is presented. With the ultimate goal of finding a list of NEAs to be captured into Sun-Earth LPOs, the trajectory is computed using a layered approach: the KM method is employed to obtain a preliminary impulsive solution, which is posteriorly refined into the CR3BP. The latter is then formulated as an optimal control problem and solved to obtain a full low-thrust trajectory.
	\item Chapter 6: One of the conservative force models based on the K3BP is used to undertake a short preliminary study on the search of periodic orbits. The process describing the computation of these orbits is described, and an alternative method to the commonly utilised grid search process is identified.
	\item Chapter 7: A summary of the results of this investigation is presented, followed by recommendations for extending the analysis of the developed models of motion and the presented applications.
\end{itemize}

\chapter{Dynamical Models and Methods}
\label{chap:dyn}
This work focuses on modelling the motion of a spacecraft in a three-body configuration. This requires developing novel perturbation methods and comparing their performance to two well-established models of motion: the two-body problem (2BP) and the circular restricted three-body problem (CR3BP). As such, \textcolor{External}{both} are here presented in detail, together with all the necessary information regarding reference frames and associated dynamical structures.

The 2BP is employed for a quick estimation of trajectories and manoeuvres of a space object. This model is used when \textcolor{Internal}{the motion of said object is simplified as having only one main celestial body governing it. Trajectories computed in the 2BP are typically represented in a fixed, \textit{inertial reference frame}.} In contrast, the CR3BP is the highest fidelity model employed in this study, acting as the baseline with respect to which all trajectory errors are computed. \textcolor{Internal}{It is utilised to determine the motion of an object subject to the gravitational influence of two other great celestial bodies, which move in circles. Trajectories computed in the CR3BP are regularly represented in a rotating, \textit{synodical reference frame}.}

\textcolor{Internal}{When computing trajectories for any given space object, the main celestial body governing it is called the \textit{primary} or \textit{central body}. If the influence of an additional great celestial body is considered, both are named the \textit{primaries}---alternatively, the bigger body is termed again \textit{primary} and the smaller \textit{secondary}. If the space object whose motion is being modelled has a negligible mass when compared to the primaries, it is referred to as the \textit{massless particle}.} 
	
\textcolor{Internal}{It is common occurrence to normalise the quantities involved in the computation of space trajectories, especially when assuming a three-body system. In this case, the unit of mass is taken to be $m_1 + m_2$ (subscripts $1$ and $2$ referring to the primary and secondary, respectively); the unit of length is selected to be the distance between the centres of these bodies; the unit of time is chosen such that the orbital period of the primary and the secondary about their common centre of mass is $2\pi$. Then, the universal constant of gravitation becomes $G$ = 1; it follows that the common mean motion of the primaries, $n$, is also unity.} 

\section{Reference Frames}
\label{sec:rf}

Several different \textcolor{Internal}{coordinate} frames of reference are going to be used throughout this document, depending on the employed motion model. \textcolor{Internal}{The main ones are the inertial, the synodic and the Local Vertical, Local Horizontal (LVLH) reference frames.} This section describes their characteristics, together with the transformation matrices that can be used \textcolor{Internal}{to transform the state vector from one to another}. 

\begin{figure}[h]
	\centering 
	\begin{minipage}[t]{0.49\linewidth}
		\includegraphics[width=\textwidth]{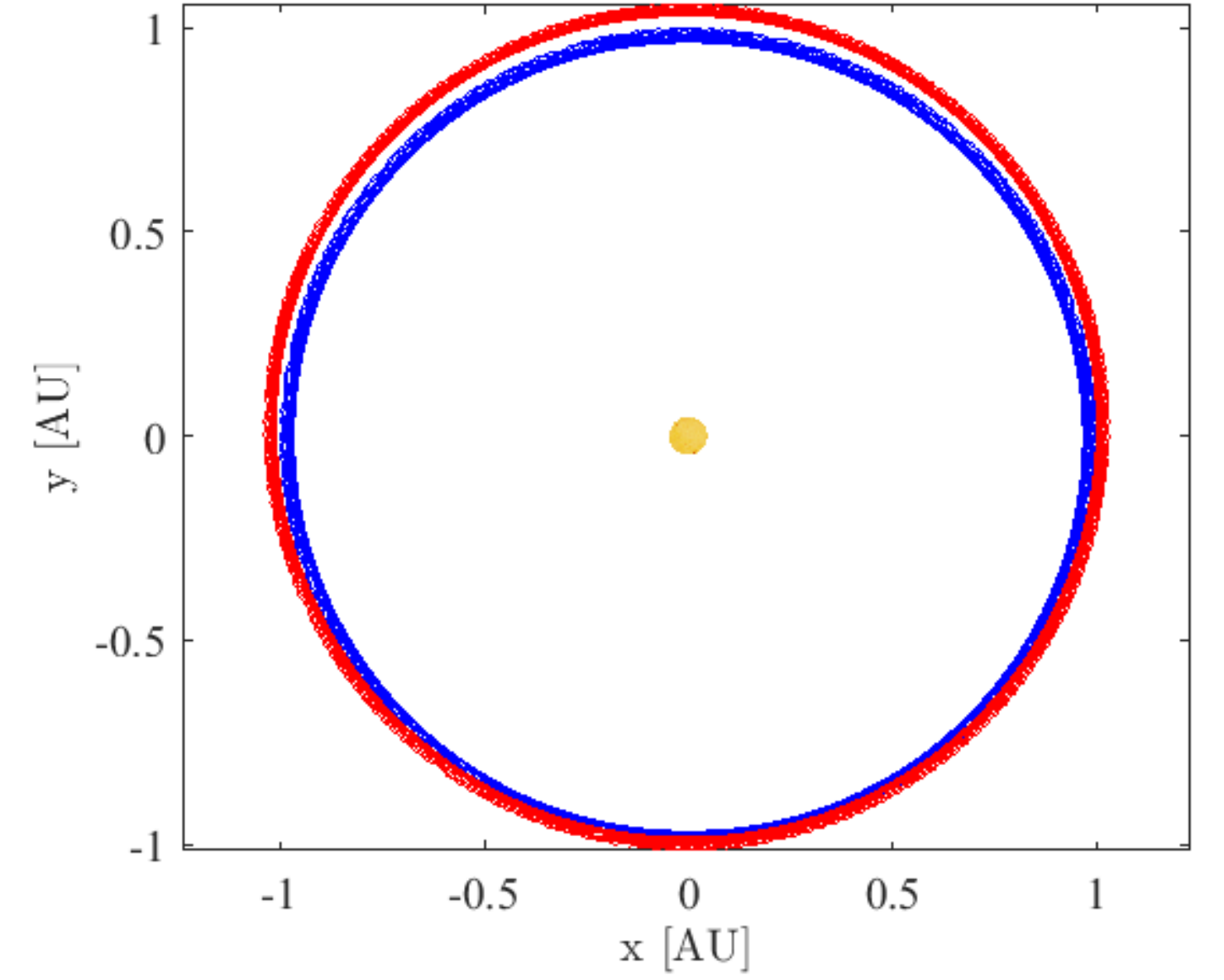}
		\caption*{\textcolor{External}{a) Inertial Cartesian frame: the red and blue lines correspond to the movement of the asteroid and the Earth, respectively}}
	\end{minipage}
	\hfill
	\begin{minipage}[t]{0.49\linewidth}
		\includegraphics[width=\textwidth]{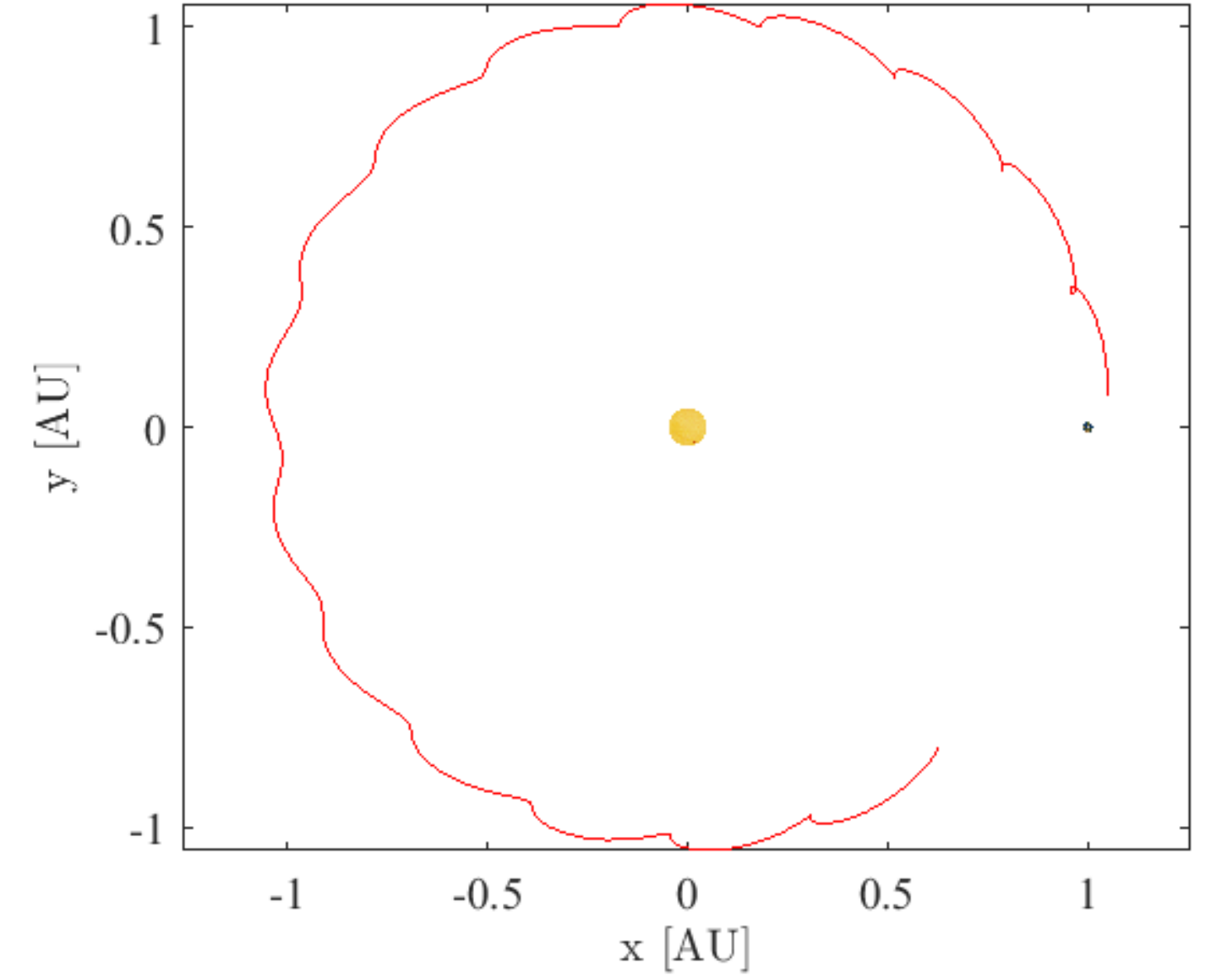}
		\caption*{\textcolor{External}{b) Synodic reference frame: the red line depicts the asteroid's motion, while the Earth appears fixed}}
	\end{minipage}
	\caption{\textcolor{Both}{Motion of asteroid 2011 MD in the Sun-Earth system.} Earth is scaled to three times the Hill radius, for visibility\label{fig:srfvscar}}
\end{figure}

\textcolor{Both}{Figure \ref{fig:srfvscar} illustrates the typical representation of the inertial and the synodic reference frames. The example motion is the one of asteroid 2011 MD, computed in the Sun-Earth system, in the CR3BP. Figure \ref{fig:srfvscar} a) depicts the inertial frame with the asteroid trajectory in red, together with the Earth's two-body motion with the Sun in blue; Figure \ref{fig:srfvscar} b) shows the same motion in the synodic reference frame, where the Earth appears fixed.}

\subsection{Inertial Reference Frame}

An inertial reference frame is one in which Newton's first law of motion is valid. It follows that the centre of mass of the system is always at rest or in uniform motion: from this point, the X-axis points towards a pre-established reference direction (e.g. Vernal equinox), the \textcolor{Internal}{Z-axis is perpendicular to a reference plane (e.g. ecliptic) and the Y-axis completes the right-hand coordinate system}. 

\textcolor{Internal}{This is the frame of reference commonly used to plot the 2BP for the motion of a massless particle around a primary. It is typically centred on the body that exerts the gravitational attraction---in the Sun-Earth system's case, the Sun. However, it is important to mention that the primary is not fixed in space, and the 2BP equations describe the motion around a fixed centre of attraction. As such, the adoption of the term \textit{Sun-centred} inertial reference frame from literature implies the assumption that the motion of the Sun is negligible.}

\subsection{Synodic Reference Frame}

The synodic or rotating frame is a non-inertial frame of reference typically employed for \textcolor{Internal}{restricted} three-body systems, in which the massless particle is deemed to have an insignificant mass when compared to the primaries. This frame is characterised by having the X-axis lying along the vector that connects the centres of mass of the two primaries. Therefore, their positions are kept constant with respect to this frame, which rotates at the same angular speed as the mean motion of the secondary around the primary. 

\textcolor{Internal}{The quantities involved in this reference frame are regularly presented in non\-/dimensional units: the distance between the primaries is equal to one, the frame rotates with unit angular velocity, the normalised gravitational parameter is determined as $\mu = \frac{\mu_2}{\mu_1 + \mu_2}$ and the two primaries are located on the X-axis at the points ($-\mu$, 0) and ($1 - \mu$, 0), with the barycentre as the origin.}

\begin{figure}[h]
	\centering
	\includegraphics[width=0.6\linewidth]{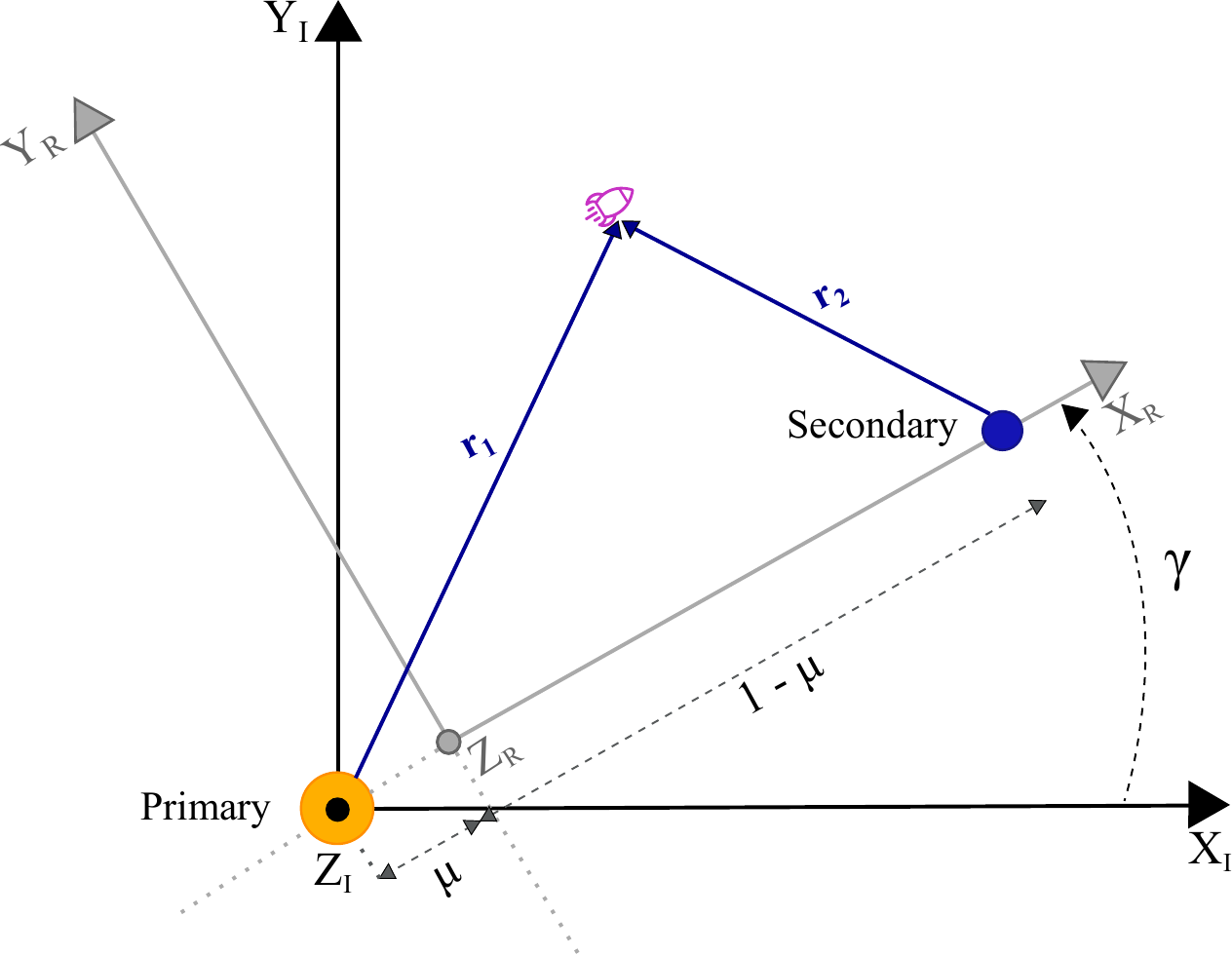}
	\caption{\label{fig:rf}Relationship between coordinate systems for the rotating and inertial frames}
\end{figure}

Figure \ref{fig:rf} illustrates the rotating relationship between the synodic frame \textcolor{Internal}{($O_{R xyz}$)} centred on the barycentre and the Sun-centred inertial reference frame \textcolor{Internal}{($O_{I xyz}$)}, again using the Sun-Earth system as example, with the spacecraft as the massless particle. The angle $\gamma$ is positive in the counter-clockwise direction and is equivalent to the dimensionless time. 

\textcolor{Internal}{In order to transform the position and velocity of a body from the inertial to the synodic frame, the first step is to move the frame's origin to the barycentre. This is shown in Eqs. \eqref{eq:trans1} and \eqref{eq:trans2}, where the superscript $(B)$ represents the transitional barycentric origin for a quantity in the inertial frame:}
\begin{align}\label{eq:trans1}
\begin{bmatrix} x_I^{(B)} & y_I^{(B)} & z_I^{(B)} \end{bmatrix}^T &= \begin{bmatrix} x_I & y_I & z_I \end{bmatrix}^T - \begin{bmatrix} \mu & 0 & 0 \end{bmatrix}^T\\
\begin{bmatrix} \dot{x}_I^{(B)} & \dot{y}_I^{(B)} & \dot{z}_I^{(B)} \end{bmatrix}^T &= \begin{bmatrix} \dot{x}_I & \dot{y}_I & \dot{z}_I \end{bmatrix}^T - \begin{bmatrix} 0 & \mu & 0 \end{bmatrix}^T\label{eq:trans2}
\end{align}

\textcolor{Internal}{The rotation of the inertial frame to the synodic frame} is done using the following \textcolor{Internal}{matrix}:
\begin{align}\label{eq:rot}
\text{\textbf{R}}^R_I = \begin{bmatrix}
\cos\gamma & \sin\gamma & 0 \\
-\sin\gamma & \cos\gamma & 0 \\
0 & 0 & 1\\
\end{bmatrix}
\end{align}

Finally, by employing Eqs. \eqref{eq:trans1} to \eqref{eq:rot}, the transformation of the position and velocity of a body from the inertial to the synodic frame becomes:
\begin{align}
\begin{bmatrix} x_R \\ y_R \\ z_R \\ \end{bmatrix} = 
\text{\textbf{R}}^R_I\begin{bmatrix} x_I^{(B)} \\ y_I^{(B)} \\ z_I^{(B)} \\ \end{bmatrix}, \hspace{1cm}
\begin{bmatrix} \dot{x}_R \\ \dot{y}_R \\ \dot{z}_R \\ \end{bmatrix} = 
\text{\textbf{R}}^R_I\begin{bmatrix} \dot{x}_I^{(B)} + y_I^{(B)} \\ \dot{y}_I^{(B)} - x_I^{(B)} \\ \dot{z}_I^{(B)} \\ \end{bmatrix}
\end{align}

\subsection{Local Vertical, Local Horizontal Reference Frame}
\label{sec:LVLH}

\begin{figure}[hbt!]
	\centering
	\includegraphics[width=0.56\linewidth]{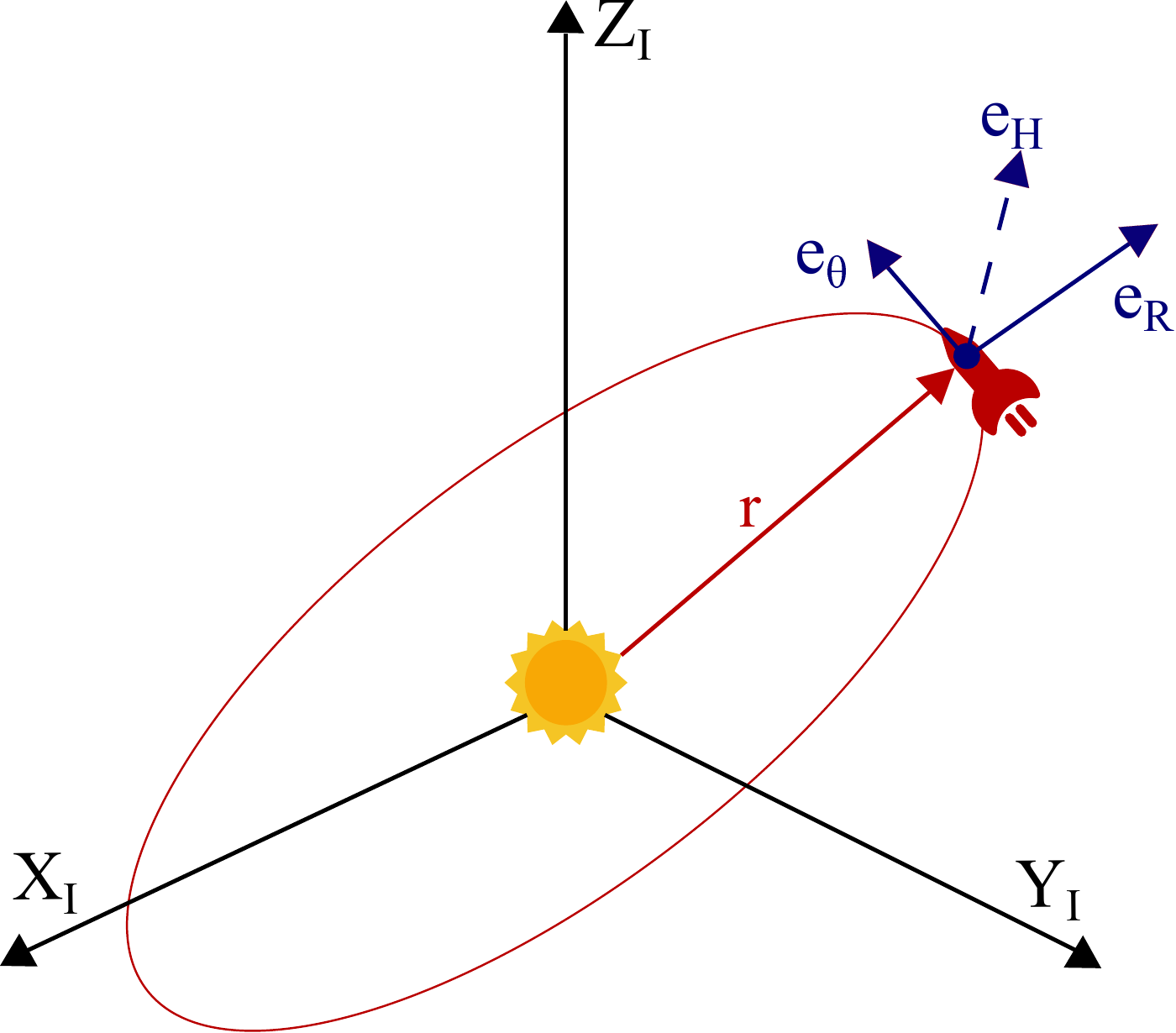}
	\caption{\label{fig:lvlh}Relationship between LVLH and inertial reference frames}
\end{figure}

\textcolor{Internal}{The LVLH reference frame is a non-inertial frame of reference whose origin is placed at the centre of mass of the spacecraft. The axes are represented as $\vec{e}_r$, $\vec{e}_{\theta}$ and $\vec{e}_h$: respectively for the radial, in-track and cross-track directions. This reference frame can be visualised in Figure \ref{fig:lvlh}, together with the spacecraft's orbit in the inertial reference frame.}

\textcolor{Internal}{The rotation of the LVLH frame ($O_{r \theta h}$) to the orbital plane frame ($O_{eph}$) is done using Eq. \eqref{eq:battin_eq}. In turn, Eq. \eqref{eq:battin_eq} is used to convert vectors from the orbital plane ($O_{eph}$) to the inertial reference frame ($O_{I xyz}$), complying with the definitions in Battin \cite{battin}.}
\begin{align}\label{eq:battin_eq2}
\text{\textbf{R}}_{r \theta h}^{eph} = 
\begin{bmatrix}
	\cos\nu & -\sin\nu & 0 \\
	\sin\nu & \cos\nu & 0 \\
	0 & 0 & 1 \\
\end{bmatrix}
\end{align}
\begin{align}\label{eq:battin_eq}                                    
\text{\textbf{R}}_{eph}^{I} = \begin{bmatrix}
\cos\Omega & -\sin\Omega & 0 \\
\sin\Omega & \cos\Omega & 0 \\
0 & 0 & 1 \\
\end{bmatrix}
\begin{bmatrix}
1 & 0 & 0 \\
0 & \cos i & -\sin i \\
0 & \sin i & \cos i \\
\end{bmatrix}
\begin{bmatrix}
\cos\omega & -\sin\omega & 0 \\
\sin\omega & \cos\omega & 0 \\
0 & 0 & 1 \\
\end{bmatrix}
\end{align}

\section{The Two-Body Problem}
\label{sub:2bp}

The mathematical problem of designing a trajectory in the Solar System, an environment composed of countless bodies, is named the \textit{N-body problem}, in which $N$ \textcolor{External}{stands for the number of bodies interacting with object whose motion is being computed, including the latter}. Using Newton's Law of Gravitation, its equations of motion can be represented by Eq. \eqref{eq:nbody}:
\begin{align}
\ddot{\bm{r}_i} &= G \sum_{\substack{j=1 \\ j \neq i}}^{N} \frac{m_j}{r^3_{ij}}(\bm{r}_j - \bm{r}_i)\label{eq:nbody}
\end{align}
\textcolor{External}{in which the subscript $i$ singles out the object whose motion is being computed, $m$ represents each bodies' mass, $\bm{r}$ is the position vector, with $r = \norm{\bm{r}}$ and $r_{ij} = \norm{\bm{r_i} - \bm{r_j}}$.}

This problem has $6N$ variables, but only 10 constraint equations are obtained using conservation laws---6 equations for the conservation of momentum of the center of mass, 3 equations for the conservation of the total angular momentum and one equation for total energy conservation. Each constraint equation defines a conserved quantity, so-called \textit{integral of motion}. Thus, there are $6N - 10$ quantities to be determined, making the general $N$-body problem analytically not integrable without performing a range of approximations. This is one reason for models of motion of the Solar System, called full ephemerides models, to require statistical models and historical data in order to reach approximate solutions \cite{ephemerides}.

The 2BP is the case when $N$ is 2; this particular scenario describes the motion of two bodies that interact only with each other. \textcolor{Internal}{In most cases, only the motion of one of the objects relative to the other is of interest. This is the situation presented throughout this thesis, where the singled-out body is treated as a massless particle and the central one is deemed fixed.} The equations of motion are the particular case of Eq. \eqref{eq:nbody} for $N = 2$:
\begin{align}
\ddot{\bm{r}} &= -\mu_C \cdot \frac{\bm{r}}{r^3}\label{eq:2bp}
\end{align}
where $\bm{r}$ is the position vector relative to the central body, $\mu_C = G m_C$ is the gravitational parameter of the central body, equal to the product of the gravitational constant and the object's mass (subscript $C$ used to contrast with the case when two primaries exist, which bears no extra notation). 

The 2BP is the basis of the patched-conics method: an approach used in trajectory design that divides space into areas in which an object is only gravitationally perturbed by one central body at a time, within each so-called \textit{sphere of influence} or Hill's sphere \cite{fundamentals}.

\section{The Circular-Restricted Three-Body Problem}

The three-body problem refers to the case where $N = 3$: three bodies move under their mutual gravitational attraction. As previously stated, this has no closed-form solutions. \textcolor{Internal}{A common} simplification can be applied: the CR3BP. In this model, the primaries move in circular, co-planar orbits around their common barycentre.

\subsection{Equations of Motion}

The full derivation of the equations of motion of the CR3BP is omitted here for simplicity's sake, since it can be found in many classical works \cite{szebehely1969theory}. Following the nomenclature defined by Koon et. al. \cite{koon2008}, the final model is depicted by Eq. \eqref{eq:cr3bp} in non-dimensional units in the synodic reference frame:
\begin{align}
\ddot{x} - 2\dot{y} &= -\frac{\partial \bar{U}}{\partial x}\nonumber\\
\ddot{y} + 2\dot{x} &= -\frac{\partial \bar{U}}{\partial y}\nonumber\\
\ddot{z} &= -\frac{\partial \bar{U}}{\partial z}
\label{eq:cr3bp}
\end{align}
where $\bar{U}$ is the effective potential function of the system, computed by Eq. \eqref{eq:pseudop}:
\begin{align}
\bar{U} &= -\frac{1}{2}(x^2 + y^2) - \frac{1 - \mu}{r_1} - \frac{\mu}{r_2}
\label{eq:pseudop}
\end{align}
\textcolor{Internal}{in which $r_1 = \norm{\bm{r_1}}$ and $r_2 = \norm{\bm{r_2}}$ are the distances from the massless particle to the primary and the secondary, respectively} (as depicted in Figure \ref{fig:rf}). These equations describe the state of the body in the synodic reference frame, generally with normalised components.

Given that the CR3BP is not analytically solvable, its solutions are obtained by employing numerical integrators. In contrast to the two-body case, the orbital elements of the trajectory will not be constant, instead changing over time. Some interesting properties and particular solutions of the motion under this model can be derived and will be detailed in the following sections.

\section{Numerical Targeting}
\label{sec:num_targ}

\textcolor{Internal}{As stated in Section \ref{sub:2bp}, there cannot be generic closed-form analytical solutions} to dynamical systems of higher complexity than the 2BP. So, when a specific orbital behaviour is sought and no analytical equations can be utilised, a two-point boundary value problem has to be solved by employing numerical targeting techniques. 

The tools here presented are not unique to any motion model, but are essential in the computation of the invariant manifolds and periodic orbits of the CR3BP and any trajectory targeting presented throughout this work.

\subsection{State Transition Matrix} 

\textcolor{Internal}{Given a set of initial conditions, the numerical integration process of a motion model will propagate the state in time. Yet, the final condition obtained by this means may not necessarily be the one that achieves the trajectory design goals (e.g. targeting of a specific point in space, rendezvous with an object, computation of a periodic orbit). Thus, in order to reach the desired target, the initial conditions of the propagated motion may require modification. To correctly and efficiently adjust a trajectory, the computation of the state transition matrix (STM) is necessary.} 

\begin{figure}[h]
	\centering
	\includegraphics[width=0.8\linewidth]{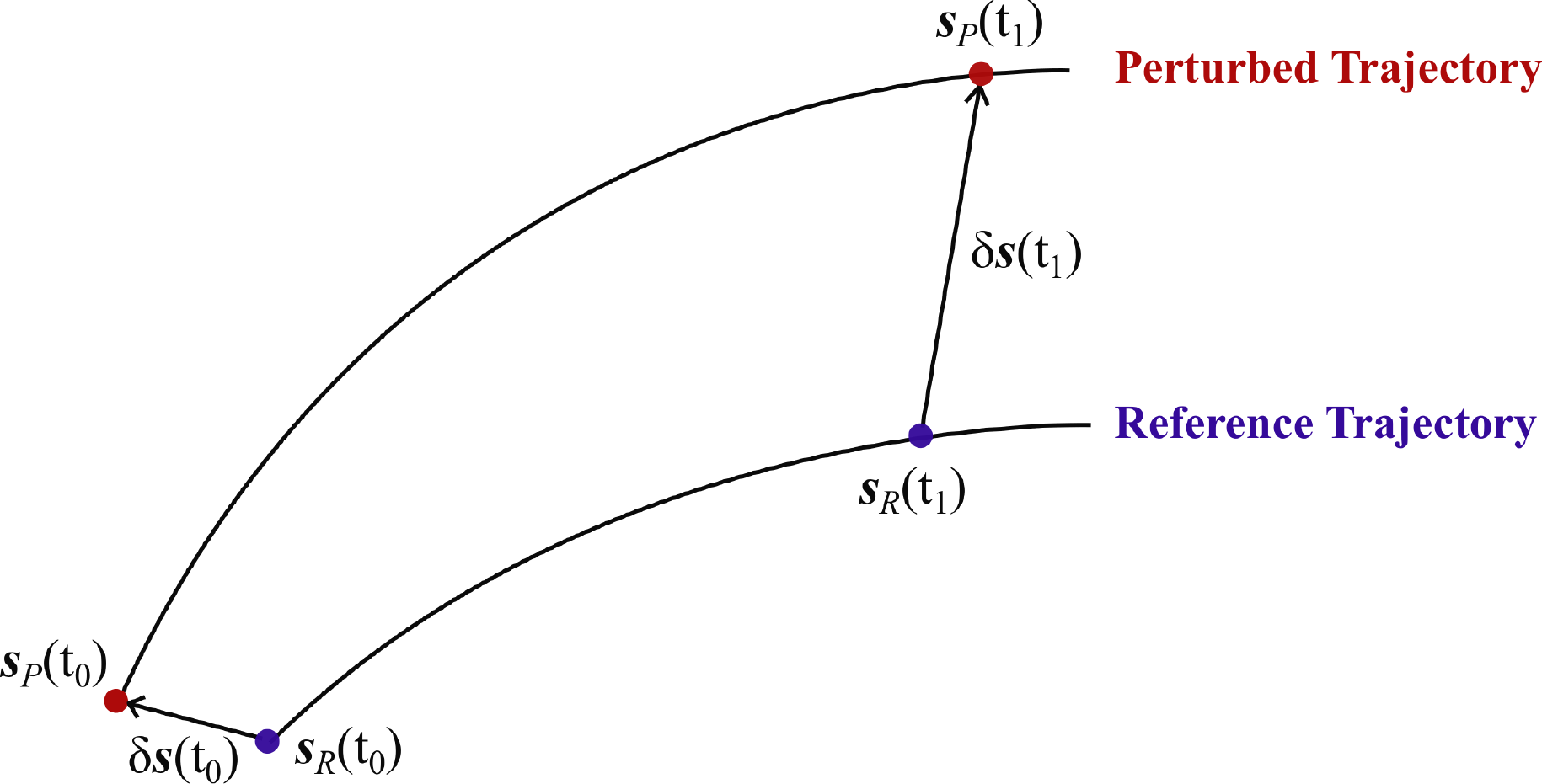}
	\caption{\label{fig:stm}Reference path and perturbed solution at times $t_0$ and $t_1$}
\end{figure}

\textcolor{Internal}{In order to illustrate how a trajectory can be adjusted, Figure \ref{fig:stm} depicts a two motions: one of them, highlighted in purple, is a naturally propagated reference trajectory $\bm{s}_R(t)$, from an initial condition $\bm{s}_R(t_0)$ to a final state $\bm{s}_R(t_1)$; the other, coloured red, is the perturbed trajectory $\bm{s}_P(t)$, computed when a perturbation $\delta \bm{s}(t_0)$ is introduced in the reference motion. After integrating this perturbed state in time, the resulting trajectory is given by:}
\begin{align}
\textcolor{Both}{\bm{s}_P(t) = \bm{s}_R(t) + \delta \bm{s}(t)}\label{eq:stm0}
\end{align}

\textcolor{Internal}{Integrating both the reference and the perturbed initial states in the CR3BP results in two distinct paths---at time $t_1$, the states along these two paths are not equal. In order to estimate their difference $\delta \bm{s}(t_1)$, the motion is linearised using a first-order Taylor series expansion about the reference trajectory.}

The perturbed arc is expressed by Eq. \eqref{eq:stm1}.
\begin{align}
\textcolor{Both}{\dot{\bm{s}}_P = \dot{\bm{s}}_R + \delta \dot{\bm{s}}  = f(\bm{s}_R + \delta \bm{s})}\label{eq:stm1}
\end{align}
where the function $f$ represents the non-linear differential equations of the motion model. The Taylor series expansion on Eq. \eqref{eq:stm1} yields the first-order terms:
\begin{align}
\dot{\bm{s}}_P &\approx f(\bm{s}_R, t) + \delta \dot{\bm{s}}\\
\delta \dot{\bm{s}} &\approx \frac{\partial f}{\partial \bm{s}} \bigg\rvert_{\bm{s}_R} \delta \bm{s}\label{eq:stm2}
\end{align}

Eq. \eqref{eq:stm2} has solutions of the form:
\begin{align}
\delta \bm{s}(t_1) &= \bm{\Phi}(t_1, t_0)\delta \bm{s}(t_0)
\end{align}
where $\bm{\Phi}(t_1, t_0)$ denotes the STM. Thus, the STM provides a linear mapping from time $t_0$ to a time $t_1$, establishing a relationship between initial and final deviations that can be used to adjust trajectories in order to match a final outcome. The propagation of the STM is described by the following equation:
\begin{align}
\textcolor{Both}{\dot{\bm{\Phi}}} &= \text{\textbf{D}}f(\bm{s}) \bm{\Phi} \label{eq:1}
\end{align}
in which $\text{\textbf{D}}f(\bm{s})$ is the Jacobian of the STM. 

\subsection{Differential Correction}

Given an initial and final set of trajectory parameters, the goal of a differential corrector is to use the STM to determine which modifications of the initial conditions are required in order to get a closer approach to the desired state at the end of the integration period. Thus, this method can be used to determine the necessary orbital manoeuvres to \textcolor{Internal}{be performed} on an initial state \textcolor{Both}{$\bm{s}_0$}, so that an adjusted transfer will end up at the desired final state \textcolor{Both}{$\bm{s}_1$} \cite{koon2008}. \textcolor{Internal}{Given a} reference trajectory governed by the non-linear dynamical equations $f$:  
\begin{align}
\textcolor{Both}{\dot{\bm{s}}} &= \textcolor{Both}{f(\bm{s}(t))} \\
\textcolor{Both}{\delta \bm{s}(t_1)} &= \textcolor{External}{\bm{\Phi}(t_1, t_0)\delta \bm{s}(t_0)}
\end{align}
in which $\bm{\Phi}$ is the state transition matrix (STM), whose propagation is described by Eq. \eqref{eq:1}.

The Jacobian \textcolor{Both}{$\text{\textbf{D}}f(\bm{s})$} is composed of the partial derivatives of the velocity and acceleration vectors, in the form of Eq. \eqref{eq:bigmatrix}:
\begin{align}
\text{\textbf{D}}f(\bm{s}) = \begin{bmatrix}
\frac{\partial \dot{x}}{\partial x} & \frac{\partial \dot{x}}{\partial y} & \frac{\partial \dot{x}}{\partial z} & \frac{\partial \dot{x}}{\partial \dot{x}} & \frac{\partial \dot{x}}{\partial \dot{y}} & \frac{\partial \dot{x}}{\partial \dot{z}}\\
\frac{\partial \dot{y}}{\partial x} & \frac{\partial \dot{y}}{\partial y} & \frac{\partial \dot{y}}{\partial z} & \frac{\partial \dot{y}}{\partial \dot{x}} & \frac{\partial \dot{y}}{\partial \dot{y}} & \frac{\partial \dot{y}}{\partial \dot{z}}\\
\frac{\partial \dot{z}}{\partial x} & \frac{\partial \dot{z}}{\partial y} & \frac{\partial \dot{z}}{\partial z} & \frac{\partial \dot{z}}{\partial \dot{x}} & \frac{\partial \dot{z}}{\partial \dot{y}} & \frac{\partial \dot{z}}{\partial \dot{z}}\\
\frac{\partial \ddot{x}}{\partial x} & \frac{\partial \ddot{x}}{\partial y} & \frac{\partial \ddot{x}}{\partial z} & \frac{\partial \ddot{x}}{\partial \dot{x}} & \frac{\partial \ddot{x}}{\partial \dot{y}} & \frac{\partial \ddot{x}}{\partial \dot{z}}\\
\frac{\partial \ddot{y}}{\partial x} & \frac{\partial \ddot{y}}{\partial y} & \frac{\partial \ddot{y}}{\partial z} & \frac{\partial \ddot{y}}{\partial \dot{x}} & \frac{\partial \ddot{y}}{\partial \dot{y}} & \frac{\partial \ddot{y}}{\partial \dot{z}}\\
\frac{\partial \ddot{z}}{\partial x} & \frac{\partial \ddot{z}}{\partial y} & \frac{\partial \ddot{z}}{\partial z} & \frac{\partial \ddot{z}}{\partial \dot{x}} & \frac{\partial \ddot{z}}{\partial \dot{y}} & \frac{\partial \ddot{z}}{\partial \dot{z}}\\
\end{bmatrix}\label{eq:bigmatrix}
\end{align}

Considering these mathematical relations, the process to obtain a differential corrector can be detailed. First of all, the objective is to find a solution $\bm{X}$ that satisfies the following expression,
\begin{align}\label{eq:GX}
\text{\textbf{G}}(\bm{X}) = \begin{bmatrix} G_1(\bm{X}) \\ ... \\ G_m(\bm{X}) \end{bmatrix} = 0
\end{align}
in which $\bm{X}$ is a vector of free variables and $\text{\textbf{G}}(\bm{X})$ is the vector of $m$ constraints to which the free variables are subject, so that the trajectory meets the required specifications. This equation can be expanded as a first-order approximation using a Taylor series about an initial guess $\bm{X}_0$:
\begin{align}
\text{\textbf{G}}(\bm{X}) = \text{\textbf{G}}(\bm{X}_0) + \frac{\partial \text{\textbf{G}}(\bm{X}_0)}{\partial \bm{X}_0}(\bm{X} - \bm{X}_0) 
\end{align}

Given that the constraint vector should be zero per the equality in Eq. \eqref{eq:GX}, Eq. \eqref{eq:lsm} is obtained:
\begin{align}
\bm{X}^{i + 1} = \bm{X}^{i } - \text{\textbf{D}}\text{\textbf{G}}(\bm{X}^{i})^{-1} \text{\textbf{G}}(\bm{X}^{i})\label{eq:lsm}
\end{align}
in which $\text{\textbf{D}} \text{\textbf{G}}(\bm{X}) = \frac{\partial \text{\textbf{G}}(\bm{X})}{\partial \bm{X}}$ and represents the Jacobian matrix of the constraint vector and superscript $i$ represents each iteration of the differential correction algorithm. Thus, an iterative process is required to cycle through the error ($|\text{\textbf{G}}(\bm{X}^{i+1})|$) until it falls below a pre-defined tolerance ($\epsilon$). The different steps of this iterative process are outlined below \cite{sanchez}:

\begin{enumerate}
	\item Define the problem to solve and identify the free variables $\bm{X}$
	\item Determine a first guess of the free variables vector $\bm{X}_0$
	\item Specify constraints to which variables are subject $\text{\textbf{G}}(\bm{X})$
	\item Calculate the Jacobian matrix of the system $\text{\textbf{D}} \text{\textbf{G}}(\bm{X})$
	\item Solve the system of equations $\bm{X}^{i+1} = \bm{X}^{i} - \text{\textbf{D}} \text{\textbf{G}}(\bm{X}^{i})^{-1} \text{\textbf{G}}(\bm{X}^{i})$
	\item Check the error associated to the new solution ($|\text{\textbf{G}}(\bm{X}^{i+1})|$). If it is larger than a convergence tolerance ($\epsilon$), repeat step 5 using $\bm{X}^{i+1}$ as the new $\bm{X}^i$. If the error falls below $\epsilon$, the method has converged.
\end{enumerate}

\subsubsection{Single Shooting Method}
\label{sec:ssm}

A single shooting method is a differential correction technique to connect the initial and final points of a trajectory with one single segment. It is suitable for the computation of simple trajectories (i.e. periodic orbits or direct transfers between two points). The algorithm can employ a fixed or variable time approach, although the mission design in this project concerns only the first case.

One commonly used approach is the targeting of the final position by changing the initial velocity. This is especially important in the computation of Lambert arcs, in order to compute the $\Delta v$ necessary to move the spacecraft to a certain position in space. 

In this scenario, the free variables vector $\bm{X}$ consists of the initial velocity components, which are the only design parameters that can be modified to reach the desired final state \textcolor{Both}{$\bm{s}_d$:}
\begin{align}
\bm{X} &= \begin{bmatrix}\dot{x}_0 & \dot{y}_0 & \dot{z}_0 \end{bmatrix}^T\\
\textcolor{Both}{\bm{s}_d} &= \begin{bmatrix} x_d & y_d & z_d \end{bmatrix}^T
\end{align}

The Jacobian of the system can be determined by computing the partial derivatives of the constraint vector $\text{\textbf{G}}(\bm{X})$. These correspond, in this case, to the section of the STM associated with the velocity of the initial state vector and the position of the final state.

\subsubsection{Multiple Shooting Method}
\label{sec:MSM}

A multiple shooting method is a differential correction technique utilised to connect the initial and final points of a trajectory with multiple segments, i.e. using several single shooting methods between intermediate points, the so called patch points. In this way, the errors associated with the long integration of segments are reduced. Plus, the shape of the trajectory can be more easily manipulated by adding constraints to the different patch points.

\begin{figure}[h]
	\centering
	\includegraphics[width=0.8\linewidth]{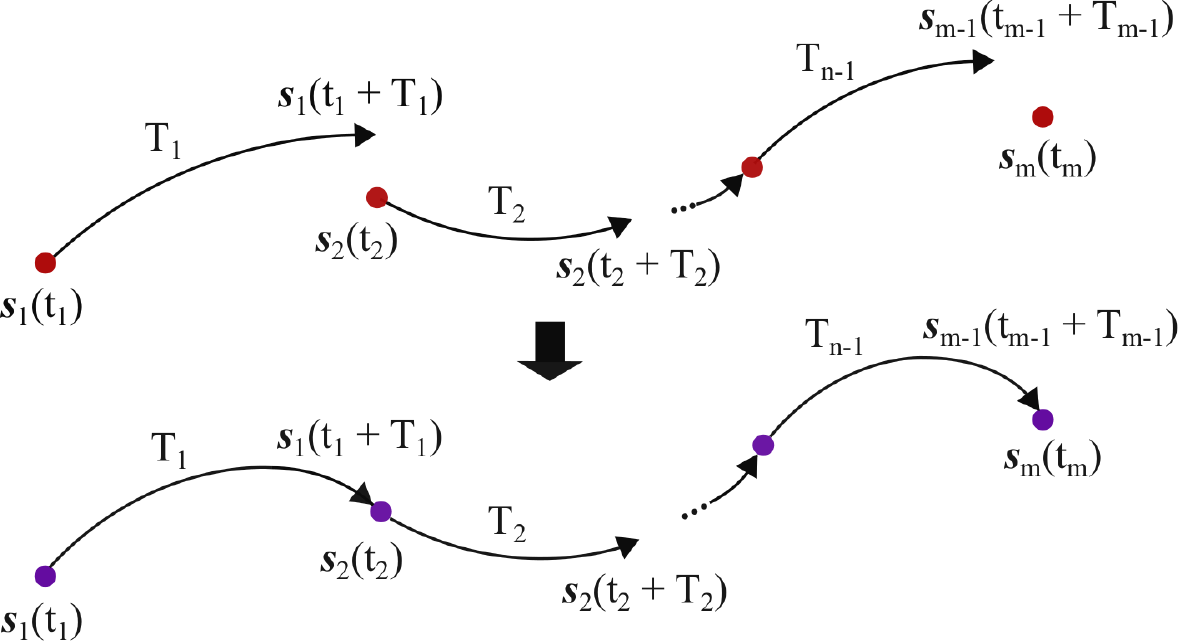}
	\caption{\label{fig:ms}Multiple shooting scheme}
\end{figure}

In the case in which the integration times are fixed, \textcolor{External}{the vector of free variables} corresponds to the linkage of each patch point's state vector \textcolor{External}{($\bm{s}_j, j = 1, ... , m$)}.
\begin{align}
\bm{X} = \begin{bmatrix}\textcolor{External}{\bm{s}_1} & ... & \textcolor{External}{\bm{s}_m} \end{bmatrix}^T
\end{align}

The free variables vector $\bm{X}$ becomes a column vector of $6m$ rows if only position and velocity are considered, or $9m$ rows if the acceleration is added (in which $m$ is the number of patch points). To ensure position or velocity continuity at the patch points, conditions have to be added to the constraint vector in the following manner:
\begin{align}
\text{\textbf{G}}(\bm{X}) = 
\begin{bmatrix}
\textcolor{External}{\bm{s}_1(t_1 + T_1) - \bm{s}_2(t_2)} \\...\\ \textcolor{External}{\bm{s}_{m-1}(t_{m-1} + T_{m-1}) - \bm{s}_m(t_m)}
\end{bmatrix}
\end{align}

Besides the named constraints, any additional ones can be added, including fixing the initial and final states of the motion.

Eq. \eqref{eq:exampleG} shows the first row of the Jacobian of the constraint vector:
\begin{align}\nonumber
\frac{\partial \text{\textbf{G}}_1(\bm{X})}{\partial \bm{X}} &= 
\begin{bmatrix}
\frac{\partial \text{\textbf{G}}_1}{\partial X_1} & ... & \frac{\partial \text{\textbf{G}}_1}{\partial X_{m}}
\end{bmatrix} = 
\begin{bmatrix}
\textcolor{External}{\frac{\partial \bm{s}_1(t_1 + T_1)}{\partial \textcolor{External}{\bm{s}_1(t_1)}}} & \textcolor{External}{-\frac{\partial \bm{s}_2(t_2)}{\partial \bm{s}_2(t_2)}} & ... & \textcolor{External}{\frac{\partial G_1}{\partial \bm{s}_m(t_m)}}
\end{bmatrix} \\ &= 
\begin{bmatrix}
\bm{\Phi}_1 & -\text{\textbf{I}} & ... & 0
\end{bmatrix}\label{eq:exampleG}
\end{align}

The same procedure applies to the remaining rows of $\text{\textbf{G}}(\bm{X})$. As such, every segment will contribute with its own STM and one identity matrix. Finally, the entire Jacobian for all the constraints of the multiple shooting method can be assembled considering all the segments:
\begin{align}\label{eq:bigJ}
\text{\textbf{DG}}(\bm{X}) = 
\begin{bmatrix}
\bm{\Phi}_1 & -\text{\textbf{I}} & \text{\textbf{0}} & 0 & 0 & 0 \\
\text{\textbf{0}} & \bm{\Phi}_2 & -\text{\textbf{I}} & 0 & 0 & 0\\
&  &  & \ddots &  & \\
0 & 0 & 0 & 0 & \bm{\Phi}_m & -\text{\textbf{I}}\\
\end{bmatrix}
\end{align}
in which each STM is computed using an adequate Jacobian matrix.

Given the following auxiliary variable:
\begin{align}
r_3 &= \sqrt{1 -2x + x^2 + y^2 + z^2}
\end{align}
and using the previously defined quantities $r$, $r_1$ and $r_2$, the Jacobian matrix of the CR3BP model \textcolor{External}{is the following}:
\begin{align}\label{eq:jac_cr}
\text{\textbf{D}}f(\bm{s}) = \begin{bmatrix}
0 & 0 & 0 & 1 & 0 & 0 \\
0 & 0 & 0 & 0 & 1 & 0 \\
0 & 0 & 0 & 0 & 0 & 1 \\
\textcolor{External}{-\partial_{xx}\bar{U}} & \textcolor{External}{-\partial_{xy}\bar{U}} & \textcolor{External}{-\partial_{xz}\bar{U}} & 0 & 2 & 0\\
\textcolor{External}{-\partial_{yx}\bar{U}} & \textcolor{External}{-\partial_{yy}\bar{U}} & \textcolor{External}{-\partial_{yz}\bar{U}} & -2 & 0 & 0\\
\textcolor{External}{-\partial_{zx}\bar{U}} & \textcolor{External}{-\partial_{zy}\bar{U}} & \textcolor{External}{-\partial_{zz}\bar{U}} & 0 & 0 & 0\\
\end{bmatrix}
\end{align}
where the values of \textcolor{External}{$\partial_{\alpha\beta}\bar{U}$ ($\alpha \in \{x, y, z\} \text{ and }\beta \in \{x, y, z\}$)} are the second partial derivatives of the effective potential presented in Eq. \eqref{eq:pseudop}. These are expanded into the following:
\begin{align}
\partial_{xx}\bar{U} &= -1 - \frac{3\mu(x - 1 + \mu)^2}{r_2^5} + \frac{\mu}{r_2^3} - \frac{3 (1 - \mu)(x + \mu)^2}{r_1^5} + \frac{1 - \mu}{r_1^3}\\
\partial_{xy}\bar{U} &= -3y\bigg(\frac{\mu(x - 1 + \mu)}{r_2^5} + \frac{(1 - \mu)(x + \mu)}{r_1^5}\bigg) \\
\partial_{xz}\bar{U} &= -3z\bigg(\frac{\mu(x - 1 + \mu)}{r_2^5} + \frac{(1 - \mu)(x + \mu)}{r_1^5}\bigg) \\
\partial_{yx}\bar{U} &= \partial_{xy}\bar{U} \\ 
\partial_{yy}\bar{U} &= -1 - \frac{3 y^2 \mu}{r_2^5} + \frac{\mu}{r_2^3} - \frac{3 y^2 (1 - \mu)}{r_1^5} + \frac{1 - \mu}{r_1^3}\\
\partial_{yz}\bar{U} &= -3yz\bigg(\frac{\mu}{r_2^5} + \frac{(1 - \mu)(x + \mu)}{r_1^5}\bigg) \\
\partial_{zx}\bar{U} &= \partial_{xz}\bar{U} \\
\partial_{zy}\bar{U} &= \partial_{yz}\bar{U} \\
\partial_{zz}\bar{U} &= -\frac{3 z^2 \mu}{r_2^5} + \frac{\mu}{r_2^3} - \frac{3 z^2 (1 - \mu)}{r_1^5} + \frac{1 - \mu}{r_1^3}
\end{align}

\section{Particular Solutions} 
\label{sub:ps}

\textcolor{Internal}{Although} the non-linear equations of motion in the CR3BP are non-integrable and autonomous (not time-dependant), some particular solutions can be derived: specifically, equilibrium points, periodic orbits and quasi-periodic orbits \cite{parker}.

\subsection{Equilibrium Points}

The libration points \textcolor{Internal}{of the CR3BP} are locations in space where an object of negligible mass, affected by the gravitational interactions between the primaries, can theoretically maintain a constant position in the synodic reference frame. This characteristic makes them very attractive for a great number of missions, e.g. to keep telescopes or other observation-type spacecraft, since the fuel consumption required to perform station-keeping is very low \cite{koon2008}.

The five libration points are the equilibrium solutions of the equations of motion of the CR3BP; they can be obtained by setting the velocity and acceleration of Eq. \eqref{eq:cr3bp} to zero:
\begin{align}
0 &= x - \frac{(1 - \mu)(x + \mu)}{r_1^3} - \frac{\mu(x + \mu - 1)}{r_2^3}\nonumber\\
0 &= y\bigg(1 - \frac{(1 - \mu)}{r_1^3} - \frac{\mu}{r_2^3}\bigg)\nonumber\\
0 &= -z\bigg(\frac{(1 - \mu)}{r_1^3} - \frac{\mu}{r_2^3}\bigg)\label{eq:lpoints}
\end{align}

\begin{figure}[!hbt]
	\centering
	\includegraphics[width=0.6\linewidth]{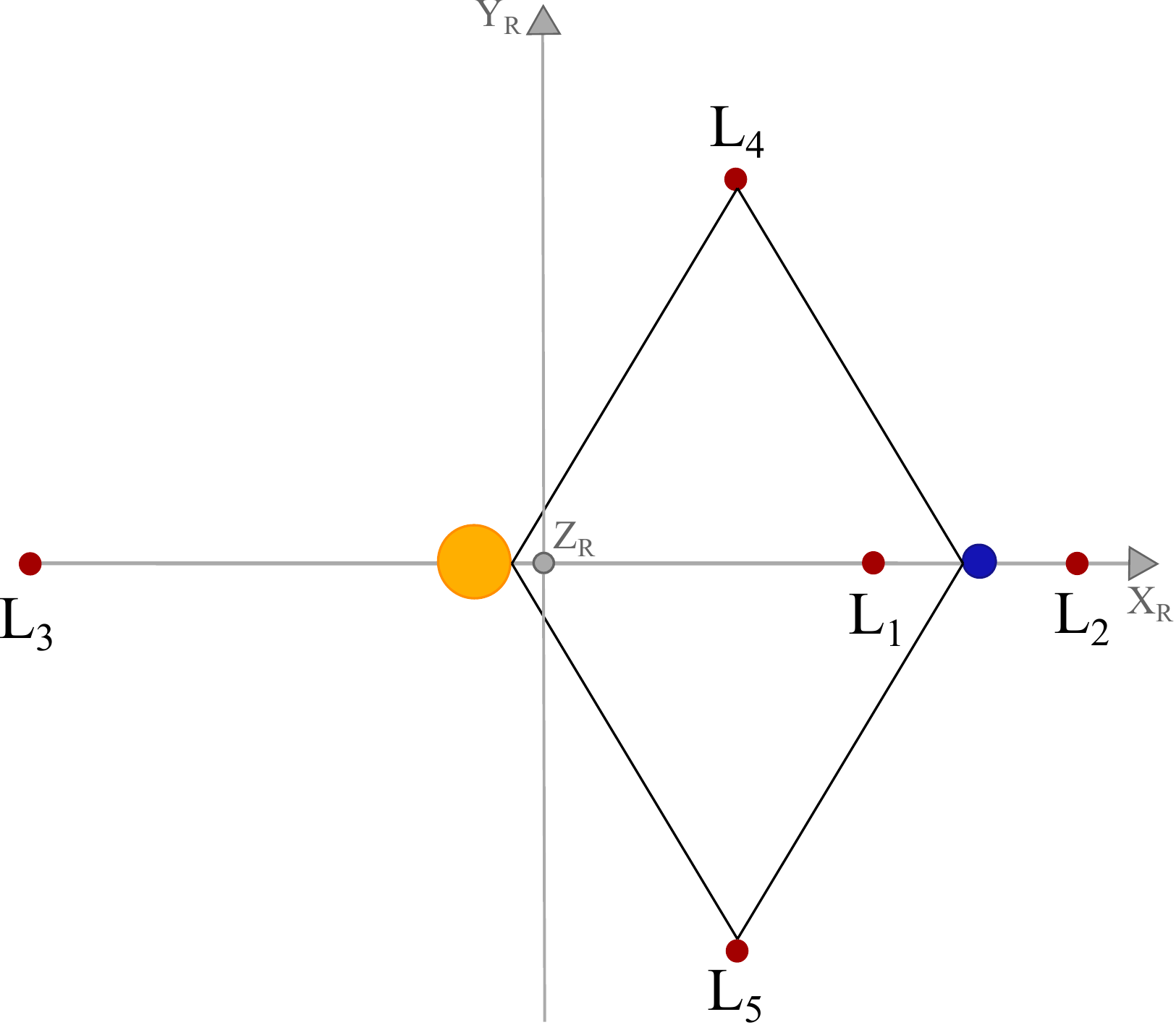}
	\caption{\label{fig:lpoints}\textcolor{Internal}{Distribution of the libration points in the CR3BP}}
\end{figure}

By analysing Eq. \eqref{eq:lpoints}, some insights can be taken. These equations imply that $z = 0$, meaning the equilibrium solutions are on the XY plane of motion. Furthermore, from setting $y = 0$, three solutions can be achieved: these represent the collinear points $L_1$, $L_2$ and $L_3$. By defining $r_1 = r_2 = 1$, the remaining solutions are obtained: points $L_4$ and $L_5$, which form an equilateral triangle with the primaries. Their position can be visualised in Figure \ref{fig:lpoints}, depicted in the synodic reference frame.

\subsection{Jacobi Constant}

The equations of the CR3BP have an energy integral of motion. Following the notation in Koon \cite{koon2008}, its formula is depicted by:
\begin{align}
E = \textcolor{Internal}{\frac{1}{2}(\dot{x}^2 + \dot{y}^2 + \dot{z}^2)} + \bar{U}\label{eq:Jacobi}
\end{align}
with $\bar{U}$ computed by Eq. \eqref{eq:pseudop}.

The quantity $-2E$ is the one generally found in literature: it is named the Jacobi constant. This is also typically represented by the variable $C$, by employing the following reformulation of Eq. \eqref{eq:Jacobi}:
\begin{align}
C = -2\bar{U} - \norm{\bm{v}}^2\label{eq:C}
\end{align}
\textcolor{Internal}{where $v$ is the norm of the spacecraft's velocity.}

The Jacobi constant can be very helpful in the characterisation of the system's dynamics, namely to determine the regions in space that are accessible to a theoretical spacecraft. Given that ${v^2 \geq 0}$ is a mathematical constraint of Eq. \eqref{eq:C}, the set of variables which obey this condition forms the accessible region, where the spacecraft can move. By setting $v = 0$, and therefore $-2\bar{U} - C > 0$, the obtained solution is a region that delimits the space in which the body can possibly move. This is also called the zero-velocity curve.

\begin{figure}[!htb]
	\centering
	\begin{minipage}[t]{0.4\textwidth}
		\centering
		\includegraphics[width=\textwidth]{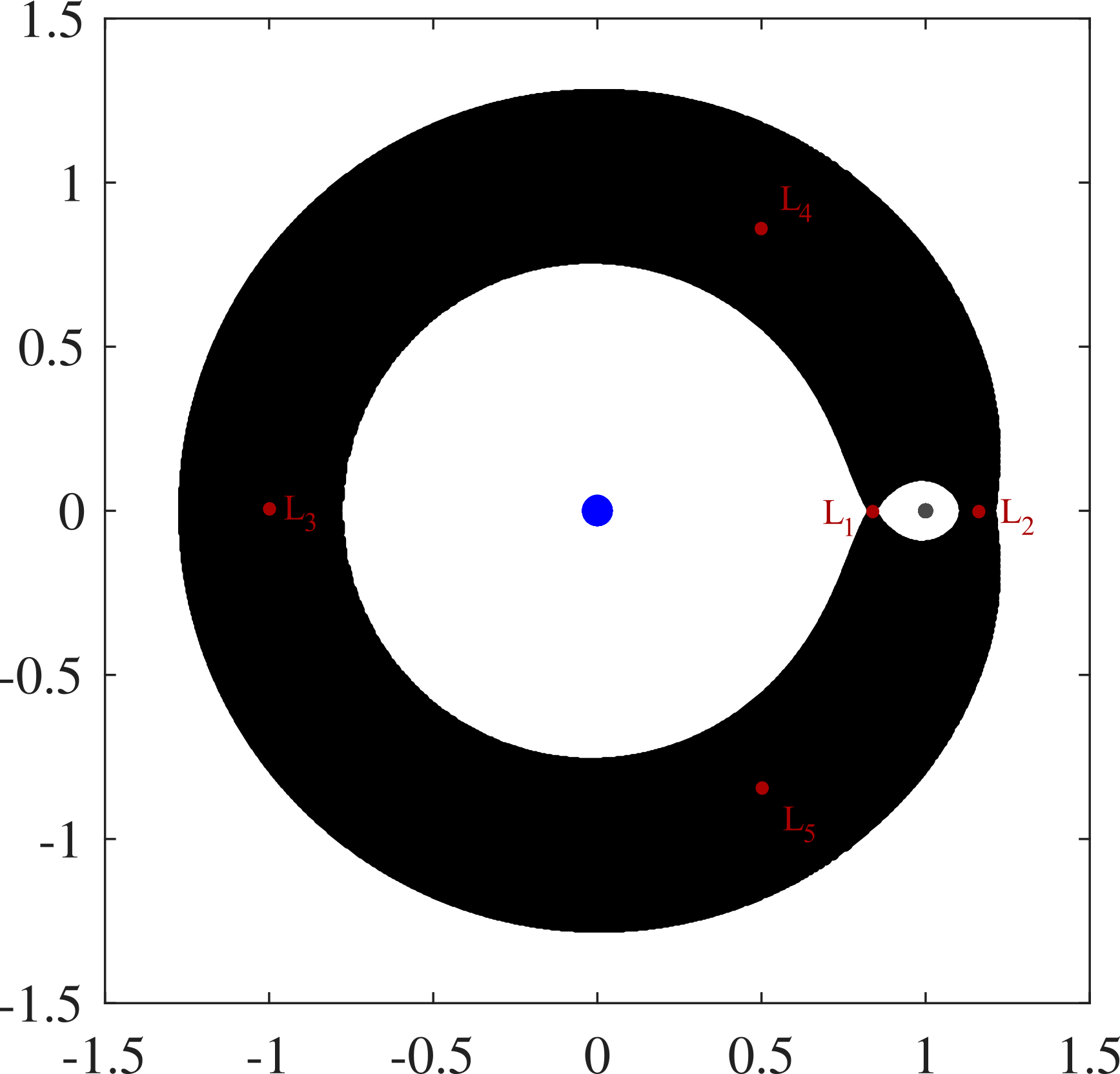}\caption*{\label{fig:L1}a) C = 3.188}
	\end{minipage}
	\begin{minipage}[t]{0.4\textwidth}
		\centering
		\includegraphics[width=\textwidth]{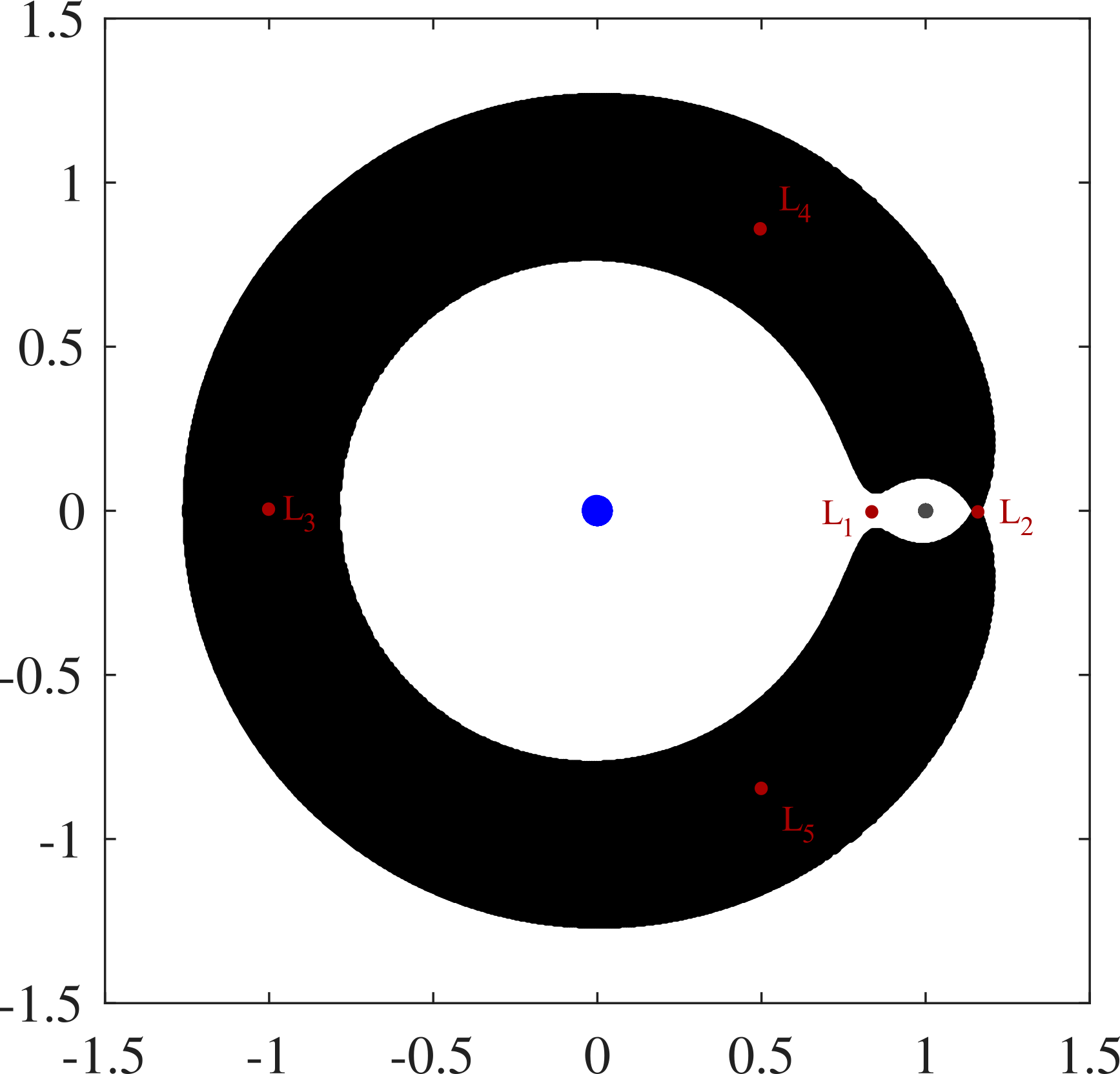}\caption*{\label{fig:L2}b) C = 3.172}
	\end{minipage}
	\begin{minipage}[t]{0.4\textwidth}
		\centering
		\includegraphics[width=\textwidth]{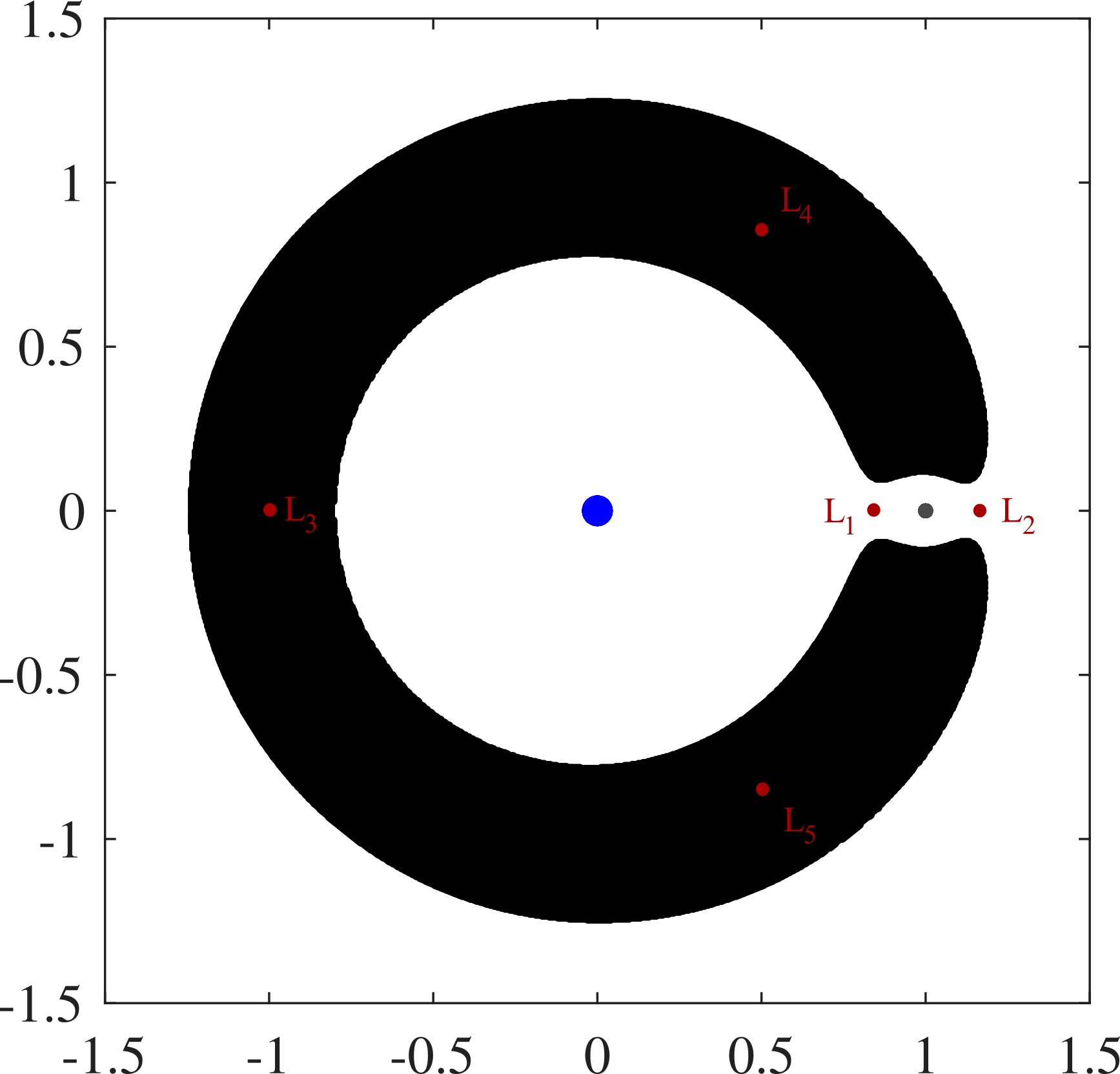}\caption*{\label{fig:L2bridge}c) C = 3.162}
	\end{minipage}
	\begin{minipage}[t]{0.4\textwidth}
		\centering
		\includegraphics[width=\textwidth]{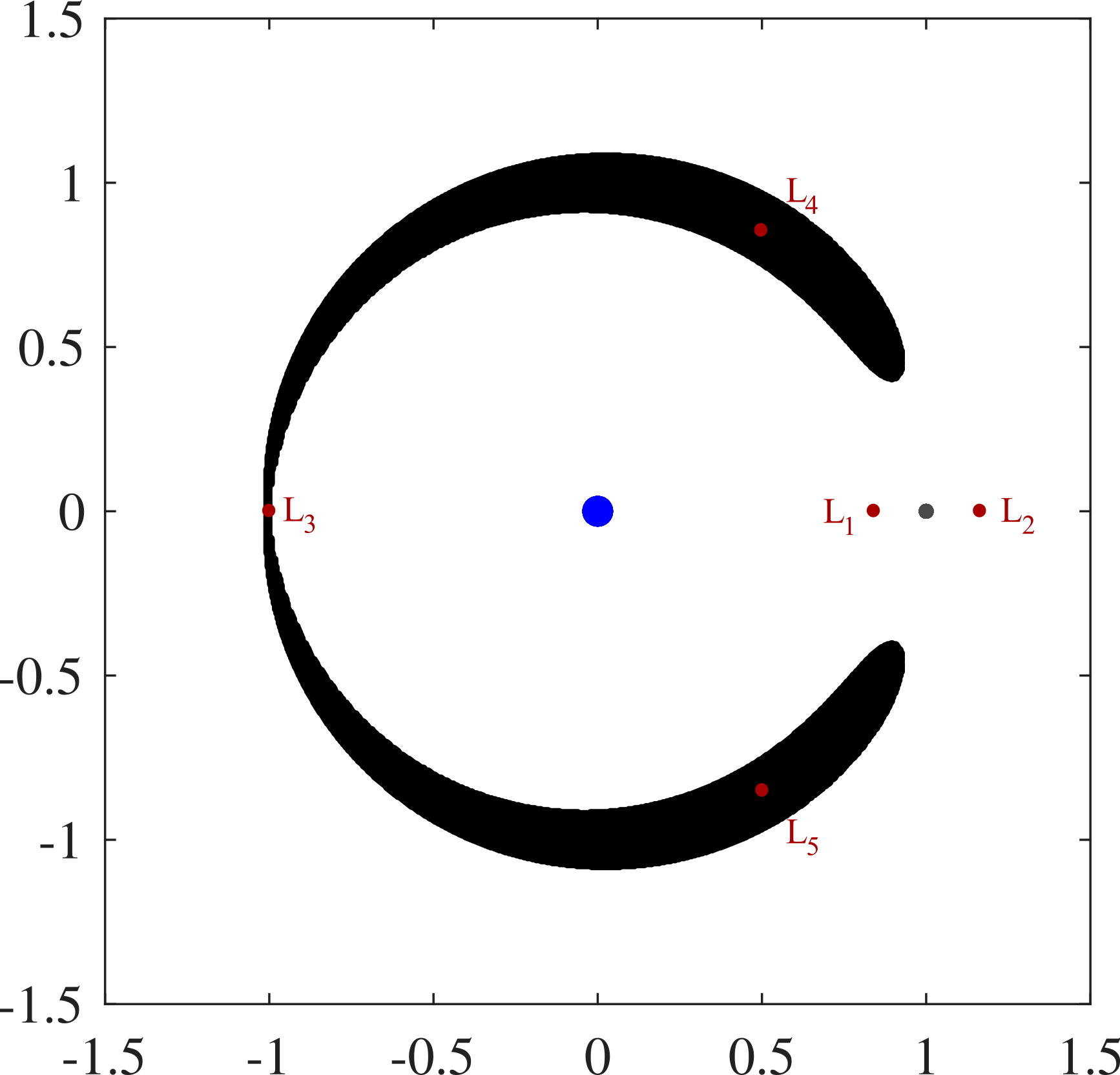}\caption*{\label{fig:L3}d) C = 3.012}
	\end{minipage}
	\begin{minipage}[t]{0.4\textwidth}
		\centering
		\includegraphics[width=\textwidth]{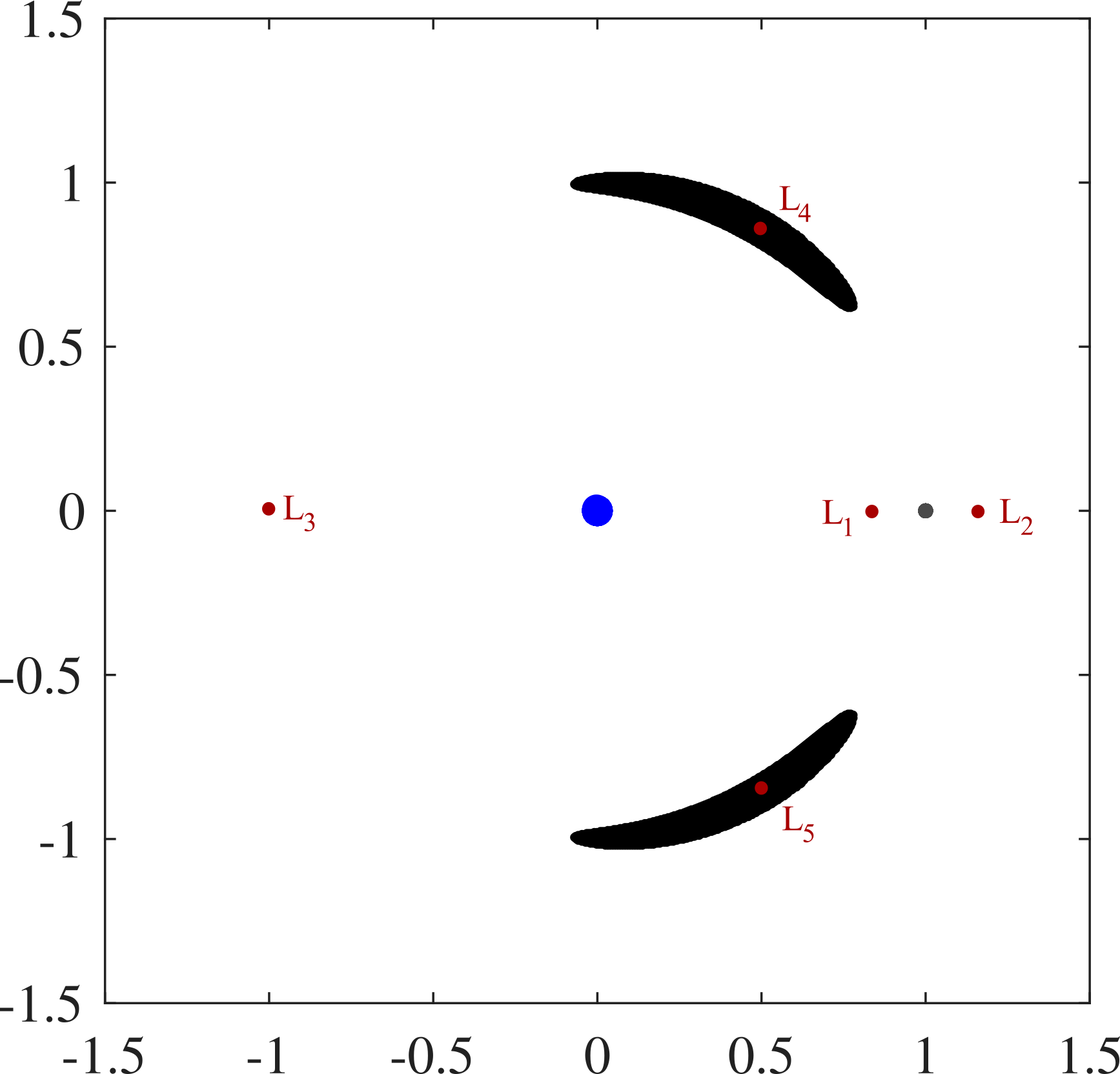}\caption*{\label{fig:L5}e) C = 2.989}
	\end{minipage}
	\begin{minipage}[t]{0.4\textwidth}
		\centering
		\includegraphics[width=\textwidth]{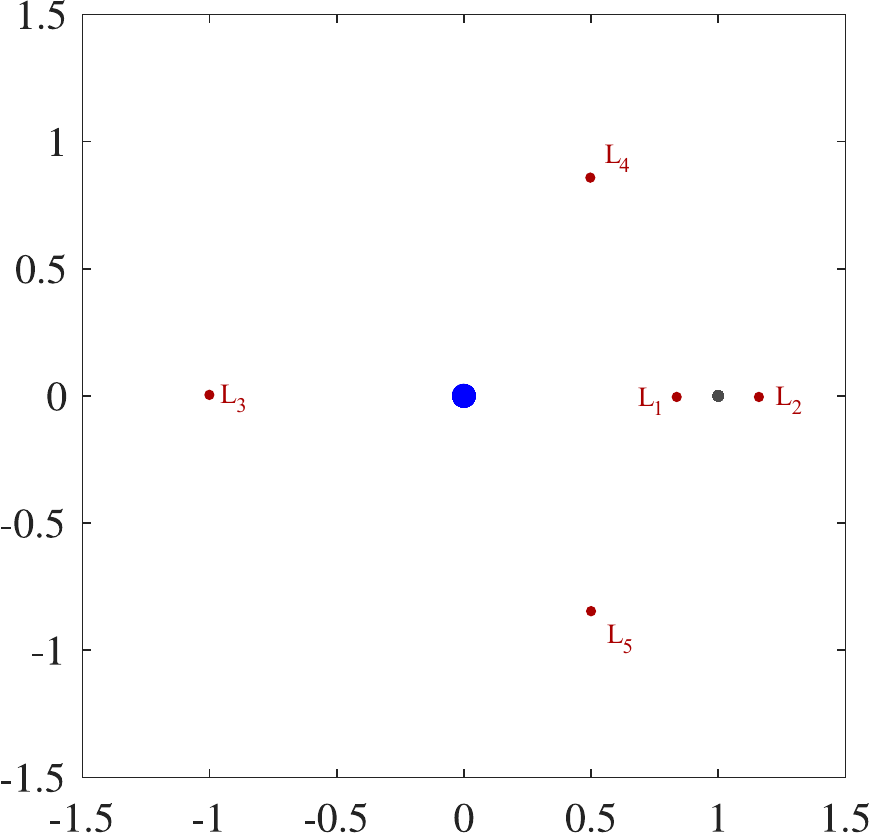}\caption*{\label{fig:L0}f) C = 2.979}
	\end{minipage} 
	\caption{Zero-velocity curves in the Earth-Moon system with decreasing Jacobi constant. Axes in \textcolor{Internal}{normalised units (unit length equal to the Earth-Moon distance)}}\label{fig:ZVC}
\end{figure}

Figure \ref{fig:ZVC} shows the zero-velocity curves for several increasing energy levels (i.e. decreasing Jacobi constant) in the Earth-Moon system. \textcolor{External}{Figure \ref{fig:ZVC} a)} shows the energy level of $L_1$, where there can be no motion between primaries. \textcolor{External}{In Figure \ref{fig:ZVC} b)} ($C = 3.172$), a trajectory from the Earth to the Moon is possible for the first time, showing that flying to the Moon via $L_1$ is energetically most favourable. The energy corresponding to \textcolor{External}{Figure \ref{fig:ZVC} c)} ($C = 3.162$) allows the test body to reach the $L_2$ point. \textcolor{External}{Figure \ref{fig:ZVC} d)} ($C = 3.012$) shows the energy of the test body increased to the level of $L_3$. If the energy is further increased, the $L_3$, $L_4$ and $L_5$ points can finally be achieved (\textcolor{External}{Figures \ref{fig:ZVC} e) and f)}, with $C = 2.989$ and $C = 2.979$).

\subsection{Libration Point Orbits}

The CR3BP contains a wide variety of periodic and quasi-periodic orbits. The former retrace their path over time: at the end of a certain time frame, the orbit's position and velocity repeat themselves. In contrast, quasi-periodic orbits have their motion confined to a particular region in space, changing slightly with each period \textcolor{Internal}{but never repeating the same orbit}. In the vicinity of the libration points, these are called the libration point orbits (LPOs).

\begin{figure}[!htb]
	\centering
	\includegraphics[width=0.7\linewidth]{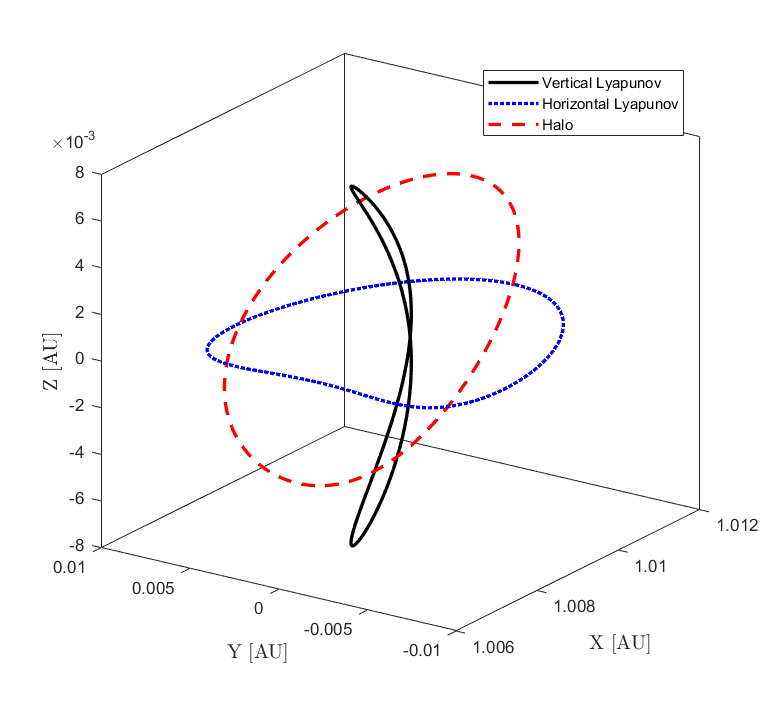}
	\caption{\label{fig:lpo}\textcolor{Internal}{Three types of LPOs in the Sun-Earth system, for C = 3.0007.}}
\end{figure}

Three types of \textcolor{Internal}{LPOs} can be highlighted: horizontal Lyapunov orbits, which are in the fundamental plane, vertical Lyapunov orbits, that are horizontally symmetric and shaped like a figure-eight, and halo orbits. Examples of these three orbital categories can be observed in Figure \ref{fig:lpo} \textcolor{Internal}{in the synodic reference frame}, for the same energy level.

\textcolor{External}{The computation of periodic orbits, as well as any trajectory with a set of desired characteristics in the CR3BP, may require the definition of a two-point boundary value problem, solved with the application of a differential corrector. The latter is} a method that, using the STM, determines how to change an initial trajectory in order to target a final set of orbital conditions. 

\subsection{Invariant Manifolds}

\textcolor{Both}{Manifold theory is essential to the understanding of the dynamical environment of the CR3BP. The invariant manifold structures connected to LPOs are particularly interesting for mission design, since they can be travelled by a spacecraft without any fuel consumption.}

\begin{figure}[h]
	\centering
	\includegraphics[width=\linewidth]{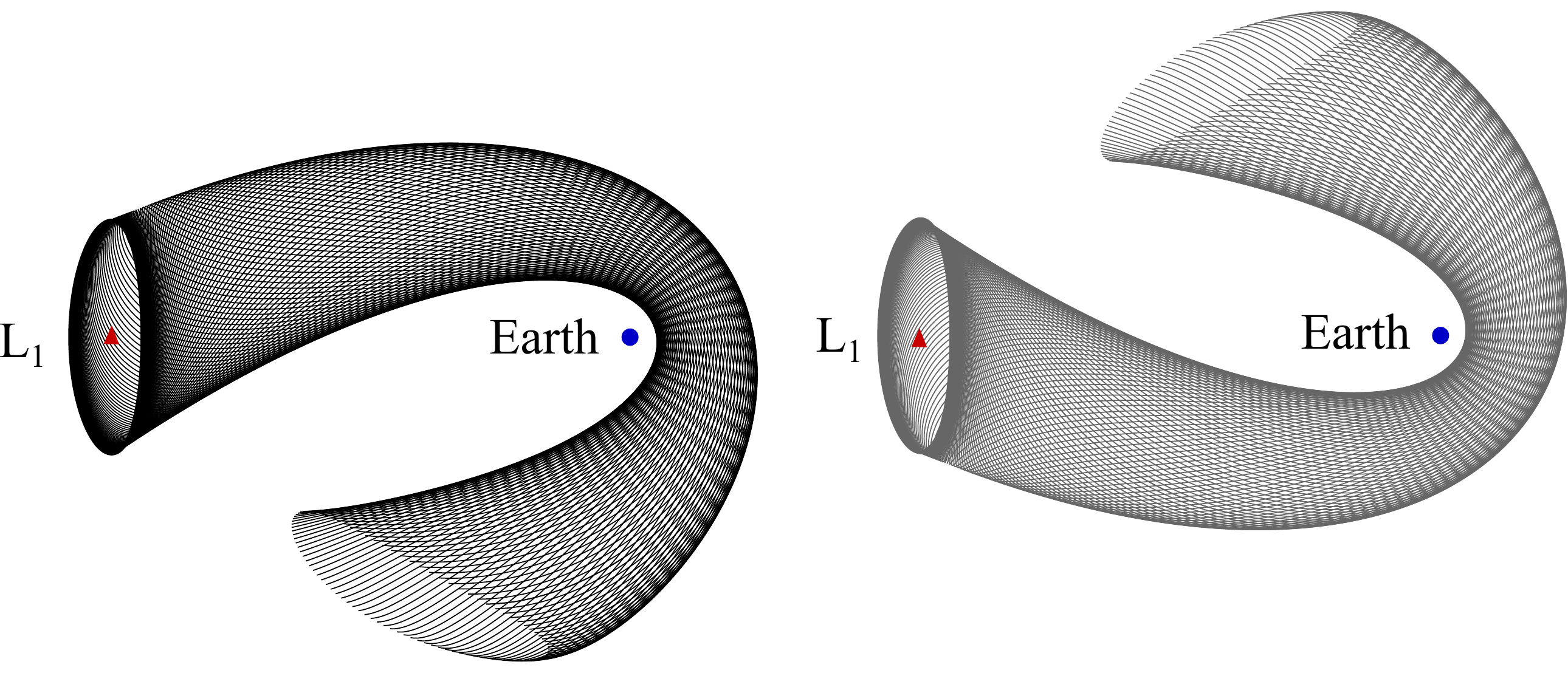}
	\caption{\label{fig:inv_man}Invariant manifold orbits connected to a halo orbit around the $L_1$ point of the Sun-Earth system, \textcolor{Internal}{C = 3.0007. On the left: stable manifold. On the right: unstable manifold}}
\end{figure}

Invariant manifolds are dynamical structures composed of countless orbits. They exist for a range of energies and form a series of 'tubes' that connect different regions around the primaries, as it can be seen on Figure \ref{fig:inv_man}. To compute them, it is necessary to acquire information about the local stability characteristics of each point along the LPO. This is obtained by computing the eigenvalues and eigenvectors of the monodromy matrix \cite{koon2008}, defined as the STM when propagated for precisely one period of the orbit.  

\textcolor{Internal}{Therefore, the computation of the invariant manifold structures of an LPO starts with the determination of its monodromy matrix. The latter possesses a certain number of distinct eigenvalues $\lambda$, with corresponding eigenvectors. The eigenvalues that have non-zero real parts are associated with the hyperbolic invariant manifolds, and the stability of the latter is determined by their real sign. If $Re(\lambda) < 0$, the manifold is stable; if $Re(\lambda) > 0$, the manifold is unstable.} 
	
\textcolor{Internal}{After the determination of the associated eigenvectors and eigenvalues, the LPO is discretised into numerous initial conditions, $\bm{s}_{P_0}$. A small perturbation step $\epsilon$ is introduced in the direction of the eigenvectors at each fixed point:}
\begin{align}\label{eq:stable}
\bm{s}_{S_0} = \bm{s}_{P_0} \pm \epsilon \bm{l}_S\\
\bm{s}_{U_0} = \bm{s}_{P_0} \pm \epsilon \bm{l}_U\label{eq:unstable}
\end{align}
where $\bm{l}_S$ and $\bm{l}_U$ are the stable and unstable eigenvectors computed from the monodromy matrix at the fixed point of interest, while $\epsilon$ is typically chosen to be equal to $10^{-6}$ (in normalised units). \textcolor{Internal}{The displacement $\epsilon$ is sufficiently small to avoid violating the linear approximation, but large enough to allow the manifold to depart the LPO after a reasonable time interval. Then, the the initial conditions of Eq. \eqref{eq:stable} are numerically integrated backwards in time to obtain the stable manifold set, while the ones of Eq. \eqref{eq:unstable} are, in contrast, integrated forwards in time to generate the unstable set. The explanations by Koon et. al. \cite{koon2001, koon_heteroclinic} are very helpful in understanding and applying the presented manifold theory.}

\section{Perturbation Methods}

Perturbation methods are a class of mathematical techniques used to generate solutions that describe the motion of a body subject to disturbing forces. \textcolor{Internal}{Depending on the orbit that is being analysed, the latter can be caused by the non-spherical shape of planets, atmospheric drag, solar radiation pressure, the gravitational attraction of other celestial bodies or other disturbing forces.} There are many techniques counted as perturbation methods, which can then be distinguished based on several different features. 

The techniques to solve perturbation equations fall into three broad categories: analytical, semi-analytical and numerical solutions \cite{vallado}. The analytical approaches were, historically, developed first due to the lack of computational power before the 20\textsuperscript{th} century. Semi-analytical and numerical methods employ computational routines to solve the differential equations. Still, even with the current computing capabilities, analytical solutions are sometimes preferable for very fast or real-time computations, provided they respect a certain level of accuracy. 

As such, perturbation methods can be divided in two main categories: \textit{special} and \textit{general} approaches. The first concerns the cases where the equations are solved using numerical methods. A specific, or \textit{special} answer is produced, only valid for the given data and initial conditions. Although special perturbation methods tend to be very accurate, they often suffer from over-specificity, since they cannot easily be extrapolated to different data. In contrast, general approaches employ anaytical and semi-analytical techniques, which can be used in several distinct case scenarios.

When describing different perturbation methods, some important characteristics should also be outlined. One refers to the difference between \textit{fast} and \textit{slow} variables. Fast variables change greatly during an orbit, even in the absence of perturbations (e.g. true, eccentric and mean anomalies). Slow variables change very little and are not modified at all if there is no perturbing effect (e.g. semi-major axis, eccentricity, inclination, argument of the periapsis and longitude of the ascending node).

The perturbed motion of a spacecraft can be computed using either \textit{mean} or \textit{osculating elements}. Mean elements are averaged over a chosen interval of time or true, eccentric, or mean anomaly, so they are relatively smoothly varying. By doing this, the perturbation is averaged throughout the motion instead of computed at each time-step. Single-averaged elements result from removing the high-frequency short-periodic motions, while double-averaged elements remove also the long periodic variations. On the other hand, osculating elements include all periodic (long and short-periodic) and secular effects. The osculating orbit is equivalent to the two-body orbit that the spacecraft would follow if the perturbing forces were suddenly removed at that instant: they are time-varying, so each trajectory point has a corresponding set of osculating elements. This representation is useful for highly accurate simulations, including real-time pointing and tracking operations. 

\subsection{Variation of Parameters}
\label{sub:VOP}

The variation of parameters (VOP) method is a mathematical tool to solve first-order differential equations, which yields a formulation of the equations of motion to use in perturbed dynamical systems. This theory is based on the premise that the solution for the unperturbed system can be used to \textcolor{Internal}{find} the perturbed one. Concretely, the unperturbed system is governed by the two-body problem discussed in Section \ref{sub:2bp}; when a small perturbative force is added, its effect is described by equations of motion that represent the time-varying, osculating orbital elements \cite{vallado}. These equations were originally developed by Euler and Lagrange as a general perturbation method.

For the perturbed problem, the equations of motion of the 2BP change from Eq. \eqref{eq:2bp} to the general form:
\begin{align}
\ddot{\bm{r}} + \mu_C \cdot \frac{\bm{r}}{r^3} = \nabla \mathpzc{R} \label{eq:pert}
\end{align}
where $\mathpzc{R}$ is the disturbing function that fully describes the perturbation. This equation can be further expanded into:
\vspace{-5mm}
\begin{align}
\ddot{x}+ \textcolor{Internal}{\mu_C}\frac{x}{r^{\,3}} = \frac{\partial {\mathpzc R}}{\partial x} = a_x\nonumber\\	
\ddot{y}+ \textcolor{Internal}{\mu_C}\frac{y}{r^{\,3}} = \frac{\partial {\mathpzc R}}{\partial y} = a_y\nonumber\\	
\ddot{z}+ \textcolor{Internal}{\mu_C}\frac{z}{r^{\,3}} = \frac{\partial {\mathpzc R}}{\partial z} = a_z\label{eq:lpe1}
\end{align}

Different VOP formulations can be distinguished depending on the forces acting on the spacecraft, which are either conservative (e.g. third-body effect) or non-conservative (e.g. pressure, thrust, drag). For the conservative case, Lagrange planetary equations (LPE) \textcolor{Internal}{can be} employed. For the non-conservative case, Gauss' variational equations (GVE) are more adequate to compute the spacecraft's motion. These methods employ different ways of describing the perturbing effect: either using a disturbing function or a disturbing force (or acceleration). The disturbing function is the difference between perturbed and unperturbed potential functions, whereas the disturbing force expresses the specific acceleration being exerted on the spacecraft.

\paragraph{Lagrange Planetary Equations:} this method was derived to describe the osculating motion of a body subject to a conservative perturbation (i.e any gravitational effect). 

The formulation of the LPE employs the first equality of Eq. \eqref{eq:lpe1}, using the disturbing function as the way to describe the perturbation. The final set of equations can be found \textcolor{Internal}{in Eq. \ref{eq:LPE} and their derivation can be followed in Battin \cite{battin}.}
\begin{align}
\nonumber\frac{da}{dt} &= \frac{2}{n a} \frac{\partial{\mathpzc{R}}}{\partial M_0}\\\nonumber
\frac{de}{dt} &= \frac{1 - e^2}{n\,a^2 e}\,\frac{\partial{\mathpzc{R}}}{\partial M_0} - \frac{\sqrt{1 - e^2}}{n\,a^2 e}\,\frac{\partial{\mathpzc{R}}}{\partial \omega}\\\nonumber
\frac{di}{dt} &= -\frac{1}{\,n a^2 \sqrt{1 - e^2} \sin i}\,\frac{\partial{\mathpzc{R}}}{\partial \Omega} - \frac{\cos i}{\,n a^2 \sqrt{1 - e^2} \sin i}\,\frac{\partial{\mathpzc{R}}}{\partial \omega}\\\nonumber
\frac{d\Omega}{dt} &= \frac{1}{n\,a^{\,2}\,\sqrt{1-e^2}\sin i}\,\frac{\partial{\mathpzc{R}}}{\partial i}\\\nonumber
\frac{d\omega}{dt} &= \frac{\sqrt{1 - e^2}}{n\,a^2 e}\,\frac{\partial{\mathpzc{R}}}{\partial e} - \frac{\cos i}{\,n a^2 \sqrt{1 - e^2} \sin i}\,\frac{\partial{\mathpzc{R}}}{\partial i}\\
\frac{dM_0}{dt} &= -\frac{2}{n\,a}\,\frac{\partial{\mathpzc{R}}}{\partial a} - \frac{1 - e^2}{\,n a^2 e}\,\frac{\partial{\mathpzc{R}}}{\partial e} \label{eq:LPE}
\end{align}
in which $a$ is the semi-major axis, $e$ is the eccentricity, $i$ is the inclination, $\Omega$ is the longitude of the ascending node, $\omega$ is the argument of the periapsis, $M$ is the mean anomaly and $n$ is the mean motion.

\paragraph{Gauss' Variational Equations:} Gauss' form of the planetary equations is perfectly equivalent to the LPE, but instead uses the second equality of Eq. \eqref{eq:lpe1}. \textcolor{Internal}{The full derivation of the equations can again be found in Battin \cite{battin}.} Since the perturbation is computed based a disturbing force, the GVE can be employed for non-conservative effects (i.e. pressure, drag or thrust). These equations can be found below:
\begin{align}
\nonumber\frac{da}{dt} &= \frac{2 a^2}{L} \Big(a_r e \sin\nu + a_{\theta}\frac{p}{r}\Big)\\\nonumber
\frac{de}{dt} &= \frac{1}{L} \Big(a_r p\sin\nu + a_{\theta}((p + r)\cos\nu + r e)\Big)\\\nonumber
\frac{di}{dt} &= a_h\frac{r \cos \upsilon}{L}\\\nonumber
\frac{d\Omega}{dt} &= a_h\frac{r \sin \upsilon}{L\sin i}\\\nonumber
\frac{d\omega}{dt} &= \frac{1}{L e} \Big(-a_r p\cos\nu + a_{\theta}(p + r)\sin\nu \Big) - a_h\frac{r \sin \upsilon \cos i}{L\sin i}\\
\frac{d\nu}{dt} &=  \frac{L}{r^2} + \frac{1}{L e} \Big(a_r p\cos\nu - a_{\theta}(p + r)\sin\nu \Big)\label{eg:gve}
\end{align}
in which \textcolor{Internal}{$a_{r}$, $a_{\theta}$ and $a_{h}$ are the acceleration components in the LVLH frame,} $L$ is the angular momentum, $p$ is the semilatus rectum, $b$ is the semi-minor axis, $r$ is the orbital position and $\upsilon = \nu + \omega$ is the argument of latitude, with $\nu$ as the true anomaly.  

\section{Flow Maps}
\label{sec:FM}

In order to study of the evolution of a perturbed orbit, the computation of the motion's dynamics at each instant may not necessarily be of interest. As an alternative, one of the main techniques to study the behaviour of complex dynamical systems is the flow map: a method that maps points from their initial location at time $t_0$ to their state at time $t$ \cite{koon2008}.

The concept of flow maps is very helpful in exploring the behaviour of numerical propagations. Many techniques that fall under this category are commonly applied in astrodynamics: a classical example is the Poincar\'e map. This method is used to replace the flow of a $N$-dimensional continuous time system by a $(N-1)$-dimensional discrete time one \cite{parker}. When used in orbital propagation, it maps the state of the orbit between two consecutive crossings of the motion with a hyperplane, i.e. a Poincar\'e section. Poincar\'e maps are frequently used to study the stability of periodic orbits and can elegantly show the orbital evolution of different trajectories in the domain of interest (see e.g. \cite{villac}).

\begin{figure}[h]
	\centering
	\includegraphics[width=0.5\linewidth]{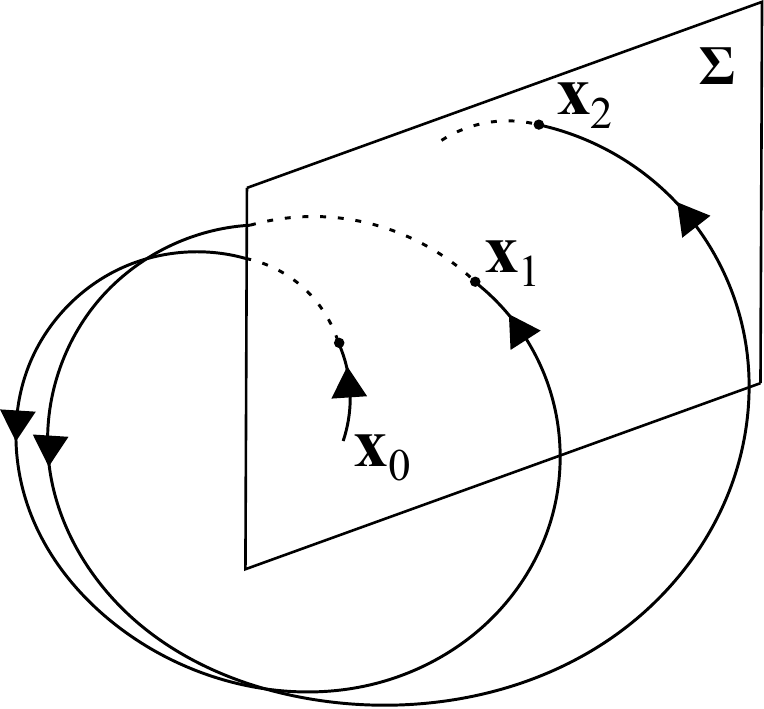}
	\caption{\label{fig:poincare2}Poincar\'e surface of section with crossings $\bm{x}_0$, $\bm{x}_1$ and $\bm{x}_2$}
\end{figure}

There are two basic types of Poincar\'e maps: the first one is the Poincar\'e surface of section \textit{in \textcolor{Internal}{phase} space}, which can be seen in Figure \ref{fig:poincare2}. This consists of a curve on a hyperplane, $\Sigma$, transverse to the flow in $\mathbb{R}^N$, which reflects the evolution of a trajectory point in a certain space region (for a 3-D system, $\Sigma$ is a 2-D plane). In other words, starting from some initial conditions $\bm{x}_0$, the $\Sigma$ plane contains a set of points that indicate the intersection \textcolor{External}{layout} of the orbit.

A second version of a Poincar\'e map is obtained by a Poincar\'e surface of section \textit{in time}. For this purpose, the time-continuous dynamical system is sampled not with respect to a constraint in phase space but at discrete times $t_m = t_0 + m\cdot T,\text{ }m \in \mathbb{N}$.  In this case, the map is called a stroboscopic sampling of the phase space---the method is then named \textit{stroboscopic map}. 

\textcolor{Internal}{The use of stroboscopic maps is common when computing motion in perturbed Keplerian dynamics \cite{gondelach}. The time sampling of orbital elements is often done by employing semi-analytical models \cite{roth, alessi_semi}. These are generally implemented by transforming the chosen equations of motion from being time-dependant to becoming a function of the fast angular variable (e.g. true anomaly) of the orbital elements of the massless particle. Since this fast variable changes by a fixed amount in a time period ($0$ to $2\pi$), orbital conditions can be updated at each time period by integrating over this parameter.}
	
\textcolor{Internal}{In order to write the equations of motion as a function of the fast variable, the former have to be multiplied by the derivative of time with respect to the fast variable $dt/dv$, where $v$ indicates the fast variable. Finally, the propagation result over one period in the fast angle $v$ returns the stroboscopic map \cite{gondelach}.}

\chapter{The Keplerian Third-Body Potential}
\label{chap:k3bp}
This work proposes to compute the third-body effect using a disturbing function obtained from the Keplerian third-body potential (K3BP). This potential function has been previously used in the Keplerian map (KM) for two and three-dimensions \cite{scheeres_multiple, alessi_semi}. This chapter describes this well-known method, as well as four novel implementations of the function. The first three are obtained using the Lagrange planetary equations (LPE), suitable for systems where only conservative forces are at play (e.g. gravitational effects). The last one makes use of Gauss' variational equations (GVE), so that non-conservative forces are also taken into account (e.g. thrusting accelerations).

The accuracy of the different methods detailed in this Chapter will be determined by comparing their propagation to the results of the circular restricted three-body problem (CR3BP). In contrast, the two-body problem (2BP) propagation will be used as the lowest-fidelity method, the probable worst performer in the cases where the disturbing effect of the secondary impacts the mission design. 

The comparison between models of motion will be done in different ways, depending on what is being analysed. When conservative methods are studied, the main concern falls on the mapping and approximation strategies to obtain the \textcolor{Internal}{orbital elements' evolution}. As such, the comparison between methods is performed by studying each relevant orbital element individually, so that its dynamical behaviour can be assessed. When non-conservative methods are analysed, the main novelty is the low-thrust trajectory propagation; consequently, the error study is done instead by comparing orbital positions in proposed mission scenarios.

The K3BP can be used to describe the third-body perturbation in planetary configurations of small gravitational parameter (e.g. Sun-Earth, Jupiter-Callisto, Saturn-Titan). Given the application scenarios in this work, the methods will be characterised using the Sun-Earth system as the main example, unless otherwise stated.

\section{The Third-Body Perturbation}
\label{sec:3BP}

The third-body perturbation is a term that refers to the added effect of an extra body to the motion of a spacecraft (here considered to be the massless particle) around a central body (e.g. the Sun). This effect is gravitational and therefore conservative, so it can be described using either disturbing functions or accelerations \cite{vallado}, as previously shown for Eq. \eqref{eq:pert}. The third-body perturbing acceleration caused by the Earth, in the Sun-Earth system, can be written as the following:
\begin{align}\label{eq:3bvallado}
\bm{\ddot{r}}_{3B} = \mu_{\Earth} \bigg(\frac{-\bm{r}_{\Earth} + \bm{r}}{\norm{-\bm{r}_{\Earth} + \bm{r}}^3}
- \frac{\bm{r}_{\Earth}}{\norm{\bm{r}_{\Earth} }^3} \bigg)
\end{align}
in which $\bm{r}_{\Earth}$ and $\bm{r}$ are respectively the Earth's and spacecraft's \textcolor{External}{position} vectors with respect to the Sun.

\begin{figure}[h!]
	\centering
	\includegraphics[width=0.65\linewidth]{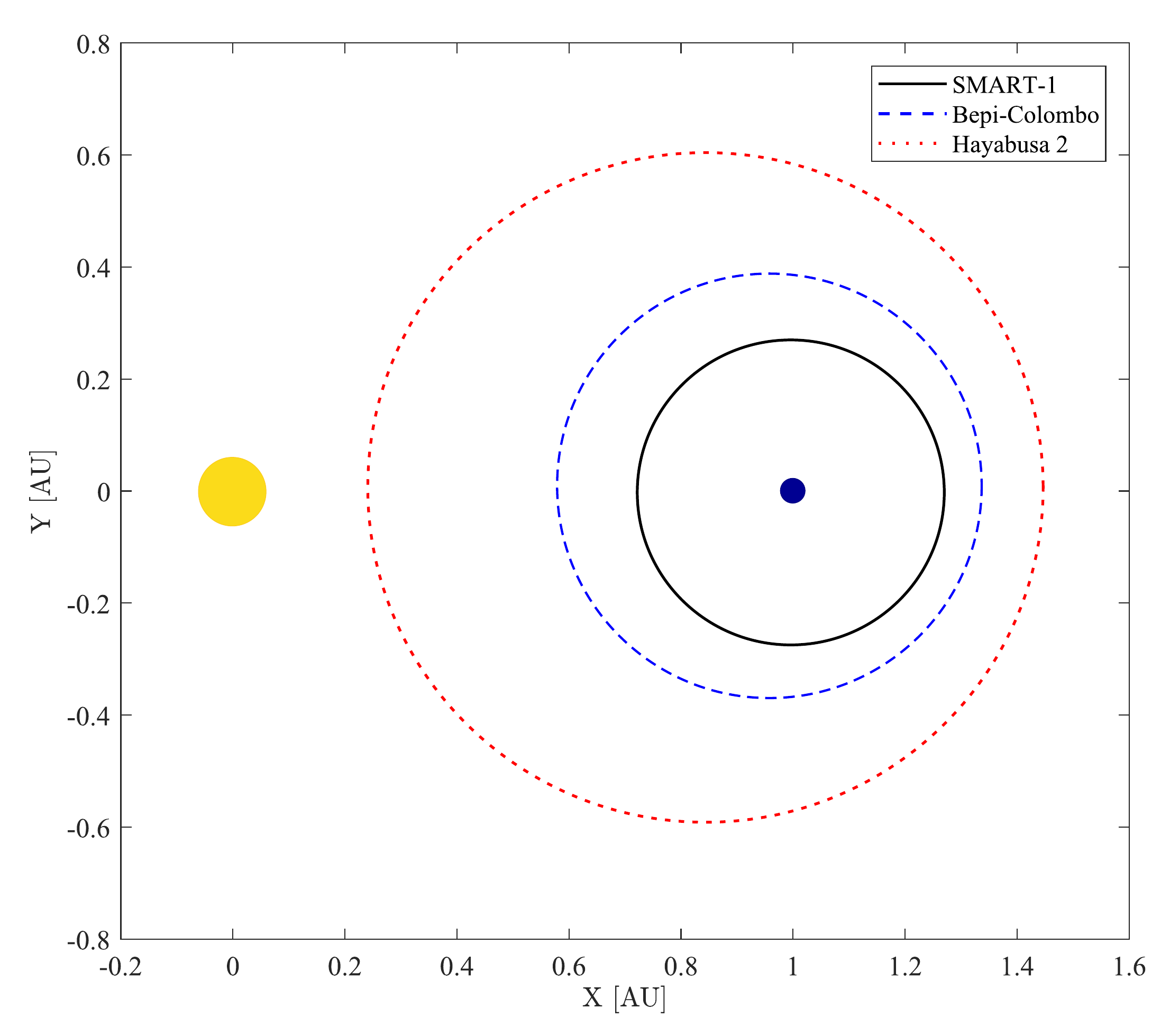}
	\caption{\label{fig:sc}Limits of the regions within which each spacecraft's acceleration is at most 1000 times greater than the one of the Earth. Earth is scaled to \textcolor{Internal}{its} Hill radius}
\end{figure}

Depending on the magnitude of this acceleration, even when outside the sphere of influence of the Earth (around 0.01 AU), the spacecraft can still be affected by its perturbation. \textcolor{Internal}{As an example of this situation,} the disturbing accelerations of the Earth on a hypothetical spacecraft were computed with Eq. \eqref{eq:3bvallado}. These were then compared to the output accelerations of the electric engines of three different spacecraft: SMART-1, Bepi-Colombo and Hayabusa 2 \cite{smart1, bepicolombo, hayabusa2}. The results can be seen in Figure \ref{fig:sc}; they depict \textcolor{Internal}{circular areas within which the Earth's perturbation is at most 1000 times smaller than the spacecraft's output acceleration.}

\textcolor{Internal}{While a perturbation 1000 times smaller than the main governing accelerations may often be considered negligible, a careful regard for such small effects can bring important benefits to the design of a trajectory. A well-known supporting example is that of Sun-synchronous orbits, which cannot be computed in the 2BP unless the $J_2$ perturbation is added into the dynamics of an Earth orbiting spacecraft. Yet, the $J_2$ effect is 1000 times smaller than the central gravitational acceleration---hence, this ratio is arbitrarily used in Figure \ref{fig:sc} to show that the Earth's third-body perturbation may still be a significant acceleration to account for. Naturally, the regions depicted in the figure will change with the low-thrust system considered; however, it is clear that the Earth's disturbing acceleration is non-negligible within a much larger space than the classical sphere of influence.} 
	
This space, where the 2BP is determined not to be accurate enough to describe the spacecraft's motion, is here termed the \textit{perturbation region}, and will be further defined and discussed later in this chapter.

\section{The Keplerian Disturbing Function}
\label{sec:dist_fcn}

In order to describe the third-body effect in the perturbation region, the K3BP and its disturbing function are here derived. \textcolor{Both}{The computation starts with the Hamiltonian of the three-body problem in an inertial reference frame, Sun-centred:}
\begin{align}
\mathpzc{H}_{3B} = \frac{1}{2}(o_x^2 + o_y^2 + o_z^2) - \frac{1 - \mu}{r_1} - \frac{\mu}{r_2}
\end{align}
\textcolor{External}{in which $o_x$, $o_y$, and $o_z$ are the generalized momenta of the massless particle and $r_1$ and $r_2$ are again the distances from the massless particle to the primary and secondary, respectively.} 

\begin{figure}[h!]
	\centering
	\includegraphics[width=0.7\linewidth]{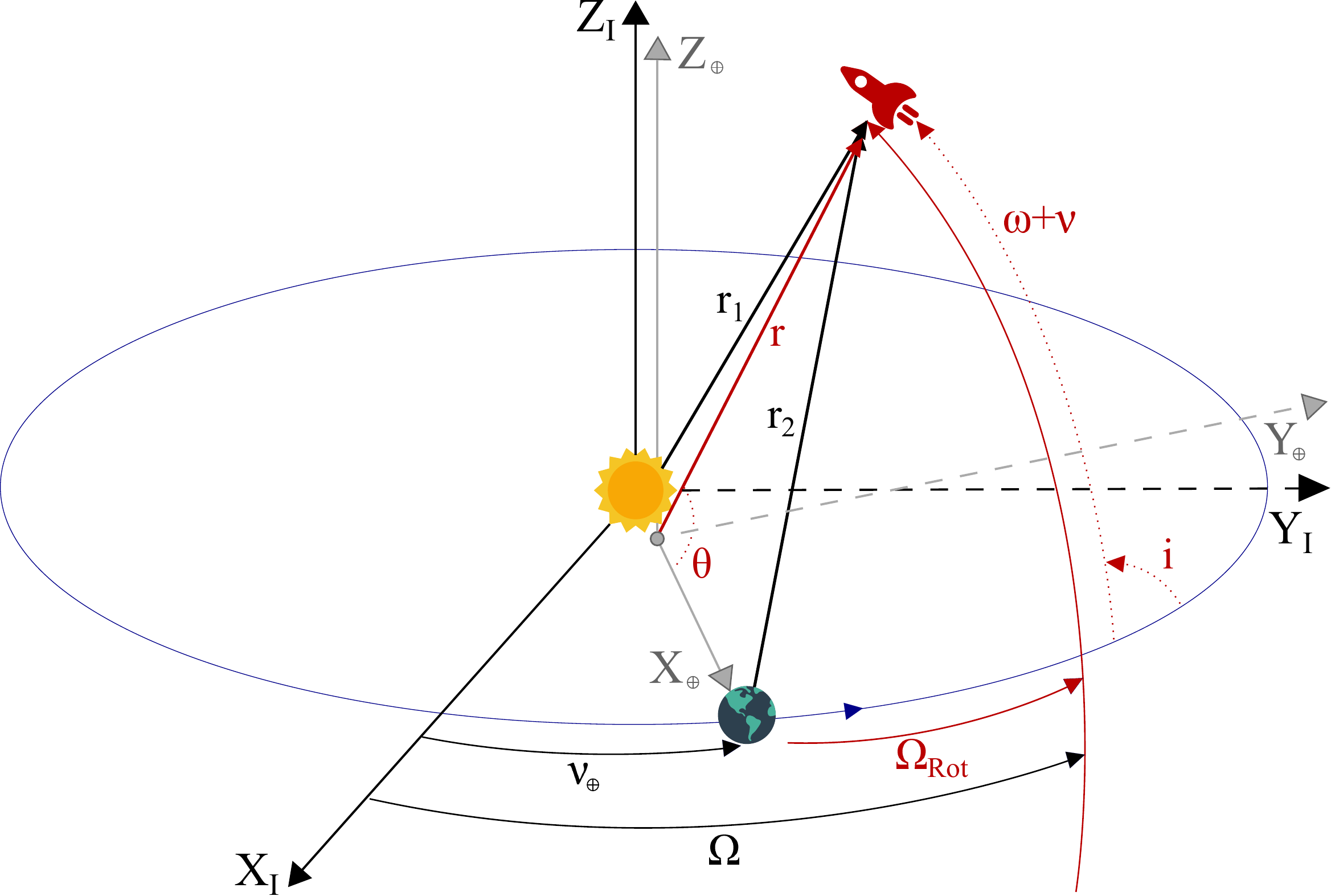}
	\caption[Three-dimensional geometry of the three-body problem in the inertial reference frame]{\label{fig:3d}Three-dimensional geometry of the three-body problem, noting the inertial and Earth-pointing reference frames\cite{neves2018}}
\end{figure}

\textcolor{Both}{This Hamiltonian can be simplified into a barycentric notation, by defining $r_1$ and $r_2$ as functions of the distance to the barycentre $r$. This is done by utilising polar coordinates $\{x = r\cos\theta; \sqrt{y^2 + z^2} =  r\sin\theta\}$:}
\begin{align}
r_1^2 &= (r\cos\theta + \mu)^2 + (r\sin\theta)^2\nonumber\\
r_2^2 &= (r\cos\theta -1 + \mu)^2 + (r\sin\theta)^2
\end{align}
where $\theta$ is the angle between $r$ and the Sun-Earth line, as it can be seen on Figure \ref{fig:3d}.

Starting with the development of $r_1$ into a function of $r$, Eq. \eqref{eq:r1inv} is obtained:
\begin{align}
r_1^2 &= r^2 + \mu^2 + 2 r \mu\cos\theta \Leftrightarrow\nonumber\\
\Leftrightarrow\frac{1}{r_1} &= \frac{1}{r}\frac{1}{\sqrt{1 + 2 \cos\theta \frac{\mu}{r} + (\frac{\mu}{r})^2}}\label{eq:r1inv}
\end{align}

Assuming a Taylor expansion around $\mu = 0$ for the terms with $r_1$ and $r_2$, Eqs. \eqref{eq:R1} and \eqref{eq:R2} are obtained. This approximation shortens the application range of the method to systems of small $\mu$.
\begin{align}
\frac{1 - \mu}{r_1} &= \frac{1}{r} + \mu\Big(-\frac{1}{r} - \frac{\cos\theta}{r^2}\Big) + \mathpzc{O}(\mu^2)\label{eq:R1}\\
\frac{\mu}{r_2} &= \frac{\mu}{\sqrt{r^2 - 2 r\cos\theta + 1}} + \mathpzc{O}(\mu^2)\label{eq:R2} 
\end{align}

It is important to note that the Taylor expansions done on the previous equations imply that $\mu \ll r$. This means that, ultimately, the location of the primary and the barycentre are nearly indistinguishable. Finally, the Hamiltonian becomes:
\begin{align}
\mathpzc{H}_{3B} = \mathpzc{T} + \mathpzc{U}_{3B} + \mathpzc{O}(\mu^2)\label{eq:hamiltonian}
\end{align}
in which:
\begin{align}
\mathpzc{T} &= \frac{1}{2}(p_x^2 + p_y^2 + p_z^2) - \frac{1}{r}\\
\mathpzc{U}_{3B} &= \mu \bigg(\frac{1}{r} + \frac{\cos\theta}{r^2} - \frac{1}{\sqrt{1 + r^2 - 2 r\cos\theta}}\bigg)\label{eq:hamil}
\end{align}

Finally, the K3BP is obtained as $\mathpzc{U}_{3B}$. The disturbing function $\mathpzc{R}$ is easily computed as $\mathpzc{R} = -\mathpzc{U}_{3B}$. 

\textcolor{Both}{With this disturbing function, another frame of reference can be introduced: a barycentric coordinate frame, with the X-axis pointing towards the Earth at all times. This will hence be named the \textit{Earth-pointing reference frame}. It can be examined in Figure \ref{fig:3d}, together with the geometry of the three-body planetary system in question: the inertial reference frame is represented by $O_{I xyz}$ and the Earth-pointing one is denoted by $O_{\Earth xyz}$.} 
	
By changing the frame of reference, it follows that a new quantity has to be introduced when computing the spacecraft motion: $\Omega_{Rot}$, which replaces the traditional $\Omega$ in the orbital elements \cite{alessi_semi}. \textcolor{Internal}{This quantity is the rotational longitude of the ascending node of the spacecraft, defined with respect to the new Earth-pointing reference frame. In this way, $\Omega_{Rot}$ is determined as:}
\begin{align}\label{eq:omrot}
\Omega_{Rot} = \Omega - \nu_{\Earth}
\end{align}
in which $\nu_{\Earth}$ is the Earth's true anomaly.

Following Alessi and S\'anchez \cite{alessi_semi}, $\nu_{\Earth}$ can be formulated as a function of the normalised integration time by expanding the following law of planetary motion:
\begin{align}
M_{\Earth} - M_{{\Earth}_0} &= n_{\Earth} (t - t_{{\Earth}_0})
\end{align}

Considering a circular Earth orbit (in non-dimensional variables, $n_{\Earth}  = 1$), a relationship between the normalised time $t$ and the Earth's true anomaly is inferred by Eq. \eqref{eq:normtime}.
\begin{align}
M_{\Earth} - M_{{\Earth}_0} &= \nu_{\Earth} - \nu_{{\Earth}_0} = t - t_{{\Earth}_0}\label{eq:normtime}
\end{align}

This can be further developed by describing Earth's mean anomaly using purely the orbital elements of the massless particle. \textcolor{Both}{The latter are represented without any kind of subscript, in contrast to the orbital elements of the Earth. The starting point is the massless particle's mean anomaly equation, where the normalised time is isolated (using $M_0 = 0$):} 
\begin{align}\label{eq:t1}
M &= n(t - t_0) \Leftrightarrow	 t = \frac{M}{n} + t_0
\end{align}

\textcolor{Both}{Given that the normalised time is the same variable for both the spacecraft and the Earth in the chosen reference frame, Eqs. \eqref{eq:normtime} and \eqref{eq:t1} can be combined to obtain the desired dependency between Earth's mean anomaly and the massless particle's elements, choosing $\nu_{{\Earth}_0}$ so that $t_{{\Earth}_0} = 0$.}
\begin{align}\label{eq:t2}
\frac{M}{n} + t_0 &= \nu_{\Earth} - \nu_{{\Earth}_0}
\end{align}

\textcolor{Both}{Since the initial reference time for the Earth's orbit ($t_{{\Earth}_0}$) and its true anomaly are interchangeable, Eq. \eqref{eq:t2} cn be further simplified into the following:}
\begin{align}
\nu_{\Earth} - \nu_{{\Earth}_0} &= \sqrt{\frac{a^3}{1 - \mu}} M + t_0\label{eq:tEarth}
\end{align}

Thus, the K3BP can be used to describe the third-body perturbation in a similar way to Eq. \eqref{eq:3bvallado}, using Keplerian elements and without having to explicitly compute the position of the secondary at all times. This comes at the cost of having a non-autonomous reference frame.

\subsection{Formulations of the Disturbing Function}
\label{sub:forms}

The disturbing function derived from the K3BP is highly dependant on two terms: $r$ and $\cos\theta$. The way these are written bears influence when computing the motion of an object under a third-body effect. Particularly, the choice of fast variable may change the description of any flow maps used to compute the perturbed motion. As such, two possible formulations are here presented: one having the true anomaly as the fast variable (the KM used by Alessi and S\'anchez \cite{alessi_semi}) and the other using the eccentric anomaly instead.

\subsubsection{Formulation in True Anomaly}
\label{subsub:nu}

\textcolor{Both}{In this formulation, the quantity $r$ remains the magnitude of the position vector with respect to the barycentre, while $\cos\theta$ is defined using a spherical trigonometric formula that considers the triangle formed by the massless particle, the Earth and the line of nodes, as it can be seen on Figure \ref{fig:3d}:}
\begin{align}\label{eq:rtheta}
r &= \frac{a(1 - e^2)}{1 + e\cos\nu}\\
\cos\theta &= \cos\Omega_{Rot}\cos(\omega + \nu) + \sin\Omega_{Rot}\sin(\omega + \nu)\cos i\label{eq:rtheta2}
\end{align}

Using the common relations between mean and true anomaly, Eq. \eqref{eq:tEarth} can be further expanded into the following:
\begin{align}\label{eq:true_t}
\nu_{\Earth} = \textcolor{Both}{\sqrt{\frac{a^3}{1 - \mu}}\bigg[ 2\arctan \bigg( \sqrt{\frac{1-e}{1+e}} \tan \bigg(\frac{\nu}{2}\bigg)\bigg) - \frac{e \sqrt{1 - e^2}\sin\nu}{1 + e\cos\nu} \bigg] - t_0}
\end{align} 

From Eq. \eqref{eq:true_t}, a singularity can be spotted for $\nu = \pi + 2 k \pi, k \in \mathbb{Z}$.

\subsubsection{Formulation in Eccentric Anomaly}
\label{subsub:ee}

The disturbing function can also be written using the eccentric anomaly $E$ as the fast variable. This solves the problem of describing $\nu_{\Earth}$ as an arctangent of a tangent function shown in Eq. \eqref{eq:true_t}, which is not smooth. In order to do so, $r$ and $\cos\theta$ are written using Eqs. \eqref{eq:theta_e} and \eqref{eq:theta_e2}, obtained using well-known formulae that can be found in Battin \cite{battin}: 
\begin{align}\label{eq:theta_e}
r &= a(1 - e\cos E)\\\nonumber
\cos\theta &= \cos i\sin\Omega_{Rot} \cdot \frac{(\sqrt{1-e^2}\cos\omega \sin E (-e + \cos E)\sin\omega)}{-1 + e\cos E} + \\
& \hspace{3cm} \frac{\cos\Omega_{Rot}((e - \cos E)\cos\omega + \sqrt{1-e^2}\sin E \sin\omega)}{-1 + e\cos E}\label{eq:theta_e2}
\end{align}

Using the relations between mean and eccentric anomaly, Eq. \eqref{eq:tEarth} as a function of the massless particle's elements becomes:
\begin{align}\label{eq:true_e}
\nu_{\Earth} = \textcolor{Both}{\frac{E - e\sin E}{n}}
\end{align}

The equations for $r$ and $\nu_{\Earth}$ become much less complex when compared to the true anomaly formulation, while the opposite happens for the term $\cos\theta$. Plus, this formulation has the aforementioned advantage of lacking singularities (except in the case of a parabolic orbit, which is out of the scope of this project).

Thus, as previously discussed, Eqs. \eqref{eq:true_t} and \eqref{eq:true_e} can be used to compute $\Omega_{Rot}$ as a function of the true or eccentric anomaly of the perturbing body per Eq. \eqref{eq:omrot}. At the same time, by observing Eq. \eqref{eq:normtime}, it can be inferred that the aforementioned equations are also useful in the mapping from the normalised time to the fast variable, which is of particular importance given the time-dependency of the reference frame of motion.

\section{Conservative Forces}
\label{sec:conservative}

This section presents and analyses the methods to compute third-body motion that are obtained using LPE. Although they all share the same general equations of motion, the way these are solved and the independent variable makes each of them unique in their suitability for different orbital computation problems.

\subsection{Equations of Motion}

As described in Section \ref{sub:VOP}, the use of LPE together with a disturbing function can be employed to describe the motion of a massless particle under a third-body perturbation (e.g. the influence of the Earth over the spacecraft in the Sun-Earth system). 

Observing the LPE of Eq. \eqref{eq:LPE}, the derivatives of the disturbing function are required in order to propagate the orbital motion. These derivatives take the general form:
\begin{align}\nonumber
\frac{\partial{\mathpzc{R}}}{\partial \mathpzc{K}} = &-\frac{1}{r^2}\frac{\partial{r}}{\partial \mathpzc{K}}  + \frac{1}{r^2}\frac{\partial{\cos\theta}}{\partial \mathpzc{K}} - \frac{2\cos\theta}{r^3}\frac{\partial{r}}{\partial \mathpzc{K}} + \\  &\frac{1}{(r^2 - 2 r \cos\theta + 1)^{\frac{3}{2}}} \bigg(r \frac{\partial{r}}{\partial \mathpzc{K}} - \cos\theta\frac{\partial{r}}{\partial \mathpzc{K}} + r \frac{\partial{\cos\theta}}{\partial \mathpzc{K}}\bigg)
\label{eq:dRdK}\end{align}
\textcolor{External}{in which $\mathpzc{K}$ is a placeholder for any of the regular Keplerian elements.} The derivatives of $r$ and $\cos\theta$ with respect to each element are trivial to compute, given their formulation in either Eqs. \eqref{eq:rtheta} to \eqref{eq:rtheta2} or Eqs. \eqref{eq:theta_e} to \eqref{eq:theta_e2}, depending on the fast variable. 

These equations can be solved in several distinct ways, depending on the application case. The different possible formulations are presented in Table \ref{tab:formulations}. One of them can already be found in literature (the KM) and the remaining ones are underlined: the E-KM, the PAP-KM and the Analytical Taylor Approximation.

\begin{table}[htb]
	\begin{center}
		\caption{\label{tab:formulations}Different formulations using the disturbing function $\mathpzc{R}$ to compute the evolution a placeholder orbital element $\mathpzc{K}$}
		\begin{tabular}{c|c|c}
		\textbf{Solution} & \textbf{Mathematical Representation} & \textbf{Method Name} \\\hline
		\textbf{Numerical} & $\frac{d\mathpzc{K}}{dt} = f\Big(\mathpzc{K}, \frac{d\mathpzc{R}}{d\mathpzc{K}}\Big)$ & \underline{\smash{E-KM}}\\
		\textbf{Semi-Analytical} & $\Delta \mathpzc{K} = \int_{t_i}^{t_f} f\bigg(\mathpzc{K}, \frac{d\mathpzc{R}}{d\mathpzc{K}}\bigg) dt$ & KM and \underline{\smash{PAP-KM}}\\
		\textbf{Analytical} & $\Delta \mathpzc{K} = \mathpzc{T}(\mathpzc{K}, t_f) - \mathpzc{T}(\mathpzc{K}, t_i)$ & \underline{\smash{Analytical Taylor Approximation}}		
	\end{tabular}
	\end{center}
\end{table}

\subsection{Analytical Solution}
\subsubsection{Taylor Approximation}

The first approach tried when solving the LPE with the disturbing function $\mathpzc{R}$ was to seek a fully analytical approximation of the orbital element change. \textcolor{External}{However, despite the number of trials and techniques tested, such an approximation was not found. The reasoning behind this relates to the time-dependency of the Earth-pointing reference frame. This adds complexity to the equations of motion in the form of additional relationships and dependencies between variables.} Namely, the mapping from time to the fast variable, using Eqs. \eqref{eq:true_t} and \eqref{eq:true_e}, cannot be derived with the remaining expression in Eq. \eqref{eq:dRdK}, having to stay unchanged. This is due to the fact that the fast variable is also the independent integration parameter (akin to the normalised time). 

Still, approximations can be tried so that an analytical solution to some of the orbital element equations is achieved. The process can be broken down in the following manner: first, a Taylor series is used to approximate Eq. \eqref{eq:dRdK} for each orbital element. Then, this expression is inserted back into the LPE in Eqs. \eqref{eq:LPE}, which may be solved analytically.

For this project, a Taylor series around $e = 0$ was successfully employed. Other orbital elements and orders of approximation were tried, without any success: from approximations on the inclination to the semi-major axis and the mean motion. Nevertheless, the factors listed above make it so that the only equation for which this works is the semi-major axis one; it is the only element for which the equation is simple enough to obtain an analytical formulation, without resorting to any other approximation.

\begin{figure}[htb!]
	\centering
		\begin{minipage}[t]{0.48\linewidth}
		\includegraphics[width=\textwidth]{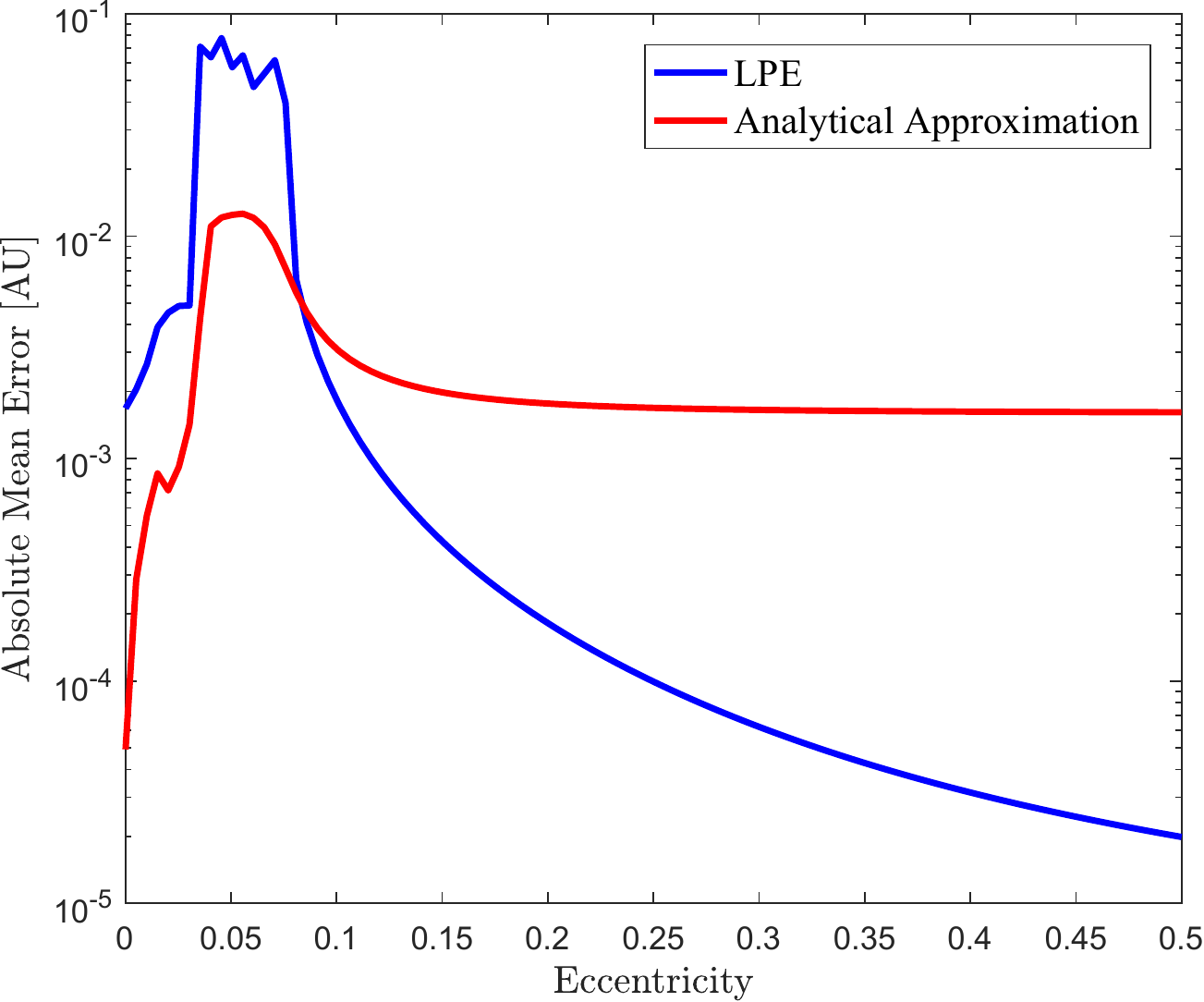}
		\caption{\label{fig:AN_error}Absolute mean error, averaged for 10,000 initial orbital conditions, in semi-major axis update as a function of the eccentricity (logarithmic scale)}
	\end{minipage}
	\hfill
	\begin{minipage}[t]{0.48\linewidth}
		\includegraphics[width=\textwidth]{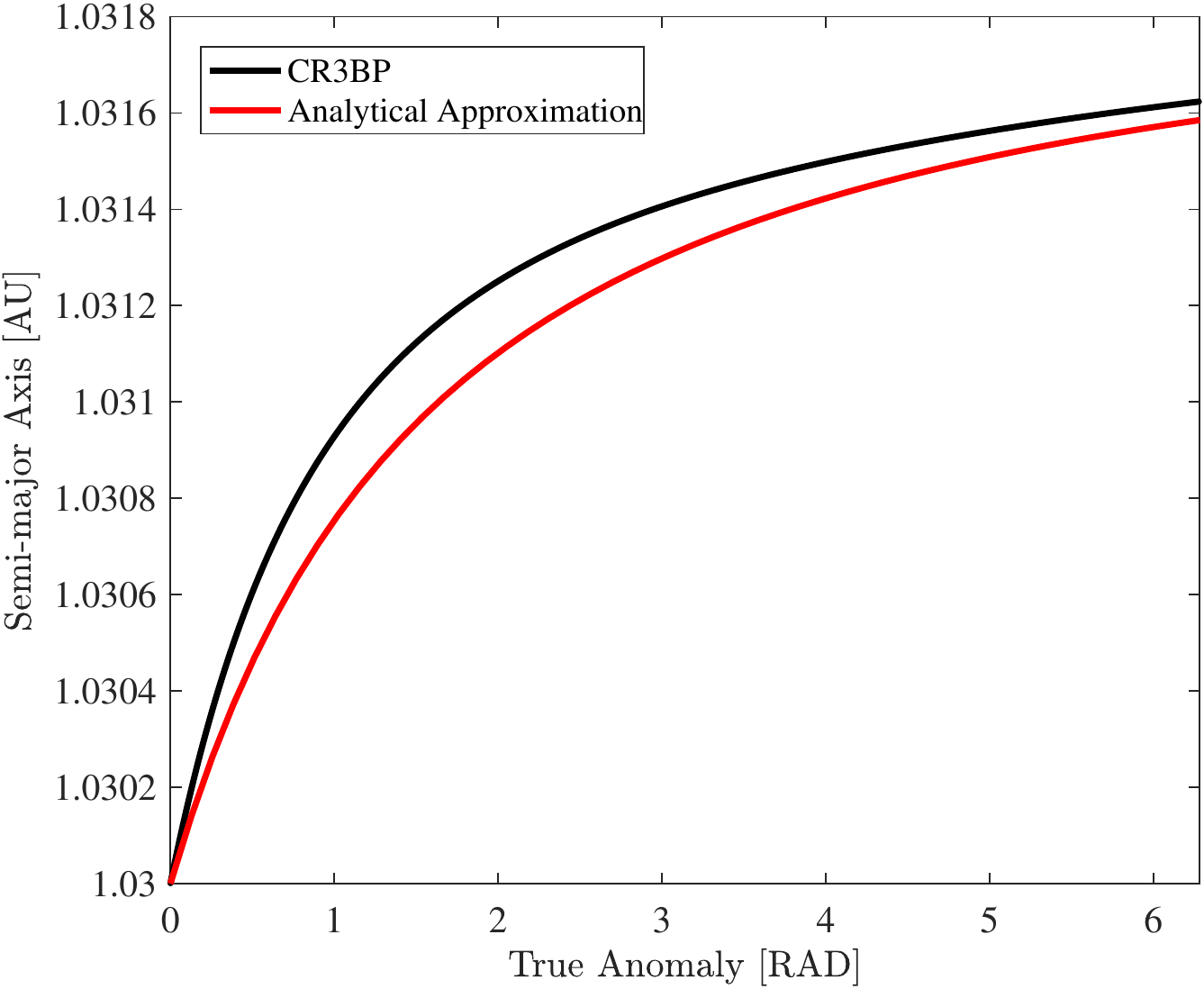}
		\caption{\label{fig:1ANCR}Semi-major axis evolution for Keplerian element set $\{ a = 1.03, e = 0.001, i = 0.001, \Omega = 5.93, \omega = 2.32, E = 0\}$}
	\end{minipage}
\end{figure}

The analytical solution found for the evolution of the semi-major axis is presented in Eq. \eqref{eq:anal}. It provides a simple expression to quickly compute the behaviour of the orbital element; nevertheless, it is limited by the fact that the orbit must have a very low eccentricity. In order to understand what kind of error would be obtained as a function of the latter parameter, 10,000 randomly sampled different initial orbital conditions are computed ($a\in \big[1.03, 1.08\big], e\in \big]0, 0.5\big], i\in \big]0, 0.5\big], \omega\in \big]0, \pi/2\big]$). The change in semi-major axis of each of these orbits after one full period, for each eccentricity value, was computed using both Eq. \eqref{eq:anal} and the original propagation of the LPE equations. \textcolor{External}{The \textit{ode45} numerical solver from the MATLAB\footnote{MATLAB and Statistics Toolbox Release 2017a} code suite \cite{matlab} is employed, with the value $10^{-10}$ for both the absolute and relative tolerances.\footnote{This solver and the tolerances indicated are the ones used for every numerical orbital propagation throughout this document.}} Figure \ref{fig:AN_error} shows the absolute error, computed as the difference between these models and the CR3BP, averaged for each eccentricity value. 
\begin{align}
a = &\frac{2\mu}{na} \Bigg[\frac{1}{(n-1)\sqrt{1+a^2-2a\cos(E - \frac{E}{n} + \omega + \Omega)}} \nonumber\\
&- \frac{1}{4 a^2(n^2-1)}\Bigg(2(n-1)\cos\Big(E + \frac{E}{n} + \omega - \Omega\Big) +(1-n)\cos\Big(E + \frac{E}{n} -i + \omega - \Omega\Big) \nonumber\\
&+(1-n)\cos\Big(E + \frac{E}{n} + i + \omega - \Omega\Big) +2(1+n)\cos\Big(E - \frac{E}{n} + \omega + \Omega\Big) \nonumber\\
&+(1+n)\cos\Big(E - \frac{E}{n} -i + \omega + \Omega\Big) +(1+n)\cos\Big(E - \frac{E}{n} + i + \omega + \Omega\Big)\Bigg)\Bigg] + \mathpzc{O}(e^1)\label{eq:anal}
\end{align}

As it can be seen, the analytical approximation actually works better than the LPE propagation up until an eccentricity of roughly 0.1. This is explained by the fact that the analytical simplification avoids the well-known singularity in eccentricity of the LPE. As such, this approximation can be very useful when computing the behaviour of low eccentricity orbits, like the ones of many near-Earth asteroids (NEAs).

In order to better visualise the behaviour of Eq. \eqref{eq:anal}, Figure \ref{fig:1ANCR} highlights one of the initial conditions for $e = 0.001$ as a representative example across the 10,000 samples. It shows the single-period propagation of the semi-major axis for both the CR3BP and the analytical approximation. It can be seen that, while the final result is slightly different, the evolution of the orbital elements is very similar. Thus, the analytical equation is shown to keep an accurate dynamical behaviour, albeit producing slightly different results.

\subsection{Semi-Analytical Solutions}
\label{sub:km}

The semi-analytical solutions here presented are inserted into the category of stroboscopic maps, as described in Section \ref{sec:FM}, since the orbital elements are computed once for each period of the motion. This is achieved by employing a first Picard iteration \textcolor{Internal}{\cite{picard}} on Eqs. \eqref{eq:LPE}, which yields a change in each orbital element after a period. \textcolor{External}{This method results in Eq. \eqref{eq:km}, which can be solved using numerical integration: in this work, the \textit{integral} function from the MATLAB code suite \cite{matlab_int} was employed, with an absolute tolerance of $10^{-12}$.\footnote{This solver and the tolerances indicated are the ones used in every semi-analytical method throughout this document.}}
\vspace{-5mm}
\begin{align}
\Delta a = &\frac{2}{n a} \int_{t_i}^{t_f}  \frac{\partial{\mathpzc{R}}}{\partial M_0} dt\nonumber\\
\Delta e = &\frac{(1 - e^2)}{n a^2 e} \int_{t_i}^{t_f} \frac{\partial{\mathpzc R}}{\partial M_0}dt - \frac{\sqrt{1 - e^2}}{n a^2 e} \int_{t_i}^{t_f} \frac{\partial{\mathpzc{R}}}{\partial \omega} dt\nonumber\\
\Delta i = &-\frac{1}{n a^2 \sqrt{1 - e^2}\sin i}\int_{t_i}^{t_f}\frac{\partial{\mathpzc R}}{\partial \Omega} dt + \frac{\cos i}{n a^2 \sqrt{1 - e^2}\sin i}\int_{t_i}^{t_f}\frac{\partial{\mathpzc{R}}}{\partial \omega} dt \nonumber\\ 
\Delta\Omega = &\frac{1}{n \sqrt{1 - e^2} a^2 \sin i} \int_{t_i}^{t_f} \frac{\partial{\mathpzc{R}}}{\partial i} dt\nonumber\\
\Delta\omega = &\frac{\sqrt{1 - e^2}}{n a^2 e}\int_{t_i}^{t_f} \frac{\partial{\mathpzc R}}{\partial e} dt - \frac{\cos i}{n a^2 \sqrt{1 - e^2}\sin i} \int_{t_i}^{t_f} \frac{\partial{\mathpzc{R}}}{\partial i} dt \nonumber\\
\Delta M_0 = &-\frac{2}{n a} \int_{t_i}^{t_f} \frac{\partial{\mathpzc R}}{\partial a} dt - \frac{1 - e^2}{n a^2 e}\int_{t_i}^{t_f} \frac{\partial{\mathpzc{R}}}{\partial e} dt \label{eq:km}
\end{align}

Given that the semi-major axis is changing throughout the motion, the orbital period is accordingly altered, making the time an unsuitable integration limit. Since the orbital element update is computed for one full orbit, it is then more convenient to perform the integration by taking the fast variable, either the true or eccentric anomaly, as the independent integration parameter. This follows \textcolor{Internal}{from} the procedure explained to generate stroboscopic maps in Section \ref{sec:FM}. 

In order to correctly modify Eqs. \eqref{eq:km} for this purpose, some relations need to be stated. For both the cases of the true and eccentric anomalies, the corresponding formulae are obtained from Chao \cite{chao}:
\begin{align}
\frac{dt}{d\nu} = \frac{r^2}{n a^2 \sqrt{1 - e^2}} &, \text{ } \frac{dt}{dE} = \frac{1 - e\cos E}{n}\\
\frac{\partial \mathpzc{R}}{\partial M_0} = \frac{a^2 \sqrt{1 - e^2}}{r^2}\frac{\partial \mathpzc{R}}{\partial \nu} &, \text{ }\frac{\partial \mathpzc{R}}{\partial M_0} dt = \frac{1}{n}\frac{\partial \mathpzc{R}}{\partial E} dE\label{eq:chao}
\end{align}

\subsubsection{The Keplerian Map}
\label{sub:km2}

The KM is an established model to measure an orbital element change caused by the third-body perturbation throughout one period of motion. Since the KM is computed as a stroboscopic map, the solution to the LPE is obtained much faster than a straightforward numerical propagation. 

The fast variable employed in the KM, coincident with the independent integration parameter, is the true anomaly: the resulting equations of motion are obtained by implementing Eqs. \eqref{eq:rtheta} to \eqref{eq:true_t} together with the LPE \textcolor{External}{of Eqs. \ref{eq:km}:}
\begin{align}
\Delta a &=\frac{2}{n^2 a} \int_{\nu_i}^{\nu_f}  \frac{\partial{\mathpzc{R}}}{\partial\nu} d\nu\nonumber\\
\Delta e &= \frac{(1 - e^2)}{n^2 a^2 e} \int_{\nu_i}^{\nu_f} \frac{\partial{\mathpzc{R}}}{\partial\nu}d\nu - \frac{(1 - e^2)^2}{n^2 a^2 e} \int_{\nu_i}^{\nu_f} \frac{1}{(1 + e\cos\nu)^2}\frac{\partial{\mathpzc{R}}}{\partial \omega} d\nu\nonumber\\
\Delta i &= -\frac{(1 - e^2)}{n^2 a^2 \sin i}\int_{\nu_i}^{\nu_f}\frac{1}{(1 + e\cos\nu)^2}\frac{\partial{\mathpzc{R}}}{\partial \Omega} d\nu + \frac{(1 - e^2)\cos i}{n^2 a^2 \sin i}\int_{\nu_i}^{\nu_f}\frac{1}{(1 + e\cos\nu)^2} \frac{\partial{\mathpzc{R}}}{\partial \omega} d\nu \nonumber\\
\Delta\Omega &= \frac{(1 - e^2)}{n^2 a^2 \sin i}\int_{\nu_i}^{\nu_f} \frac{1}{(1 + e\cos\nu)^2} \frac{\partial{\mathpzc{R}}}{\partial i} d\nu \nonumber\\
\Delta\omega &= \frac{(1 - e^2)^2}{n^2 a^2 e}\int_{\nu_i}^{\nu_f} \frac{1}{(1 + e\cos\nu)^2} \frac{\partial{\mathpzc{R}}}{\partial e} d\nu - \frac{(1 - e^2)\cos i}{n^2 a^2 \sin i} \int_{\nu_i}^{\nu_f} \frac{1}{(1 + e\cos\nu)^2} \frac{\partial{\mathpzc{R}}}{\partial i} d\nu \nonumber\\
\Delta M_0 &= -\frac{2(1 - e^2)^{3/2}}{n^2 a}\int_{\nu_i}^{\nu_f} \frac{1}{(1 + e\cos\nu)^2} \frac{\partial{\mathpzc{R}}}{\partial a} d\nu - \frac{(1 - e^2)^{5/2}}{n^2 a^2 e}\int_{\nu_i}^{\nu_f} \frac{1}{(1 + e\cos\nu)^2} \frac{\partial{\mathpzc{R}}}{\partial e} d\nu \label{eq:km_true}
\end{align}

The integral in the original KM is computed at each periapsis passage, with $\nu_i = -\pi$ and $\nu_f = \pi$ \cite{alessi_semi}. These changes are then added to the previously known orbital elements to obtain the updated motion. 

\begin{figure}[h!]
	\centering
	\includegraphics[width=0.7\linewidth]{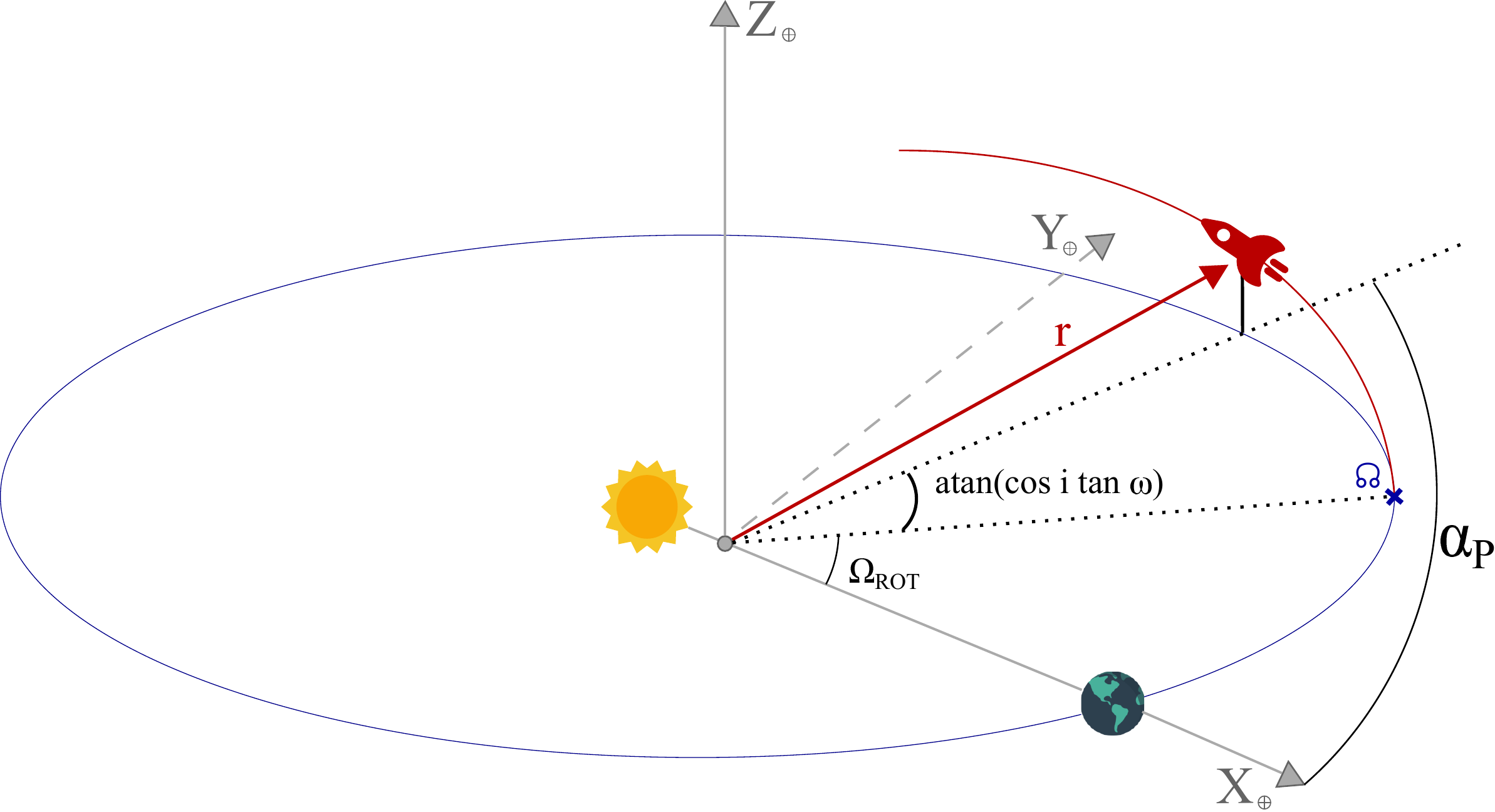}
	\caption{\label{fig:alpha}\textcolor{Both}{Geometry of the angle $\alpha_P$, concerning a spacecraft's trajectory (red line) in the Earth-pointing reference frame}}
\end{figure}

When using the KM, another physical quantity can be introduced: the parameter $\alpha_P$, representing the phasing of the massless particle with the disturbing body, while the first is at periapsis (hence the subscript $P$). \textcolor{External}{In other words, it is the angle between the Sun-Earth axis and the projection to the ecliptic plane of the line connecting the barycentre to the massless particle, when the latter is at the periapsis position.} This can be easily visualised in Figure \ref{fig:alpha}. 

After one full period, the new phasing can be computed with an update in $\alpha_P$, using Eq. \eqref{eq:alpha}.
\begin{equation}\label{eq:alpha}
\Delta\alpha_P = 2\pi - 2\pi \sqrt{\frac{a^3}{1-\mu}}
\end{equation} 

As described by Alessi and S\'anchez \cite{alessi_semi}, the variable $\alpha_P$ can also be computed as a function of the Keplerian elements:
\begin{equation}\label{eq:alpha_omega}
\alpha_P = \Omega_{Rot}(t_P) + \arctan(\cos i \tan \omega)
\end{equation} 
in which $t_P$ is the time at periapsis passage.

Given that $\alpha_P$ is a function of the longitude of the ascending node, its update given by Eq. \eqref{eq:alpha} can replace the change in $\Delta\Omega$ in Eqs. \eqref{eq:km}. Finally, gathering Eqs. \eqref{eq:km} and \ref{eq:alpha}, the action of the KM can be represented by the mapping $\mathpzc{M}$:
\begin{align}\label{eq:map}
\mathpzc{M}: &\{a, e, i, \omega, M_0 | \alpha_{P}\} \mapsto \{\Delta a, \Delta e, \Delta i, \Delta \omega,  \Delta M_0| \Delta\alpha_{P}\}
\end{align} 

These equations do not need to be computed throughout the entire motion; instead, they are used inside a neighbouring region of the Earth, within which its gravitational perturbation is non-negligible. This has been defined as the perturbation region detailed in Section \ref{sec:3BP}. This region is here defined by an interval of $\alpha$ values, shown to capture the extent of the disturbing effect; outside it, the gravitational perturbation of the Earth is so small that can be effectively neglected. Alessi and S\'anchez \cite{alessi_semi} have determined this region to be ${\abs{\alpha} \leq \frac{\pi}{8}}$; throughout this project, this is defined as $\abs{\alpha} \leq \frac{\pi}{8} + \frac{\Delta \alpha_{P_0}}{2}$, in which $\Delta \alpha_{P_0}$ is the change in phasing corresponding to one full period of the motion, using the initial orbital elements of the massless particle.

\begin{figure}[h]
	\centering
	\begin{minipage}[t]{0.49\linewidth}
		\includegraphics[width=\textwidth]{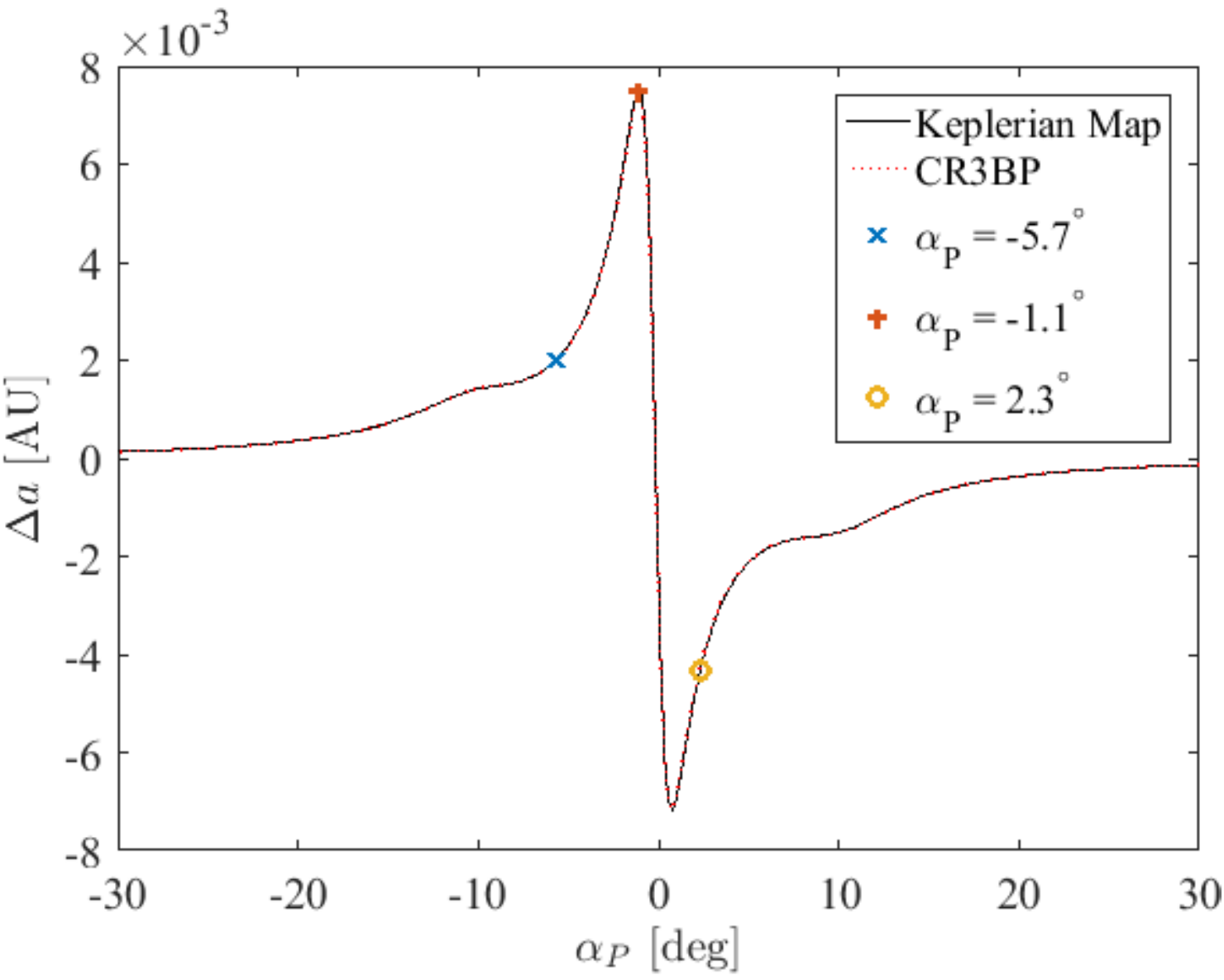}
		\caption[Kick-map: change in $\Delta a$ with $\alpha_P$ for asteroid 2016 RD34]{\label{fig:km_angle}Kick-map: change in $\Delta a$ with $\alpha_P$ for asteroid 2016 RD34 \cite{neves2018}}
	\end{minipage}
	\hfill
	\begin{minipage}[t]{0.48\linewidth}
		\includegraphics[width=\textwidth]{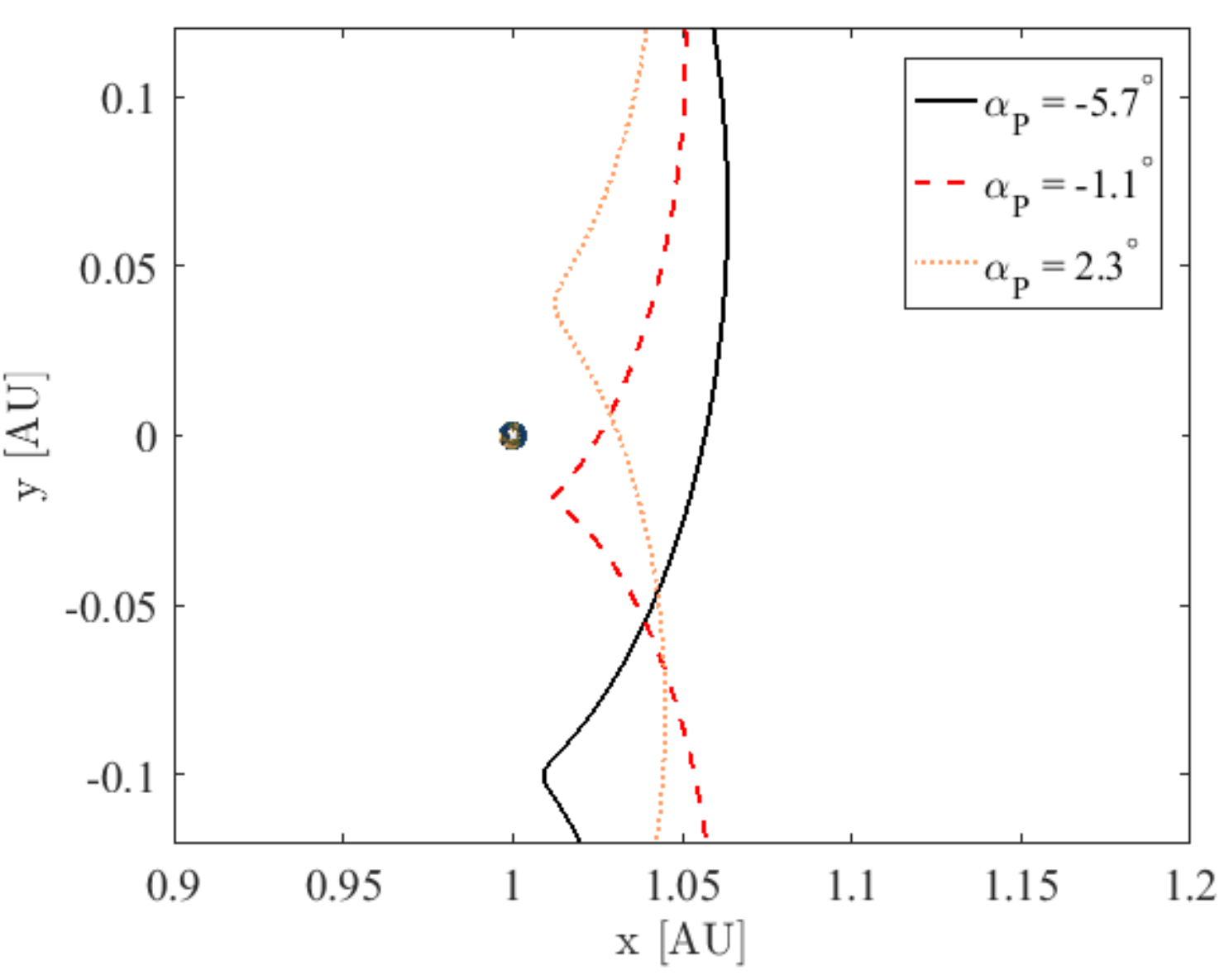}
		\caption[Three different Earth encounters of asteroid 2016 RD34]{\label{fig:grid}Three different Earth encounters of asteroid 2016 RD34 \cite{neves2018}}
	\end{minipage}
\end{figure}

An interesting application of the KM is the kick-map, a visual representation of the orbital elements' variation as a function of $\alpha_P$. As an example, Figure \ref{fig:km_angle} shows the kick-map representing the semi-major axis change as a function of the phasing $\alpha_P$, for asteroid 2016 RD34. Three points are highlighted, corresponding to the particular changes undergone by three different possible encounters shown in Figure \ref{fig:grid} (Earth's radius is scaled to the size of Hill radius, for visibility). It is worth noting that the semi-major axis function is also computed using the CR3BP, as shown on Figure \ref{fig:km_angle}. The corresponding plot overlaps with the kick-map computed with the KM, reporting a very good accuracy for the latter model.

\subsubsection{The \textcolor{External}{Periapsis-Apoapsis-Periapsis Keplerian Map}}

\begin{figure}[htb!]
	\centering
	\includegraphics[width=0.65\textwidth]{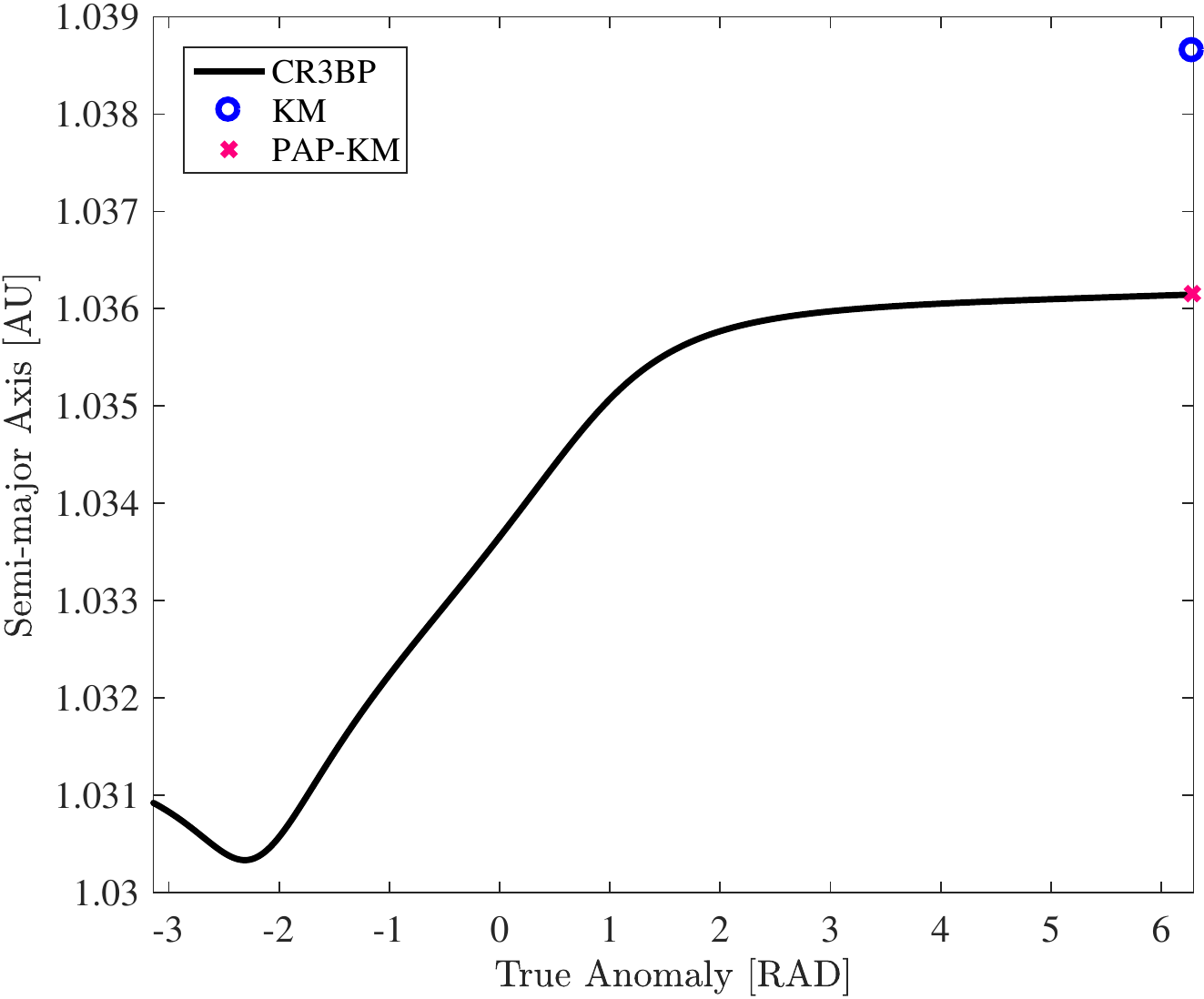}
	\caption{\label{fig:1per}Semi-major axis evolution for asteroid 2018 AV2}
\end{figure}

As detailed in the previous section, the original KM is obtained using the true anomaly as the independent variable for the integration. The interval ${\big[-\pi, \pi\big]}$ is chosen in order to avoid numerical errors, and it also bypasses the non-smoothness of the mapping shown on Eq. \eqref{eq:true_t}. As such, the new conditions at periapsis are taken from the orbital changes that happen between apoapses. 

\textcolor{External}{The PAP-KM is a semi-analytical mapping technique akin to the KM but, instead of having the true anomaly as the fast and independent variable, the eccentric anomaly is selected. In contrast to the KM, given that this formulation has no singularities of importance, the propagation interval can be changed to $\big[0, 2\pi\big]$. Using Eqs. \eqref{eq:km} and \eqref{eq:chao}, the obtained PAP-KM equations are the following:}
\begin{align}
\Delta a &=\frac{2}{n^2 a} \int_{E_i}^{E_f}  \frac{\partial{\mathpzc{R}}}{\partial E} dE\nonumber\\
\Delta e &= \frac{1 - e^2}{n a^2 e} \int_{E_i}^{E_f} \frac{\partial{\mathpzc{R}}}{\partial E} dE -  \frac{\sqrt{1 - e^2}}{n^2 a^2 e} \int_{E_i}^{E_f} (1 - e\cos E)\frac{\partial{\mathpzc{R}}}{\partial \omega} dE\nonumber\\
\Delta i &= -\frac{1}{n a^2 \sqrt{1 - e^2} \sin i}\int_{E_i}^{E_f} (1 - e\cos E) \frac{\partial{\mathpzc{R}}}{\partial \Omega} dE + \frac{\cos i}{n^2 a^2 \sqrt{1 - e^2} \sin i}\int_{E_i}^{E_f} (1 - e\cos E) \frac{\partial{\mathpzc{R}}}{\partial \omega} dE \nonumber\\
\Delta\Omega &= \frac{1}{n^2 a^2 \sqrt{1 - e^2} \sin i} \int_{E_i}^{E_f} (1 - e\cos E) \frac{\partial{\mathpzc{R}}}{\partial i} dE \nonumber\\
\Delta\omega &= \frac{\sqrt{1 - e^2}}{n^2 a^2 e}\int_{E_i}^{E_f} (1 - e\cos E) \frac{\partial{\mathpzc{R}}}{\partial e} dE - \frac{\cos i}{n^2 a^2 \sqrt{1 - e^2} \sin i} \int_{E_i}^{E_f} (1 - e\cos E) \frac{\partial{\mathpzc{R}}}{\partial i} dE \nonumber\\
\Delta M_0 &= -\frac{2}{n^2 a} \int_{E_i}^{E_f} (1 - e\cos E) \frac{\partial{\mathpzc{R}}}{\partial a} dE - \frac{1 - e^2}{n^2 a^2 e}\int_{E_i}^{E_f} (1 - e\cos E) \frac{\partial{\mathpzc{R}}}{\partial e} dE \label{eq:km_ecc}
\end{align}

\textcolor{External}{In order to better understand the advantages of using the PAP-KM instead of the KM, an example trajectory propagation can be studied. Figure \ref{fig:1per} depicts the change in semi-major axis of asteroid 2018 AV2 in its closest pass near the Earth, starting in 2036.} The plot highlights the evolution of this orbital element as computed by the CR3BP from one apoapsis to the periapsis after next ($-\pi$ to $2\pi$ in the X-axis). The point representing the KM update is computed using \textcolor{External}{${a_{\nu = 0} + \Delta a_{KM}}$}, where \textcolor{External}{$\Delta a_{KM}$ is determined using Eq. \ref{eq:km_true},} integrated from $-\pi$ to $\pi$. In a similar fashion, the value corresponding to the PAP-KM employs \textcolor{External}{${a_{\nu = 0} + \Delta a_{PAP-KM}}$}, where \textcolor{External}{$\Delta a_{PAP-KM}$ is calculated using Eq. \ref{eq:km_ecc},} integrated from $0$ to $2\pi$.

It can be seen that \textcolor{External}{$\Delta a_{KM}$} corresponds quite well to the change between apoapses; however, given this particular orbital evolution, this change is quite different from the one between periapses, \textcolor{External}{leading to a large discrepancy in the final semi-major axis. Thus, it can be inferred that the KM method will naturally yield significant accuracy errors, especially in cases where the orbital elements are noticeably altered between periapses.} In contrast, the updated element as computed with the PAP-KM matches the CR3BP motion quite clearly.

\begin{figure}[h]
	\begin{minipage}{.5\linewidth}
		\centering
		\includegraphics[width=0.95\textwidth]{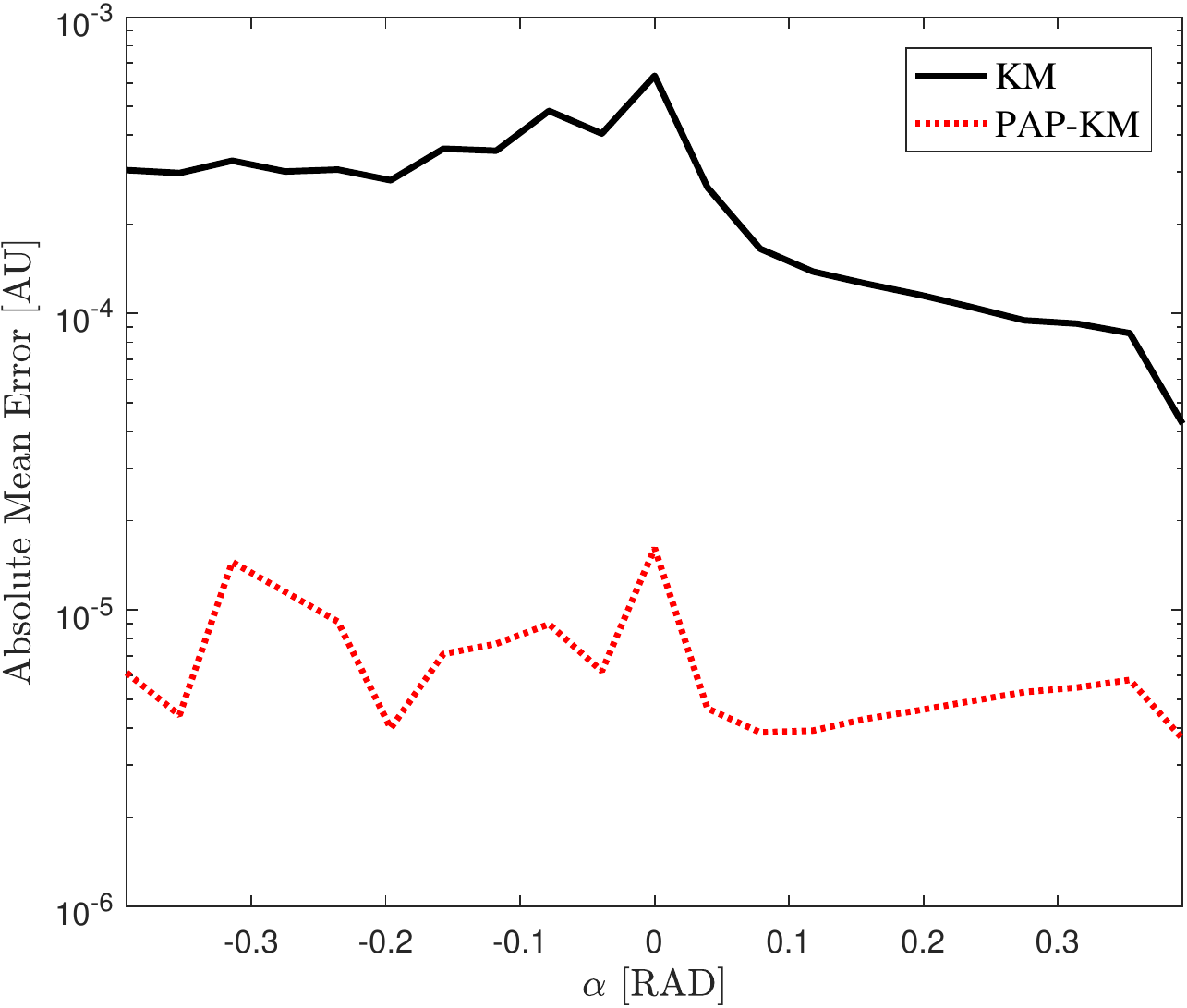}
		\caption*{\textit{a) Semi-major Axis}}
	\end{minipage}%
	\begin{minipage}{.5\linewidth}
		\centering
		\includegraphics[width=0.95\textwidth]{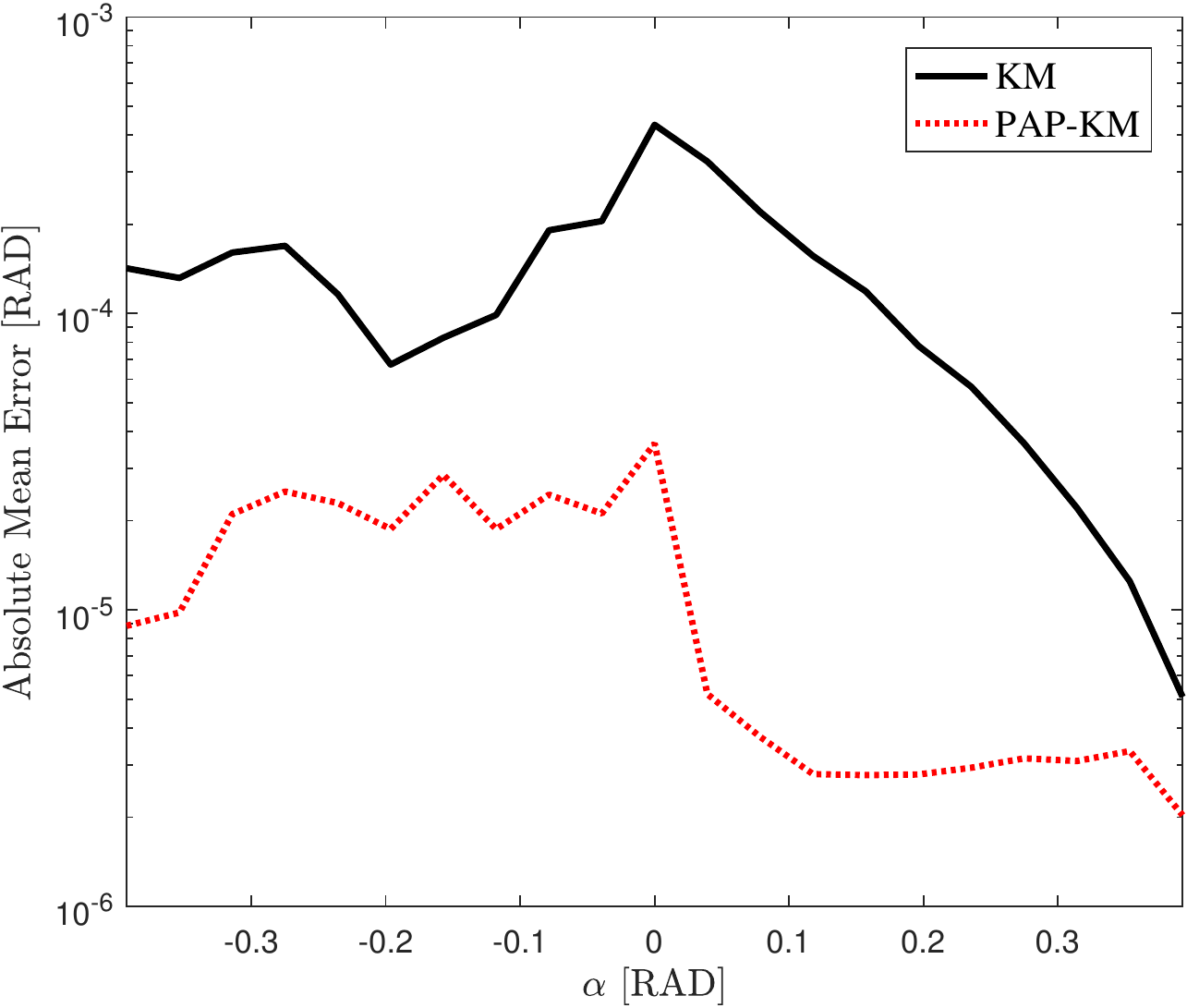}
		\caption*{\textit{b) Eccentricity}}
	\end{minipage}
	\begin{minipage}{.5\linewidth}
		\centering
		\includegraphics[width=0.95\textwidth]{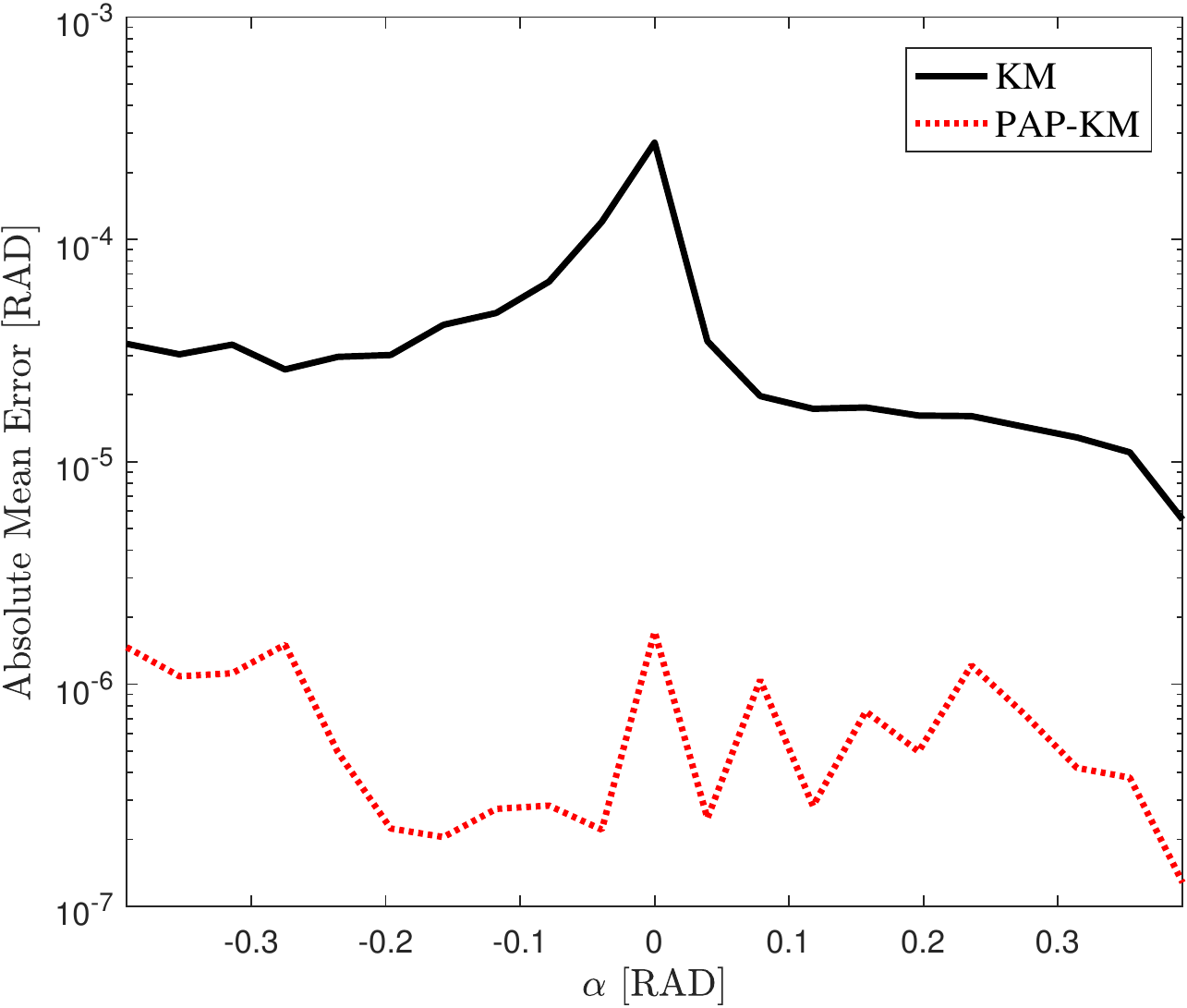}
		\caption*{\textit{c) Inclination}}
	\end{minipage}%
	\begin{minipage}{.5\linewidth}
		\centering
		\includegraphics[width=0.95\textwidth]{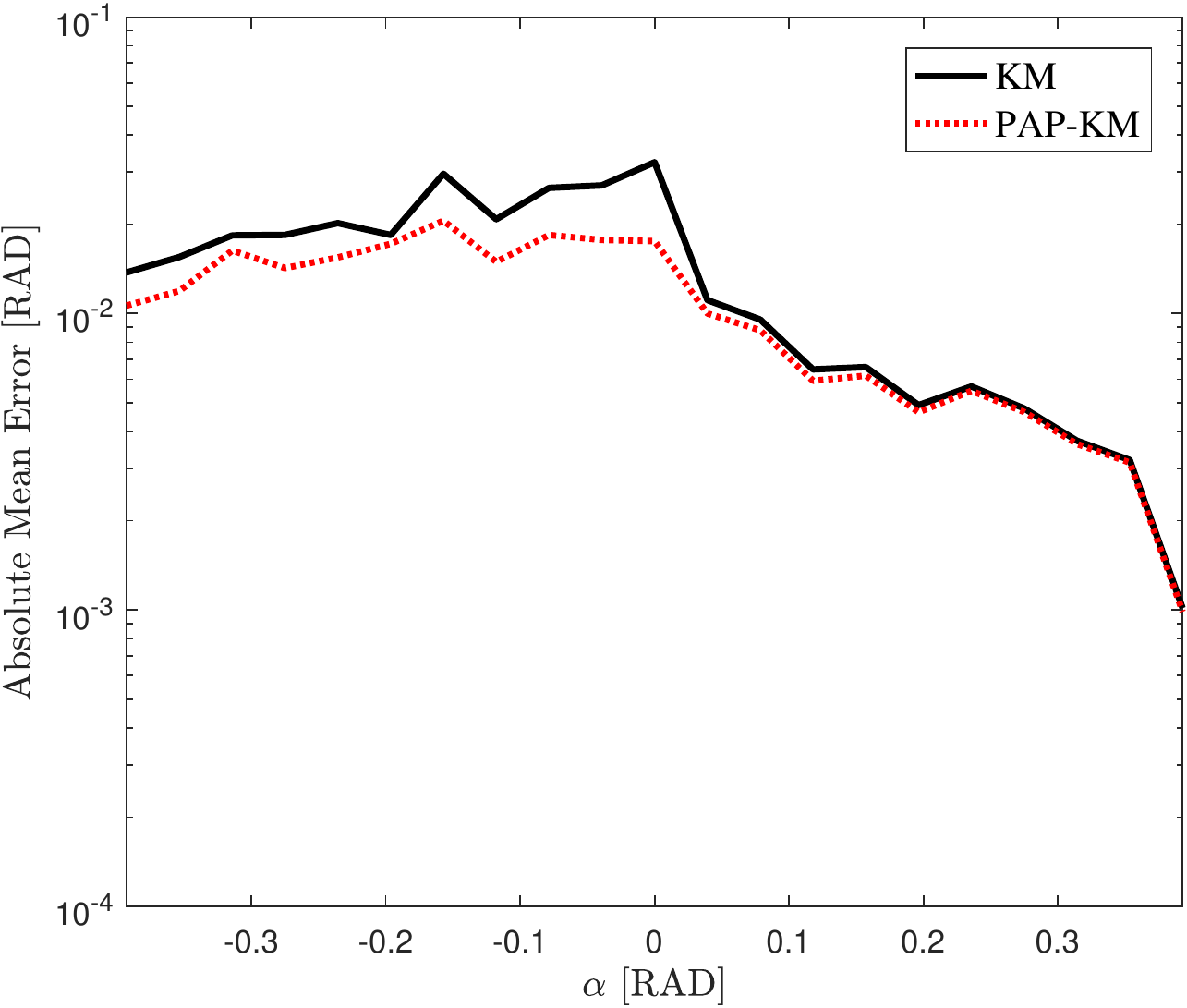}
		\caption*{\textit{d) Argument of Periapsis}}
	\end{minipage}\caption{\label{fig:all}Absolute mean error, averaged for 10,000 initial orbital conditions, for one period as a function of initial $\alpha_P$ (logarithmic scale)}
\end{figure}

To judge the performance of the KM and the PAP-KM after one period of motion, these two methods were computed for the same previously described 10,000 randomly sampled initial conditions. The averaged absolute error compared to the CR3BP was determined for one period of motion for each starting $\alpha_P$, and plotted in Figures \ref{fig:all}. As it can be observed, the PAP-KM case shows a much smaller error after just one period, which confirms it as a more accurate alternative to the original KM for a stroboscopic map to compute the third-body effect.

\subsection{Numerical Solutions}

The most straightforward manner to solve the LPE equations is to employ a numerical propagation algorithm. From the options presented, \textcolor{Internal}{given the solvers and tolerances used}, this is the costliest in terms of computational expenditure, but it is also the most rigorous one. This is due to the fact that, as opposed to the previously described methods, the instantaneous evolution of each element is taken into account in the computation of the others, at each time step. In order to keep this characteristic but decrease the computational time demand, the E-KM was developed.

\subsubsection{\textcolor{External}{The Euler-Keplerian Map}}
\label{sub:EK}

An obvious drawback of the previously described methods is that the parameter update is made using only its initial value, regardless of its evolution throughout the orbit. In order to account for this effect, the LPE can be \textcolor{External}{integrated} with an Euler method. This will yield an approximation of the instantaneous element change and also avoid possible integration errors from the PAP-KM or the original KM.

The number of \textcolor{External}{step sizes} was chosen in order for the method to have the same computational time of the KM: this yielded the value 20. The algorithm's pseudo-code is the following:	

\begin{algorithm}[h]
	\textbf{Initialization:} $\mathpzc{K}$ = $\mathpzc{K}_0$; $E$ = 0\;
	\While{$E < E_{final} $}{
		slope = Eq. \eqref{eq:LPE}\;
		$\mathpzc{K}$ = $\mathpzc{K}$ + slope*step\;
		$E$ = $E$ + step\;
	}
	\caption{\label{alg}Euler method for computing the evolution of the orbital element $\mathpzc{K}$}
\end{algorithm}

\paragraph{Phasing Update:} As previously discussed, Eq. \eqref{eq:alpha} shows that the orbital period, and more concretely the initial semi-major axis of the orbit, is used to estimate the $\alpha_P$ phasing of the motion at the next periapsis. However, the semi-major axis may be changing steeply throughout the orbit, rendering unsuitable the approximation to only use its initial value.

In order to solve this issue, the contribution of an instantaneous semi-major axis to Eq. \eqref{eq:alpha} can improve the chosen model's accuracy. This can be obtained by applying Euler's method on the $\alpha_P$ computation, using the semi-major axis evolution as generated from Algorithm \ref{alg}. The obtained value for each discretised time interval is given by $a_{list}$. This vector is then used for the $\alpha_P$ update in Algorithm \ref{alg2}: 

\begin{algorithm}[h]
	$a_{list}$ = Algorithm 1 for $\mathpzc{K} = a$ $^*$\\
	\textbf{Initialization:} $\alpha = 0$, $a_0 = 0$\\
	\For{$i = 1:len(a_{list})$}{
		$\textcolor{Internal}{\zeta_i = \zeta_{i-1}} + \sqrt{\frac{\big(a_{list}\big)_i^3}{1 - \mu}}$
	}
	$\Delta \alpha \textcolor{External}{\rightarrow} 2\pi - \textcolor{Internal}{2\pi\frac{\zeta}{len(a_{list})}}$\\
	\begin{flushleft}
		$^*$ $a_{list}$ contains the semi-major axis value at each timestep\\
	\end{flushleft}
	\caption{\label{alg2}Computing the update in $\alpha_P$ using Algorithm \ref{alg} for the semi-major axis}
\end{algorithm}
It is important to denote that Algorithm \ref{alg2} can also be implemented for the PAP-KM method, just by adding the extra computation of the instantaneous semi-major axis with Algorithm \ref{alg}.

\begin{figure}[htb!]
	\centering
	\includegraphics[width=0.65\textwidth]{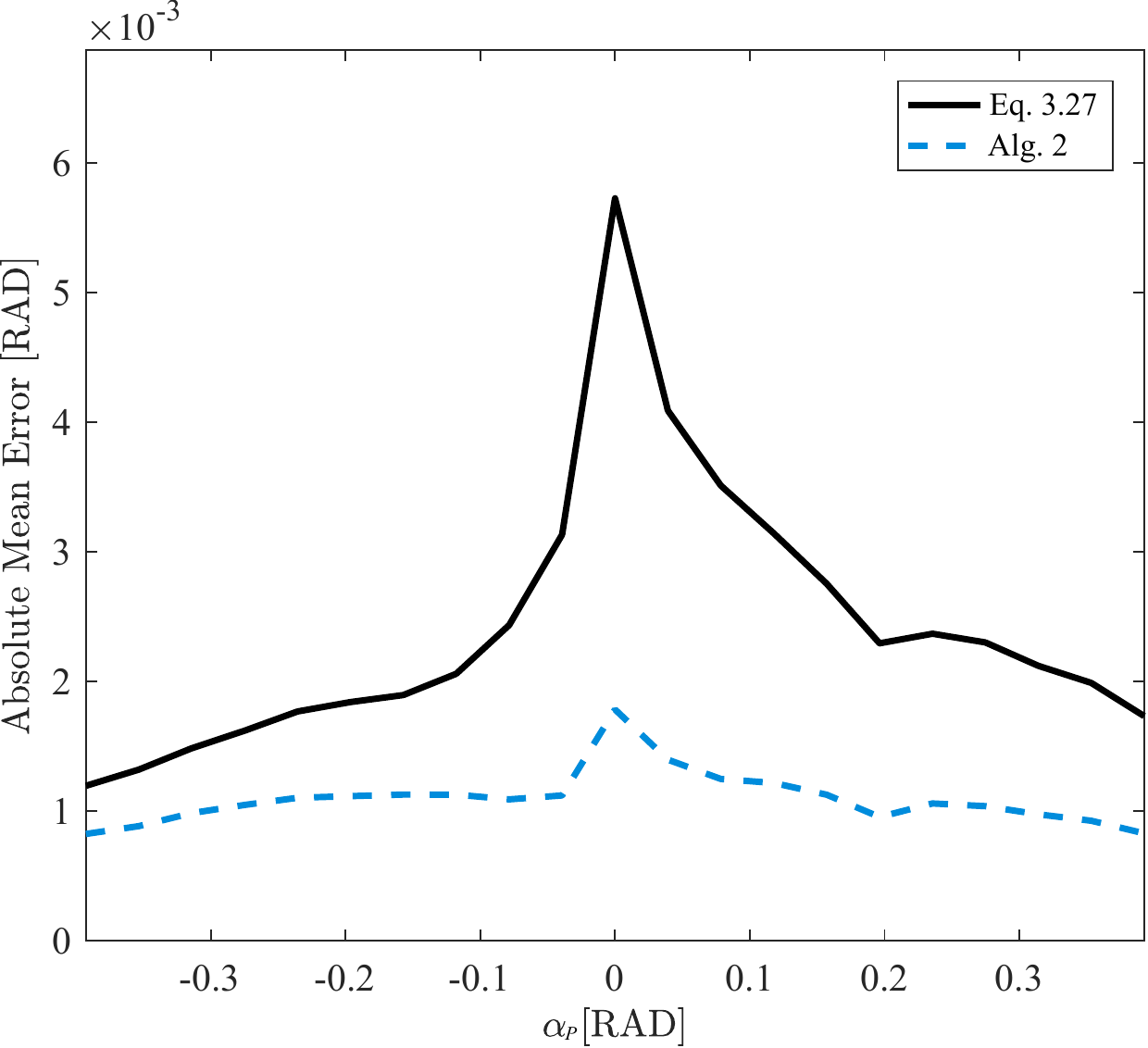}
	\caption{\label{fig:mean_alpha}Absolute mean error after one period as a function of initial $\alpha_P$.} 
\end{figure}

In order to further verify how much the usage of Algorithm \ref{alg2} can improve the overall orbital element update, the resulting $\alpha_P$ from this method and the regularly computed $\alpha_P$ in Eq. \eqref{eq:alpha} were compared with the phasing yielded by the CR3BP for 10,000 different initial conditions, in the same fashion as Figures \ref{fig:AN_error} and \ref{fig:all}. The absolute error is again shown in Figure \ref{fig:mean_alpha} as the average of each initial $\alpha_P$. 

It can be seen that the instantaneous update from Algorithm \ref{alg2} performs much better. This is more obvious in the region closer to the Earth ($\alpha_P \approx 0$), where the semi-major axis is predicted to change more drastically. \textcolor{External}{It is important to remember that this gain in accuracy comes at the cost of an increased computational time, depending on the method's step size. In contrast, the computation provided by Eq. \eqref{eq:alpha} is explicit.}

\subsection{Long-Term Propagation}

\begin{figure}[hbt!]
	\begin{minipage}{.5\linewidth}
		\centering
		\includegraphics[width=0.95\textwidth]{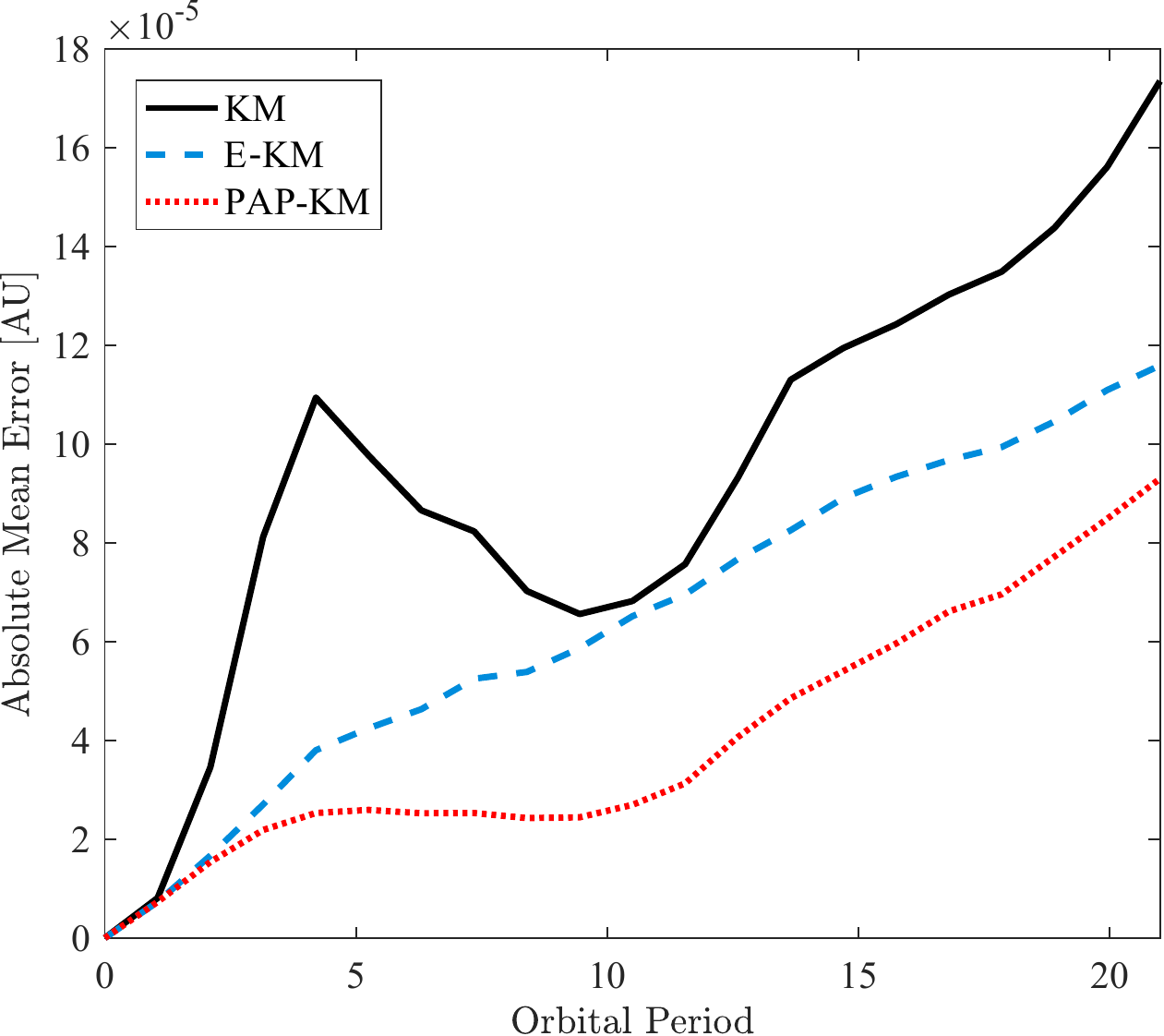}
		\caption*{\textit{a) Semi-major Axis}}
	\end{minipage}%
	\begin{minipage}{.5\linewidth}
		\centering
		\includegraphics[width=0.95\textwidth]{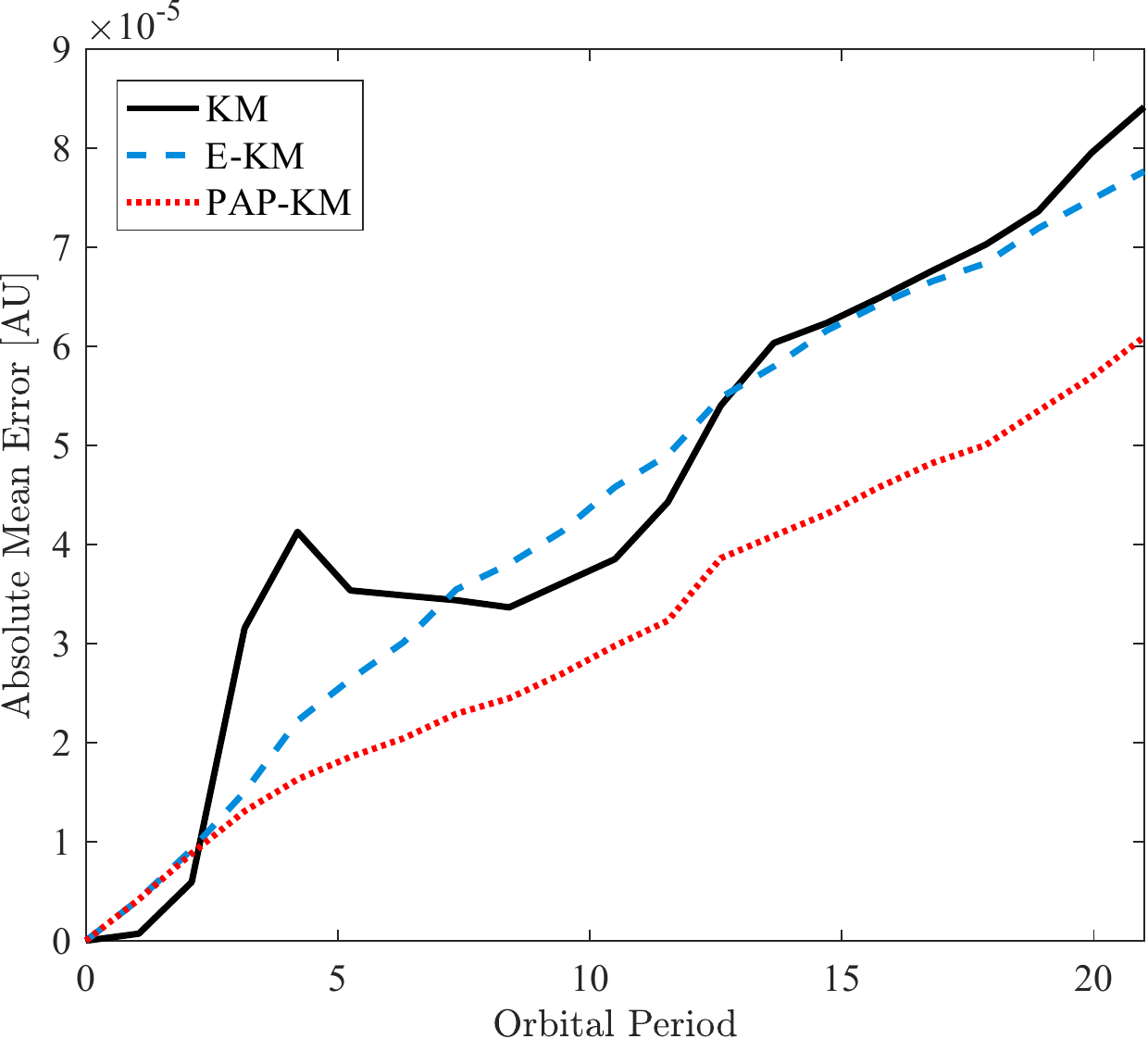}
		\caption*{\textit{b) Eccentricity}}
	\end{minipage}
	\begin{minipage}{.5\linewidth}
		\centering
		\includegraphics[width=0.95\textwidth]{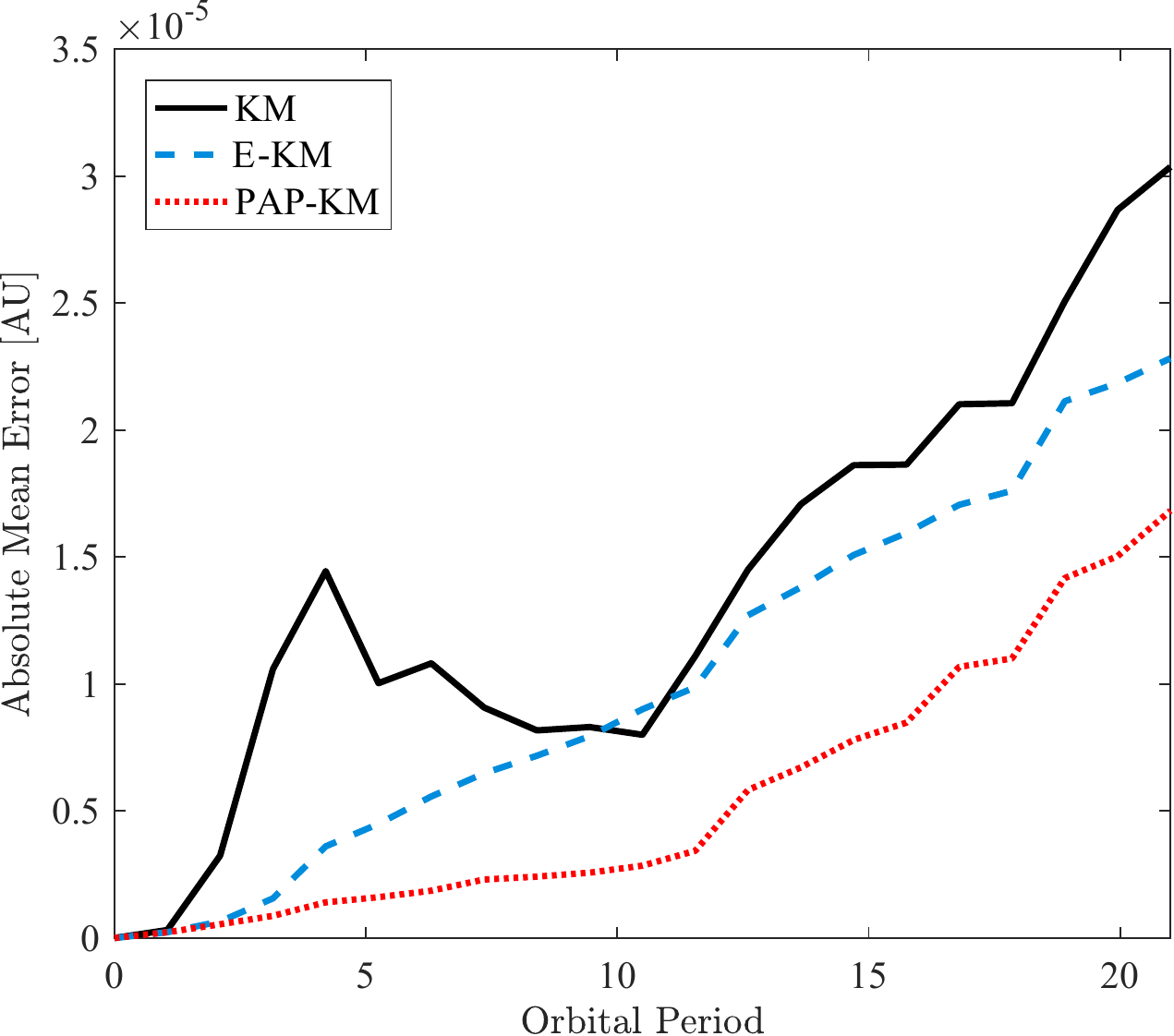}
		\caption*{\textit{c) Inclination}}
	\end{minipage}%
	\begin{minipage}{.5\linewidth}
		\centering
		\includegraphics[width=0.95\textwidth]{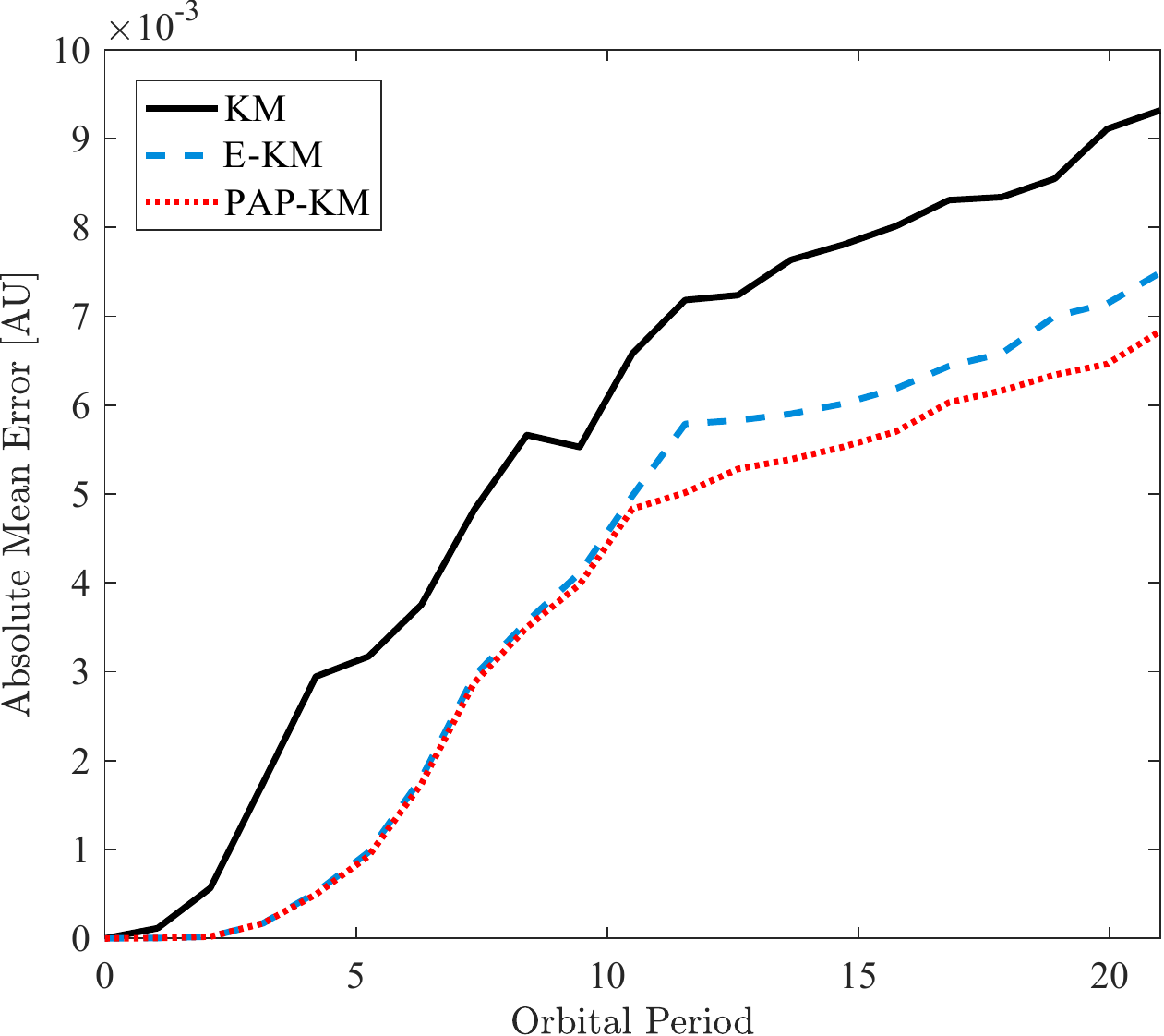}
		\caption*{\textit{d) Argument of Periapsis}}
	\end{minipage}\caption{\label{fig:long_term} Average absolute error for orbital element update in the the KM, PAP-KM and E-KM for 21 periods}
\end{figure}

A short analysis of the long term propagation error of three of the models discussed in this Chapter is made on Figure \ref{fig:long_term}. The KM (with the $\alpha$ update of Eq. \eqref{eq:alpha}), the PAP-KM and the E-KM (both employing the $\alpha$ update computed with Algorithm \ref{alg2}) are compared to the CR3BP: the same 10,000 initial conditions are propagated over 21 periods of motion (i.e. about 21 years). At each periapsis, \textcolor{Internal}{the absolute error for each model is averaged over all the initial conditions}, using the CR3BP propagation as the baseline. This could possibly be improved by increasing the number of integration intervals in the E-KM, at the cost of a longer computational time.

As expected, the error accumulates with the number of periods for all of the orbital elements. For both methods presented in this section, this happens much slower than for the original KM. The expectation would be for the E-KM to perform the best, since it takes into account the behaviour of the osculating orbital elements, at each time step; however, the error provided by the PAP-KM with the $\alpha$ update of Algorithm \ref{alg2} is clearly the smallest. This indicates that the small errors when calculating the instantaneous orbital elements accumulate faster in the numerical computation setting of the E-KM.

\textcolor{External}{In order to further validate the dynamical behaviours of the PAP-KM and the E-KM, a similar approach to the one presented by Ross and Scheeres \cite{scheeres_multiple} was adopted. In that paper, the KM is compared to the CR3BP by propagating a series of initial conditions in the Jupiter-Callisto system. The values of the semi-major axis were taken from the interval $a \in \big[1.1,\text{ }1.8\big]$, avoiding the islands corresponding to stable mean motion resonances of the massless particle's orbit with Callisto's: these are the white ovals depicted in the plots of the paper's Figure 5.}
	
\begin{figure*}[h!]
	\centering
	\begin{minipage}{.32\textwidth}
		\centering
		\includegraphics[width=1.1\linewidth]{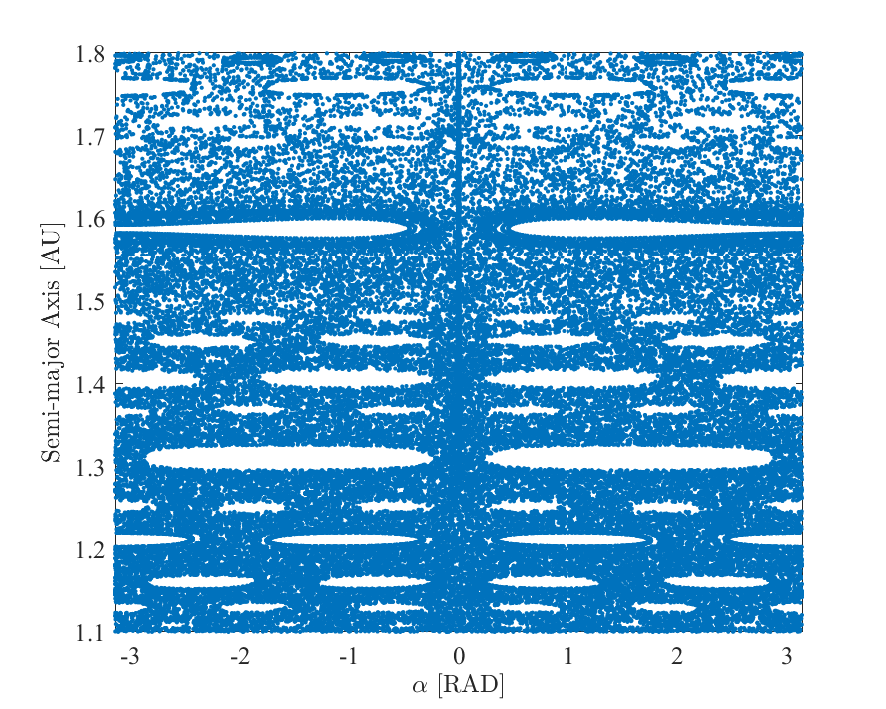}
		\caption*{\label{fig:S1}a) CR3BP}
	\end{minipage}%
	\begin{minipage}{.32\textwidth}
		\centering
		\includegraphics[width=1.1\linewidth]{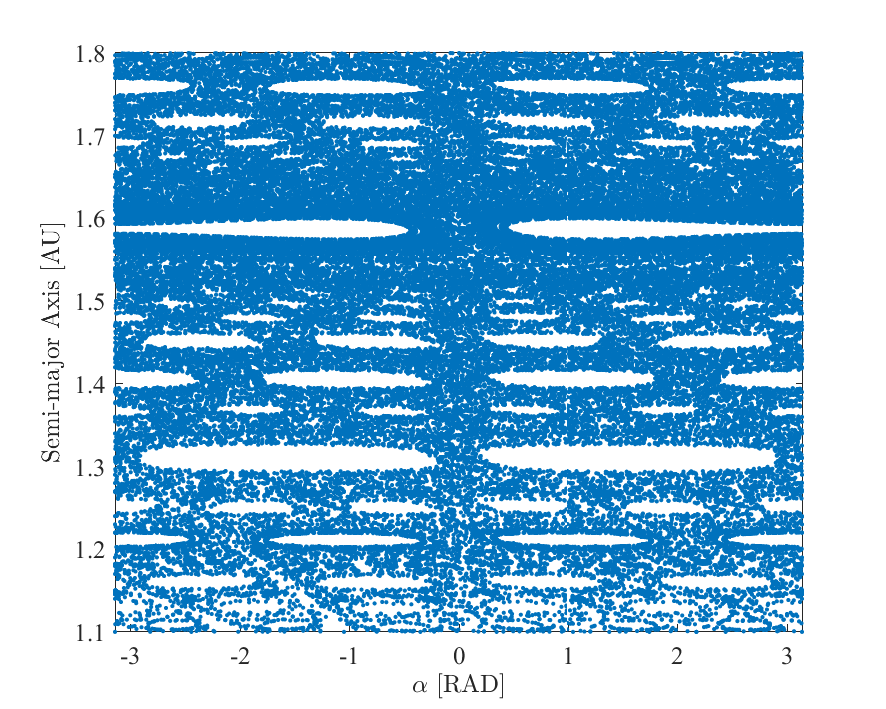}
		\caption*{\label{fig:S2}b) E-KM}		
	\end{minipage}
	\begin{minipage}{.32\textwidth}
		\centering
		\includegraphics[width=1.18\linewidth]{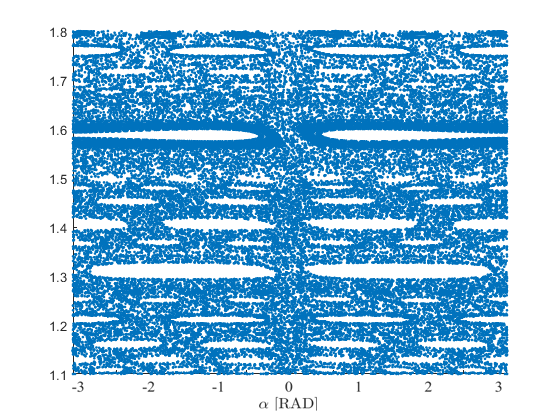}
		\caption*{\label{fig:S22}c) PAP-KM}		
	\end{minipage}
	\caption{\label{fig:S3}Plot of the stable resonances in the Jupiter-Callisto system \textcolor{Both}{($\mu = 5.667 \times 10^{-5}$)}}	
\end{figure*}

\textcolor{External}{In this work, Figure \ref{fig:S3} depicts the same methodology. It shows the evolution of the semi-major axis in a long term propagation, as a function of the initial $\alpha_P$---this time, for the E-KM and PAP-KM to be compared with the CR3BP. Each point corresponds to the new semi-major axis after one orbit, computed for about 300 periods. All three figures were generated using the same initial conditions, with the energy values and resonances found in the mentioned paper \cite{scheeres_multiple}.}

\textcolor{External}{As expected, Figure \ref{fig:S3} a) replicates the left-hand side of Figure 5 in the paper by Ross and Scheeres \cite{scheeres_multiple}, representing the CR3BP propagation.} It is possible to see that the distribution of points is not the same between figures a), b) and c); thus, a starting condition may have different updates in all of these three models. Nevertheless, the behaviour of the resonances and islands is maintained in all the plots, which ultimately confirms the validity of the E-KM and PAP-KM models.

\subsection{Computational Time Analysis}
\label{sub:ctime}

Considering the low-fidelity of the methods developed so far, it was considered important to have a decreased computational cost in contrast to the CR3BP. Without any further code optimisations for time, the original KM and the PAP-KM were determined to be respectively $13\%$ and $22\%$ faster than the latter for one period, with the E-KM having the same computational cost as the KM, as reasoned in Section \ref{sub:EK}. 

Some points remain to be stated. First, the usage of stroboscopic maps is a very interesting solution to speed up the calculation of long-term propagations without a remarkable loss in accuracy. Second, the integration methods currently used to compute both the KM and PAP-KM can still be improved in terms of efficiency, rendering the computation speed even faster. \textcolor{Internal}{Third, as detailed in Section \ref{sub:km}, the described methods need only to be used in the perturbation region.} For the remaining motion, the model used is the 2BP, which can decrease the overall computational cost immensely. Fourth, the computation time can be further reduced by taking advantage of the kick-maps. In cases when the orbit does not change very drastically, kick-maps can be computed to show the evolution of the Keplerian elements as a function of $\alpha_P$. Then, for long-term propagation, the orbital update can be made by interpolating on these kick-maps, significantly reducing the number of needed computations and the overall cost. This method will be later considered in the application scenario of Chapter \ref{chap:ast2}.

\section{Non-Conservative Forces}
\label{sec:GVE}

This section details the equations of motion generated using the GVE with the accelerations obtained from the K3BP in order to obtain the GVE third-body (GVE-3B) framework. The accuracy of the model is also analysed, as well as its suitability for mission design.

\subsection{Equations of Motion}

GVE have been extensively used in astrodynamics to compute motion perturbed by a disturbing force. In contrast to the LPE, they can also account for non-conservative accelerations. Consequently, they are especially useful to model a low-thrust spacecraft moving in three-body configurations. The accelerations in that particular scenario are represented by Eq. \eqref{eq:a_sum}.
\begin{equation}\label{eq:a_sum}
	\bm{a}_{GVE} = \{a_r, a_{\theta}, a_h\} = \bm{a}_{LT} + \bm{a}_{3B}
\end{equation}

Per Newton's first law, the low-thrust acceleration $\bm{a}_{LT}$ is easily computed as the quotient of the thrust vector and the system's mass. The formulation of \textcolor{External}{the third body acceleration} $\bm{a}_{3B}$ is more complex: in order to account for the third-body perturbation, it is here based on the disturbing function derived in Section \ref{sec:dist_fcn}.

\subsection{Disturbing Accelerations}

The derivation of the disturbing accelerations for the GVE-3B is done using the K3BP ($\mathpzc{U}_{3B}$), derived in Section \ref{sec:dist_fcn}. Following Hamiltonian mechanics, the accelerations are computed as in Eq. \eqref{eq:a1}:
\begin{align}\label{eq:a1}
	a_x = -\frac{\partial\mathpzc{U}_{3B}}{\partial x}, a_y = -\frac{\partial\mathpzc{U}_{3B}}{\partial y}, a_z = -\frac{\partial\mathpzc{U}_{3B}}{\partial z}
\end{align}
which, together with Eqs. \eqref{eq:hamiltonian} to \eqref{eq:hamil}, yield the final output (in the instantaneous Earth-pointing reference frame): 
\begin{align}
	\nonumber 
	\bm{a}_{3B} &= \{a_x, a_y, a_z\}\\\nonumber
	a_x &= - \mu\Bigg(\frac{-1 + x}{(1 - 2x + x^2 + y^2 + z^2)^\frac{3}{2}} - \frac{3 x^2}{(x^2 + y^2 + z^2)^\frac{5}{2}} + \frac{1 - x}{(x^2 + y^2 + z^2)^\frac{3}{2}}\Bigg)\\\nonumber	
	a_y &= -y\Bigg(\frac{\mu}{(1 - 2x + x^2 + y^2 + z^2)^\frac{3}{2}} - \frac{\mu(3 x + x^2 + y^2 + z^2)}{(x^2 + y^2 + z^2)^\frac{5}{2}}\Bigg)\\
	a_z &= -z\Bigg(\frac{\mu}{(1 - 2x + x^2 + y^2 + z^2)^\frac{3}{2}} - \frac{\mu(3 x + x^2 + y^2 + z^2)}{(x^2 + y^2 + z^2)^\frac{5}{2}}\Bigg)\label{eq:acc}
\end{align}

\subsection{Gauss' Variational Equations Third-Body Framework}

The accelerations used in the GVE (Eqs. \eqref{eg:gve}) have to be written in the LVLH frame. However, $\bm{a}_{3B}$ is described in the previously detailed Earth-pointing reference frame, while the low-thrust accelerations $\bm{a}_{LT}$ may be depicted in any reference frame, depending on the setting of the problem. For the sake of reaching a formula for $\bm{a}_{GVE}$, $\bm{a}_{LT}$ is here described in a barycentric, inertial Cartesian frame. Thus, some transformations have to be taken into account. The first one is determined by matrix $\text{\textbf{R}}_I^{\Earth}$, computed using Eqs. \eqref{eq:normtime} and \eqref{eq:tEarth} and used to convert a vector in the inertial Cartesian frame \textcolor{Internal}{($O_{Ixyz}$)} to the Earth-pointing reference frame \textcolor{Internal}{($O_{\Earth xyz}$)}:
\begin{align}
\bm{a}_{\Earth xyz} &= \text{\textbf{R}}_I^{\Earth} \text{ } \textcolor{Internal}{\bm{a}_{I xyz}}, \text{  }
\text{\textbf{R}}_I^{\Earth} = \begin{bmatrix}
\cos (t + \nu_{\Earth_0}) & -\sin (t + \nu_{\Earth_0}) & 0 \\
\sin (t + \nu_{\Earth_0}) & \cos (t + \nu_{\Earth_0}) & 0 \\
0 & 0 & 1 \\
\end{bmatrix}
\end{align}

The remaining necessary matrices are described in Section \ref{sec:LVLH}: matrix $\text{\textbf{R}}_{eph}^{I}$ is used to convert from the orbital plane ($O_{eph}$) to the inertial Cartesian reference frame, and $\text{\textbf{R}}_{r \theta h}^{eph}$ rotates the vector from the LVLH frame ($O_{r \theta h}$) to the orbital plane one:
\begin{align}
\textcolor{Internal}{\bm{a}_{I xyz}} &= \text{\textbf{R}}_{eph}^{I} \text{ }\bm{a}_{eph}\\
\bm{a}_{eph} &= \text{\textbf{R}}_{r \theta h}^{eph} \text{ }\bm{a}_{r \theta h}
\end{align}

Finally, the accelerations to use in Eqs. \eqref{eg:gve} in the LVLH frame are computed as follows:
\begin{align}
\textcolor{Internal}{\bm{a}_{GVE} = \bigg(\text{\textbf{R}}_{eph}^{I} \text{\textbf{R}}_{r \theta h}^{eph}\bigg)^{-1} \bigg(\text{\textbf{R}}_I^{\Earth^{-1}} \text{ } \bm{a}_{3B} + \bm{a}_{LT}\bigg)}
\end{align}

\begin{figure}[bht!]
	\centering
	\includegraphics[width=\linewidth]{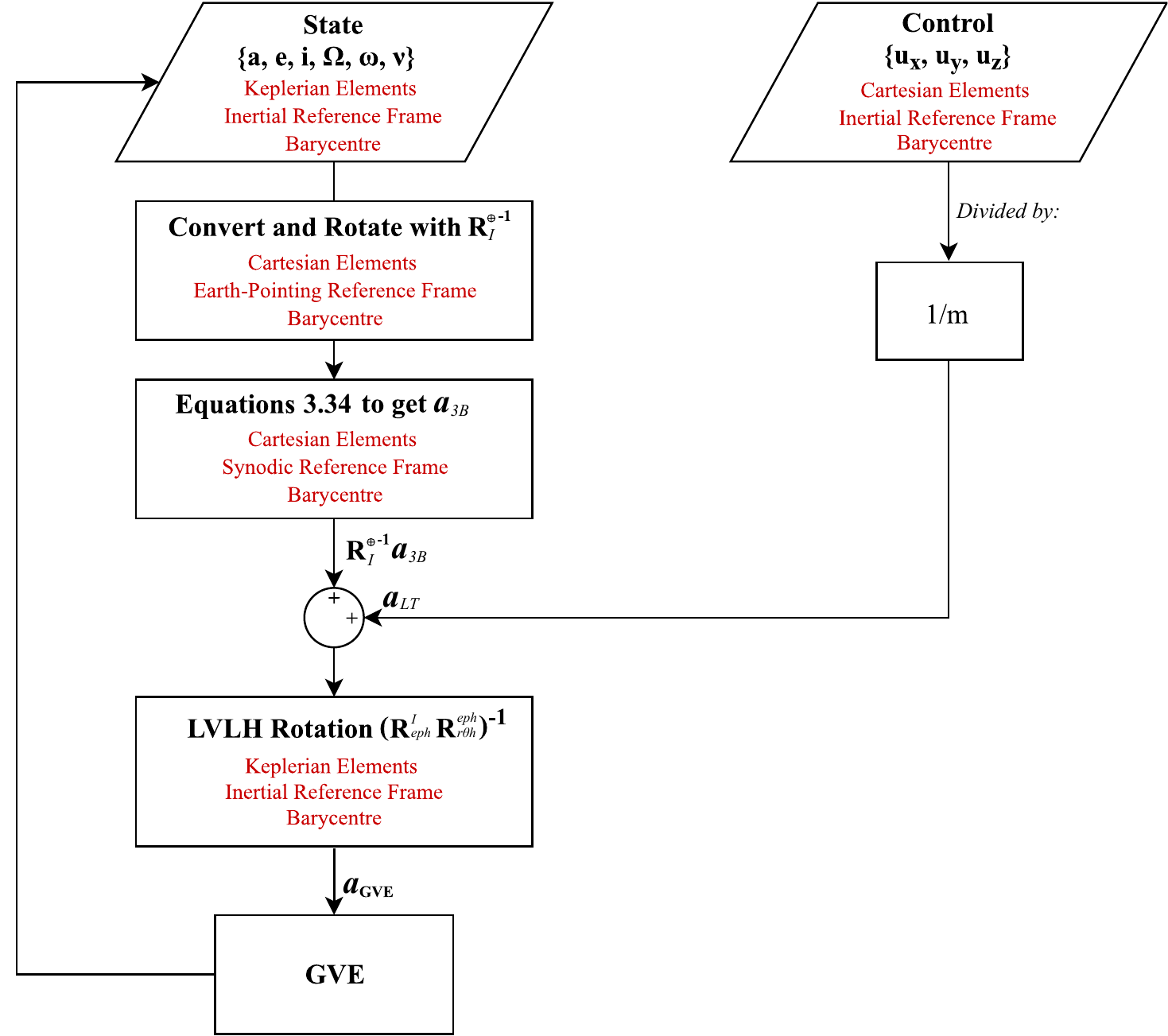}
	\caption{\label{fig:flow} State propagation using the GVE Framework}
\end{figure}

The state propagation is done in barycentric orbital elements. For a better understanding of the transformations involved and the overall framework, the flowchart of Figure \ref{fig:flow} can be analysed. \textcolor{Internal}{It is important to highlight that $\bm{a}_{LT}$ may be described in a reference frame that is not the one indicated---}in that case, the appropriate transformations have to be implemented beforehand. 

\subsection{Accuracy of the Model}
\label{sec:Accuracy}

The GVE-3B equations are formulated as a low-fidelity tool for the computation of third-body perturbations in a system \textcolor{Internal}{with} small gravitational parameter. It is then important to figure out their performance as a function of the distance to the perturbing body. As such, in the following analysis, the modelled motion will have no thrusting acceleration: the influence of the latter will be analysed in the asteroid mission trajectories of Chapter \ref{chap:ast1}. 

For the purpose of analysing the GVE-3B accuracy, the propagation error is defined as the distance, at each time step, between the positions computed with this model and with the CR3BP. In order to further highlight the importance of taking the third-body perturbation into account, the defined GVE-3B error is contrasted to the similarly computed 2BP error. This work claims that the accuracy of the 2BP is lacking in the vicinity of the Earth and up to very distant regions from its sphere of influence (the so-called perturbation region), making it unsuitable for the proposed asteroid trajectory design and presenting the GVE-3B framework as a much better alternative.

\begin{figure}[h!]
	\centering
	\begin{minipage}[t]{0.465\linewidth}
		\includegraphics[width=\textwidth]{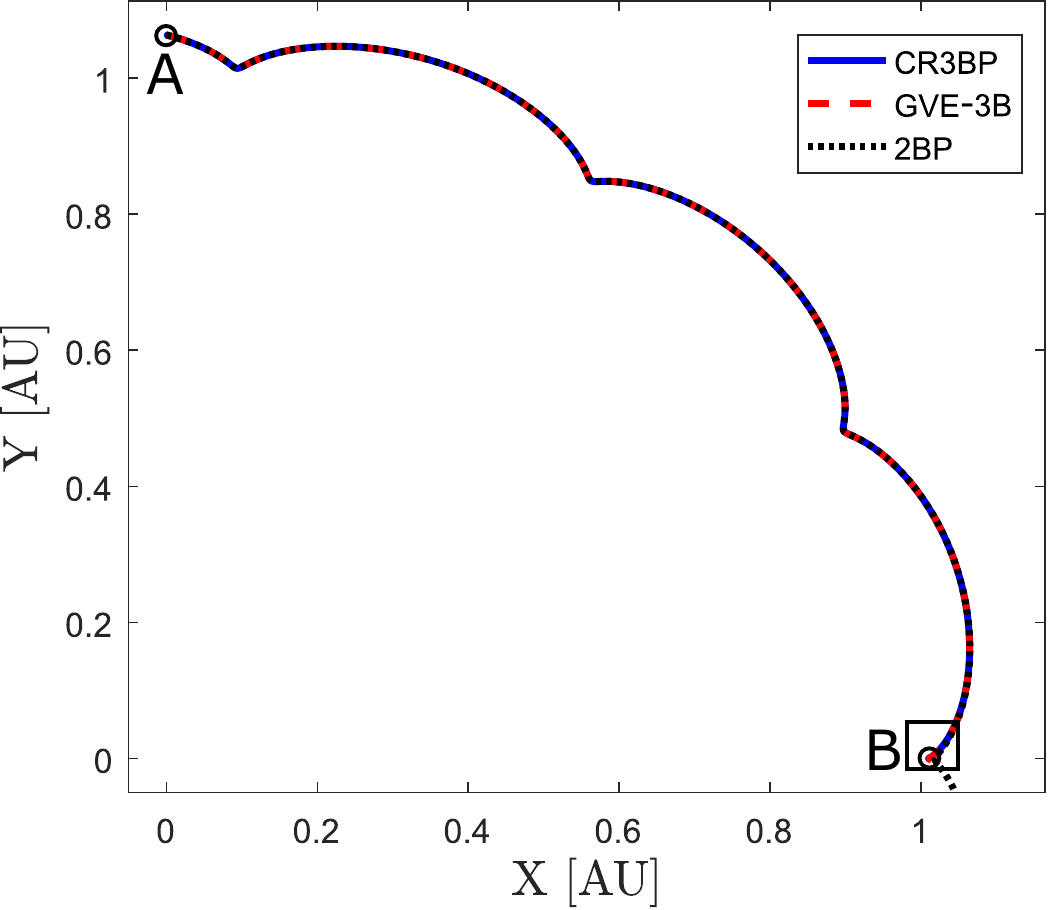}
		\caption*{\textcolor{External}{a) Full propagation from A to B}}
	\end{minipage}
	\hfill
	\begin{minipage}[t]{0.52\linewidth}
		\includegraphics[width=\textwidth]{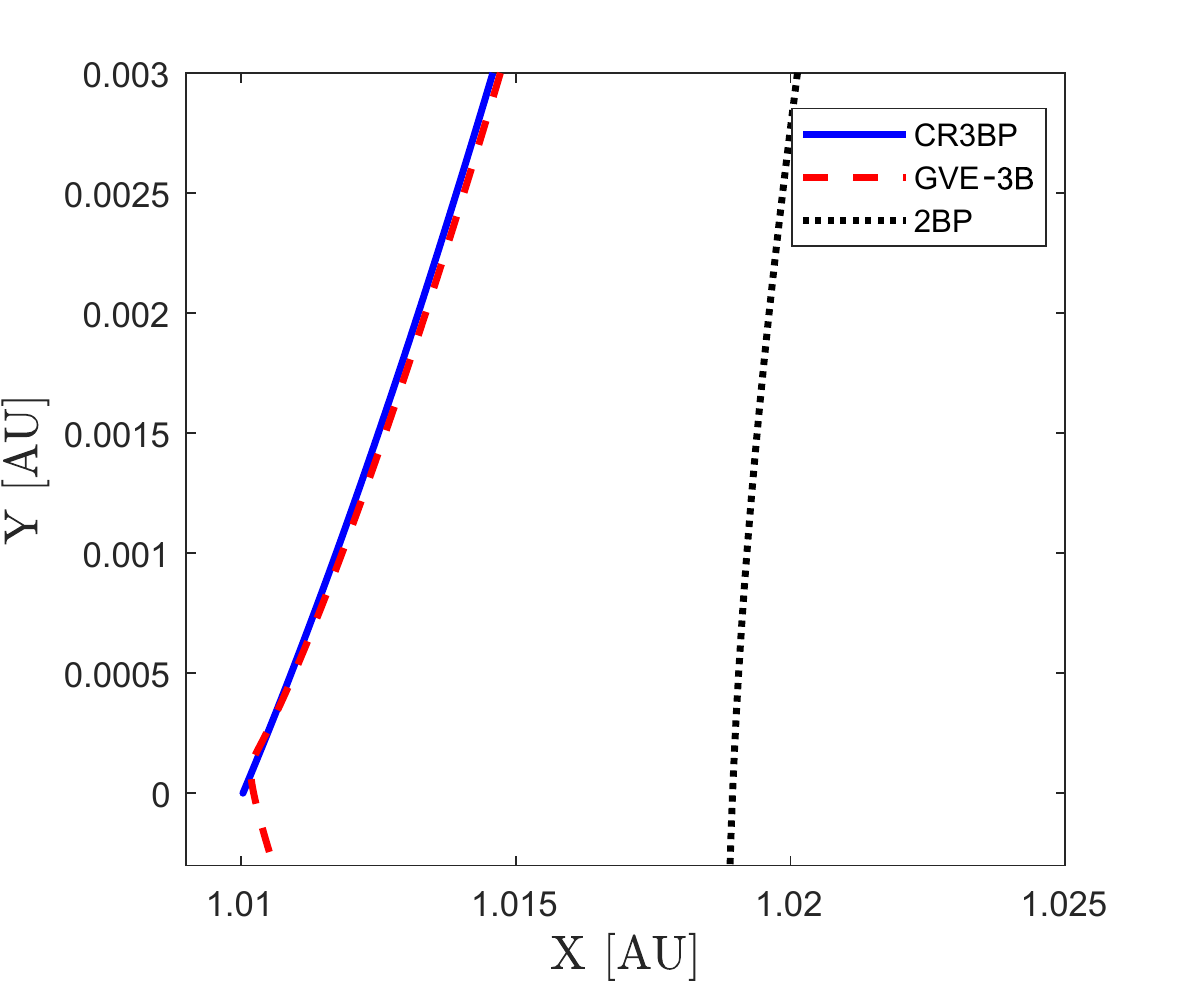}
		\caption*{\textcolor{External}{b) Zoom-in at point B}}
	\end{minipage}
	\caption{\textcolor{External}{Propagation of the CR3BP, the GVE-3B and 2BP from point A to point B\label{fig:traj_e}}}
	\centering
	\begin{minipage}[t]{0.49\linewidth}
		\includegraphics[width=\textwidth]{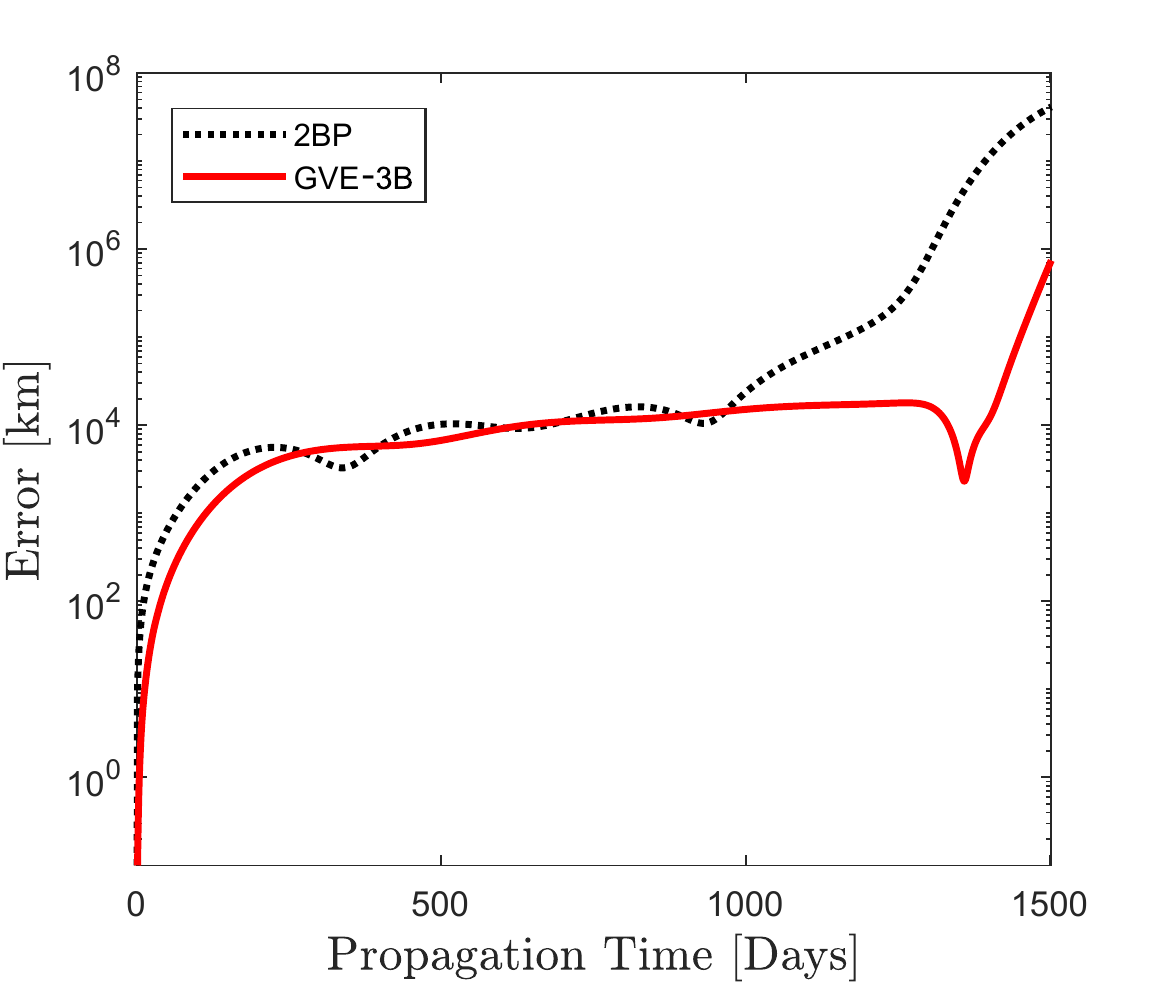}
		\caption*{\textcolor{External}{a) Error as a function of time}}
	\end{minipage}
	\hfill
	\begin{minipage}[t]{0.49\linewidth}
		\includegraphics[width=\textwidth]{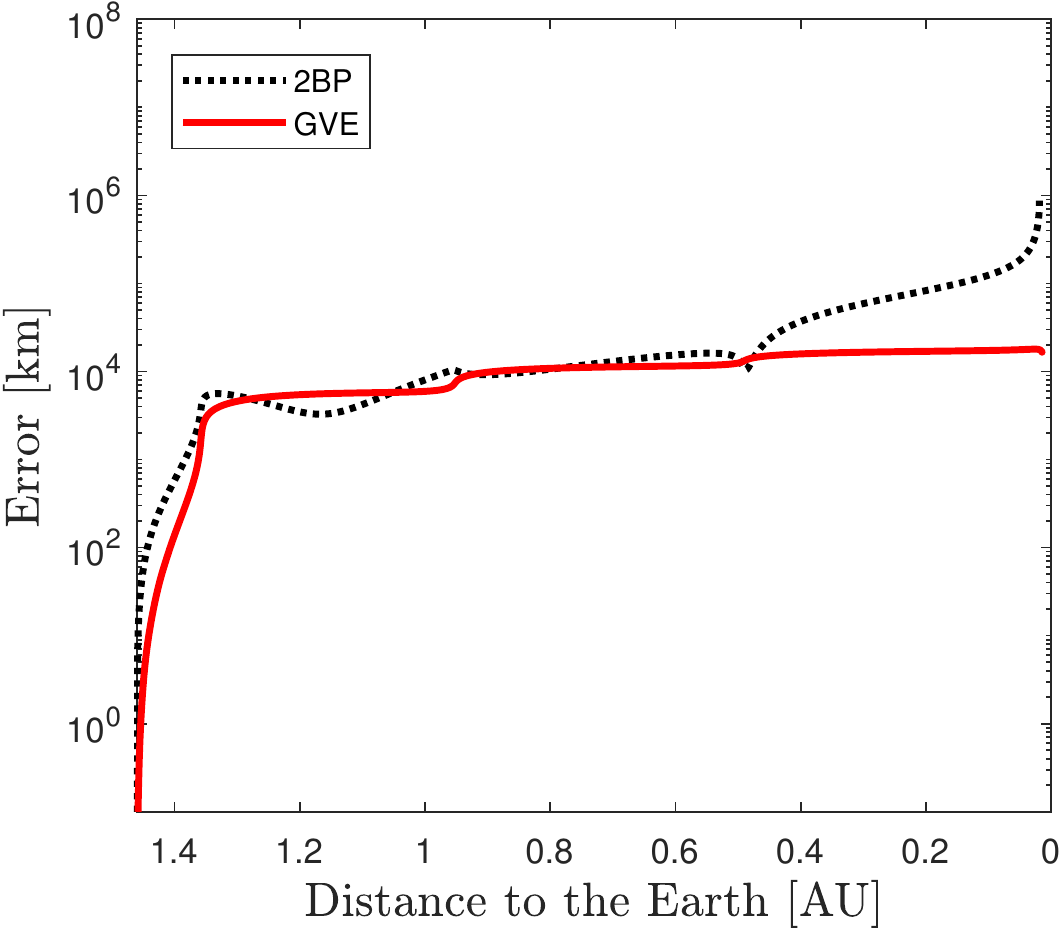}
		\caption*{\textcolor{External}{b) Error as a function of distance to the Earth,} \textcolor{Internal}{shown until the crossing of the X-axis}}
	\end{minipage}
	\caption{Propagation error for the 2BP and the GVE-3B for the trajectory in Figure \ref{fig:traj_e}\label{fig:e_time}}
\end{figure}

The first baseline trajectory for error comparison is found in Figure \ref{fig:traj_e}: the full propagation from initial point A to end point B can be seen in Figure \ref{fig:traj_e} a), while Figure \ref{fig:traj_e} b) highlights a zoom-in of the final state, as propagated by each of the models. This trajectory was constructed in the following manner: first, the stable invariant manifold of the $L_2$ point is backpropagated in the CR3BP for a period of 1500 days, ending up at point A. Then, starting from this point, all three models were propagated forward for the same time period. It follows that the end state for the CR3BP, the 2BP and the GVE-3B should be the $L_2$ point; however, when getting closer to the Earth, the difference between the models increases, and the error provided by the 2BP is shown to be especially large.

This error can be better analysed in Figure \ref{fig:e_time}: while both plots show the same error on a logarithmic scale, \textcolor{External}{Figure \ref{fig:e_time} a)} is plotted as a function of time, while \textcolor{External}{Figure \ref{fig:e_time} b)} is a function of the distance to the Earth. It can be seen that the GVE-3B error remains quite stable in the $10^4$ km value, which stays the same for the majority of the motion, until the X-axis of the synodic reference frame is reached. As expected, when the trajectory approaches the region of the Earth's sphere of influence ($\sim$ 0.01 AU), there is a clear spike in the GVE-3B error. In contrast, the 2BP error increases faster, starting to do so much earlier. 

\begin{figure}[htb!]
	\centering
	\includegraphics[width=0.65\textwidth]{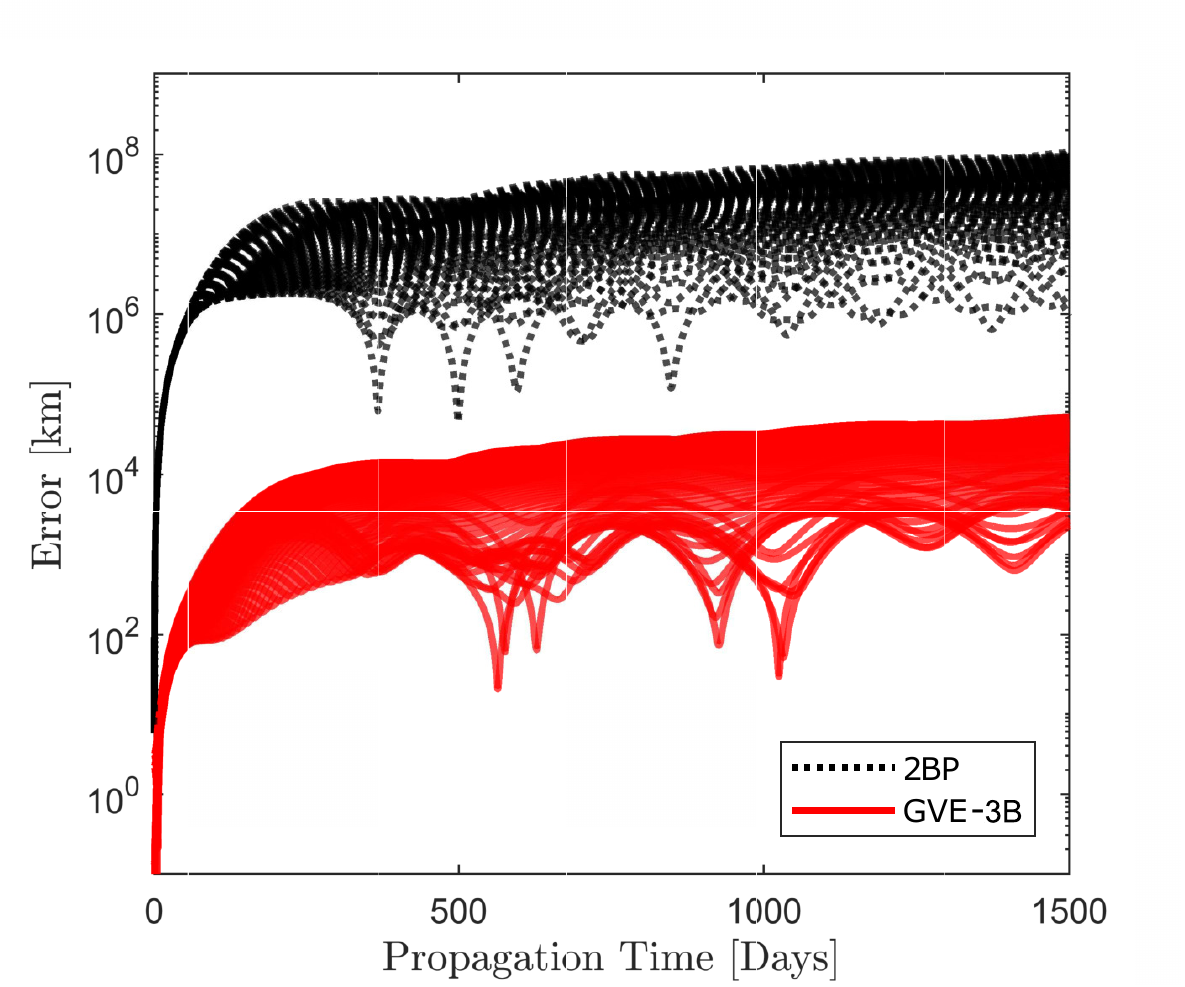}
	\caption{Propagation error for the 2BP and the GVE-3B for 100 initial states in the Earth's vicinity}\label{fig:50prop}
\end{figure}

In order to further verify these findings, the performance of the three models is compared by propagating 100 different trajectories backwards in time for 1500 days. Each of them starts, with zero velocity, very close to the Earth's sphere of influence and never crosses it throughout the propagation time. For each trajectory, the corresponding error when comparing the models can be seen in Figure \ref{fig:50prop}. The value yielded by the 2BP propagations is much higher overall, since the trajectories begin close to the disturbing body. In contrast, the GVE error remains similar to the previously shown case, demonstrating its consistency in application scenarios. 

From these plots, it can be concluded that the CR3BP and GVE-3B model behave in a very similar way in the feasible region, which clearly evidences the quality of the framework presented in this paper and validates its use up until a very close region to the secondary. However, akin to the KM method \cite{alessi_semi}, the GVE-3B model is not adequate for use inside the sphere of influence of the Earth or for trajectories that remain for a long time in this vicinity.

\chapter{Application: Gauss' Variational Equations for Low-Thrust Design}
\label{chap:ast1}
\textcolor{Internal}{With the pursuit of increasingly innovative and complex space missions, the focus of the space industry has been turning towards low-thrust technologies. Electric propulsion systems provide large savings in propellant mass, which can be decisive for the mission's feasibility. Since the first spacecraft using low-thrust was successfully flown in 1998 \cite{DS1}, this technology has allowed the planning of a range of missions that would otherwise be infeasible, including visits to the outer planets, comets and asteroids \cite{noton}.}

Designing a low-thrust trajectory is a more complex task than doing so for a high-thrust one. For the latter case, the few short thrust phases can be approximated by singular instantaneous $\Delta v$'s. On the contrary, low-thrust missions require the propulsion system to operate for a significant part of the transfer. Consequently, the thrust vector is a continuous function of time and the trajectory optimisation problem has to find the optimal control law \cite{dachwald}. This is an extremely complex problem that has no closed-form solutions, except for some very specific cases \cite{dachwald_cite}. Thus, the optimal control problem has to be carefully conceived in order to achieve convergence, which includes an attentive definition of bounds and constraints for the trajectory. 

Following the process described in Section \ref{sec:opt_traj_design}, one of the main steps in the formulation of the optimal control problem is choosing the model of motion in which the trajectory is developed. Certain design applications, like missions to near-Earth asteroids (NEAs) require models of motion of higher complexity than the classical two-body problem (2BP), since third-body perturbations have a non-negligible effect. This is due to the fact that NEAs \textcolor{Internal}{usually} move in \textit{low-energy regimes}. These are here defined as regimes of motion in which ballistic capture is theoretically possible \cite{ballistic}, which may occur for objects whose orbital energy does not differ much from that of the third-body perturbation, e.g. nearly co-orbital bodies \cite{TCO} or, in the case of the Sun-Earth system, spacecraft departing from or arriving to Earth with a low excess velocity ($v_{\infty}$). \textcolor{External}{However, the utilisation of an alternative higher-fidelity method (e.g. the circular restricted three-body problem (CR3BP)) may bring difficulties related to the definition of the optimal control problem. For instance, boundary conditions are not trivial to set, since the coordinates are presented in the synodic Cartesian reference frame.}

\textcolor{External}{As an alternative, the Gauss' variational equations third-body (GVE-3B) framework can describe third-body motion in terms of the Keplerian elements that define the osculating orbit of the spacecraft, in a barycentric coordinate system. This is advantageous when devising a control strategy near global optima, since it provides a better intuitive understanding of the trajectory. Given that the solution to an optimal control problem typically relies on the setting of parameters by an expert in astrodynamics, outside of the actual optimisation process \cite{dachwald}, an intuitive understanding of the trajectory's evolution until a solution is reached is crucial. Furthermore, using Keplerian elements, the bounds and boundary values of the optimal control problem can be more easily assessed, facilitating the convergence of the non-linear process to generate the control law.} 

Thus, this chapter presents the usage of the GVE-3B framework, as presented in Section \ref{sec:GVE}, in the low-thrust trajectory design of missions to NEAs. The latter are now considered the easiest celestial bodies to reach from the Earth \textcolor{Internal}{and may represent a potential impact threat to our planet \cite{RN1}.}

A test example of the usability of the GVE-3B framework is presented with the design of two trajectories: an asteroid capture and a rendezvous mission. The first case entails having a spacecraft attaching itself to an asteroid and moving it---from its nominal orbit to the stable invariant manifold of the $L_2$ point of the Sun-Earth system. Mirroring this mission, the rendezvous scenario consists on a spacecraft departing \textcolor{Internal}{$L_2$} through its unstable manifold orbit and matching its motion to the one of another NEA. 

\textcolor{External}{Some choices were made to simplify the trajectory design and maximise the chances of having an easy convergence for the optimal control problem. This is in line with the main focus of this chapter: to demonstrate the application of the GVE-3B framework with fully fledged trajectories that meet the initial requirements, which may not be the overall best solutions. The next sections analyse and explain the design process in detail.}

\section{Asteroid Mission Trajectory Design}
\label{sec:traj_design}

This section presents the results from each phase of the trajectory design approach for the asteroid capture and rendezvous missions. The benefits of using the GVE-3B framework as the model of motion for low-thrust design are highlighted, and \textcolor{External}{fully} optimised transfers are shown for asteroids 2018 AV2 and 2017 SV19.

\subsection{Mission Summary}

The trajectory design for two asteroid missions is here presented: one for capture, another for rendezvous. Asteroid missions are invariably linked to the use of low-thrust propulsion systems, since the trajectory will benefit from their high exhaust velocity \cite{frontiers}. Furthermore, both missions consider a spacecraft departing from (the rendezvous case) or arriving to (the capture case) \textcolor{External}{one of the libration points of the Sun-Earth system}. \textcolor{External}{The close-range proximity operations that precede or conclude a capture or rendezvous mission are not considered in this work.} 
	
\textcolor{External}{The asteroids were chosen using the Accessible NEAs NASA database\footnote{cneos.jpl.nasa.gov/nhats, Accessed 01-11-2018}, which was searched to find objects whose estimated mission costs do not surpass 5 km$\cdot$s$^{-1}$ (defined in the database as the total $\Delta v$ for departing a notional 400 km altitude circular Earth parking orbit, matching the NEA's velocity at arrival, departing the NEA and controlling the atmospheric entry speed at Earth return).}
	
\textcolor{Both}{The choice fell on asteroids 2018 AV2 and 2017 SV19 as the capture and rendezvous targets. At the time of search, they were the most recently discovered ones with a clearly defined optical opportunity, i.e. a set calendar date in which the asteroid will next be observable from the Earth. Since both asteroids have semi-major axes greater than the one of the Earth, the endgame was set for the $L_2$ point. This location is connected to several past missions (e.g. Herschel and Planck in 2009 \cite{herschel, planck}) and benefits from} the existence of invariant manifold orbits that can be travelled without spending any fuel, both arriving to and departing from it. It is important to denote that the mission $\Delta v$ costs may be further minimized by picking out optimal libration point orbits (LPOs) \textcolor{Internal}{and} connected manifold orbits (i.e. closer to the asteroids' energy), instead of the generic $L_2$ point. Since this component would bring an additional choice to the design process, it was not included in order to keep the focus on showcasing the methodology in a straightforward manner. 

\subsubsection{Asteroid Capture}

Asteroid 2018 AV2 is an Apollo asteroid discovered in January, 2018. Currently, there is little data regarding its composition and nature: it is known that its diameter is on the \textcolor{Internal}{small} range (3.2 - 14 m) and, assuming an average material composition, its mass is of about 318 tonnes \cite{mass, mass2}. It is postulated that this object may likely be artificial \cite{mini-moon}; still, its dimensions make it a feasible notional body to study a capture mission.

The mission design entails the following: the spacecraft meets the asteroid at a certain point of the asteroid's nominal orbit and grabs it by any chosen means \cite{ARRM, artemis}. Then, the coupled system uses its propulsive capabilities to insert itself into a stable invariant manifold trajectory of the $L_2$ point, the target destination. This happens around the asteroid's next close approach with the Earth, starting in 2036.

\subsubsection{Asteroid Rendezvous}

Asteroid 2017 SV19 is an Amor asteroid discovered in September, 2017. Its diameter is in the 17-78 m range, with an estimated mass of 52,850 tonnes. This makes the body very heavy for capture, but adequate for a rendezvous mission with a spacecraft. 

While the capture trajectory takes the spacecraft-asteroid system to $L_2$, the rendezvous one takes the spacecraft away from this point. The remaining mission mirrors the previously described one: after the spacecraft departs $L_2$ into an unstable invariant manifold orbit, it changes its trajectory to meet the nominal motion of the asteroid. The rendezvous happens during the asteroid's next close approach with the Earth, in 2040. 

\begin{figure}[h]
	\centering
	\includegraphics[width=\linewidth]{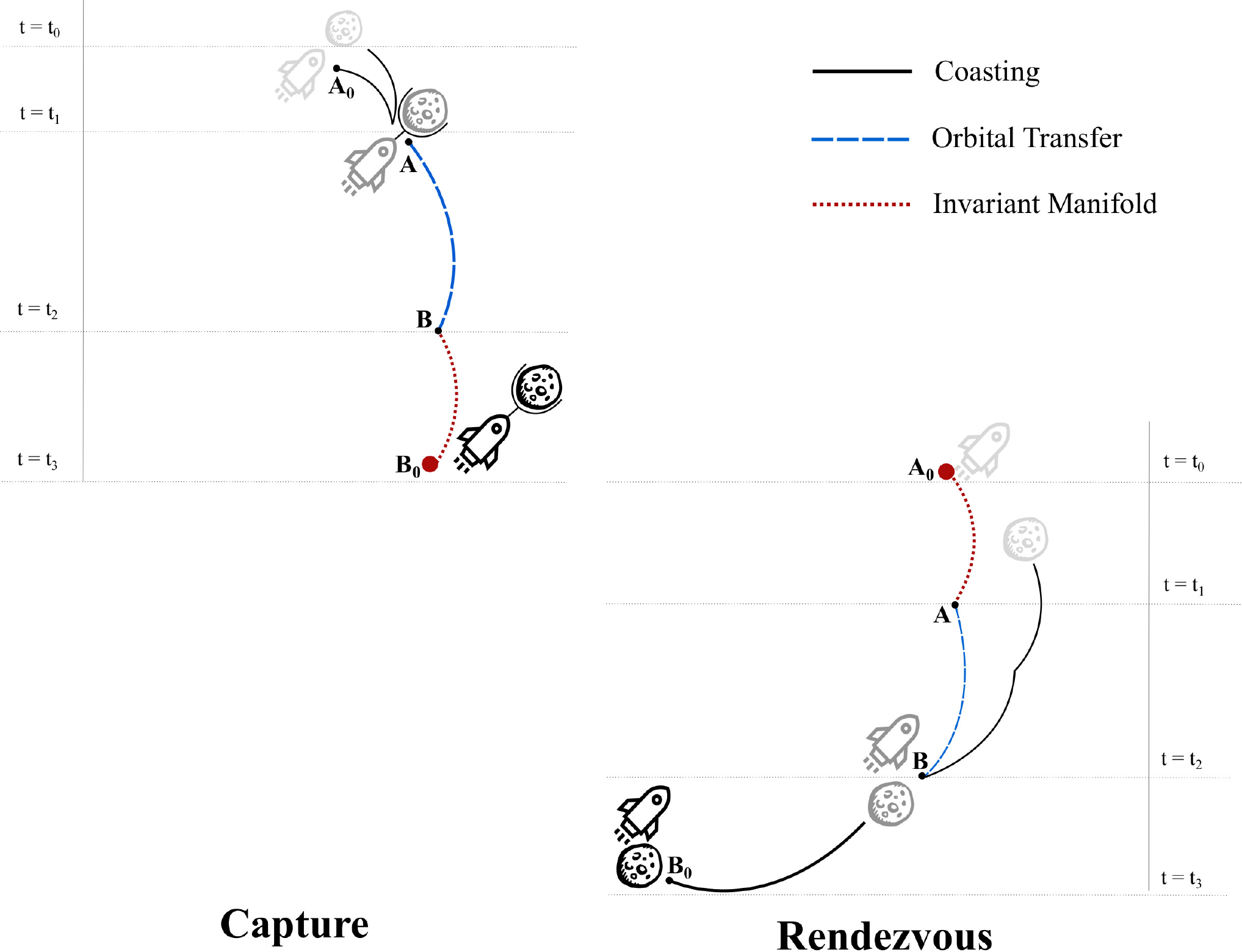}
	\caption{\label{fig:caperend}Trajectory diagram for the asteroid capture case (left) and the asteroid rendezvous mission (right)}
\end{figure}

As previously stated, both bodies are moving in a very similar energy regime to the one of the Earth, making them great candidates to test the use of the GVE-3B framework. In order to better visualize and compare both mission scenarios, Figure \ref{fig:caperend} can be observed. The orbital transfer that will be designed in this chapter concerns the blue dashed segment from $\mathrm{t_1}$ to $\mathrm{t_2}$. The required steps in order to determine its initial and final points (A and B, respectively) will be explained in the following sections.

\subsection{Low-Thrust Trajectory Design}
\label{sub:ltdesign}

The problem of designing a spacecraft's trajectory can be simply stated as the determination of a transfer that satisfies some initial and final conditions, while minimizing a chosen parameter \cite{conway}. Its mathematical formulation is given in Chapter \ref{chap:intro}, together with a detailed overview of the particular problem of low-thrust modelling.

Several methods exist to solve optimal control problems, as surveyed by several authors \cite{ltreview, betts}. Currently, direct methods are the ones most employed for mission design \cite{conway}. While robust, these still require the generation of a reasonable initial guess of the solution parameters, i.e. a set of state and control time-histories that yield a sub-optimal trajectory to be improved by the NLP problem solver. 

\textcolor{Internal}{In short, feasible initial guesses that can be posteriorly transcribed and solved using an optimal control software (e.g GPOPS-II \cite{gpops}) have to be generated.} This computation is achieved by implementing the following sequential approach, which will be detailed in the following sections:

\begin{description}
	\item{\textbf{\textcolor{External}{Step 1: Transfer Optimisation.}}} The optimal initial and final dates for the orbital transfer manoeuvre ($\mathrm{t_1}$ and $\mathrm{t_2}$) are determined by computing several possible Lambert arcs with different boundary conditions and choosing the one with the lowest $\Delta v$.
	\item{\textbf{\textcolor{External}{Step 2: Sims-Flanagan Approach.}}} The set transfer dates are used to compute an initial guess trajectory in the style of a Sims-Flanagan approach \cite{sims}. The result will be a transfer divided into equal-time segments, with an impulsive $\Delta v$ applied in each of them.
	\item{\textbf{\textcolor{External}{Step 3: Continuous Low-Thrust Transfer.}}} The multiple-impulse transfer is transcribed to a continuous one, by converting the $\Delta v$'s and segment times determined in the previous step into continuous accelerations and uniting the trajectory with a multiple shooting method.
\end{description}

\subsubsection{Step 1: Transfer Optimisation}
\label{sub:LA}

Finding the best possible trajectory for both asteroid missions requires defining several potential starting and ending points and calculating the transfer costs between them. Looking at Figure \ref{fig:caperend}, one can see that this means choosing the best Lambert arc out of different values of $\mathrm{t_1}$ and $\mathrm{t_2}$ and related asteroid ephemerides. In this case, the best arc is chosen as the one with the lowest $\Delta v$: the corresponding values of $\mathrm{t_1}$ and $\mathrm{t_2}$ are the initial and final transfer times and the related asteroid ephemerides are fixed to the state of points A and B. \textcolor{External}{These ephemerides remain fixed in time in the trajectory design, which may be sub-optimal for the final solution, as low-thrust trajectories are generally slower than impulsive ones.}

More concretely, looking first to the capture case: the first step is to obtain the real ephemerides of the asteroid 2018 AV2 in a far-away position from the Earth. Referring back to Figure \ref{fig:caperend}, this position corresponds to point $\mathrm{A_0}$. Then, this state is propagated forward in time to $\mathrm{t_1}$ (point A). In a similar fashion, the invariant manifold of the $L_2$ point is backpropagated to $\mathrm{t_2}$ (point B). Posteriorly, the states corresponding to $\mathrm{t_1}$ and $\mathrm{t_2}$ are connected via a Lambert arc, obtaining the impulsive transfer cost. This is done for many different values of $\mathrm{t_1}$ and $\mathrm{t_2}$, yielding a plot of transfer $\Delta v$'s as a function of the initial and final dates: a \textit{porkchop plot}.

\textcolor{External}{An analogous methodology is implemented for the rendezvous case: the ephemerides of asteroid 2017 SV19 are retrieved in a far-away position from the Earth (at $\mathrm{t_3}$, point $\mathrm{B_0}$). These are then backpropagated to $\mathrm{t_2}$ (point B). Simultaneously, the spacecraft's position is propagated forward from the $L_2$ point (at $\mathrm{t_0}$) to $\mathrm{t_1}$. The ephemerides corresponding to $\mathrm{t_1}$ and $\mathrm{t_2}$ are connected with a Lambert arc.}

\begin{figure}[hbt!]
	\centering
	\begin{minipage}[t]{0.49\linewidth}
		\includegraphics[width=\textwidth]{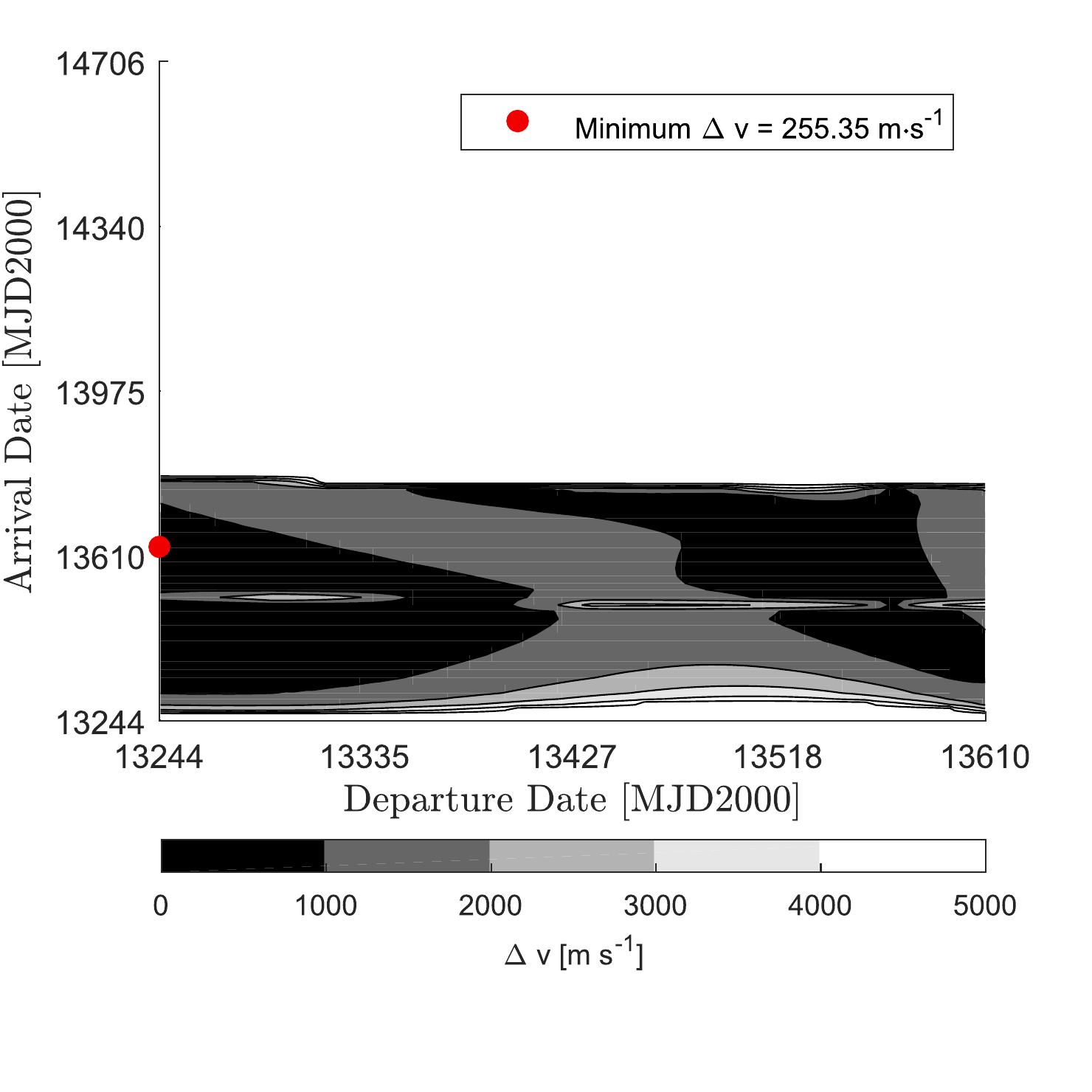}
		\caption*{a) Capture of Asteroid 2018 AV2 in the GVE-3B model}
	\end{minipage}
	\hfill
	\begin{minipage}[t]{0.49\linewidth}
		\includegraphics[width=\textwidth]{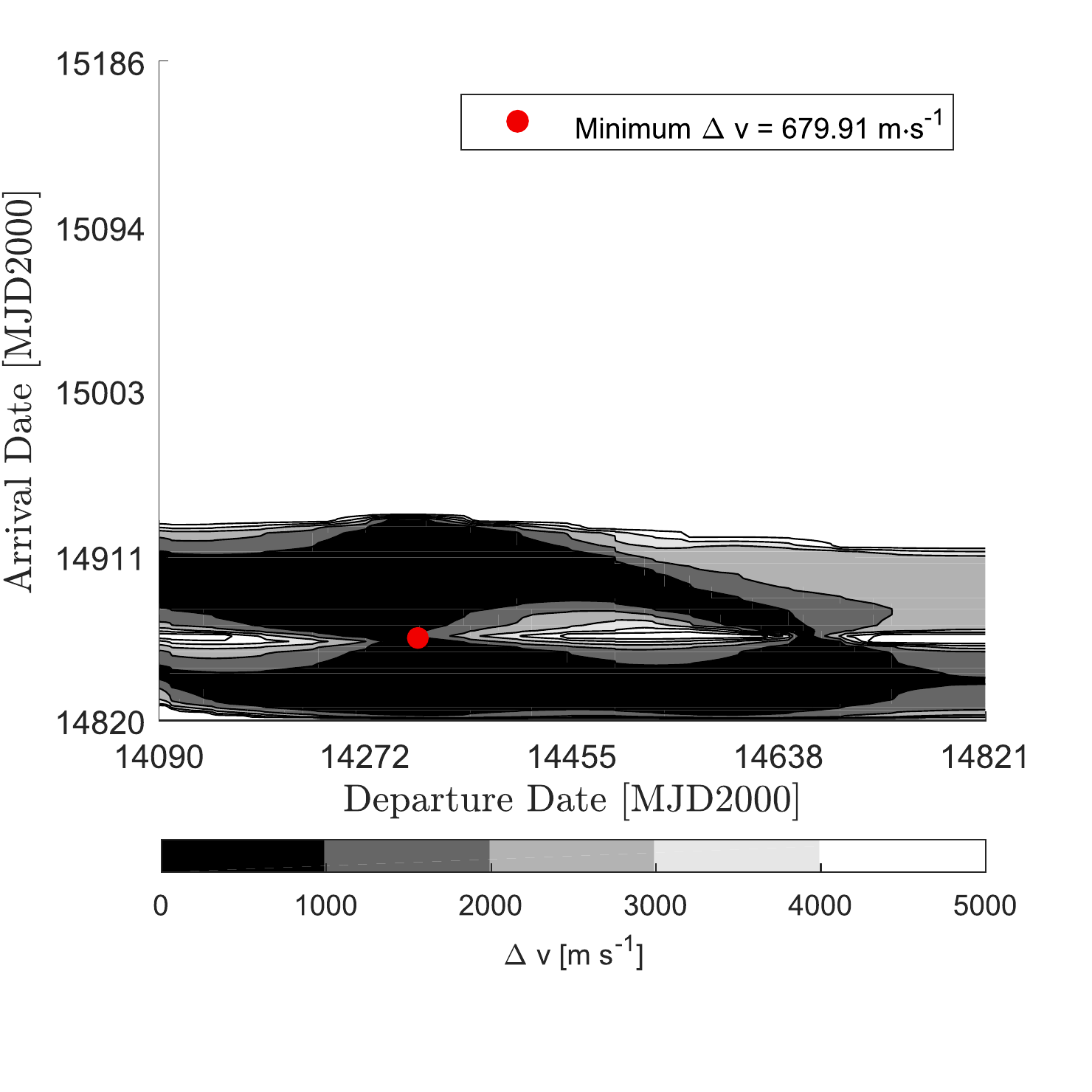}
		\caption*{b) Rendezvous with Asteroid 2017 SV19 in the GVE-3B model}
	\end{minipage}
	\centering
	\begin{minipage}[t]{0.49\linewidth}
		\includegraphics[width=\textwidth]{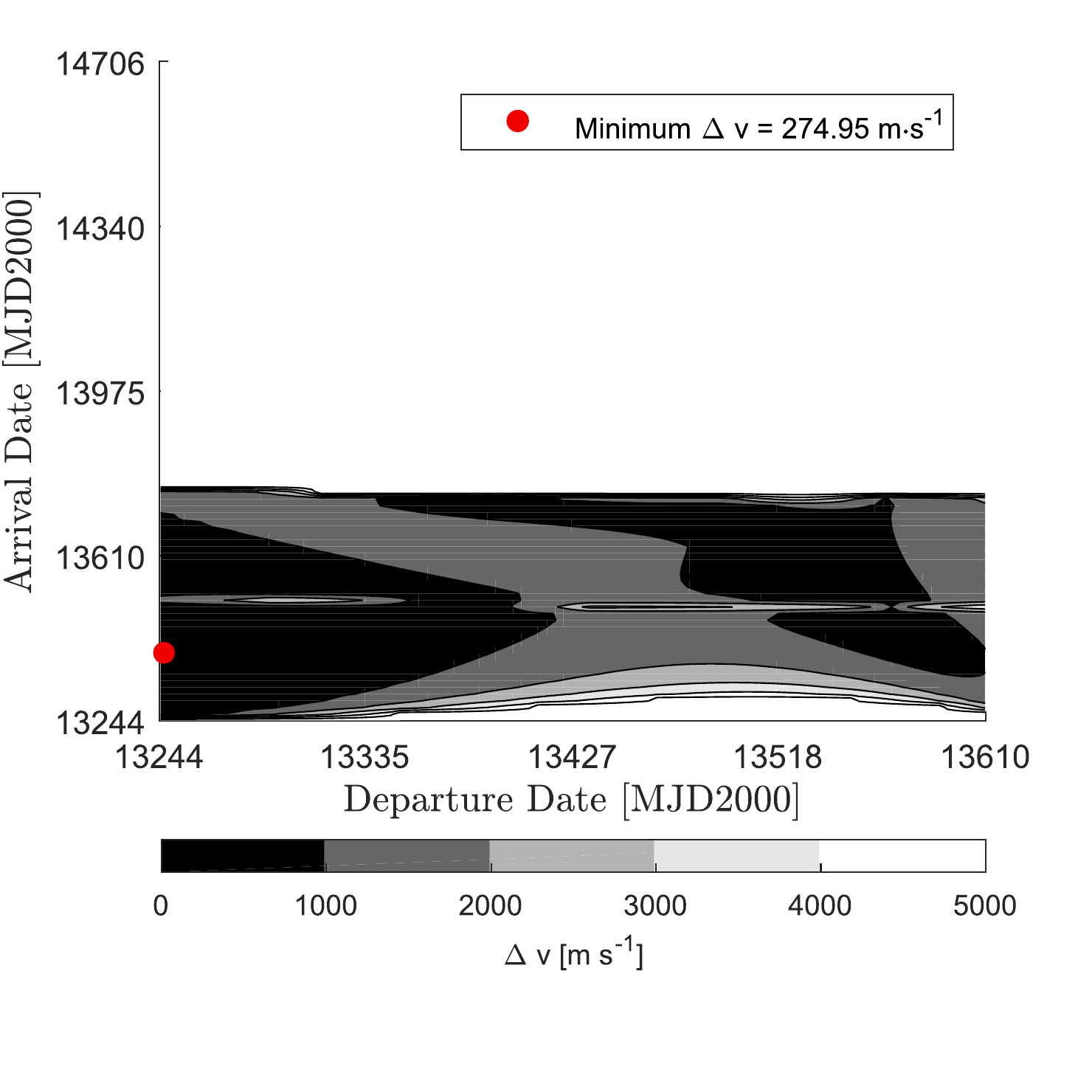}
		\caption*{c) Capture of Asteroid 2018 AV2 in the 2BP model}
	\end{minipage}
	\hfill
	\begin{minipage}[t]{0.49\linewidth}
		\includegraphics[width=\textwidth]{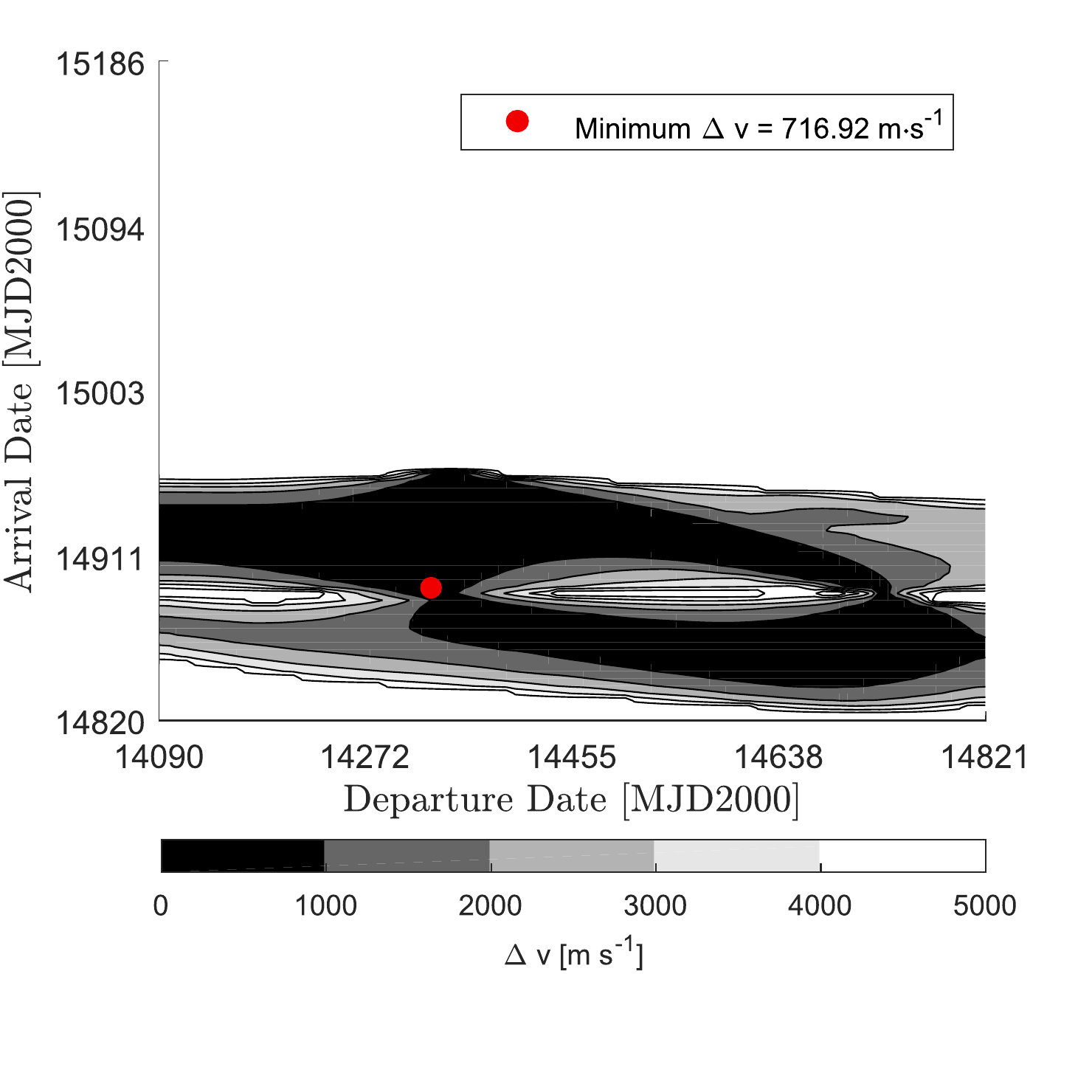}
		\caption*{d) Rendezvous with Asteroid 2017 SV19 in the 2BP model}
	\end{minipage}
	\caption{Contour plots of the $\Delta v$ cost as a function of departure and arrival transfer dates\label{fig:PC}}
\end{figure}

\textcolor{External}{In short, the initial asteroid positions were retrieved from the Horizons JPL database\footnote{https://ssd.jpl.nasa.gov/?horizons, Accessed 01-12-2018}. In this step of the process, the Lambert arcs connecting A and B are computed in the 2BP. The propagations from $\mathrm{A_0}$ to A and $\mathrm{B_0}$ to B are done either in the GVE-3B model or in the 2BP, so as to further highlight their different results. The contour plots indicating the transfer $\Delta v$ cost as a function of the initial and final manoeuvre dates can be found on Figure \ref{fig:PC}: images a) and b) depict the capture and rendezvous cases in which the propagations from $\mathrm{A_0}$ to A and $\mathrm{B_0}$ to B are done in the GVE-3B framework, while c) and d) show the computation using the 2BP.} 

The initial conditions for asteroids 2018 AV2 and 2017 SV19 were taken to be April, 2036 ($\mathrm{t_0}$ of Figures \ref{fig:caperend} a) and c)) and July, 2041 ($\mathrm{t_3}$ of Figures \ref{fig:caperend} b) and d)). \textcolor{External}{The minimum and maximum transfer durations are set to zero and three years, respectively (the former condition being obviously infeasible, but being employed to establish the limit). The departing times range from the initial condition dates up to one year after these.} In total, 10,000 Lambert arcs were computed, with 100 departure and 100 arrival dates. The optimal $\Delta v$ manoeuvres are marked with a red circle; in the case of the GVE-3B framework propagation, the capture trajectory yields a cost of about 255 m$\cdot$s$^{-1}$, while the rendezvous requires 680 m$\cdot$s$^{-1}$. The optimal values obtained with the 2BP propagation are slightly different, \textcolor{Internal}{while} the optimal dates obtained in the two models are quite distinct. For instance, the optimal departure and arrival dates of the GVE-3B rendezvous porkchop yield, in the 2BP plot, a $\Delta v$ of 1502 m$\cdot$s$^{-1}$. This demonstrates that ignoring the Earth's influence in these trajectories could clearly yield sub-optimal transfers.

\subsubsection{Step 2: Sims-Flanagan Approach}

\begin{figure}[h]
	\centering
	\includegraphics[width=0.8\linewidth]{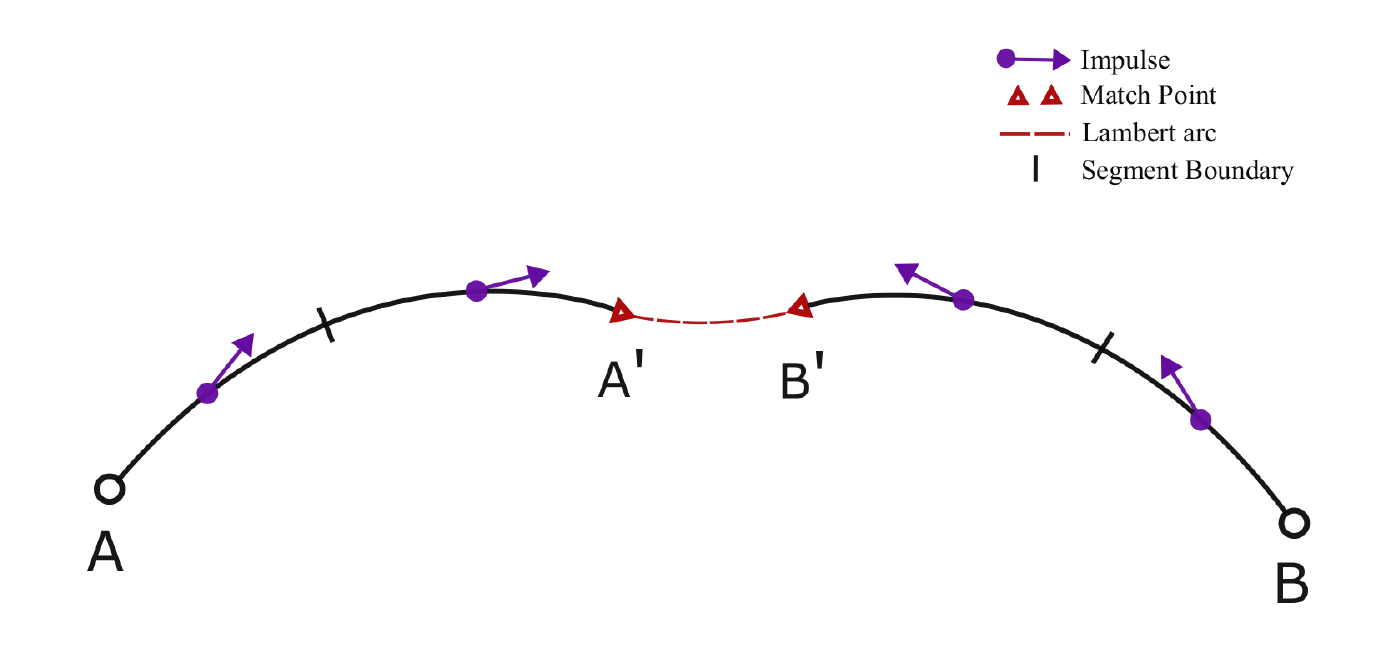}
	\caption{\label{fig:SF}Sims-Flanagan trajectory scheme}
\end{figure}

The Sims-Flanagan method has been extensively used to further refine impulsive thrust first guesses into low-thrust trajectories \cite{sims}. This approach consists on adding several short impulses along the path, making it more similar to a low-thrust trajectory and, thus, more likely for an optimal control solver to converge easily \cite{hargraves}. The trajectory is divided into segments \textcolor{External}{with pre-defined durations}, as it can be seen on the adapted Sims-Flanagan schematic of Figure \ref{fig:SF}. In the middle of each segment, a small impulsive manoeuvre is implemented. Each transfer (called a leg, defined from point A to B) is propagated forward and backward to a match point, located usually halfway through the leg. \textcolor{External}{This process can be further optimised by not fixing each segments' duration, at the cost of a more complex optimisation problem.}

In other applications of the method, the problem is inserted into an optimiser with a constraint on the match point distance \cite{sims2}. In this work, the match points are connected with a Lambert arc, and a genetic algorithm is used to minimise the sum of each segments' $\Delta v$. As such, the method is merely inspired on the Sims-Flanagan approach, with the named fundamental differences. 

The application of the Sims-Flanagan method can be further detailed in the following manner: first, points A and B are obtained from the optimal impulsive trajectory found via the porkchop plot. \textcolor{External}{One thing to look out for when defining these transfer conditions is not to choose points A and B with very close ephemerides---the connecting Lambert arc would become very short, leading to an increased sensitivity of the $\Delta v$ parameter in the optimisation problem.} Then, a genetic algorithm is employed: \textcolor{External}{it is part of the MATLAB code suite \cite{matlab_ga} and was implemented using an initial population of 1000 parameters, with the same number of maximum generations to convergence.} Following the characterisation of a typical optimisation problem, the design variable is the array of $\Delta v$'s that represent the mid-segment impulses (purple arrows of Figure \ref{fig:SF}). The objective function propagates the motion forward from A and backwards from B using the GVE-3B model, implementing the impulsive $\Delta v$'s in each segment. This yields match points A' and B', which are then connected using a Lambert arc. Finally, the cost function can be computed as the sum of this connecting Lambert arc $\Delta v$ with the $\Delta v$'s provided by the design variable array. 

\subsubsection{Step 3: Continuous Low-Thrust Transfer}

After obtaining the Sims-Flanagan solution, a low-thrust trajectory can finally be determined. Since the guiding motivation is to showcase the capabilities of the GVE-3B framework, the optimal control problem was simplified in a way that only the acceleration is optimised, while assuming a constant mass. Thus, the controlled trajectory design is computed in the following manner:

\begin{enumerate}
	\item Each impulsive $\Delta v$ from the Sims-Flanagan approach is converted into continuous accelerations using the formula $\bm{a} = \frac{\bm{v}_1 - \bm{v}_0}{\Delta t}$, \textcolor{External}{in which $\bm{v}_0$ is the velocity at the beginning of the segment, $\bm{v}_1$ is the velocity at the end of the segment and $\Delta t$ is its time duration.}
	\item A is propagated forwards and B backwards, using the respective accelerations for each segment. The final points of the trajectory will be the new match points A'' and B'', \textcolor{Internal}{different from the previous A' and B' since they are achieved with continuous accelerations}.
	\item A'' and B'' are linked using the multiple shooting method \textcolor{External}{shown in Section \ref{sec:MSM} using the GVE-3B model, which} changes the thrusting accelerations to match the initial and final states, achieving a fully connected trajectory.
\end{enumerate}

A multiple shooting method \textcolor{External}{is} used to connect two ephemerides by computing several intermediate points in between them and, finally, linking them with individual segments. \textcolor{External}{This multiple shooting method was first described in Section \ref{sec:MSM}:} the solution to this scheme involves iterating through Eq. \eqref{eq:lsm} until the error ($\text{\textbf{G}}(\bm{X}^{i+1})$) is smaller than a convergence tolerance ($\epsilon$).

The convergence tolerance was chosen to be $\epsilon = 10^{-8}$. In this particular scenario of a multiple shooting scheme with pre-determined initial and final points and free accelerations, the beginning and ending position and velocities are fixed. No acceleration constraints are imposed for any of the patch points. 

The particular matrices required for this multiple shooting scheme are described by Eqs. \eqref{eq:J1} to \eqref{eq:J2}, taking into consideration the Jacobian of the GVE-3B framework in Eq. \ref{eq:jac_gve}. This is here represented as taking into account the system's accelerations, yielding a $9\times9$ \textcolor{Internal}{matrix}. If these are not considered, the matrix may be reduced to a $6\times6$ one.
\begin{align}\label{eq:jac_gve}
\text{\textbf{D}}f(\bm{s})_{3B} &= \begin{bmatrix}
& & & & & & & & \\
& \text{\textbf{0}}_{3\text{x}3} & & & \text{\textbf{I}}_{3\text{x}3} & & & \text{\textbf{0}}_{3\text{x}3} &\\
& & & & & & & & \\
& & & & & & & & \\
& \text{\textbf{R}}_I^{\Earth^{-1}} \text{\textbf{A}}_{3B} & & & \text{\textbf{0}}_{3\text{x}3} & & &\text{\textbf{I}}_{3\text{x}3} &\\
& & & & & & & & \\
& & & & & & & & \\
& \text{\textbf{0}}_{3\text{x}3} & & & \text{\textbf{0}}_{3\text{x}3} & & & \text{\textbf{0}}_{3\text{x}3} &\\
& & & & & & & & \\
\end{bmatrix}\text{, where }
\text{\textbf{A}}_{3B} = \begin{bmatrix}
\textcolor{External}{\partial_x a_x} & \textcolor{External}{\partial_y a_x} & \textcolor{External}{\partial_z a_x}\\
\textcolor{External}{\partial_x a_y} & \textcolor{External}{\partial_y a_y} & \textcolor{External}{\partial_z a_y}\\
\textcolor{External}{\partial_x a_z} & \textcolor{External}{\partial_y a_z} & \textcolor{External}{\partial_z a_z}\\
\end{bmatrix}
\end{align}
\textcolor{External}{where $\partial_\beta a_{\alpha}$ ($\alpha \in \{x, y, z\} \text{ and }\beta \in \{x, y, z\}$) are the first partial derivatives of the accelerations $\{a_x, a_y, a_z\}$ presented in Eq. \eqref{eq:acc}}. These are expanded into the following:
\begin{align}\nonumber
\partial_x a_{x} &= \mu\bigg(\frac{3(x - 1)^2}{r_3^5} - \frac{1}{r_3^3} - \frac{15 x^3}{r^7} + \frac{9x - 3x^2}{r^5} + \frac{1}{r^3}\bigg)\\
\partial_y a_{x} &= - 3 y\mu \bigg(\frac{1 - x}{r_3^5} + \frac{5 x^2}{r^7} - \frac{1- x}{r^5}\bigg)\nonumber\\
\partial_z a_{x} &= -3 z \mu \bigg(\frac{1 - x}{r_3^5} + \frac{5 x^2}{r^7} - \frac{1- x}{r^5}\bigg)\nonumber\\
\partial_x a_{y} &= y\mu \bigg(\frac{3(x - 1)}{r_3^5} + \frac{3 + 2x}{r^5} - \frac{5x(3x + r^2)}{r^7}\bigg)\nonumber\\
\partial_y a_{y} &= \mu\bigg(\frac{3x + r^2}{r^5} -\frac{1}{r_3^3} + 3 y^2 \Big(\frac{1}{r_3^5} - \frac{5x + r^2}{r^7}\Big)\bigg)\nonumber\\
\partial_z a_{y} &= 3 y z \mu \bigg(\frac{1}{r_3^5} - \frac{5x + r^2)}{r^7}\bigg)\nonumber\\
\partial_x a_{z} &= z\mu \bigg(\frac{3(x - 1)}{r_3^5} + \frac{3 + 2x}{r^5} - \frac{5x(3x + r^2)}{r^7}\bigg)\nonumber\\
\partial_y a_{z} &= 3 y z \mu \bigg(\frac{1}{r_3^5} - \frac{5x + r^2)}{r^7}\bigg)\nonumber\\
\partial_z a_{z} &= \mu\bigg(\frac{3x + r^2}{r^5} - \frac{1}{r_3^3}  + 3 z^2 \Big(\frac{1}{r_3^5} - \frac{5x + r^2}{r^7}\Big)\bigg)
\end{align}

The number of patch points for the multiple shooting method was chosen to be 3. This can clearly be seen on the Jacobian of the constraint vector in Eq. \eqref{eq:J1}, where the first 3 conditions represent the patching requirements and the remaining ones establish that \textcolor{Internal}{A'' and B''} are fixed. This number was chosen in order to simulate the conditions of an on-off control, which will be later validated.

\begin{align}\label{eq:J1}
\text{\textbf{DG}}(\bm{X}) &= 
\begin{bmatrix}
\bm{\Phi}_{1, 6\text{x}9} & -\text{\textbf{IO}}_{6\text{x}9} & \text{\textbf{0}}_{6\text{x}9} & \text{\textbf{0}}_{6\text{x}9}\\
\text{\textbf{0}}_{6\text{x}9} & \bm{\Phi}_{2, 6\text{x}9} & -\text{\textbf{IO}}_{6\text{x}9} & \text{\textbf{0}}_{6\text{x}9}\\
\text{\textbf{0}}_{6\text{x}9} & \text{\textbf{0}}_{6\text{x}9} & \bm{\Phi}_{3, 6\text{x}9} & -\text{\textbf{IO}}_{6\text{x}9}\\
\text{\textbf{IO}}_{6\text{x}9} & \text{\textbf{0}}_{6\text{x}9} & \text{\textbf{0}}_{6\text{x}9} & \text{\textbf{0}}_{6\text{x}9}\\
\text{\textbf{0}}_{6\text{x}9} & \text{\textbf{0}}_{6\text{x}9} & \text{\textbf{0}}_{6\text{x}9} & \text{\textbf{IO}}_{6\text{x}9}\\
\end{bmatrix}, \text{  } 
\text{\textbf{G}}(\bm{X}) = 
\begin{bmatrix}
\bm{s}_1(t_1 + T_1) - \bm{s}_2(t_2) \\
\bm{s}_2(t_2 + T_2) - \bm{s}_3(t_3)\\
\bm{s}_3(t_3 + T_3) - \bm{s}_4(t_4)\\
\bm{s}_1(t_1) - \bm{A''}\\
\bm{s}_{4}(t_4) - \bm{B''}\\
\end{bmatrix}\\\nonumber\\
\label{eq:J2}
\text{\textbf{IO}} &= \begin{bmatrix}
1 & 0 & 0 & 0 & 0 & 0 & 0 & 0 & 0 \\
0 & 1 & 0 & 0 & 0 & 0 & 0 & 0 & 0 \\
0 & 0 & 1 & 0 & 0 & 0 & 0 & 0 & 0 \\
0 & 0 & 0 & 1 & 0 & 0 & 0 & 0 & 0 \\
0 & 0 & 0 & 0 & 1 & 0 & 0 & 0 & 0 \\
0 & 0 & 0 & 0 & 0 & 1 & 0 & 0 & 0 \\
\end{bmatrix}
\end{align}

\section{Results and Discussion}

\begin{figure}[p!b]
	\centering
	\begin{minipage}[t]{0.424\linewidth}
		\includegraphics[width=\textwidth]{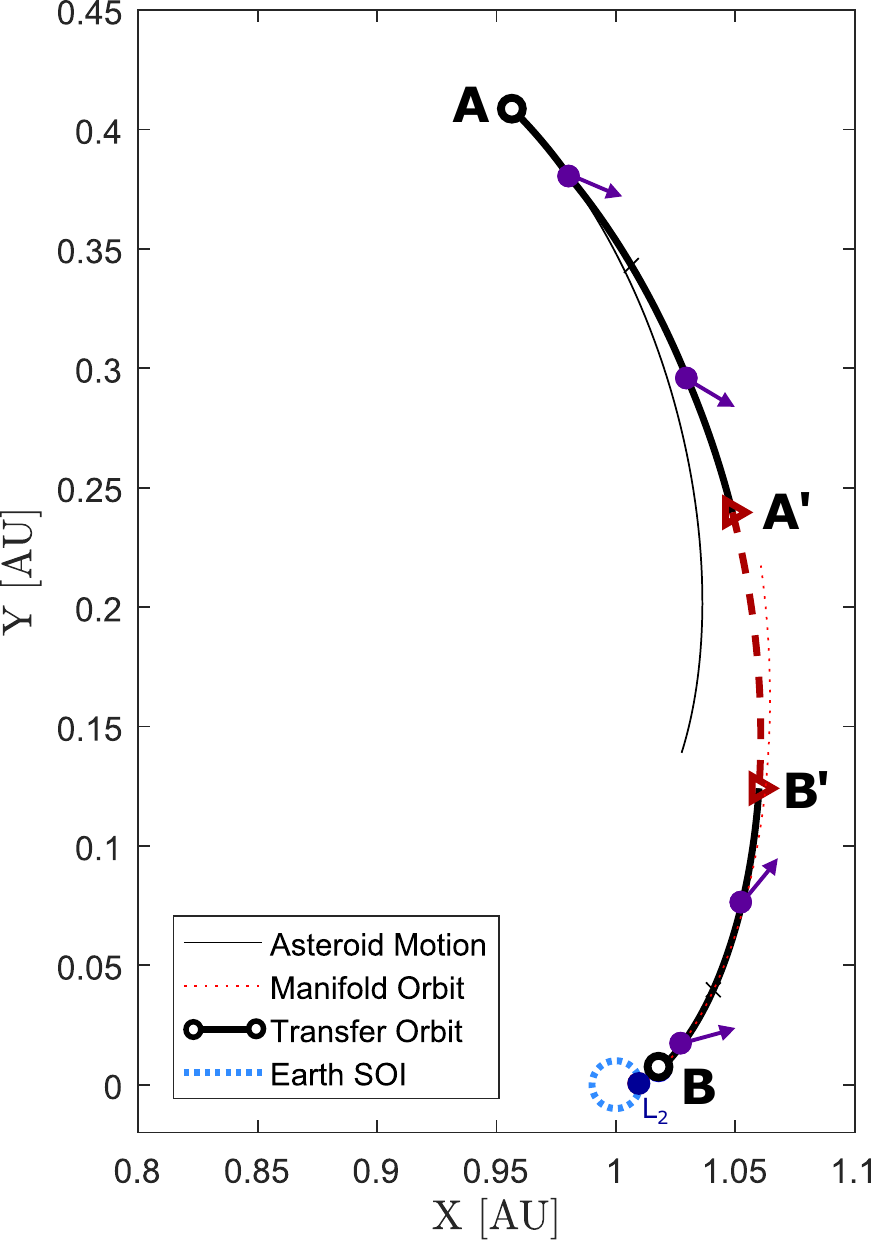}
		\caption*{a) Capture of Asteroid 2018 AV2}
	\end{minipage}
	\hfill
	\begin{minipage}[t]{0.42\linewidth}
		\includegraphics[width=\textwidth]{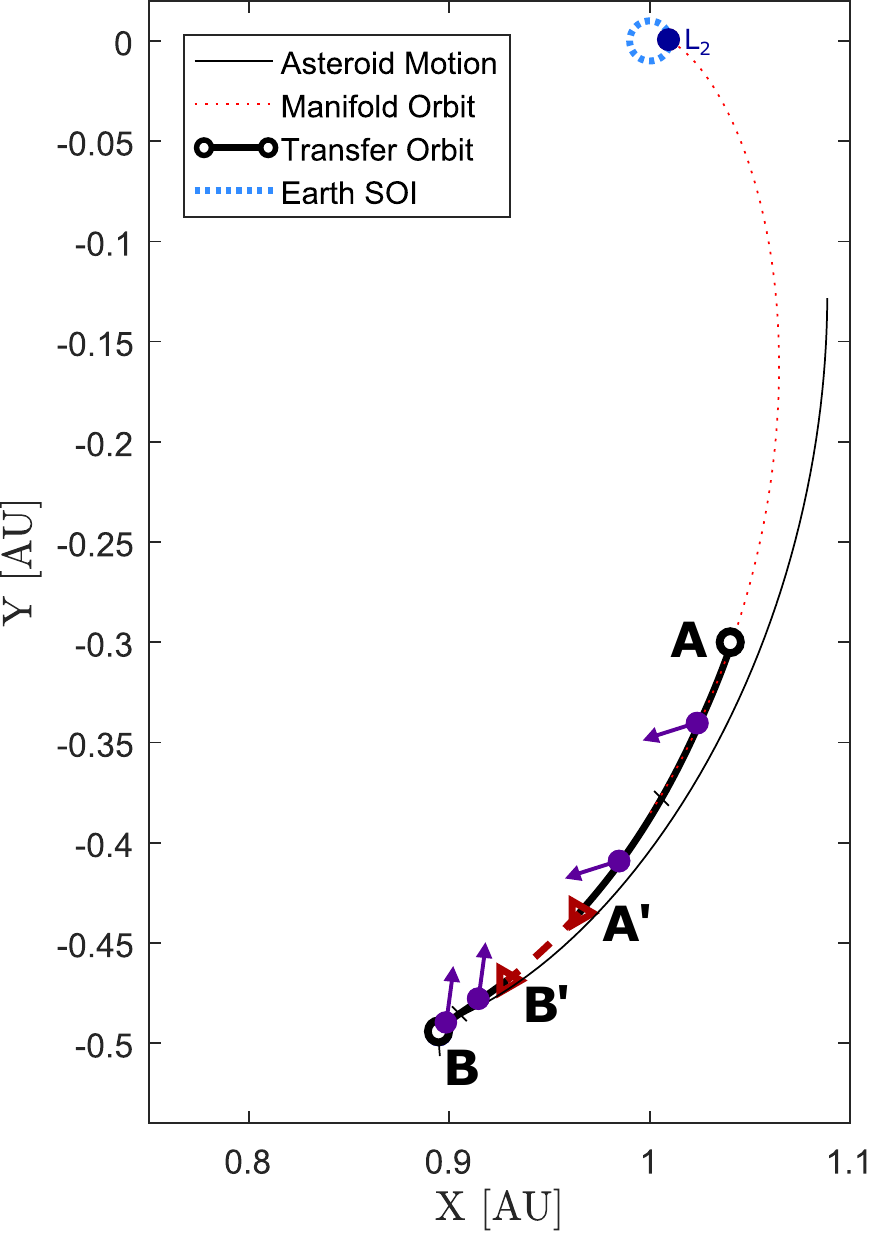}
		\caption*{b) Rendezvous with Asteroid 2017 SV19}
	\end{minipage}
	\caption{Trajectory segments in the Sims-Flanagan approach\label{fig:SFRes}}
	\vspace{4mm}
	\begin{minipage}[t]{0.42\linewidth}
		\includegraphics[width=\textwidth]{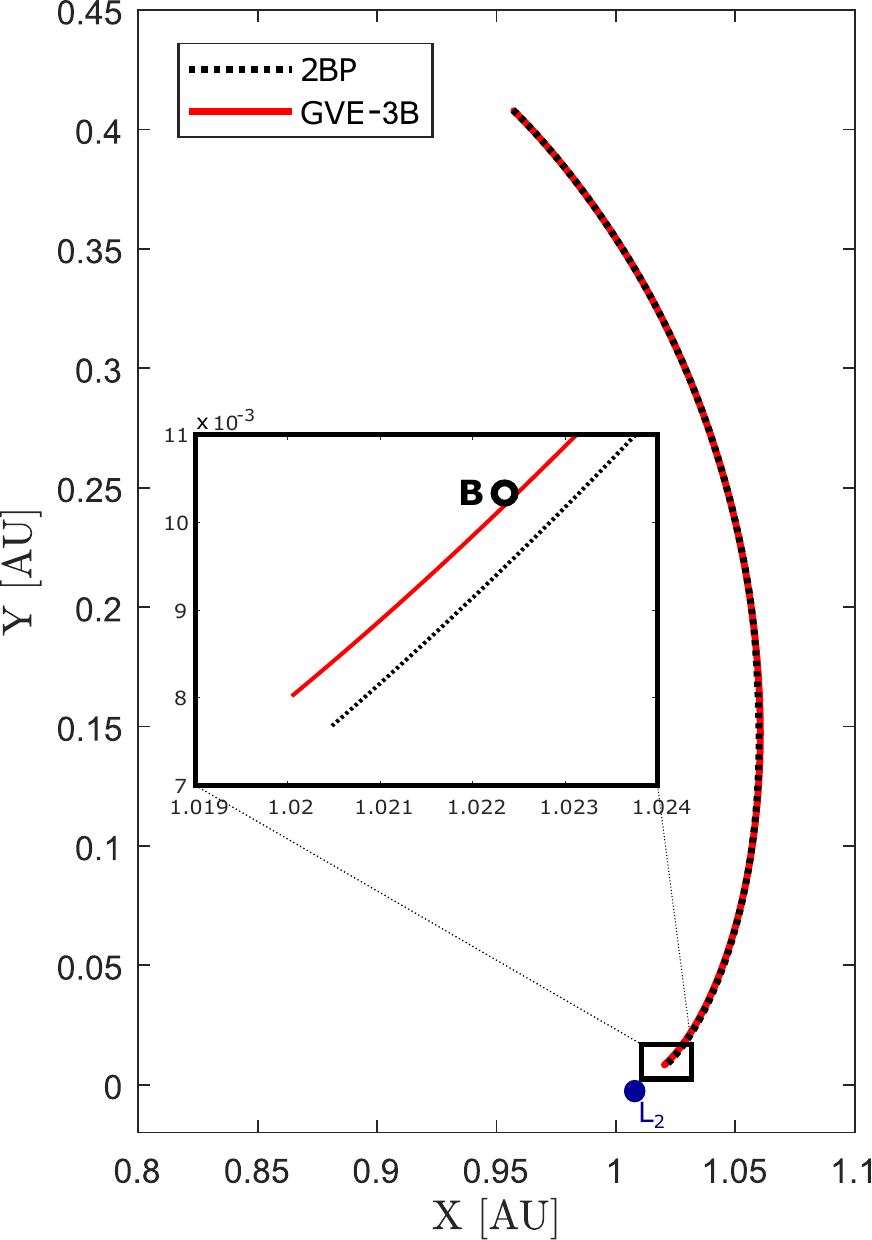}
		\caption*{a) Capture of Asteroid 2018 AV2}
	\end{minipage}
	\hfill
	\begin{minipage}[t]{0.422\linewidth}
		\includegraphics[width=\textwidth]{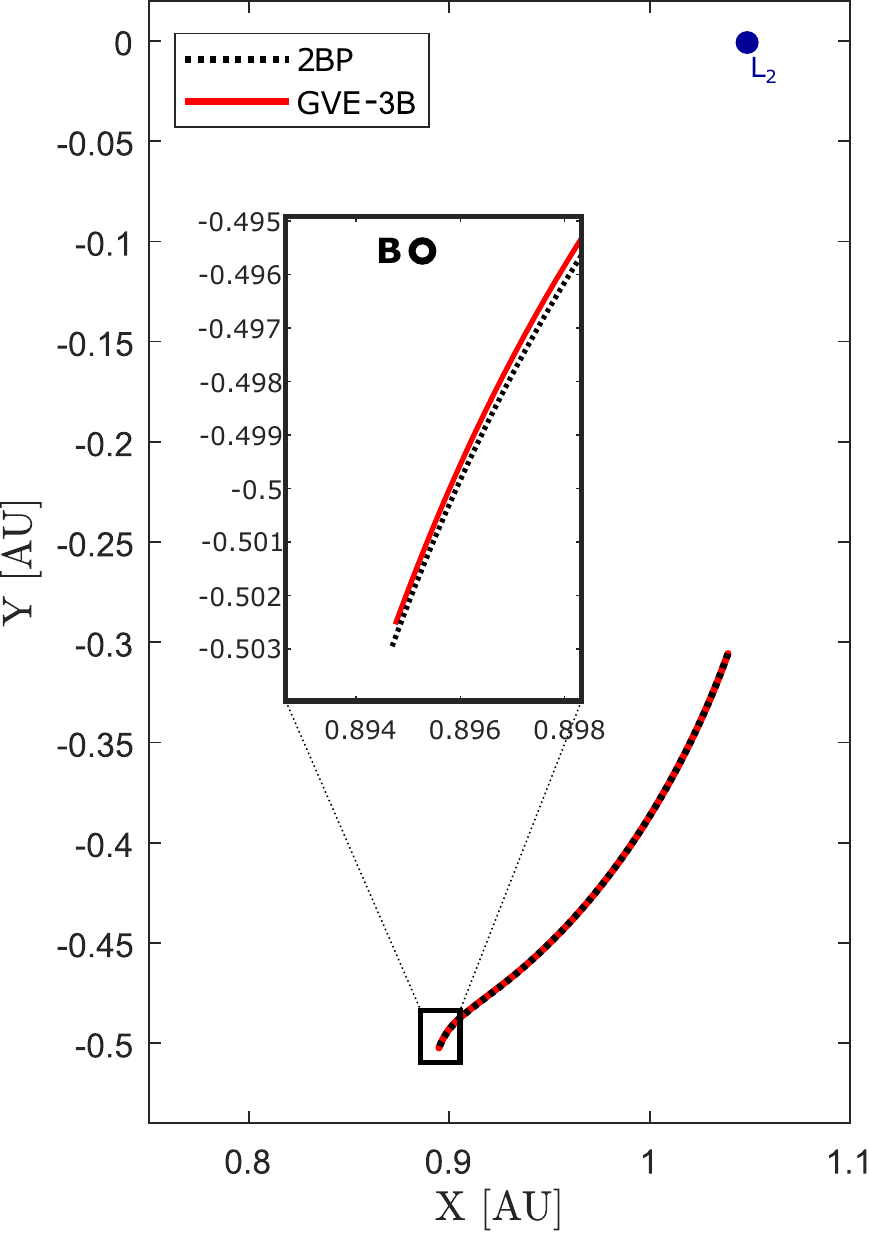}
		\caption*{b) Rendezvous with Asteroid 2017 SV19}
	\end{minipage}
	\caption{Final \textcolor{External}{trajectories with control laws determined with the GVE-3B framework (in red) and with the 2BP (in black), propagated with the CR3BP}\label{control_law}}
\end{figure}

\textcolor{External}{After determining the ephemerides of points A and B for both mission scenarios, the Sims-Flanagan approach was implemented using 4 trajectory segments,} having the extra Lambert arc to link points A' and B'. The total $\Delta v$ for the optimal solutions was 380 m$\cdot$s$^{-1}$ and 875 m$\cdot$s$^{-1}$, respectively for the capture and rendezvous cases. The obtained trajectory can be found in Figure \ref{fig:SFRes}. Figure \ref{fig:SFRes} a) depicts the capture transfer: starting from point A in the nominal motion of 2018 AV2 and finishing at point B of the invariant manifold orbit, the purple arrows indicate the small manoeuvres of the Sims-Flanagan approach, with the red dashed line outlining the Lambert arc that connects the patch points A' and B'. Figure \ref{fig:SFRes} b) contains the rendezvous trajectory from point A in the invariant manifold orbit to point B of the nominal motion of 2017 SV19, using the same notation. 

The low-thrust design is computed as detailed in Section \ref{sub:ltdesign}. Two control laws (sequence of accelerations and times to implement them) were devised, using the Sims-Flanagan approach and the multiple shooting method. One is computed entirely using the GVE-3B framework. The other is modelled with the same multiple shooting method, but using the 2BP model. Figure \ref{control_law} shows the propagation of both control laws using the CR3BP model. 

By comparing the propagations, done with the CR3BP, of the GVE-3B and 2BP control laws, one can see that the two trajectories finish at distinct end-points. It is noteworthy that the desired target position is not reached, which is particularly evident in the asteroid rendezvous case. Considering that this scenario demands a much higher amount of $\Delta v$ in a considerably shorter mission time, the translation from impulsive to low-thrust trajectory is more prone to accumulating errors. \textcolor{External}{In the capture case, the minimum distance to target for the trajectory computed with the GVE-3B control law is of about 20,000 km, while the one calculated in the 2BP is about four times that length. Similarly, the minimum distance in the rendezvous case computed with the 2BP control law exceeds the one of the GVE-3B by over 10,000 km. While it is clear that the GVE-3B control law performs much better with respect to the target, there is likely a timing problem in the trajectory propagation.}

\begin{figure}[ht!]
	\centering
	\begin{minipage}[t]{0.48\linewidth}
		\includegraphics[width=\textwidth]{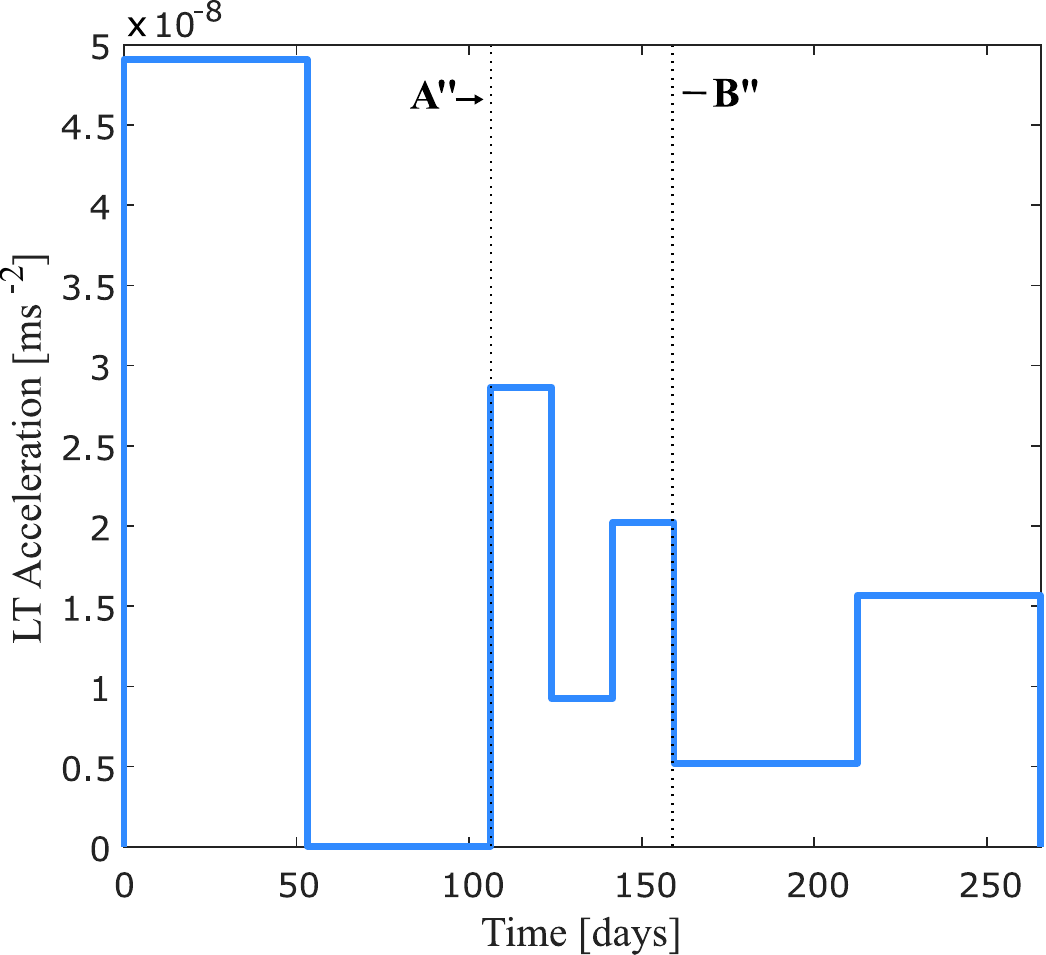}
		\caption*{a) Capture of Asteroid 2018 AV2}
	\end{minipage}
	\hfill
	\begin{minipage}[t]{0.48\linewidth}
		\includegraphics[width=\textwidth]{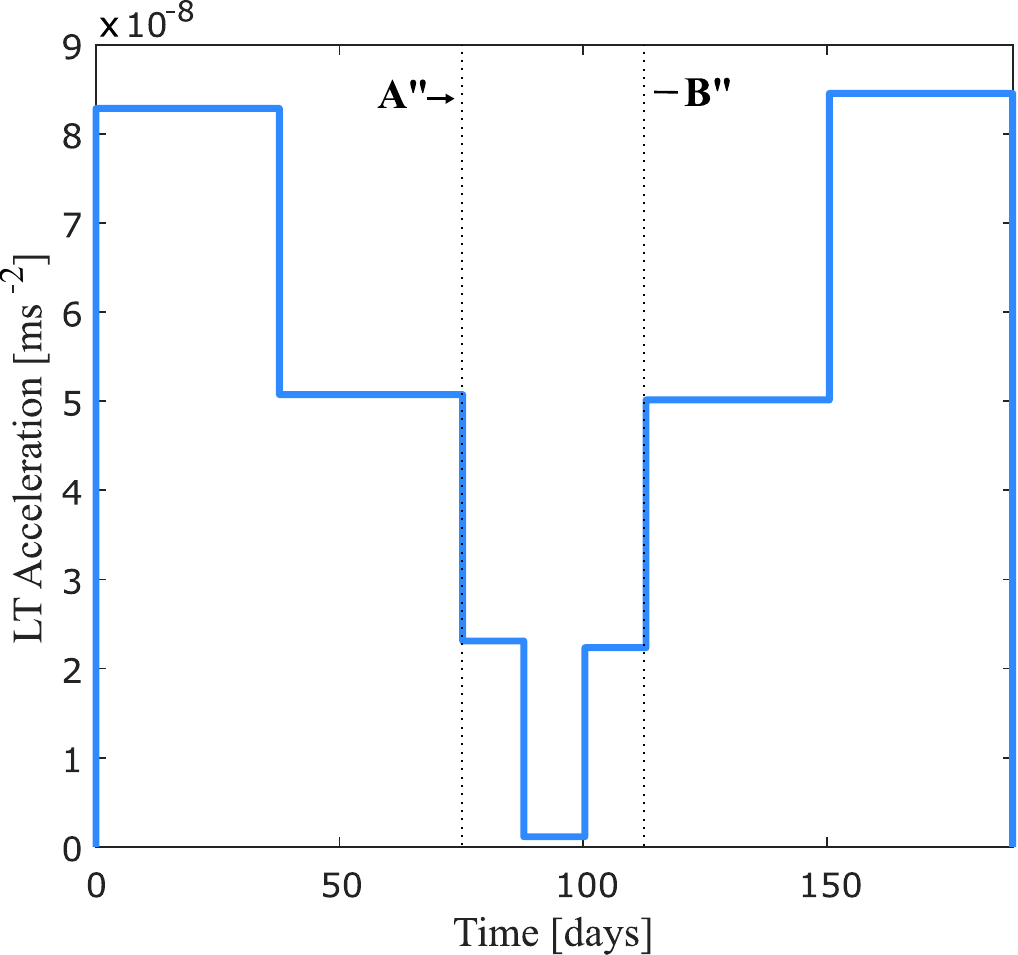}
		\caption*{b) Rendezvous with Asteroid 2017 SV19}
	\end{minipage}
	\caption{Control history for the GVE-3B model, as depicted in Figure \ref{control_law}\label{fig:cl_time}}
\end{figure}

A different analysis involves considering \textcolor{External}{solely trajectories propagated using the GVE-3B control law.} The corresponding control history is detailed in Figure \ref{fig:cl_time}, where the different trajectory phases can be distinguished. The solution is close to a on-off control pattern, but there is a clear disparity in between phases. Future implementations could have the multiple shooter be used throughout the entire motion, so that a smoother and more consistent on-off control can be attained.
	
\begin{figure}[ht!]
	\centering
	\begin{minipage}[t]{0.48\linewidth}
		\includegraphics[width=\textwidth]{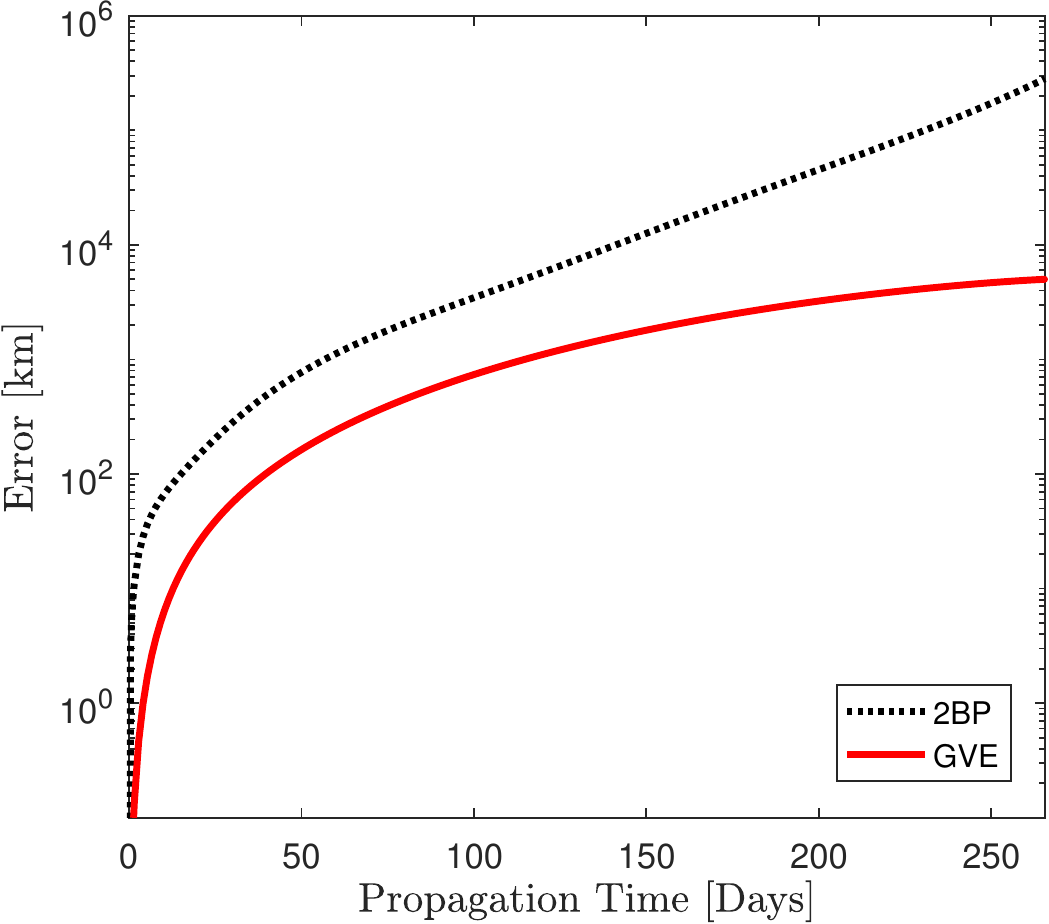}
		\caption*{a) Capture of Asteroid 2018 AV2}
	\end{minipage}
	\hfill
	\begin{minipage}[t]{0.48\linewidth}
		\includegraphics[width=\textwidth]{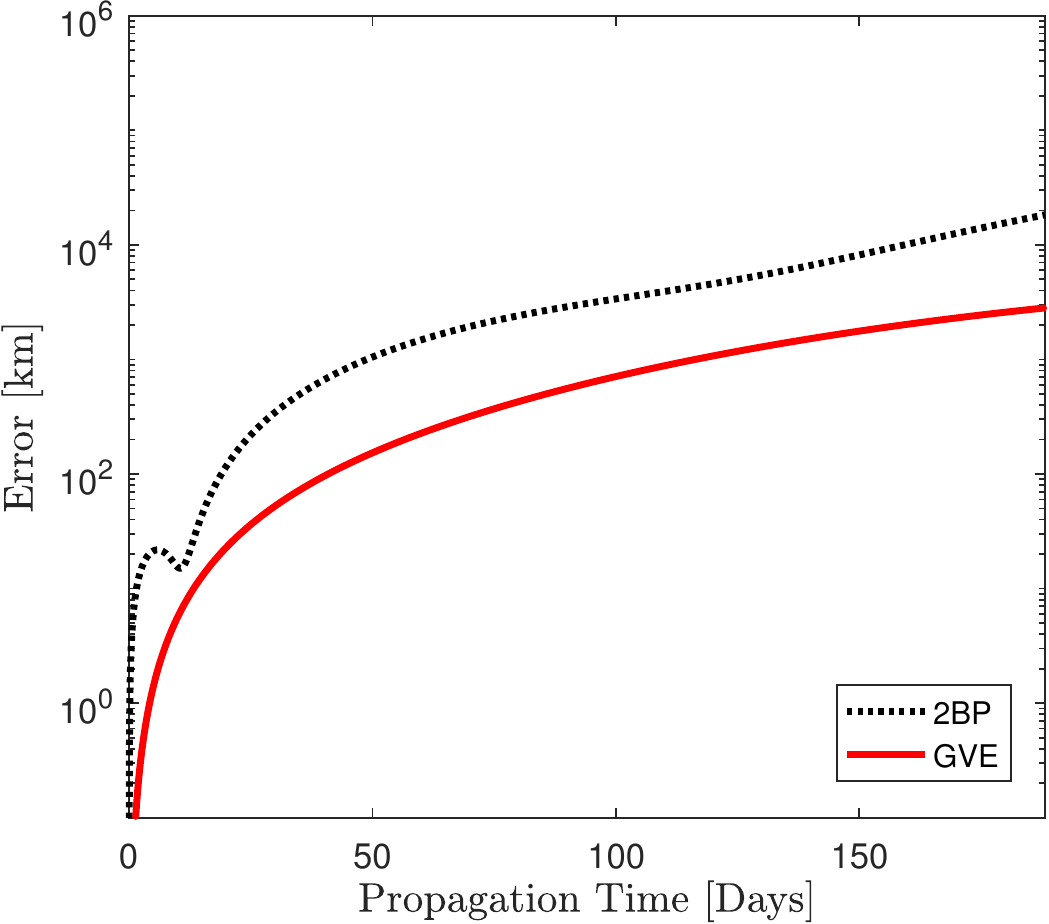}
		\caption*{b) Rendezvous with Asteroid 2017 SV19}
	\end{minipage}
	\caption{\textcolor{External}{Absolute error, with respect to the CR3BP baseline, for the propagation of the 2BP and the GVE-3B. Error corresponding to the trajectory of Figure \ref{control_law} using the GVE-3B control law}\label{fig:cl_compare}}
\end{figure}

\textcolor{External}{Considering only this GVE-3B control law, each mission's trajectory can be propagated using the GVE-3B, the 2BP and the CR3BP models. The absolute errors of the propagations with the GVE-3B and the 2BP are computed, using the CR3BP model as the reference solution.} The results can be found on Figure \ref{fig:cl_compare}: the error yielded by the 2BP propagation is much higher for both mission scenarios, reaching around $10^6$ and $10^4$ km for capture and rendezvous, respectively. Plus, the 2BP model error grows very fast in the capture case, where the motion gets very close to the Earth. The rendezvous trajectory also proves to have a smaller error when using the GVE-3B, despite the fact that the motion starts at a distance of 30 times the Earth's Hill radius. This further demonstrates that the planet's impact on the mission design is noteworthy, even when quite far from its sphere of influence. Therefore, it can be inferred that the GVE-3B framework is more accurate \textcolor{External}{than the 2BP when compared} to the CR3BP. 

\textcolor{Both}{It is now established that, in terms of accuracy, the GVE-3B far surpasses the 2BP. In terms of convergence of the optimal control problem, the situation is similar. A preliminary study using the optimal control solver GPOPS-II was performed for the capture of asteroid 2018 AV2. Under the initial and final time and ephemerides conditions and using the CR3BP model, the solver takes about 50\% more iterations to converge with the 2BP control law than under the GVE control law. The fact that the initial guess is already a fully fledged trajectory also contributes to the easiness in establishing bound and boundary constraints---however, there is clearly still much room for improvement in the search of optimal solutions.}

\section{Summary}
\label{sec:GVEconclusions}

This chapter \textcolor{Internal}{demonstrates the application of a novel formulation of the third-body perturbation, using a GVE framework, to the propagation of trajectories within low-energy regions and subject to low-thrust propulsive accelerations.}

The nature of the GVE-3B equations makes them very advantageous for low-thrust trajectory design. Particularly, the intuitive observation of the orbital elements evolution and the easy definition of boundaries and constraints for the optimal control problem make the GVE-3B framework straightforward to set up and solve. \textcolor{External}{In summary, its usage facilitates the set up and convergence of the optimal control problem solver.} Thus, its utilisation in preliminary mission design can be extremely valuable, providing important insights into the time evolution of the orbital elements. 

The presented framework can be utilised in many different \textcolor{External}{applications for space mission design}. One such concept entails Jovian or Saturnian moon tours---as shown by Alessi and S\'anchez \cite{alessi_semi}, both of these planetary systems have small enough gravitational parameters that the developed equations of motion remain accurate (as opposed to, for example, the Earth-Moon system). However, computing trajectories for moon tours would demand a higher complexity in the mission design, since this would require the integration of the third-body framework with multiple planetary systems.

\chapter{Application: Trajectory Design Framework for Earth-Resonant Trajectories}
\label{chap:ast2}
\textcolor{Internal}{The Solar System is swarmed by millions of minor bodies, including asteroids and comets. Near-Earth asteroids (NEAs), in particular, have been brought into attention because of two important aspects: they are among the easiest celestial bodies to reach from the Earth and may also represent a threat to our planet \cite{RN1}. Furthermore, they are appealing for their potential resources, since they contain a plethora of useful materials \cite{RN4, RN2}.}

Following these trends, asteroid capture or retrieval missions have come under the spotlight. The concept envisages a spacecraft that performs a rendezvous with an asteroid, lassos it and then hauls it back to the Earth's neighbourhood, so that it can be more easily accessed. The mission has clear synergies with all three of the main aspects of asteroid missions: science, planetary defence and mining. However, a high percentage of NEAs have masses several orders of magnitude larger than that of the typical interplanetary spacecraft ($\sim 10^{3}$ kg). As such, for a capture to be feasible, it may have to be done using extremely low-energy and low-cost transfers. 

As discussed in Chapter \ref{chap:intro}, in order to achieve low-energy, fuel efficient transfers, the exploitation of encounters with other bodies in the Solar System has been previously studied, particularly swing-bys of the Moon \cite{O,R} and Earth \cite{RN18}, using the patched conic approximation. However, the most easily retrievable objects \cite{pau_eros} of the NEA population move on such a regime of motion that they rarely get close enough to the Earth for a swing-by manoeuvre. Nevertheless, even if the asteroid undergoes an encounter with the Earth outside its sphere of influence, its gravitational interaction will still be noticeable and may substantially modify the asteroid's trajectory. This was shown in Chapter \ref{chap:k3bp}, with the definition of the perturbation region and the introduction of the Keplerian third-body potential (K3BP) to compute the third-body effect.

This chapter presents a design of retrieval trajectories to transfer asteroids into Sun-Earth libration point orbits (LPOs). The work includes multiple Earth encounters along the capture trajectory, occurring outside the sphere of influence of the Earth in a low-energy regime; then, the optimal control problem is solved to obtain a final low-thrust transfer. The trajectory design is pursued using a layered approach: first, the Keplerian map (KM) method is used to obtain a preliminary solution. Despite being a lower-fidelity model, the KM captures well the underlying dynamics of the Earth perturbation, being a sensible procedure to explore the infinite space of possible trajectories \footnote{The KM has been shown to be a worse performer than the comparable model PAP-KM, in Chapter \ref{chap:k3bp}. However, since the trajectory design case was studied previous to the perturbation methods' analysis, the more established model was here selected.}. Then, a multi-fidelity framework is used to refine the solution from an impulsive approximation, designed in a low-fidelity dynamical model, into a low-thrust transfer in the circular-restricted three-body problem (CR3BP).

\section{\textcolor{Both}{Design of Earth-Resonant Captures}}
\label{sec:design}

This section \textcolor{Both}{describes the} Earth-resonant capture trajectory \textcolor{Both}{that is the final goal of the mission design}, \textcolor{Internal}{bearing in mind that the word \textit{resonant} is here used to convey two consecutive encounters with the Earth.} 

\begin{figure}[bh!]
	\centering
	\includegraphics[width=\textwidth]{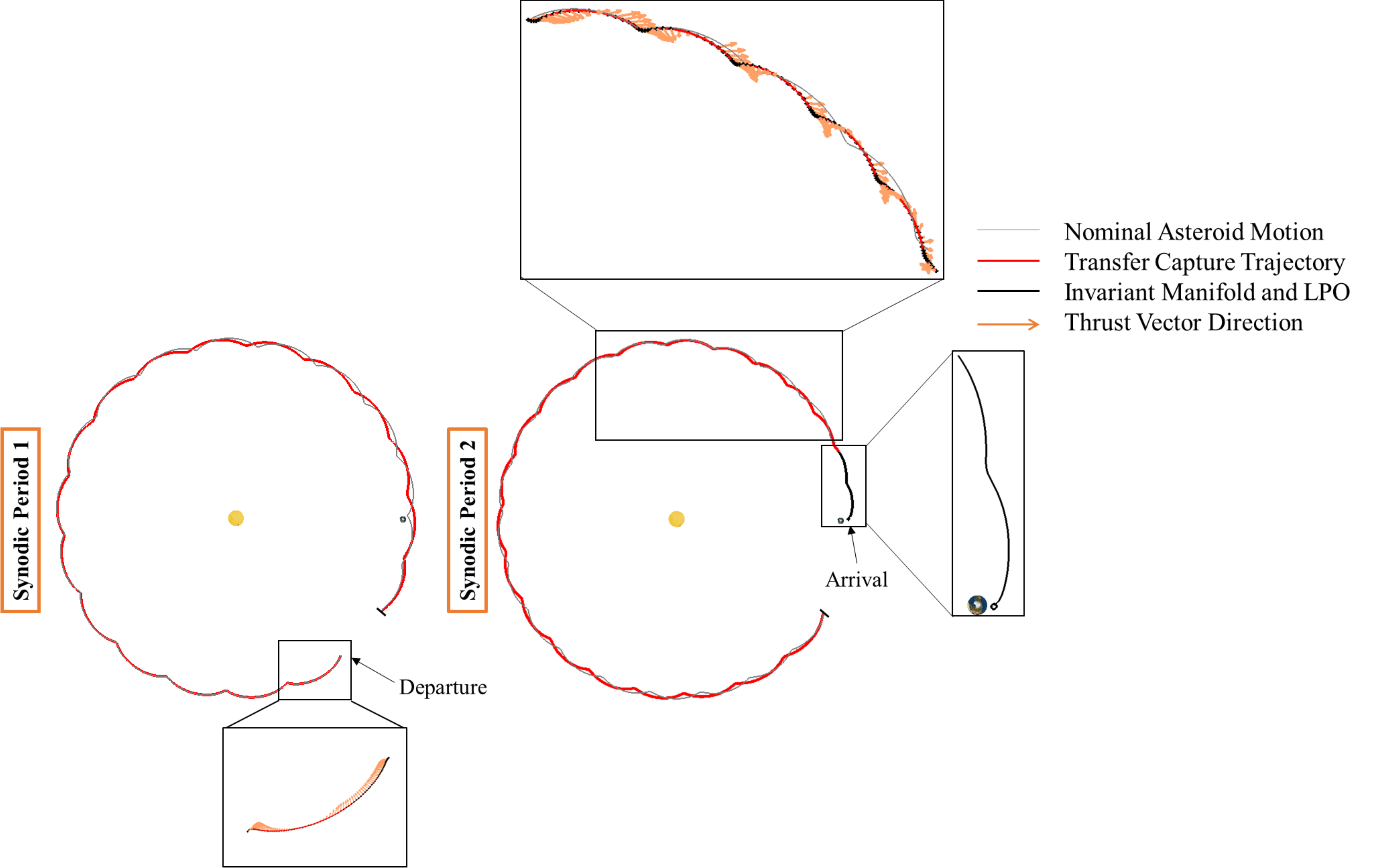}
	\caption{Earth-resonant capture trajectory for asteroid 2011 MD in the synodic reference frame, divided in two sequential synodic periods for clarity. Earth is increased to the size of the Hill radius for better visualisation}\label{fig:entire_traj}
\end{figure}

Figure \ref{fig:entire_traj} illustrates the end product of the multi-fidelity design: a low-thrust resonant Earth capture. \textcolor{Internal}{The two encounters with the Earth, which correspond to the asteroid's closest approaches to the planet, occur outside its} sphere of influence. Their geometry is ultimately defined by a strict relation between the orbital periods of the Earth and the asteroid\footnote{For the sake of simplicity, the system formed by the coupling of the spacecraft with the asteroid will be henceforth referred to as \textit{the asteroid.}}. 

The trajectory arrives to a stable invariant manifold orbit connected to an LPO after two sequential synodic periods, wherein each includes one Earth encounter. Two \textcolor{External}{highlighted} arcs of Figure \ref{fig:entire_traj} depict the thrusting segments (the control vector is coloured orange); for the remaining orbits, the asteroid is coasting.

\textcolor{Both}{It is important to highlight the difference between the two trajectory designs that will be compared from now on: the \textit{Earth-resonant} design and a \textit{direct capture} trajectory. For the latter, the asteroid is moved from its initial orbit directly into the invariant manifold connected to the LPO, without exploiting previous Earth's approaches. As such, a direct capture requires only one manoeuvre: the manifold insertion, henceforth referred to as $\Delta v_I$.}

\section{\textcolor{Both}{Multi-Fidelity Dynamical Framework}}
\label{sec:frame}

In order to choose the dynamical model in which to compute the motion of a body in space, a trade-off has to be done between accuracy and computational cost. Besides, the chosen method has to take into consideration the perturbative accelerations caused by nearby celestial bodies. 

Instead of choosing only one model, a multi-fidelity framework for the asteroid motion in a capture mission scenario is proposed. In this way, a sequential pruning of the NEA population can be made with increasingly higher fidelity models. From stage to stage of the framework, the number of analysed asteroids decreases, as the ones whose capture costs are deemed too high are pruned out. Thus, with adequate accuracy, it is possible to compute an extensive number of trajectories with lower computational cost, and refine only the interesting ones.

As such, the asteroid's capture cost is first estimated using a simple manoeuvre in the two-body problem (2BP). Then, the Earth-resonant trajectory is computed with a low-fidelity model---the KM. Finally, the motion is then refined into the CR3BP. These models were described separately in Chapters \ref{chap:dyn} and \ref{chap:k3bp}, with their computational costs preliminarily compared in Section \ref{sub:ctime}. The framework leaves the possibility of further refinement beyond the CR3BP, using for instance homotopy continuation methods \cite{zamaro}.

\begin{figure}[h!]
	\centering
	\includegraphics[width=1\textwidth]{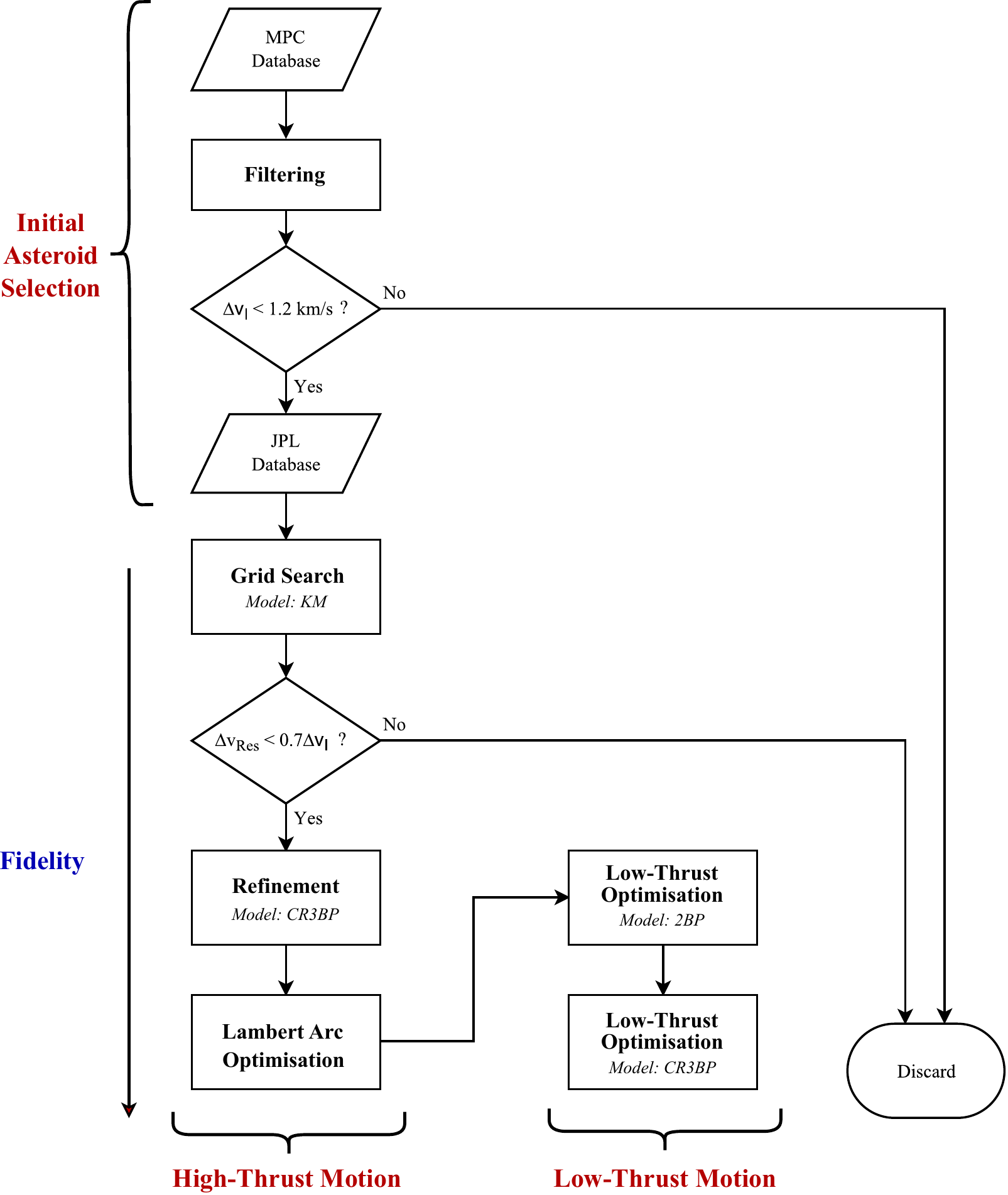}
	\caption{Flowchart detailing the proposed multi-fidelity design of resonant captures}\label{fig:flowchart_mf}
\end{figure}

The entire multi-fidelity dynamical framework, trajectory design and asteroid selection processes can be observed in the flowchart of Figure \ref{fig:flowchart_mf}, from the initial asteroid selection to the multi-fidelity high-thrust and low-thrust trajectory designs. Each of the actions is then explained in its specific section.

Following Figure \ref{fig:flowchart_mf}, the asteroid selection process \textcolor{External}{(\textit{Initial Asteroid Selection})} starts by obtaining each NEA's ephemerides from the Minor Planet Centre (MPC) database\footnote{http://www.minorplanetcenter.net/iau/TheIndex.html, Accessed 01-06-2018}. An estimate of the capture cost $\Delta v_I$ is made using a filter \textcolor{External}{(\textit{Filtering})}, and all asteroids whose capture cost does not comply with an imposed threshold are discarded. Subsequently, detailed orbital parameters are obtained from the Horizons JPL database\footnote{https://ssd.jpl.nasa.gov/?horizons, Accessed 01-06-2018} to start the multi-fidelity high-thrust design, as expanded in Section \ref{sub:ht}. First, the computation of the resonant capture is made for the current asteroid pool, using the KM as the model of motion \textcolor{External}{(\textit{Grid Search})}. Then, the trajectory for the asteroids that were not pruned out is refined in the CR3BP \textcolor{External}{(\textit{Refinement})}. Lastly, the filter estimate of the capture cost $\Delta v_I$ is replaced by a \textcolor{External}{\textit{Lambert Arc Optimisation}}. The obtained trajectory is used as a first guess for the \textcolor{External}{\textit{Low-Thrust Optimisation}}.

\subsection{\textcolor{Both}{Initial Asteroid Selection}}
\label{sub:pruning}

In order to select the asteroids for which the proposed Earth-resonant trajectory would show the greatest advantages, the entire population of registered NEAs was considered for the initial asteroid pool. Their orbital elements were first obtained from the MPC website. 

From this pool, the targets deemed unfit were pruned out. In order to do so, a filter was employed, \textcolor{External}{which is described later in this section.} It computes the $\Delta v$ cost of hypothetical bi-impulsive transfers linking the asteroid's orbit to the invariant manifolds connected to LPOs of the $L_1$ and $L_2$ points. \textcolor{External}{The filter returns the minimum cost and corresponding orbit out of several possible $\Delta v$'s, depending on energy level, type of LPO and libration point of the target trajectory.} 

The filter has been shown to act as a lower threshold to a Lambert arc capture cost, while still being a good approximation for it \cite{pau_retrieval}. Thus, its main purpose is to work as a fast decision-maker to rule out clearly unfit targets. As indicated by the flowchart of Figure \ref{fig:flowchart_mf}, it is only used in a preliminary stage, being later replaced by a Lambert arc optimisation.

\textcolor{External}{A rough limit of $\Delta v_I = $1.2 km$\cdot$s$^{-1}$ was imposed as the maximum estimated capture cost,} as it left already a large pool of asteroids for study, while pruning out those whose capture fuel would clearly be too high for a feasible mission. Posteriorly, the detailed ephemerides from this reduced list were taken from the Horizons JPL database, representing the period between 2020 and 2100. There are, in this time-frame, several synodic periods which can be selected for capture, and a careful choice yields the lowest $\Delta v_I$. The preferred period is the one in which the asteroid experiences the greatest change in orbital elements when encountering the Earth. This decision comes from the assumption that this change corresponds to a greater sensitivity to the Earth's perturbation and, therefore, to the highest optimisation sensitivity. 

In fact, the greatest change in orbital elements can be related to the biggest impact caused by the Earth perturbation on the asteroid. Furthermore, this impact can, generally, be related to the geometry of the encounter. A closer encounter will feel a greater disturbing effect, which can be correlated to the phasing of the asteroid with the Sun-Earth line (variable $\alpha_P$). As previously shown, Figures \ref{fig:km_angle} and \ref{fig:grid} depict three different Earth encounters and the resulting change in semi-major axis. It can be observed that, the closer $\alpha_P$ is to zero, the bigger the impact the perturbation has on the orbital elements. Accordingly, the chosen period for the trajectory design is the one in which $\alpha_P$ is closest to zero.

\paragraph{\textcolor{Both}{Filtering:}} the filter tool was first described by S\'anchez et al. \cite{pau_gravitational} and later expanded upon \cite{pau_retrieval}. It is utilised to establish simple heuristic rules that distinguish asteroids that may become good candidates, from others that can be completely discarded. The filter performs an approximation of a bi-impulsive manoeuvre with one burn on the perihelion and one on the aphelion; only one of the two is responsible for an inclination correction ($\Delta v_i$), but both include a semi-major axis change ($\Delta v_a$). From well-known orbital mechanics, the inclination change manoeuvre can be computed as:
\begin{equation}\label{eq:inc_change}
\Delta v_i = 2\sqrt{\bigg(\frac{\mu_{\Sun}}{a_0}r^*\bigg)} \sin\bigg(\frac{\norm{i_t-i_0}}{2}\bigg)
\end{equation}
in which $\mu_{\Sun}$ is the Sun's gravitational parameter, $i_t$ is the inclination of the target hyperbolic trajectory, $i_0$ is the initial asteroid inclination and $r^*$ corresponds to the ratio of perihelion and aphelion distance if the inclination change is done at aphelion, or its inverse if performed at perihelion. 

The semi-major axis change manoeuvre obeys Eq. \eqref{eq:sma_change}:
\begin{equation}\label{eq:sma_change}
\Delta v_a = \sqrt{\mu_{\Sun}\bigg(\frac{2}{r} - \frac{1}{a_1}\bigg)} - \sqrt{\mu_{\Sun}\bigg(\frac{2}{r} - \frac{1}{a_0}\bigg)}
\end{equation}
in which $r$ is the distance to the Sun at which the burn is made (perihelion or aphelion distance) and $a_0$ and $a_1$ are respectively the initial and final semi-major axis. 

Finally, the total $\Delta v$ of the manifold insertion manoeuvre can then be estimated by:
\begin{equation}\label{eq:bi-impulse}
\Delta v_I = \sqrt{\Delta v_{a1}^2 + \Delta v_{i1}^2} + \sqrt{\Delta v_{a2}^2 + \Delta v_{i2}^2}
\end{equation}

Eq. \eqref{eq:bi-impulse} yields four values, depending on whether the perihelion or aphelion burn is the first and which will include the inclination correction; the minimum result out of these will be the filter output.

This filter provides only a very rough approximation, intended for pruning unlikely candidates from the large asteroid database. For example, it implicitly assumes that the line of nodes coincides with the line of apsides, so that the inclination change can be performed at one of the apsides. Plus, these formulae only take into consideration the shape and inclination of the orbits (semi-major axis, eccentricity and inclination), ignoring any phasing with the Earth \cite{pau_retrieval}.

\subsection{\textcolor{External}{High-Thrust Design}}
\label{sub:ht}

\begin{figure}[h]
	\centering
	\begin{minipage}[t]{0.48\textwidth}
		\includegraphics[width=\textwidth]{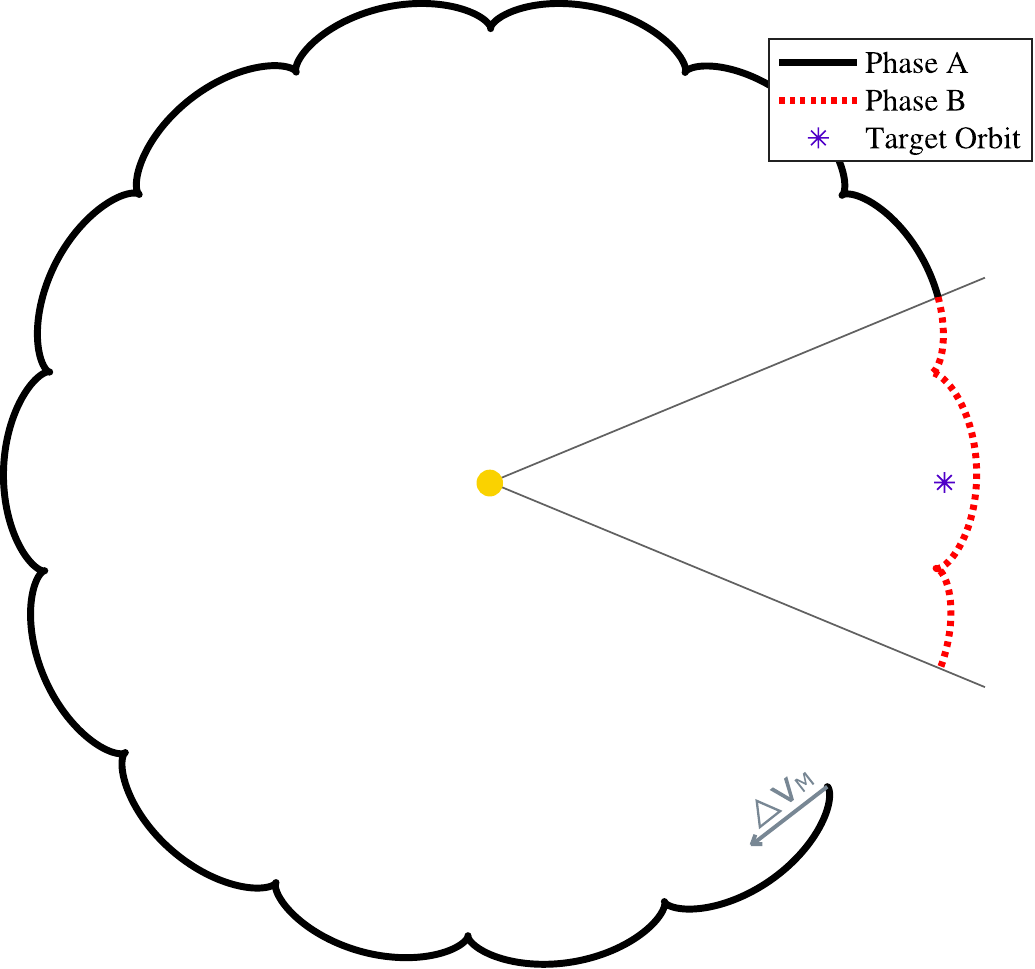}
	\end{minipage}
	\hfill
	\begin{minipage}[t]{0.48\textwidth}
		\includegraphics[width=\textwidth]{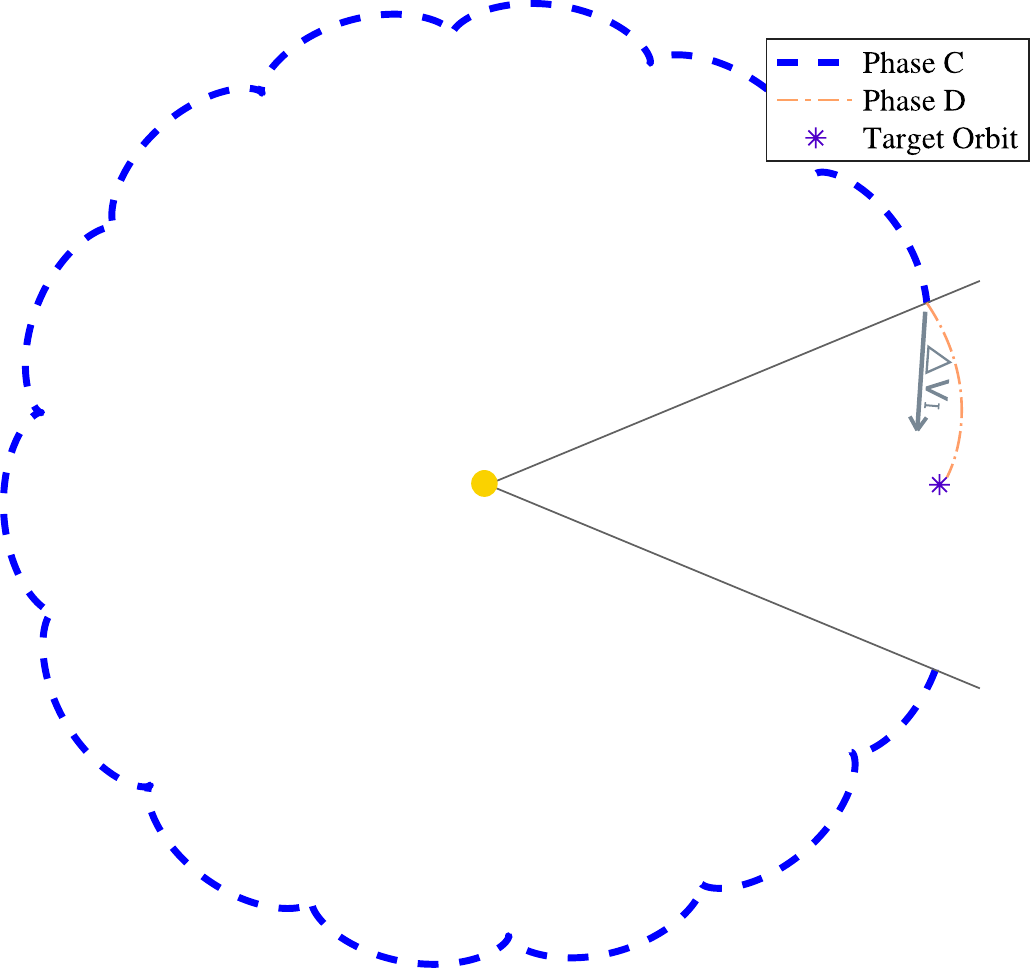}
	\end{minipage}
	\caption{\textcolor{External}{Phases of the Earth-resonant capture trajectory in sequence: first and second synodic periods respectively on the left and right-hand sides. The initial manoeuvre $\Delta v_M$ (Phase A) and the first Earth-encounter (Phase B) can be seen on the first synodic period. The coasting (Phase C), posterior second encounter with the Earth and manifold insertion (Phase D) are found on the second synodic period. The direction of the motion represents an asteroid of semi-major axis $>$ 1 in the synodic barycentric reference frame}}\label{fig:phases}
\end{figure}

After the reduced asteroid list is obtained, the process to develop the multi-fidelity high-thrust trajectory is applied to each candidate. This motion is used as an initial guess for the low-thrust optimisation of the final trajectory. The full motion can be divided in four different phases, which are highlighted in Figure \ref{fig:phases} and further explained here:

\begin{description}
	\item[\textit{Phase A}] starts when the asteroid is at the first periapsis right outside the perturbation region of the Earth; at this point, the asteroid's velocity is changed by $\Delta v_M$, altering its path. In the final low-thrust trajectory, this manoeuvre will correspond to a short thrusting period \textcolor{External}{at} the beginning of the motion. This can be seen in Figure \ref{fig:entire_traj}, in the highlighted section in Synodic Period 1.
	\item[\textit{Phase B}] corresponds to the first encounter with the Earth, inside the perturbation region. As previously stated, this region is defined by $\abs{\alpha} = \frac{\pi}{8} + \frac{\Delta \alpha_{P_0}}{2}$, which delimits a sufficiently large zone to encompass all $\alpha_P$ in which the object's motion is noticeably perturbed. In Figure \ref{fig:phases}, this area is marked by the diagonal lines forming a conical region in space.
	\item[\textit{Phase C}] consists \textcolor{External}{of} the motion in between Earth encounters, including the manoeuvre to insert the asteroid into an invariant manifold connected to an LPO, with a cost given by the variable $\Delta v_I$. In the low-thrust trajectory, this manoeuvre will equate to a short thrusting period before insertion into the LPO. This can be seen on Figure \ref{fig:entire_traj}, in the highlighted section in Synodic Period 2.
	\item[\textit{Phase D}] culminates with the second Earth encounter, where the asteroid moves from the insertion invariant manifold into the LPO. 
\end{description}

The dynamics in Phases A and C are modelled with the 2BP, with the Sun exerting the central gravitational attraction; the asteroid's path is only altered by an initial $\Delta v_M$, and a $\Delta v_I$ at the end of the latter. However, due to the close proximity to the Earth, this planet's perturbation cannot be neglected in Phase B, and both the KM and CR3BP will be used to model its dynamics. Finally, the invariant manifold dynamics in Phase D are modelled using the CR3BP, as described in previous publications \cite{pau_retrieval}.

The final capture $\Delta v_C$ is the added total of the two different manoeuvres, shown in Eq. \eqref{eq:sum} for the resonant and direct cases---the latter of which, as previously mentioned, only consists of Phases C and D. \textcolor{External}{The quantity $\Delta v_C$ is used as a scalar value, since the velocity change is defined as tangential. Since the point of the resonant manoeuvre is to change the phasing of the asteroid with the Earth, this requires only a modification in semi-major axis. The evolution of this orbital element depends only on the tangential component of the acceleration, according to the Gauss' variational equations (GVE) form in Battin's Problem 10.7 \cite{battin}.}

\begin{equation}\label{eq:sum}
\Delta v_C = \begin{cases} \Delta v_M + \Delta v_I, & \mbox{Resonant Capture} \\ \Delta v_I, & \mbox{Direct Capture} \end{cases}
\end{equation}

The multi-fidelity framework for the high-thrust motion is then divided into three steps, distinctive due to their different dynamical models and increasing complexity. The resulting trajectory acts as an initial guess to reach a low-thrust controlled solution. 

For the \textit{Grid Search} and the \textit{Refinement} steps, the trajectory in Phase B is modelled respectively in the KM and in the CR3BP. The $\Delta v_I$ cost in Phase D is estimated by the previously described filter, which is later replaced by a \textit{Lambert Arc Optimisation}. As argued in Section \ref{sub:pruning}, this filter is useful to obtain a quick estimate of the transfer costs, so that the initial optimisation focuses on exploiting Earth's perturbation. Nevertheless, future work could consider more accurate approaches to obtain $\Delta v_I$, such as a fast Lambert arc estimator \cite{an_lambert}.

It is important to note that one of the targets to bear in mind when developing this high-thrust trajectory is the achievement of the smallest possible $\Delta v_C$; according to the Tsiolkovsky equation, this will correspond to the highest retrievable mass for a fixed amount of propellant.

\subsubsection{Step 1: \textit{Grid Search}}

In this step, the modelling of the Earth encounter in Phase B is made using the KM, with the capture cost $\Delta v_C$ computed as the sum of $\Delta v_M$ with the filter estimate for $\Delta v_I$. The purpose of implementing $\Delta v_M$ in the beginning of the transfer is to ensure that the orbital changes occurring during the Earth encounter due to its perturbation are optimal: the KM is used for a quick assessment of how the asteroids' positioning relative to the Earth will impact their orbits.  

\begin{figure}[h!]
	\centering
	\includegraphics[width=0.65\linewidth]{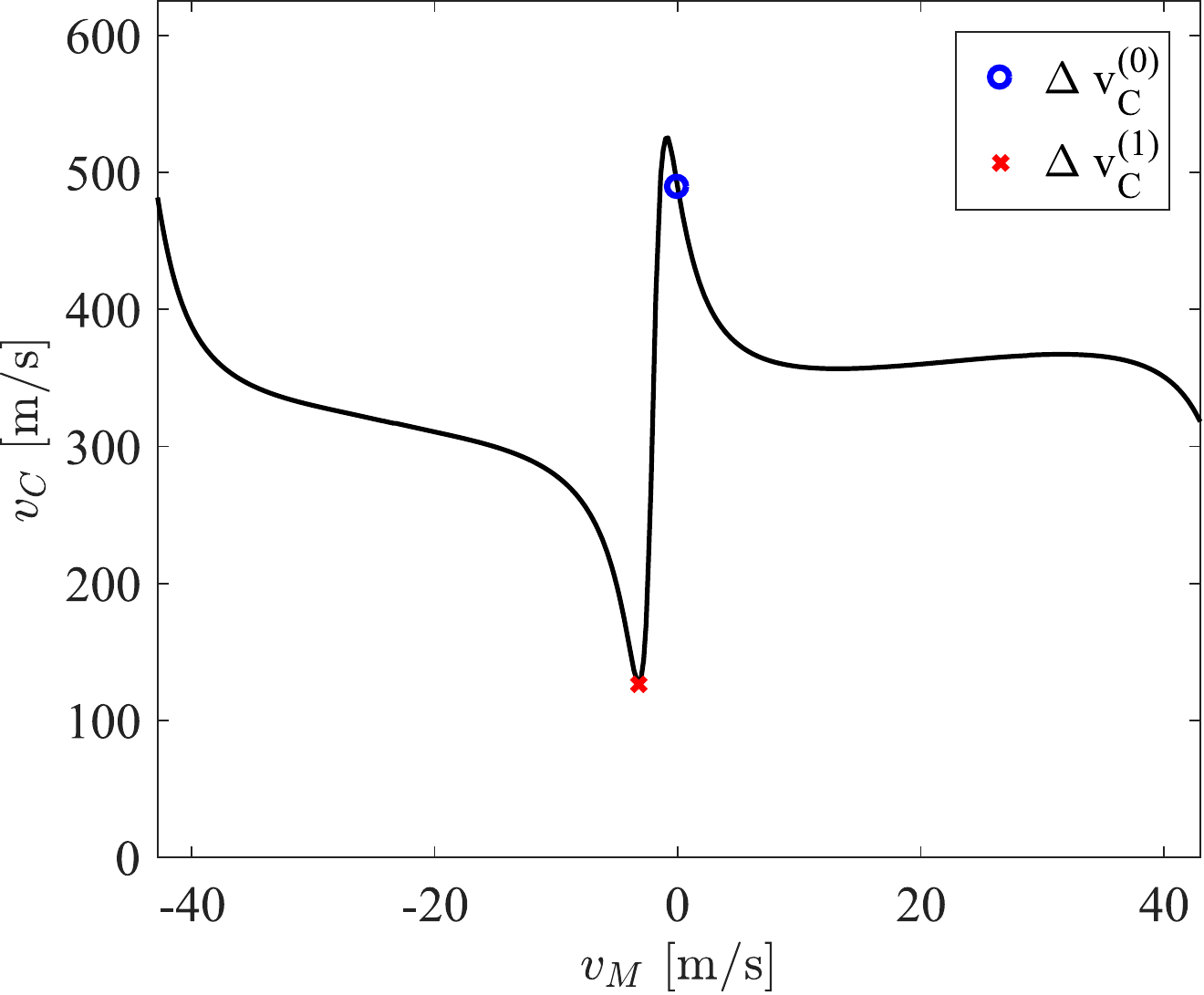}
	\caption{Grid search for asteroid 2016RD34}\label{fig:dv_comp}
\end{figure}

As such, the grid search is performed by computing $\Delta v_C$ for values of $\Delta v_M$ from approximately -50 m$\cdot$s$^{-1}$ to 50 m$\cdot$s$^{-1}$, with a step of 0.02 m$\cdot$s$^{-1}$: values that cause significant change in the final capture cost, while being comparatively very small. This is done in the following order: each computed motion starts with a change in orbital elements provoked by their respective $\Delta v_M$. Then, the KM equations are used to calculate the $\{\Delta a, \Delta e, \Delta i, \Delta \omega \}$ set of orbital element changes for each case. Finally, a grid is obtained, with the $\Delta v_C$ cost computed using Eq. \eqref{eq:sum} as a function of $\Delta v_M$. This grid, together with the minimum value for $\Delta v_C$, can be seen in Figure \ref{fig:dv_comp} for asteroid 2016 RD34.

Using this grid, the smallest capture cost value and its corresponding initial manoeuvre are retrieved. On Figure \ref{fig:dv_comp}, this value is highlighted with a red cross. Throughout this chapter, these have the respective denominations of $\Delta v_{C}^{(1)}$ and $\Delta v_{M}^{(1)}$. These are the values to pass on to the next section, \textit{Refinement}.

On Figure \ref{fig:dv_comp} it can be seen that, by performing no manoeuvre (i.e. $\Delta v_M = 0$, represented by $\Delta v_{C}^{(0)}$), the capture $\Delta v_C$ would be much higher. Thus, the case of asteroid 2016 RD34 is a good example of a situation in which a very small $\Delta v_M$ decreases immensely the final capture cost, demonstrating the value of optimizing the Earth encounter.

One simplification must be mentioned: although $\Delta v_M$ will cause a change in the orbital elements before the first Earth encounter, this change is so small that makes a negligible difference when computing the KM perturbation as a function of the asteroid's phasing. In this way, the KM mapping shown on Eq. \eqref{eq:map} needs only to be determined for the initial set of orbital elements in an adequate range of $\alpha$. The $\alpha_P$ of each motion in the grid can be interpolated out of this computation.

Another point to be made is that, if the absolute value of $\Delta v_M$ is increased without any bounds, the asteroid's orbit is eventually going to move forwards or backwards one epicycle, causing $\alpha_P$ to be the same. In order to address this, the analysis is restricted to $\Delta v_M$ inside limits that correspond to the asteroid moving backwards or forwards only one epicycle. These limits are defined more strictly after the previous rough estimate of 50 m$\cdot$s$^{-1}$. For the actual computation, the semi-major axis formula of the GVE according to Battin's Problem 10.7 \cite{battin} is again employed, in the following form:
\begin{align}\label{eq:dvm}
\Delta v_M = \frac{\mu_{\Sun} \Delta a}{2 a^2 v_P}
\end{align}
where $v_P$ is the velocity at the periapsis and $\Delta a$ represents the variation in initial semi-major axis corresponding to the addition of one epicycle to the asteroid's motion.

\subsubsection{Step 2: \textcolor{Both}{\textit{Refinement}}}

While the KM is able to accurately represent the underlying gravitational disturbance caused by the Earth, resulting in a reduction of capture costs, other non-linear effects may completely distort the sought outcome when passing the solution to the CR3BP. In this way, the solution yielded by \textit{Step 1}, using the previously optimal $\Delta v_M$, needs to be refined to achieve a similar behaviour and capture cost. 

In general terms, the closer the asteroid is to the perturbing body, the greater the latter's influence on it becomes. While the asteroid may undergo several periapsis passages in the region of Earth's perturbation, it can be roughly assumed that the one exerting the most significant impact on its motion will have the $\alpha_P$ closest to zero. As such, in order to get a similar orbital change using the CR3BP as the one obtained in the previous step with the KM, the closest of its periapsis passages---$\alpha_{closest}$---should be the same in both models. 

In order to target the same $\alpha_{closest}$, a differential corrector based on a single shooting method was developed: it computes the value of $\Delta a$, and consequently $\Delta v_M$, that yields the encounter with the intended $\alpha_{closest}$. The manoeuvre value will be henceforth referred to as $\Delta v_{M}^{(2)}$, while the resulting capture cost will be, analogously, $\Delta v_{C}^{(2)}$.

\paragraph{Differential Corrector:} the shooting method developed to solve this problem was designed for the case in which the state vector is expressed in orbital elements ($\bm{\kappa} = \{a, e, i, \Omega,$ $\omega, M\}$). Instead of considering the correction of all six orbital elements, the computation is based on the premise to change only the initial semi-major axis which, assuming the dependencies to other variables to be negligible, will provide a very fast computation to manipulate the final mean anomaly $M_1$. The mean anomaly change as a function of the initial semi-major axis is represented by:
\begin{equation}
\delta M_1 = \bm{\Phi}\delta a_0\label{eq:mf}
\end{equation}
where $\dot{\bm{\Phi}} = \textcolor{External}{\text{\textbf{D}}} f(\bm{\kappa}(t))\bm{\Phi}$.

The system's Jacobian, $\text{\textbf{D}} f(\bm{\kappa}(t))$, is determined by Eq. \eqref{eq:annex}.
\begin{align}\label{eq:annex}
\text{\textbf{D}} f(\bm{\kappa}(t)) =& \frac{\delta \dot{M}}{\delta a}
\end{align}  

\textcolor{External}{Expanding the Lagrange Planetary Equations (LPE) in Eqs. \eqref{eq:LPE} with the disturbing function $\mathpzc{R}$ in Eq. \eqref{eq:dRdK}, $\dot{M}$ becomes:} 
\begin{align}\nonumber
\dot{M} =& \frac{d}{dt}n(t_f - t_0) + \frac{2\mu}{na}\Bigg[\frac{1 - e^2}{1 + e\cos\nu}\Bigg(-\frac{1}{r^2} - \frac{2\cos\theta}{r^3} + \frac{r - \cos\theta}{\big(1+r^2-2r\cos\theta\big)^{3/2}}\Bigg) \\\nonumber
& + \frac{3M\Theta}{2n}\Bigg(\frac{1}{r^2} - \frac{r}{\big(1+r^2-2r\cos\theta\big)^{3/2}}\Bigg)\Bigg] \\\nonumber
& + \frac{\mu(1 - e^2)}{nea^2}\Bigg[\Bigg(\frac{1}{r^2} + \frac{2\cos\theta}{r^3} - \frac{r - \cos\theta}{\big(1+r^2-2r\cos\theta\big)^{3/2}}\Bigg)\Bigg(\frac{2ea + r\cos\nu}{1 + e\cos\nu}\Bigg)\\
& + \frac{\Theta\beta}{n}\Bigg(\frac{1}{r^2} - \frac{r}{\big(1+r^2-2r\cos\theta\big)^{3/2}}\Bigg)\Bigg]\label{eq:mdot}
\end{align}  
where $\cos \theta$ is given by Eq. \eqref{eq:rtheta2}, $\Theta = \sin\Omega_{rot}\cos(\omega + \nu) + \cos\Omega_{rot}\sin(\omega + \nu)\cos i$ and $\beta = -\sin E (1 + \frac{1 - e\cos E}{1 - e^2})$. \textcolor{External}{To achieve the formulation in Eq. \eqref{eq:annex}, Eq. \eqref{eq:mdot} is derived with respect to $a$, considering this quantity independent from the other orbital elements. This derivation is here omitted due to its length and triviality.} 

The algorithm's goal is to change the final mean anomaly $M_1$ so that the $\alpha_P$ value of the final propagation point is equal to $\alpha_{closest}$---in other words, the periapsis of the orbit has to be moved so that its phasing falls in $\alpha_{closest}$.

The process is done in the following manner: first, the target $\alpha_{closest}$ is obtained. The motion is then propagated in the CR3BP, from its initial conditions, to reach this geometry. If the final orbital position is not at the periapsis, the difference between the mean anomaly $M_1$ and the desired target ($M_{Target} = 0$) is computed. The necessary change in mean anomaly is obtained from the expression $\Delta M_1 = \min\{M_1, 2\pi - M_1\}$. 

Finally, Eq. \eqref{eq:mf} is used to determine the appropriate change in $\Delta a_0$. By employing Eq. \eqref{eq:dvm}, this value will be used to compute the new orbital manoeuvre $\Delta v_M$ at each \textcolor{External}{iteration (superscript $i$)} of the algorithm, as per Eq. \eqref{eq:dvm_iter}. When the change in $\Delta a_0$ is smaller than a threshold ($\epsilon = 10^{-8}$), the algorithm returns the obtained $\Delta v_M$, refined for the CR3BP. 
\begin{align}\label{eq:dvm_iter}
\Delta {v_{M}}^{i+1} = \Delta {v_{M}}^{i} + \frac{\mu_{\Sun} \Delta {a_{0}}^{i}}{2 {a^i}^2 {v_{P}}^{i}}
\end{align}

\subsubsection{Step 3: \textit{Lambert Arc Optimisation}}
\label{sub:lambert}

Subsequently, the $\Delta v_I$ guess established by the filter for Phase D is replaced by the computation of a Lambert arc, adding another fidelity level to the framework. The Lambert arc solution will yield an orbital transfer that connects the asteroid to one of the manifold orbits, replacing the bi-impulsive approximation defined by the filter.

In order to devise the trajectory between asteroid and target orbit, the initial and final states of the transfer can be fixed, with their respective ephemerides taken at the chosen calendar dates to compute the connecting Lambert arc. However, an optimiser can be used to determine the set of dates when the transfer is the most cost-effective, together with the optimal destination manifold orbits.

\paragraph{Considerations on the Lambert Arc Optimiser:}the chosen optimiser to tackle this problem is EPIC \cite{epic}. EPIC is a global trajectory software that performs domain decomposition, where each domain is evaluated based on the evolution of a population of agents; its purpose is to generate a series of very good local optima instead of a global one, such that there is more flexibility to the mission design.

The Lambert arc connects one of the invariant manifold trajectories to the asteroid's trajectory after the first encounter with the Earth. EPIC is then employed to compute several arcs, depending on the number of orbital revolutions chosen---in this case, the maximum value was fixed as 4. Lambert's problem is computed by EPIC using Battin's formulation \cite{battin}. The solution with lowest $\Delta v_I$ is the one chosen for the design, except in cases where the time of flight of the computed arc was too small for the low-thrust trajectory to be feasible. 

\subsection{\textcolor{External}{Low-Thrust Optimisation}}
\label{sec:lt}

A low-thrust propulsive system is characterized by commonly having an $I_{\text{SP}}$ at least ten times higher than a chemical-thrust one, making it less propellant-consuming for the same amount of spacecraft mass. This makes the use of low-thrust engines quite attractive for asteroid retrieval, where the combined spacecraft-asteroid system's mass can be very high.  

The computation of the low-thrust trajectory required for asteroid hauling starts with the definition of the system variables. The considered initial setting is akin to the one used by NASA's ARRM concept \cite{ARRM}: a spacecraft of 5500 kg dry mass and 10 tonnes of propellant, using a high power solar electric propulsion system of roughly 40 kW and $I_{\text{SP}}$ of 3000 s, yielding a maximum thrust capability of 2 N.

For each phase of the optimisation, the asteroid's mass that can be hauled by the low-thrust engine is heavily dependent on the thrust vector, which is the decision variable of the optimisation. The entire system's mass can only be computed after solving the optimal control problem. An initial guess for this can be estimated, using the high-thrust $\Delta v_C$ solution and the Tsiolkovsky equation. However, this mass will prove to be quite far from the actual value; in order to correct it, a mass homotopy is performed, which will be explained later in this section. Since the spacecraft's fuel and dry masses are both fixed, maximising the total system's mass is equivalent to doing so for the carried asteroid mass.

\subsubsection{Optimal Control Problem} 

The conversion from the high-thrust trajectory into a low-thrust motion is done by formulating an optimal control problem, where the thrust is the control variable and the objective is to maximize the final system's mass (after the propellant is depleted), while constrained by the motion's boundaries and position targets. Once all the bounds, boundary conditions and necessary problem parameters are set, the optimal control problem is solved by a chosen optimal control solver software.

The optimal control problem is defined in the general form characterised in Chapter \ref{chap:intro}: to determine the state $\bm{s}(t)$, control $\bm{u}(t)$, initial time and final time that minimise the cost function $J$. The general problem is then adapted to fit the low-thrust trajectory case. As already stated, the decision variable will be the thrust vector throughout the motion: starting with the initial time and state, this variable is optimised in such a way that the final system's mass is the highest. Each state of the trajectory is computed using the equations of motion of the problem coupled with the acceleration provided by the thrust, forming the dynamical constraints. 

A multi-fidelity framework, going from a two-body \textcolor{Internal}{modified} equinoctial system to the CR3BP model, was implemented to find a solution to the optimal control problem. The selection of the two-body modified equinoctial system, which employs GVE to compute the motion, allows a straightforward definition of the states' bounds, leading to an improved convergence. Furthermore, this description avoids the singularities that could be encountered when dealing with easily retrievable NEAs \cite{pau_eros}, whose orbits are frequently quasi-circular and quasi-planar. The \textcolor{External}{dynamics} for the equinoctial two-body system \cite{hintz2008survey} are thus presented in Eqs. \eqref{eqs:equinoctial}. 
\begin{align}\label{eqs:equinoctial}
\frac{dp}{dt} &= \frac{2p}{w}\sqrt{\frac{p}{\mu}}\frac{u_{\theta}}{m}\nonumber\\
\frac{df}{dt} &= \sqrt{\frac{p}{\mu}}\bigg[\sin l\frac{u_{r}}{m} + \frac{1}{w}\bigg((w + 1)\cos l + f\bigg)\frac{u_{\theta}}{m} - \frac{g}{w}\bigg(h\sin l - k\cos l\bigg)\frac{u_{n}}{m}\bigg]\nonumber\\
\frac{dg}{dt} &= \sqrt{\frac{p}{\mu}}\bigg[-\cos l\frac{u_{r}}{m} + \frac{1}{w}\bigg((w + 1)\sin l + g\bigg)\frac{u_{\theta}}{m} - \frac{f}{w}\bigg(h\sin l - k\cos l\bigg)\frac{u_{n}}{m}\bigg]\nonumber\\
\frac{dh}{dt} &= \sqrt{\frac{p}{\mu}}\frac{s^2}{2w}\cos l\frac{u_{n}}{m}\nonumber\\
\frac{dk}{dt} &= \sqrt{\frac{p}{\mu}}\frac{s^2}{2w}\sin l\frac{u_{n}}{m}\nonumber\\
\frac{dl}{dt} &= \sqrt{\mu p}\bigg(\frac{w}{p}\bigg)^2 + \frac{1}{w}\sqrt{\frac{p}{\mu}}\bigg(h\sin l - k\cos l\bigg)\frac{u_{n}}{m}
\end{align}
\textcolor{Internal}{where the set $\{p, f, g, h, k, l\}$ corresponds to the commonly known modified equinoctial elements \cite{equinoctial}, the control vector $\bm{u}$ is written in spherical coordinates, $w = \frac{p}{r} = 1 + f\cos l + g\sin l$ and $s^2 = 1 + h^2 + k^2$.}

Once a solution is computed with the aforementioned dynamical model, it is used as a first guess for a final optimisation in the CR3BP. As discussed in Chapter \ref{chap:ast1}, employing a simplified model as a first solution makes it simpler and faster for the program to converge, as these types of problems are extremely sensitive to the set up and the process is very time consuming. Each of the models has a few more differences, detailed on Table \ref{tab:model_comparison}. The computation of the objective function is common for both cases, as shown in Eq. \eqref{tab:m_dot}.
\begin{equation}\label{tab:m_dot}
\dot{m} = -\frac{\norm{\bm{u}}}{g_0 I_{SP}}
\end{equation}

\begin{table*}
	\caption{\label{tab:model_comparison}Definition of parameters and functions for the two-body equinoctial and CR3BP optimal control problems}
	\begin{tabular*}{\textwidth}{c @{\extracolsep{\fill}}ccc}
		\toprule
		Problem Parameters & 2BP & CR3BP \\
		\midrule
		State Vector $\bm{y}$ & $\{p, f, g, h, k, l\}$ & $\{x, y, z, \dot{x}, \dot{y}, \dot{z}\}$\\
		Control Vector $\bm{u}$ & $\{u_r, u_{\theta}, u_n\}$ & $\{u_x, u_y, u_z\}$\\	
		Dynamical Constraints & Eqs. \eqref{eqs:equinoctial} & Eqs. \eqref{eq:cr3bp}$^*$ \\	
		Reference Frame & Inertial & Synodic \\\hline		
		Objective Function $J$ & \multicolumn{2}{c}{$-m(t_f)$}  \\
		\bottomrule
	\end{tabular*}
	\begin{flushleft}
		$^*$ Including the added thrust component\\
	\end{flushleft}
\end{table*}

\paragraph{Considerations on the Mass:}In order to establish the feasibility of the mission, the spacecraft must be able to carry its own weight, the fuel mass and the asteroid's. This is described by: 
\begin{equation}\label{eq:mass_ineq}
m_\text{ast} + m_\text{dry} + m_\text{fuel} = m_\text{opt}
\end{equation}
where $m_\text{dry}$ and $m_\text{fuel}$ are already established as respectively 5.5 and 10 tonnes, and $m_\text{opt}$ is the output result of the optimal control problem, representing the entire system's mass. The asteroid mass that can be hauled by the spacecraft, $m_\text{ast}$, can be computed by solving Eq. \eqref{eq:mass_ineq} once $m_\text{opt}$ is known. 

The goal of the optimal control problem is to utilise the propellant in the most efficient way, starting from a given initial system mass (as in Eq. \eqref{eq:mass_ineq}). This does not, however, change the initial mass. For the first iteration, the mass is computed such that the total thrust time is equal to 1\% of the time of flight. Yet, this value is very far from the capabilities of the low-thrust system engine. As such, the initial mass can be increased after each iteration of the optimal control solver software using a continuation method, here described as a \textit{mass homotopy}.

At each iteration, the initial mass value is raised in small steps. This increase has to be small enough for the optimal control problem to converge quickly, with the solution of the previous iteration as a new initial guess. As such, the initial mass at each iteration is estimated by means of the Tsiolkovsky equation, utilising the still available thrust time and propellant mass as tuning parameters.

Finally, the entire low-thrust trajectory is developed as follows: starting from the high-thrust guess, the optimisation is performed until a solution is reached. If the difference between final and initial mass is smaller than 10 tonnes (the total propellant mass), and the total thrust time is shorter than the total transfer time, then the mass homotopy is performed and the optimisation occurs again for the new solution; otherwise, the process is terminated.  

\paragraph{Considerations on the Optimal Control Solver:} GPOPS-II is the chosen solver software. It works by formulating a non-linear programming problem (NLP) \cite{gpops}, using orthogonal collocation to transform the continuous problem into a discrete one. This is posteriorly tackled by an NLP solver---IPOPT, an interior point optimiser \cite{ipopt}.

The applied collocation method is a Legendre-Gauss-Radau (LGR) scheme: the solution is given by a polynomial that can fit the continuous problem in selected points, called the \textit{collocation points}: the dynamical equations (as defined on Table \ref{tab:model_comparison}) are solved for these, and they are evaluated against the candidate Legendre polynomials. 

Since GPOPS-II is an hp-adaptive method, both the number of mesh intervals (p) and the degrees of the polynomials in question (h) can be chosen for optimal performance. The way in which GPOPS-II is set up to manipulate these parameters is found in \cite{patterson-rao,darby-rao}.

The input obtained by GPOPS-II is a discretisation of the initial trajectory guess. The bounds for each of the state, control and mass variables have to be defined, as well as the boundary conditions, corresponding to the initial and final ephemerides of the trajectory. GPOPS-II works with user-supplied dynamical equations and cost function. Since this solver is extremely sensitive to the initial conditions given, in preliminary runs, the cost function was defined so as to minimise the distance between the software-computed end trajectory point and the actual user-defined manifold target. In posterior runs, once the guess trajectory was nearer to the defined conditions, the cost function was changed to maximise the overall mass.

However, the sensitivity of the software to the initial conditions led to the decision of optimising the final low-thrust trajectory in several consecutive phases. Considering that each piece of the low-thrust trajectory has its own counterpart in the initial high-thrust guess and using the notation found in Figure \ref{fig:phases}, the division was made in the following manner: segment 1 corresponds to Phase A, segment 2 to the junction of Phases B and C, and segment 3 to Phase D. This split was manually implemented to obtain separate solutions for each segment, which were posteriorly patched together. An alternative, more automated way to proceed with this implementation would be to use the multi-phase structure of GPOPS-II.

The division is made based on thrusting demand: segments 1 and 3 are the most difficult ones for the solver to converge to an optimal solution, since they require a greater variance in the control vector throughout time to reach the boundary conditions. As such, they are solved first, with segment 2 left as a final connecting trajectory. Thus, the target states of segments 1 and 2 are respectively the starting states of segments 2 and 3, yielding a final fully connected low-thrust trajectory. 

\section{Results and Discussion}
\label{sec:results}
This section presents the obtained values at each step of the multi-fidelity design process (see Figure \ref{fig:flowchart_mf}). The benefits of employing an Earth-resonant transfer for asteroid retrieval are analysed, and the different fidelity models that make up the framework are compared in terms of accuracy and computational cost.

\subsection{Asteroid Candidate List}

From the MPC database, over 18,000 asteroids were considered. The filter described in Section \ref{sub:pruning} was used to prune the ones whose $\Delta v$ with direct capture was higher than 1.2 km$\cdot$s$^{-1}$. \textcolor{External}{A total of 3000 target conditions per energy level are computed: 500 invariant manifold trajectories per each halo, vertical and horizontal Lyapunov orbits of both libration points.} Then, the ephemerides of the remaining 88 asteroids were retrieved from the JPL database. For each candidate, the synodic period with the greatest orbital changes was selected, as described in Section \ref{sub:pruning}.

To retrieve the minimum $\Delta v$ for a resonant capture, the grid search is performed on each of these NEA. The values are posteriorly compared with the direct capture $\Delta v$ for each asteroid. The selection of candidates for refining and further analysis is based on whether or not the resonant capture $\Delta v$ is smaller than the direct one by at least 30\%; this resulted in a total of 12 asteroids. Captures with lesser improvements than this threshold were simply deemed not interesting enough to pursue, given how much longer the transfer time of a resonant capture would become, as compared with the direct option.

Table \ref{tab:dates} shows the list of candidate asteroids' designations, the times of flight and the date when each transfer begins.

\begin{table*}
	\caption{\label{tab:dates}Dates, times of flight (TOF) and final capture orbit for each asteroid in the candidate list}
	\begin{tabular*}{\textwidth}{c @{\extracolsep{\fill}}ccccc}
		Asteroid & Initial Date & TOF\textsubscript{Direct} [y] & TOF\textsubscript{Resonant} [y] & Capture LPO \\\hline	
		2016 RD34 & 05-22-2033 & 12.92 & 35.40 & VL2$^a$\\
		2012 TF79 & 01-02-2020 & 19.76 & 37.38 & VL2   \\
		2011 MD & 01-08-2070 & 14.73 & 36.73 & VL2   \\
		2011 BL45 & 23-12-2072 & 17.62 & 42.76 & VL2\\
		2014 BA3 & 09-09-2035 & 10.48 & 25.22 & VL1$^b$\\
		2000 SG344 & 22-01-2033 & 41.48 & 81.04 & PL2$^c$\\
		2017 FJ3 & 10-03-2047 & 4.46 & 11.21 & PL2 \\
		2017 BN93  & 17-04-2034 & 13.86 & 47.30 & VL2 \\
		2010 VQ98 & 15-01-2063 & 16.66 & 48.77 & VL2 \\
		2008 UA202 & 01-01-2020 & 29.39 & 40.53 & PL2\\
		2006 JY26 & 01-01-2020 & 44.05 & 72.86 & HL2$^d$\\
		2006 BZ147 & 21-01-2039 & 38.26 & 58.11 & HL2 \\\hline
	\end{tabular*}
	\begin{flushleft}
		$^a$ Vertical Lyapunov in L$_2$\\
		$^b$ Vertical Lyapunov in L$_1$\\
		$^c$ Planar Lyapunov in L$_2$\\
		$^d$ Halo in L$_2$\\
	\end{flushleft}
\end{table*}

\subsection{Trajectory Design}

For each of these 12 asteroids, the values found for every step detailed in Section \ref{sub:ht} are depicted on Table \ref{tab:asts_dv}, as well as the $\alpha_{closest}$ parameters to be targeted by the differential corrector. It is important to note that the direct $\Delta v_{C}$ is the optimised Lambert arc cost, as detailed in Section \ref{sub:lambert}. Again, the resonant $\Delta v$'s have different denominations depending on the step in the multi-fidelity design: $\Delta v_{C}^{(1)}$ corresponds to the KM grid search, $\Delta v_{C}^{(2)}$ to the CR3BP refinement and $\Delta v_{C}^{(3)}$ includes the Lambert arc optimisation.

Six asteroids were excluded from the candidate list, for two different reasons. First, the cases of asteroids 2008 UA202, 2006 JY26, 2000 SG344 and 2014 BA3 are particular ones in which the dynamics are too sensitive for an adequate capture trajectory to be computed by the suggested method, as the motion easily enters the Hill radius. These do not have the $\Delta v$ values for each step on Table \ref{tab:asts_dv}, as the framework did not tackle motions so close to the Earth. The appropriate way to work these cases would be to use a model that adequately approximates motions close to the Earth, such as {\"O}pik's method \cite{opik1, opik2, sanchez2015} or the pseudostate technique \cite{pseudostate, pseudostate_JGCD}.

Second, asteroids 2011 BL45 and 2006 BZ147 are outliers in the sense that, although the capture $\Delta v_I$ determined with the filter was shown to fit the criteria, the value resulting from the Lambert arc computed with EPIC was too great to achieve a high capture mass. It was understood that both were cases in which the filter severely underestimated the capture $\Delta v$, and therefore there was no point in pursuing their study. This can possibly be avoided by using a different estimate of $\Delta v_I$, such as a Lambert arc estimator or a similar fast transfer manoeuvre \cite{an_lambert}.

Even though the filter proves to be an under-estimator for most cases, the efficiency of the resonant trajectory compared to the direct one is kept for most asteroids. After this process, six asteroids remained for the computation of the low-thrust transfer.

\begin{table*}
	\caption{\label{tab:asts_dv}Analysis of $\alpha_{closest}$ and $\Delta v$ values on each step for the resulting asteroids. The excluded asteroids are pinpointed by reasons $^a$ or $^b$}	
	\begin{tabular*}{\textwidth}{c @{\extracolsep{\fill}}cccccc}\hline 
		\multirow{2}{*}{Asteroid} & \multirow{2}{*}{$\alpha_{closest}$ [$^{\circ}$]} & Direct $\Delta v_C$ & \multicolumn{3}{c}{Resonant $\Delta v_C$ [m$\cdot$s$^{-1}$]} \\ 
		& & & $\Delta v_{C}^{(1)}$ & $\Delta v_{C}^{(2)}$ & $\Delta v_{C}^{(3)}$ \\\hline
		2016 RD34 & 0.84 & 323.36 & 136.28 & 62.14 & 84.90 \\
		2012 TF79 & 0.86 & 272.81 & 94.27 & 83.46 & 72.85 \\ 
		2011 MD & 0.57 & 206.22 & 64.94 & 54.34 & 138.94 \\
		2011 BL45$^b$& 2.24 & 377.28 & 47.94 & 60.30 & 339.00\\
		2014 BA3$^a$ & -16.11 & - & - & - & - \\
		2000 SG344$^a$ & 2.60 & - & - & - & - \\
		2017 FJ3 & 1.52 & 1002.3 & 718.26 & 475.08 & 659.31 \\
		2017 BN93 & 1.18 & 539.70 & 145.06 & 125.76 & 274.90 \\
		2010 VQ98 & 2.03 & 265.01 & 118.77 & 148.89 & 130.89 \\
		2008 UA202$^a$ & -4.37 & - & - & - & -  \\
		2006 JY26$^a$ & -3.63 & - & - & - & -  \\
		2006 BZ147$^b$ & -26.72 & 1412.10 & 874.29 & 991.70 & 1239.50 \\\hline
	\end{tabular*}
	\begin{flushleft}
		$^a$ Too close to the Earth\\
		$^b$ Filter underestimation\\
	\end{flushleft}
\end{table*}

\subsection{Mass Comparison}

As described on Section \ref{sec:lt}, the high-thrust guess was used for the optimisation of the low-thrust solution. By iterating through several steps on the mass homotopy process, GPOPS-II achieved the final retrievable masses for the direct and resonant captures. Table \ref{tab:mass} shows the comparison between these two values, and also includes some previously published results \cite{pau_retrieval, RN11} that assume the usage of similar low-thrust propulsion systems \textcolor{External}{in their designs}. 

\begin{table*}
	\caption{\label{tab:mass} \textcolor{External}{Retrievable masses and TOF values (rounded to the nearest unit) of the candidate asteroids for direct and Earth-resonant captures, together with the state-of-the-art (SOA) results. TOF\textsubscript{Res.} repeated from Table \ref{tab:dates} for clarity. Balance computed between m\textsubscript{Direct} and m\textsubscript{Res}}}
	\begin{tabular*}{\textwidth}{c @{\extracolsep{\fill}}ccccccc}
		Asteroid & $ $TOF\textsubscript{SOA} [y] & TOF\textsubscript{Res.} [y] & m\textsubscript{SOA} [t] & m\textsubscript{Direct} [t] & m\textsubscript{Res.} [t] & Balance [t] \\\hline	
		2016 RD34 & - & 35 & - & 357 & 1227 & +870 \\
		2012 TF79 & 7.30 & 37 & 739$^{a}$ & 705 & 3161 & +2456 \\
		2011 MD & 6 & 37 & 800$^{b}$ & 784 & 1496 & +712 \\
		2017 FJ3 & - & 11 & - & 193 & 293 & +100 \\
		2017 BN93 & - & 47 & - & 322 & 521 & +199 \\
		2010 VQ98 & 8.86 & 49 & 727$^{c}$ & 493 & 1515 & +1022 \\\hline
	\end{tabular*}
\end{table*}

As it can be seen, the increase in retrievable mass between the direct and resonant trajectories was over 30\% for all cases. When comparing the computed low-thrust motion for both models used (2BP with equinoctial elements and CR3BP), the difference in values presented is less than 0.01\%, so there was no need to present a distinction.

\textcolor{External}{As expected, the times of flight of the resonant trajectories are much higher than the ones found in literature, since the computed trajectories are longer by at least one synodic period. Still, the obtained mass values are also much greater and, in some cases, allow for the retrieval of over 1000 tonnes of material back to Earth.} 

\begin{figure}[h!]
	\centering
	\includegraphics[width=0.65\linewidth]{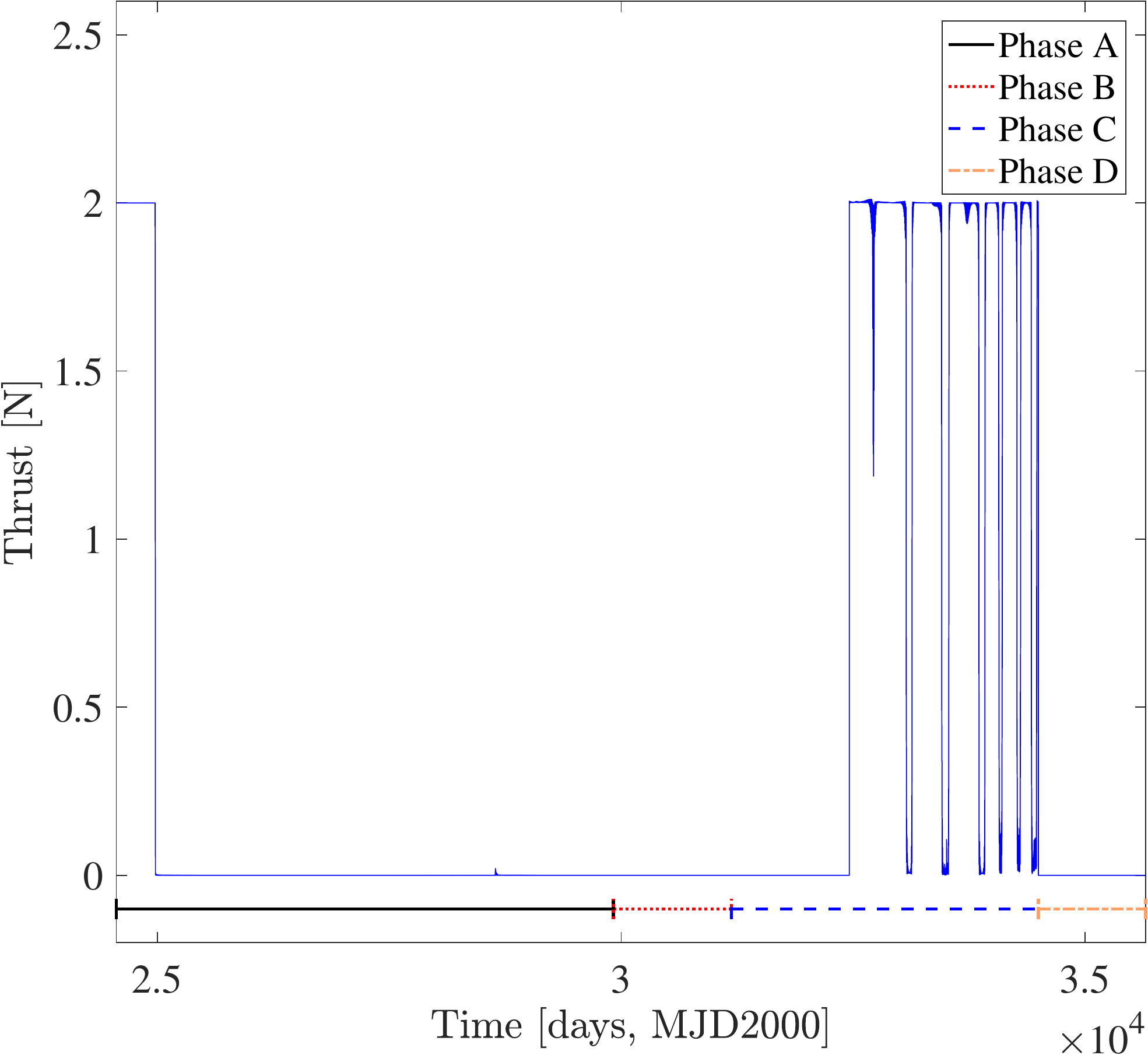}
	\caption{Thrust profile for the full low-thrust trajectory in the CR3BP as a function of time for asteroid 2011 MD}\label{fig:full_thrust}
\end{figure}

Figure \ref{fig:full_thrust} shows the complete thrust profile of the Earth-resonant capture trajectory for asteroid 2011 MD in the CR3BP. It depicts the thrust history throughout time, starting with a small control output corresponding to the $\Delta v_M$ manoeuvre. Then, the asteroid continues coasting until the trajectory reaches the Lambert arc segment, which proves to be the hardest to optimise as it is associated with the $\Delta v_I$ manoeuvre.

Despite the higher complexity of the CR3BP low-thrust optimal control problem, the final solution features a substantially neat on-off control, indicative that optimality conditions are well satisfied. 

\subsection{Model Comparison}
\label{sub:comparison}

The KM model was shown to have a similar accuracy to the CR3BP, coupled with a shorter computational time---both throughout this thesis, and in the work of Alessi and S\'anchez \cite{alessi_semi}. In order to further support these claims, the results of both models are studied in this particular case scenario of asteroid capture. They are compared for 88 different sets of initial asteroid orbital elements, and the time spent on each is analysed.

\begin{figure}[hb!]
	\centering
	\includegraphics[width=0.65\textwidth]{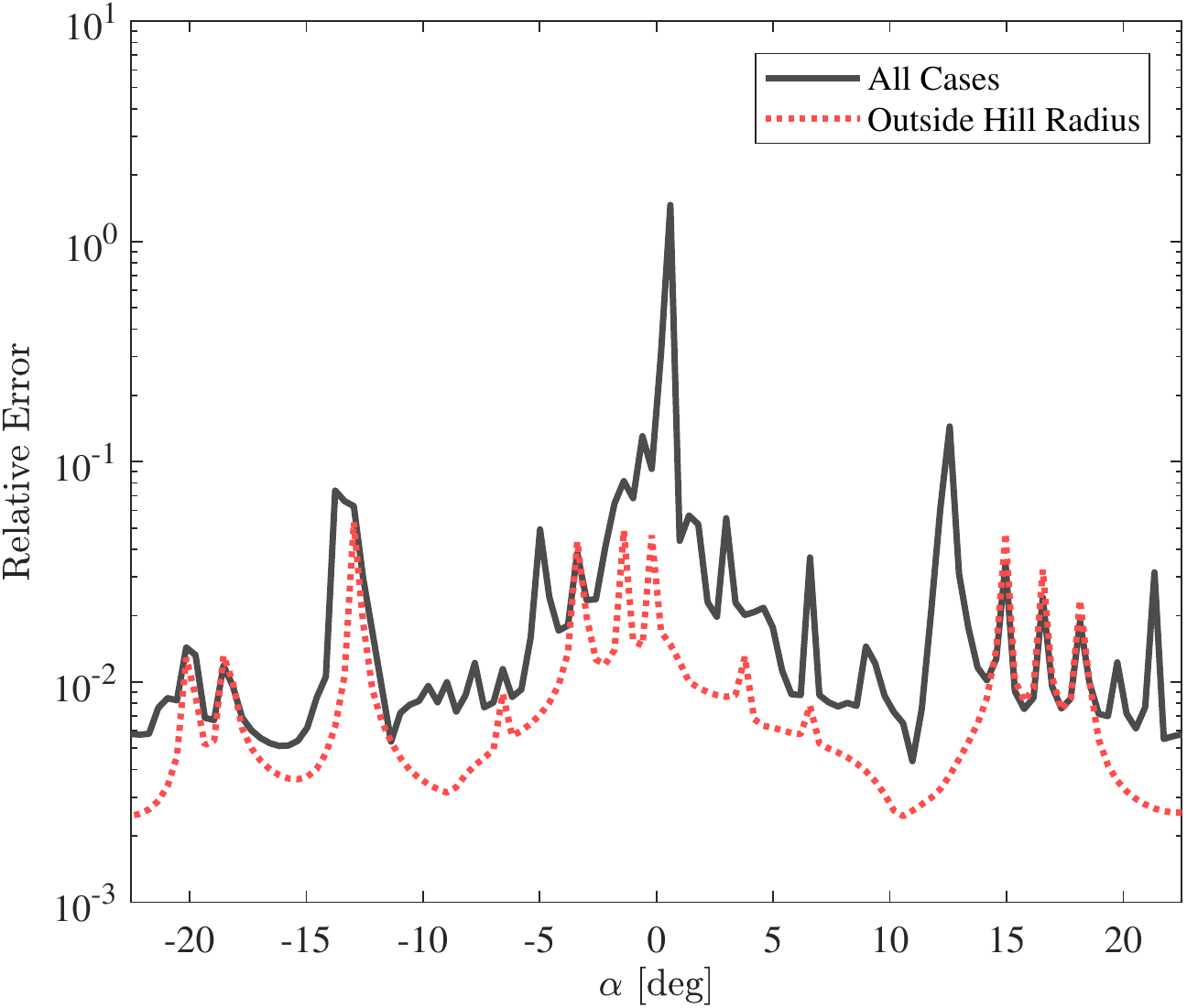}
	\caption{Mean relative error in logarithmic scale for semi-major axis update as a function of $\alpha$}\label{fig:error}
\end{figure}

On Figure \ref{fig:error}, the mean value of the relative error for the semi-major axis updated with the KM is shown for 300 different $\alpha$ values. The black \textcolor{External}{solid lines} display the mean error over the 88 asteroids regardless of their closest approach to the Earth, while the red dotted \textcolor{External}{lines} consider only the ones that never cross the sphere of influence of the planet, totalling 36 bodies. As discussed in Section \ref{sub:km}, the relative error is much smaller when the cases that cross the Hill radius are removed: the KM model works best when outside Earth's sphere of influence. Also, the error increase for values close to $\alpha = 0$ is correlated to the closest proximity to the Earth, where the Earth's perturbation is the most influential.

In terms of computational time, both models were used to compute 60 different initial $\Delta v_M$ manoeuvres for the 88 previously mentioned asteroids. \textcolor{External}{Using the MATLAB \textit{Profiler} to single out function usage, under the same situational conditions with the same computer specifications}, the KM was shown to be roughly 30 times faster than the CR3BP. This is an even better result than the one found in Section \ref{sub:ctime}, mainly caused by the then conjectured improvement due to the interpolation of the kick-maps.

Considering the multi-fidelity trajectory design, it is observed that the $\Delta v_M$ to reach the same $\alpha_{closest}$ is very similar for both the KM and the CR3BP. The latter yields slightly lower $\Delta v_C$ results, which can be seen on Table \ref{tab:asts_dv}. This allows the inference that the former, being a lower-fidelity framework, models the trajectory nearly as accurately as the higher-fidelity one. The $\Delta v_M$ for the KM grid search and the CR3BP refinement are found on Table \ref{tab:dvm}: when compared to the overall capture $\Delta v_C$, they are extremely small.

This comparison allows for the validation of the KM as a relatively accurate model that can provide a good stepping stone for an increase in fidelity in the current framework. The lower computational cost allows for a feasible analysis of the amount of NEA described on this thesis, while still presenting very similar results to higher complexity models of motion.

\begin{table*}
	\caption{\label{tab:dvm} Comparison between the manoeuvre values in the KM (Step 1) and CR3BP refinement (Step 2) models}
	\begin{tabular*}{\textwidth}{c @{\extracolsep{\fill}}ccc}
		Asteroid & $\Delta v_{M}^{(1)}$ [m$\cdot$s$^{-1}$] & $\Delta v_{M}^{(2)}$ [m$\cdot$s$^{-1}$] \\\hline	
		2016 RD34 & -3.60 & -2.86 \\
		2012 TF79 & 6.42 & 6.58 \\
		2011 MD & 1.02 & 1.60 \\
		2017 FJ3 & -16.47 & -13.24 \\
		2017 BN93  & 6.46 & 7.06 \\
		2010 VQ98 &  -5.11 & -3.98 \\\hline
	\end{tabular*}
\end{table*}

\section{Summary}
\label{sec:RESconclusions}
This chapter presents a framework to design nearly resonant trajectories in a multi-fidelity model. The design is carried out in several steps, starting from a high-thrust motion using the KM model and ending in a fully developed low-thrust trajectory in the CR3BP.

This trajectory is applied to the concept of asteroid capture missions, exploiting the orbital perturbations occurring in an Earth-resonant motion to increase the retrievable mass for a list of candidate NEA. These are hauled into LPOs in less than two synodic periods, encountering the Earth twice in their motion. 

The retrieved asteroid mass from the resonant trajectory is compared to the one obtained by direct capture and previously published results. Six of them (asteroids 2016 RD34, 2012 TF79, 2011 MD, 2017 FJ3, 2017 BN93 and 2010 VQ98) showed a huge increase with respect to the direct capture and the state of the art. This can be mainly attributed to two decisions: first, the selection of the synodic period with the most advantageous Earth encounter---since each passage will have the asteroid in a different configuration with the planet, distinct results will be obtained; second, the careful exploitation of this encounter with an initial manoeuvre, to put the asteroid into an optimal insertion orbit. In the end, the cost of this initial manoeuvre is negligible compared to the insertion cost, but both the former and the Earth encounters are performed mainly to place the asteroid in the best possible orbital configuration at the manifold insertion. Thus, the highly increased amount of retrievable mass presented by Earth-resonant trajectories is an extremely good case for their usefulness in asteroid capture missions. These trajectories may become valuable options within a future portfolio of asteroid capture missions.

The main drawback of this method is the increased mission time as opposed to a direct capture. This comes down to a mission design trade-off problem, in which flexibility, time and cost have to be managed. This method proves to be the most advantageous when the asteroid's mass is too great for any other type of capture to be feasible. Still, mission time can be decreased by scheduling the spacecraft's rendezvous with the asteroid to happen closer to the Earth encounter. Actually, Figure \ref{fig:full_thrust} contains a large coasting period (Phase A) that could be easily removed by performing the phasing manoeuvre much later. However, the cost of this manoeuvre would grow which, in consequence, would slightly decrease the amount of retrievable mass. 

The multi-fidelity framework presented in this research is very flexible in terms of the desired accuracy, as the user can choose not to undertake all the steps in the process, but instead to stop at wherever it is the most convenient. Plus, the framework can be expanded on fidelity with different tools, depending on the intended application. Future options to include in the framework would be the option to compute trajectories going inside the Hill radius using an alternative to the KM, or a continuation of the design into a full ephemerides model. Nevertheless, the KM proves to be a very good model to approximate the third-body effects while maintaining a low-computational cost, which is very useful for the assessment of multiple trajectory designs. Hence, the framework presented can be adapted to other planetary systems (i.e. Saturn-Titan, Jupiter-Europa), and be used for other purposes than asteroid capture, such as moon tours or end-of-life disposal strategies, showing a range of applications worthy of further study.

\chapter{Application: Computation of Periodic Orbits}
\label{chap:dro}
The computation of periodic orbits for a given planetary system has always been a topic of interest in astrodynamics. In 1969, H\'enon was the first to consolidate the nomenclature regarding their classification \cite{henon_periodic}. He proposed the existence of four natural families of symmetrical, \textcolor{External}{two-dimensional} simple-periodic orbits (only one period, without collisions with the secondary): \textit{a}, \textit{c}, \textit{f} and \textit{g}. Families \textit{a} and \textit{c} originate in the libration points $L_1$ and $L_2$: its orbits are commonly known as libration point orbits (LPOs), and they are discussed in Chapter \ref{chap:dyn}. Families \textit{f}, \textit{g} and their branches (\textit{g'}, \textit{g3}) are composed of orbits around the secondary body: they are respectively called distant retrograde orbits (DROs) and distant prograde orbits (DPOs).

Periodic orbits have been getting increasing attention in mission design. They represent a useful alternative to gravitationally bounded orbits when a space probe needs to remain in the neighbourhood of a celestial body for a long time. DROs and DPOs are particularly interesting due to their relative positioning with the secondary body. However, as opposed to most DPOs, single-periodic DROs are stable in the long-term \cite{lam, ARRM}, being therefore the most sought-after orbital type. 

Several authors have remarked on the possible utility of DROs for missions related to planetary defence \cite{camillaDRO, martaDRO} and exploration \cite{wallace, ARRM}. Thus, these orbits have been computed and defined for many different planetary configurations, ranging from the Sun-Earth-Moon system to the vicinity of near-Earth asteroids (NEAs) and the outer planets \cite{lam, capdevila_paper, andreasDRO}.

While there are many ways to obtain these orbits, their computation usually begins with a set of initial orbital conditions defined in a particular model of motion, as provided by H\'enon using the Hill's problem \cite{henon_periodic, henon_2003}. These are generated by trying several different starting points and testing them to see if a periodic orbit is achieved, in the way of a grid search. Posteriorly, the successfully obtained orbits are corrected into the desired model, and continuation methods can be applied to acquire more orbits of the same family. 

\textcolor{External}{This work contains} a first study on a novel way to obtain the \textcolor{External}{set parameters for a grid search} of DROs, DPOs and their related branching families. Instead of employing the Hill's problem as the model of motion, it employs the PAP-KM previously characterised in Chapter \ref{chap:k3bp}. Given that this is a low-computational cost method, the time undertaken by the grid search remains low. The relative accuracy of the PAP-KM with respect to the CR3BP makes it so that the differential correction of the orbits should take a very short time. Plus, the use of the PAP-KM as opposed to Hill's problem may present different solutions or reveal new \textcolor{External}{theoretical} periodic orbital families. 

\textcolor{External}{In summary, this chapter details a preliminary study on the search and computation of periodic orbits, using the models of motion developed throughout this thesis. Given that the work here presented is still on a developmental stage, the methods and results leave room for expansion. Ideas for future developments will be presented later in the chapter.}

\section{Orbital Characterization}

The definition of a \textit{periodic orbit} was established by Markellos \cite{markellos_1974} as the solution to the equations of motion of a given dynamical model when Eq. \eqref{eq:periodic} holds true for any value of $t_0$.
\begin{align}
\bm{x}(t_0, \mu) = \bm{x}(t_0 + P\cdot T, \mu)\label{eq:periodic} 
\end{align}
in which $\textbf{x}$ is the state vector, $\mu$ is the gravitational parameter of the planetary system, $P$ is an integer and $T$ is the orbital period of the motion. In this way, a solution is said to be $P-$periodic when the initial orbital state repeats itself after $P$ periods.

As stated in Chapter \ref{chap:dyn}, periodic orbits are particular solutions of systems in which \textcolor{External}{at least two celestial} bodies interact. Their computation is typically performed in the Hill's problem, a simplification on the three-body problem in which the gravitational parameter tends to zero ($\mu \rightarrow 0$). The reference frame of motion in which Hill's equations are written \textcolor{External}{is set in the orbital plane of the primaries} and uses the same time variable as the synodic reference frame, with $\xi$ as the horizontal axis and $\eta$ as the vertical axis. Furthermore, the Jacobi constant $C$ \textcolor{Internal}{is subject to} a change of scale to the variable $\Gamma$. The full derivation  of these equations can be found with more detail in literature \cite{szebehely1969theory}.

\begin{figure}[hbt!]
	\centering
	\begin{minipage}[b]{0.43\textwidth}
		\includegraphics[width=\textwidth]{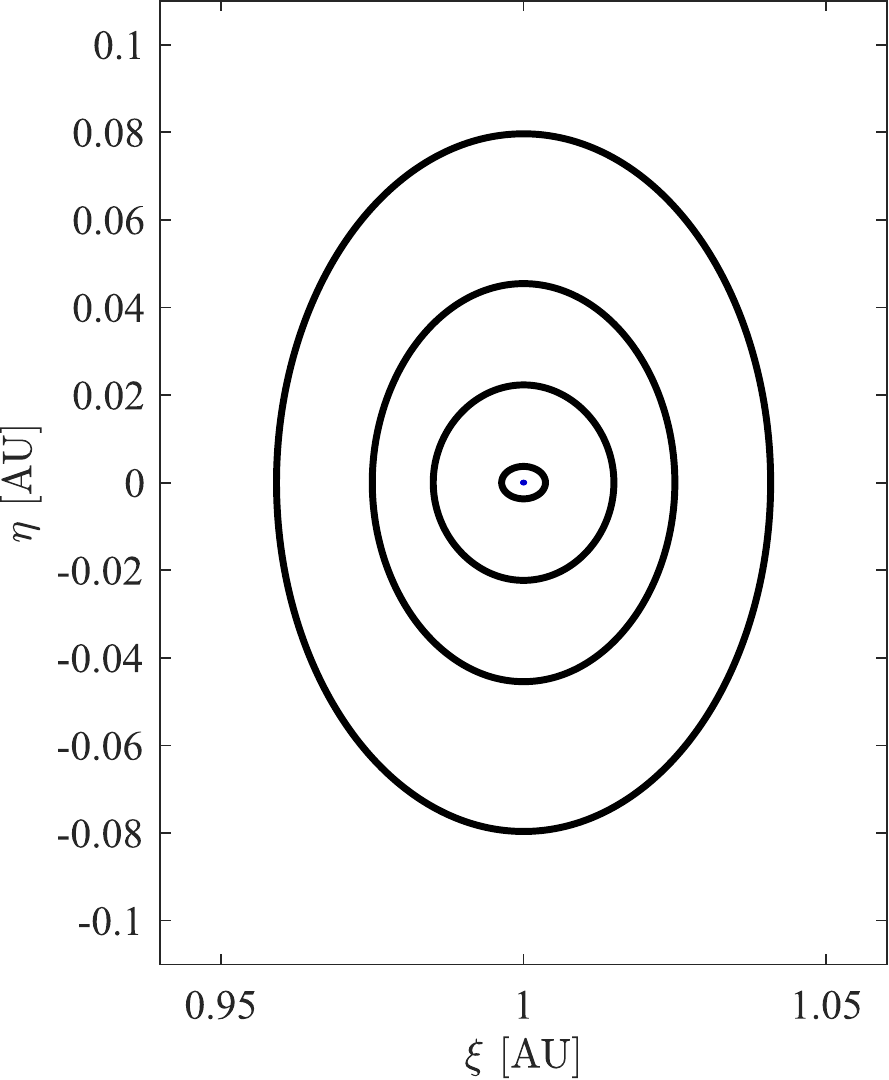}
		\caption*{a) Family \textit{f}, DROs}
	\end{minipage}
	\begin{minipage}[b]{0.43\textwidth}
		\includegraphics[width=\textwidth]{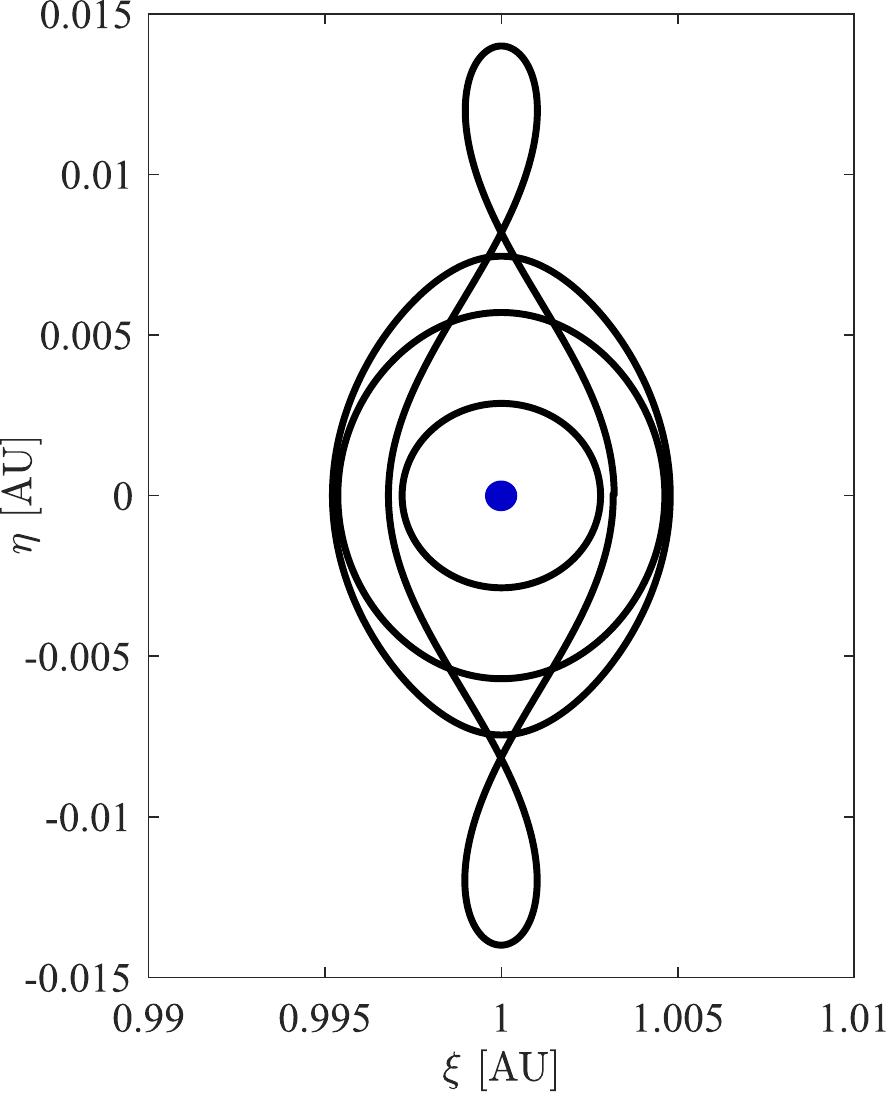}
		\caption*{b) Family \textit{g}, DPOs}
	\end{minipage}
	\begin{minipage}[b]{0.425\textwidth}
		\includegraphics[width=\textwidth]{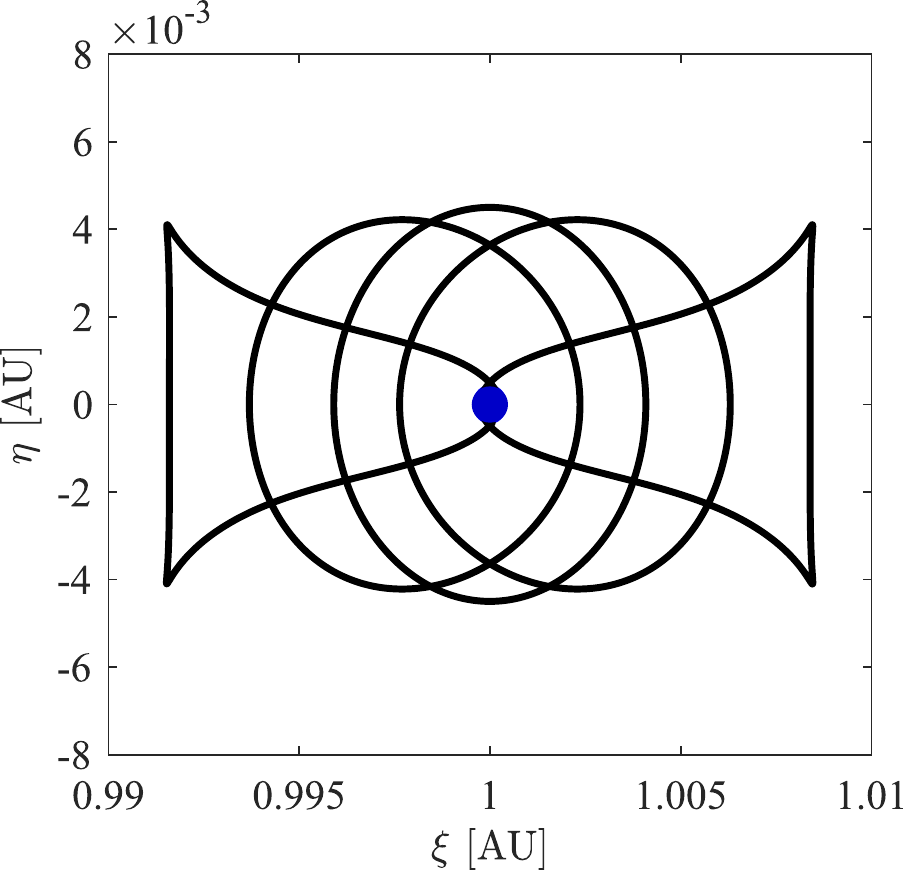}
		\caption*{c) Family \textit{g'}, DPOs}
	\end{minipage}
	\begin{minipage}[b]{0.41\textwidth}
		\includegraphics[width=\textwidth]{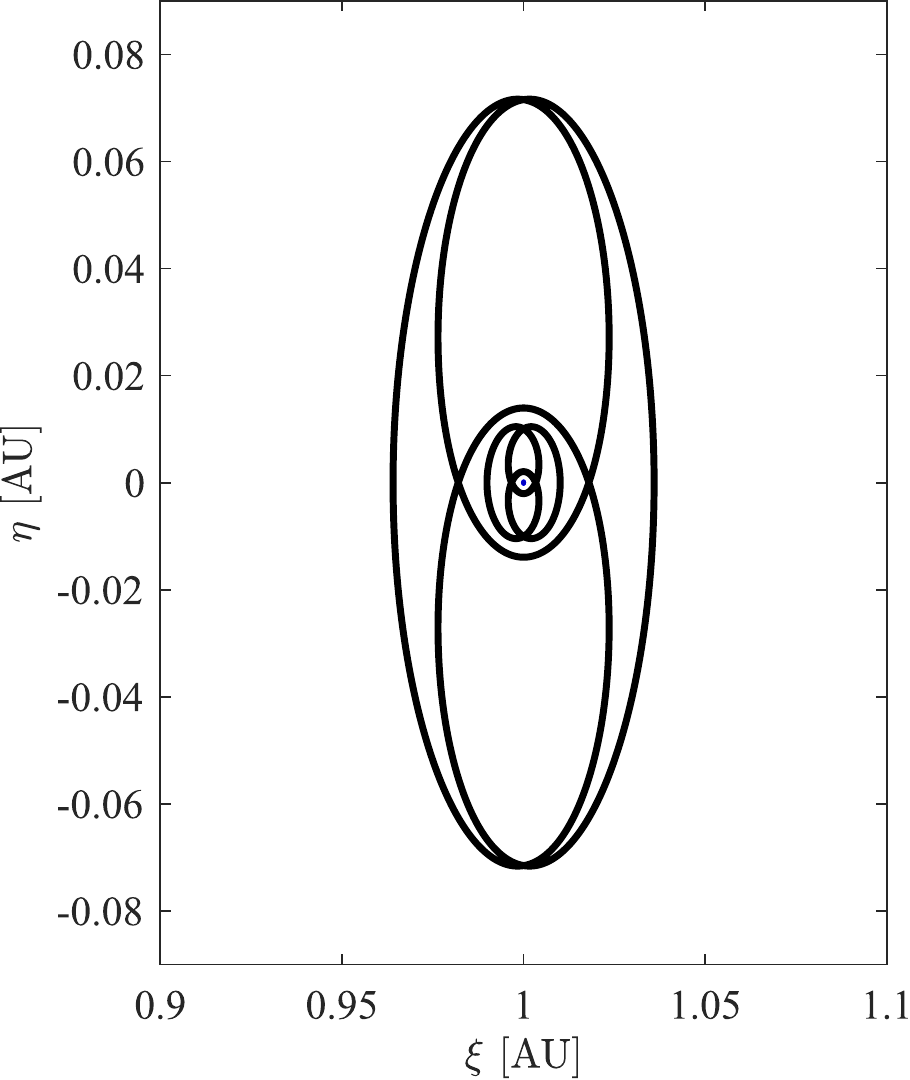}
		\caption*{d) Family \textit{g3}, P3DROs}
	\end{minipage}
	\caption{\label{fig:henon_dro}Periodic orbits of different energies in the Hill's problem. Earth's radius shown as 10 times bigger for visibility, in all cases}
\end{figure}

As an example of the implementation of Hill's problem to compute the periodic orbit families sought after in this project, Figure \ref{fig:henon_dro} was obtained for the Sun-Earth system, using values tabled by H\'enon. It highlights all natural families of periodic orbits around the secondary, together with their branches, for different energy values. It is important to denote that the depicted orbits are merely examples amongst many, as an infinite number of periodic orbits per family exists.

The simple-periodic DROs (family \textit{f}) can be seen on Figure \ref{fig:henon_dro} a): they are stable and symmetrical with respect to their centre point. Family \textit{g} of DPOs is depicted on Figure \ref{fig:henon_dro} b): their stability depends on their energy value, since the motion can get very close to the secondary. Orbits of the branching family \textit{g'} can be seen on Figure \ref{fig:henon_dro} c): most of them are stable, but with decreasing energy values collisions may be obtained (intersections with the secondary), as it can be seen for one of the represented orbits. Finally, on Figure \ref{fig:henon_dro} d), family \textit{g3} is depicted: these orbits correspond to the category of P3DROs (Period-3 DROs). They are triple-periodic DROs but, since they are unstable, they branch out of the \textit{g} family.

\section{Computation of Periodic Orbits}

The generation of periodic orbits around the secondary is typically done in three steps: first, a grid search is performed using the Hill's problem. Second, a differential correction method is implemented to achieve the intended fidelity. Finally, a continuation method is used to obtain orbits of the same family. As such, the stability of the resulting orbit depends on how much the real system \textcolor{External}{resembles the} Hill's problem \cite{henon_book}.

In this section, the grid search segment of the computation of periodic orbits is performed by employing the PAP-KM. The Sun-Earth system is picked as the main example for the computation, although the results can be extrapolated to systems with similar gravitational constants \cite{henon_periodic}. Subsequent implementations of differential correction and continuation methods are not tackled in this work, but they are well documented in literature \cite{henon_book}.

\subsection{Grid Search}   

In H\'enon's study \cite{henon_periodic}, the computed orbits are simple-periodic ($P = 1$), \textcolor{External}{two-dimensional} and symmetrical with respect to the horizontal axis ($\dot{\xi}_0 = 0$). As such, the orbital propagation is done for only half a period, so as to obtain a perpendicular crossing with the horizontal axis. Each planar orbit is then fully parametrised by its initial horizontal position and its energy, in the variable set $[\xi_0, \Gamma]$.

When adopting the PAP-KM, the initial orbital conditions have to be stated as a function of Keplerian elements. Following similar rules to H\'enon's work, only planar orbits starting in the horizontal axis will be considered: this makes both the inclination $i$ and phasing $\alpha$ equal to zero, while the longitude of the ascending node $\Omega$ becomes meaningless. Plus, by the nature of the PAP-KM, this initial orbital state represents the periapsis. As such, the parameters $[\xi_0, \Gamma]$ can be replaced by $[a, e]$, with $i = \omega = \alpha = 0$.

The grid search using the PAP-KM will be computed as a \textit{Low-Cost Likelihood Map} (LCLM), showcasing the \textcolor{External}{chance} that a point representing an initial condition can be easily converted into a periodic orbit of the \textit{f} or \textit{g} families. \textcolor{External}{The term low-cost is here used to highlight that the model of motion used, the PAP-KM, has a lower-computational cost and fidelity when compared to the CR3BP.} 

\subsubsection{Low-Cost Likelihood Maps}

The LCLM is a tool that explores the likelihood of a given initial condition $[a, e]$ being able to represent a periodic orbit around the secondary. For each $[a, e]$ pair, the LCLM yields a \textcolor{External}{likelihood}, presented in a contour map. In order to obtain this value, one must first define the requirements needed to obtain the periodic orbit. In the PAP-KM, three conditions can be inferred:

\begin{enumerate}
	\item $\sum_{i=1}^{P} \Delta a_i = 0$	
	\item $\sum_{i=1}^{P} \Delta e_i = 0$
	\item $\sum_{i=1}^{P} \Delta\alpha_i = 0$
\end{enumerate}

These conditions define that, after $P$ periods, the orbit should have the same Keplerian elements and the same phasing with the secondary as in the beginning.

Using the update in $\alpha$ of Eq. \eqref{eq:alpha} together with these conditions, an easy analytical solution can be obtained for the case of $P = 1$, showing an entire class of orbits, with varying eccentricity, inclination and argument of the periapsis for a fixed semi-major axis that is characteristic of the planetary system. This can be demonstrated by \textcolor{External}{the following equation}:
\begin{align}
	2\pi \bigg(\sqrt{\frac{a}{1 - \mu}} - 1\bigg) = 0 \Rightarrow a = 1 - \mu\label{eq:derive}
\end{align}

However, given the orbital characteristics of a periodic orbit, the $\alpha$ update of Eq. \eqref{eq:alpha} may not be a very accurate estimator for the change in phasing, considering that the semi-major axis changes drastically throughout the motion (as previously discussed in Section \ref{sub:EK}). \textcolor{External}{As such, the LCLM should consider the final change in $\alpha$ together with the updates in $a$ and $e$.}

The grid search for the LCLM starts with a varied set of initial conditions: $a \in [0.96, 1.04]$ and $e \in [0, 0.5]$, with a step in semi-major axis of 0.001 AU and 0.005 in eccentricity. These intervals were deemed coarse enough for the search not to be too cumbersome and for each differentially corrected orbit to be unique, while still being sufficiently fine to show an adequate range of orbits. 

For each initial condition, the PAP-KM is used to update each orbital element for $P$ periods. After the updated orbit is obtained, the desired likelihood is computed by applying a \textit{figure of merit} ($FM$), based on the conditions previously enumerated:
\begin{align}
	FM = \frac{\sum_{i=1}^{P} \Delta a^i}{4} + \frac{\sum_{i=1}^{P} \Delta e^i}{4} + \frac{\sum_{i=1}^{P} \Delta \alpha^i}{2} \label{eq:fm}
\end{align}

\textcolor{External}{The weighting was here chosen so that the cumulative changes in $\alpha$ would have the exact same impact on the computation as the final difference in both orbital elements ($a$ and $e$, whose influence was not distinguished in between them).}
	
Naturally, the higher $FM$ is, the more the orbit changed throughout $P$ periods. Thus, the less likely the orbit can be easily corrected into periodic motion. The $FM$ value is \textcolor{External}{mapped into a likelihood} using the following sigmoid function:
\begin{align}
	\textcolor{External}{S}(FM) = 1 - \tanh(FM), \text{  }\textcolor{External}{S}(FM) \in ]\text{ }0, 1]
\end{align}
which maps increasingly high values of the figure of merit to zero. This simply means that, the smaller the change to the initial orbital elements and relative position, \textcolor{External}{the higher the likelihood of} the orbit being periodic. However, a theoretical figure of merit equal to zero will guarantee an orbit closed in the PAP-KM model, but not necessarily in the CR3BP. 

\begin{figure}[htb!]
	\centering
	\begin{minipage}[b]{0.49\textwidth}
		\includegraphics[width=\textwidth]{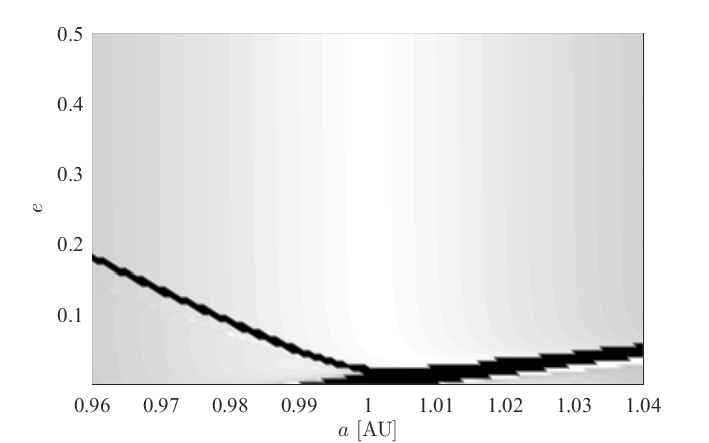}
		\caption*{a) P = 1, CR3BP}
	\end{minipage}
	\begin{minipage}[b]{0.49\textwidth}
		\includegraphics[width=\textwidth]{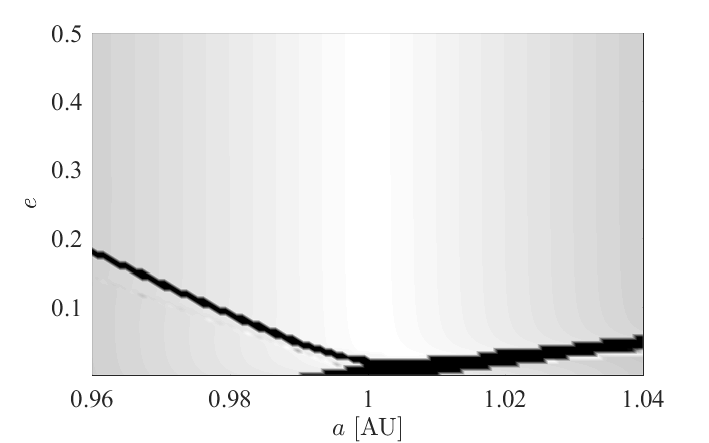}
		\caption*{b) P = 1, PAP-KM}
	\end{minipage}
	\begin{minipage}[b]{0.49\textwidth}
		\includegraphics[width=\textwidth]{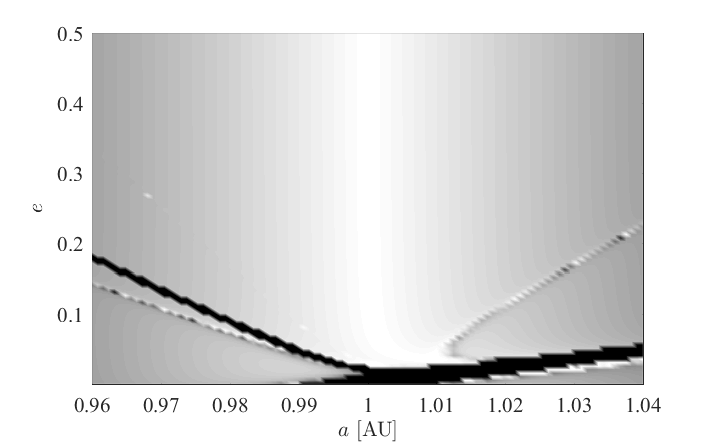}
		\caption*{c) P = 2, CR3BP}
	\end{minipage}
	\begin{minipage}[b]{0.49\textwidth}
		\includegraphics[width=\textwidth]{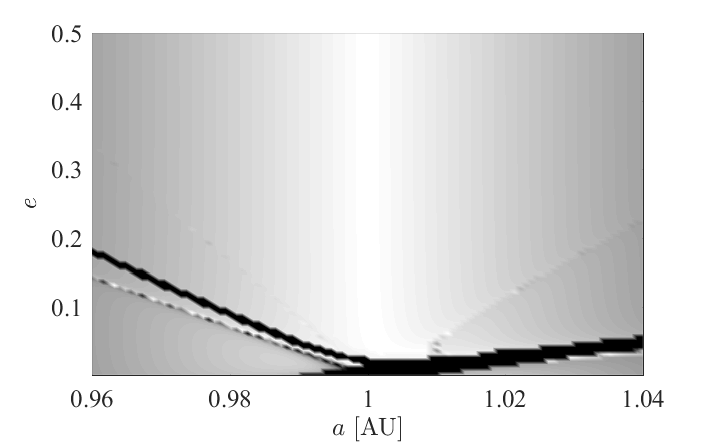}
		\caption*{d) P = 2, PAP-KM}
	\end{minipage}
	\begin{minipage}[b]{0.49\textwidth}
		\includegraphics[width=\textwidth]{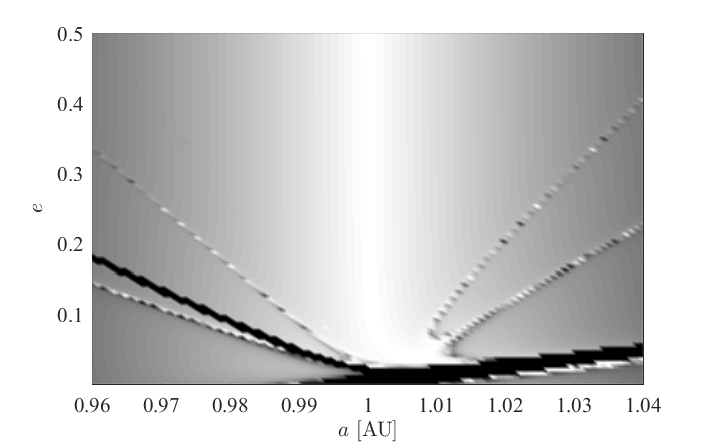}
		\caption*{e) P = 3, CR3BP}
	\end{minipage}
	\begin{minipage}[b]{0.49\textwidth}
		\includegraphics[width=\textwidth]{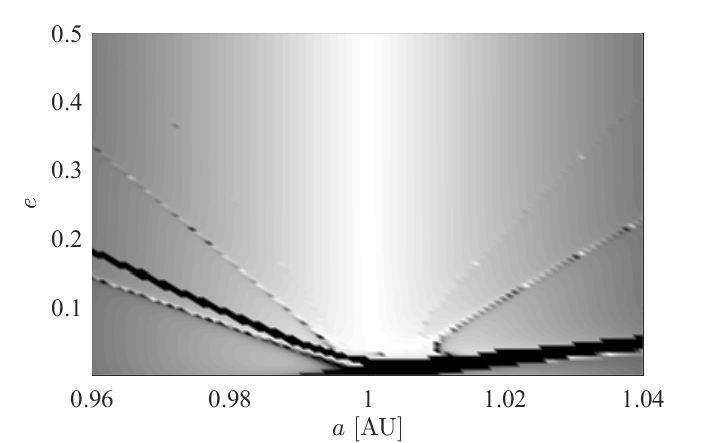}
		\caption*{f) P = 3, PAP-KM}
	\end{minipage}
	\centering
	\begin{minipage}[b]{0.91\textwidth}
		\includegraphics[width=\textwidth]{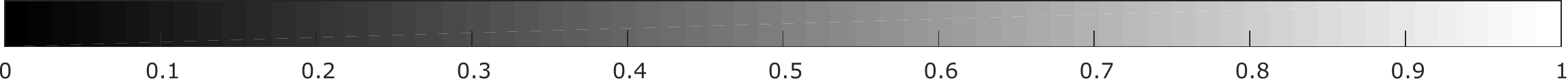}
	\end{minipage}
	\caption{\label{fig:LCLM}Low-Cost Likelihood Map using the CR3BP and the PAP-KM for the search of $P$-Periodic DROs}
\end{figure}

\section{Results and Discussion}

In order to validate the usage of the PAP-KM, the LCLM was first computed in both this model and \textcolor{External}{then adapted to the CR3BP, by converting the Keplerian elements into Cartesian synodic coordinates}. Considering that the previously mentioned families discovered by H\'enon have at most three periods, the plots were made for $P = \{1, 2, 3\}$: these can be seen in Figure \ref{fig:LCLM}. Given that the PAP-KM is not accurate inside the Earth's sphere of influence (as discussed in Section \ref{sec:Accuracy}), the zones of the LCLM corresponding to this region are painted black.

\begin{figure}[hbt!]
	\centering
	\includegraphics[width=0.7\textwidth]{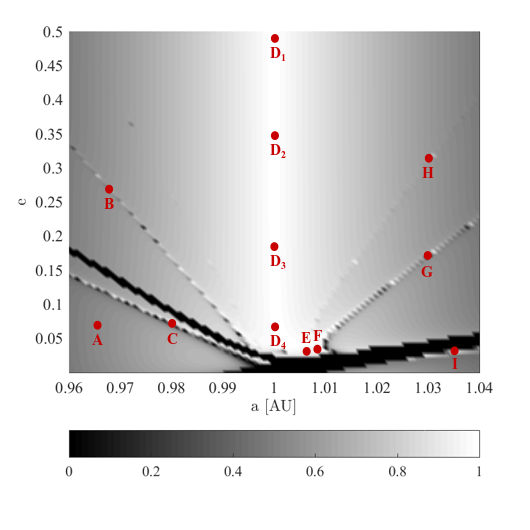}
	\caption{\label{fig:LCLMParticular}Low-Cost Likelihood Map for $P = 3$, highlighting important orbits}
\end{figure}

The first thing to be inferred is that the behaviour of the maps looks very similar for both models of motion, validating the use of the PAP-KM in this scenario. When the value of $P$ increases, two things become evident: first, the overall \textcolor{External}{likelihood} of finding a periodic orbit decreases, as the plots become darker. This happens since an orbit does not change as much in one period as it does for a higher value of $P$. Consequently, the $FM$ will accumulate greater changes in $a$, $e$ and $\alpha$ when propagating non-periodic orbits for an increasing number of periods. Second, some small lines of higher \textcolor{External}{likelihood} will branch out from the central light column: these are more evident in the CR3BP than the PAP-KM. In order to better understand their meaning, the plot for $P = 3$ in the PAP-KM is highlighted in Figure \ref{fig:LCLMParticular}.

\begin{figure}[t!]
	\centering
	\begin{minipage}[b]{0.3\linewidth}
		\includegraphics[width=\textwidth]{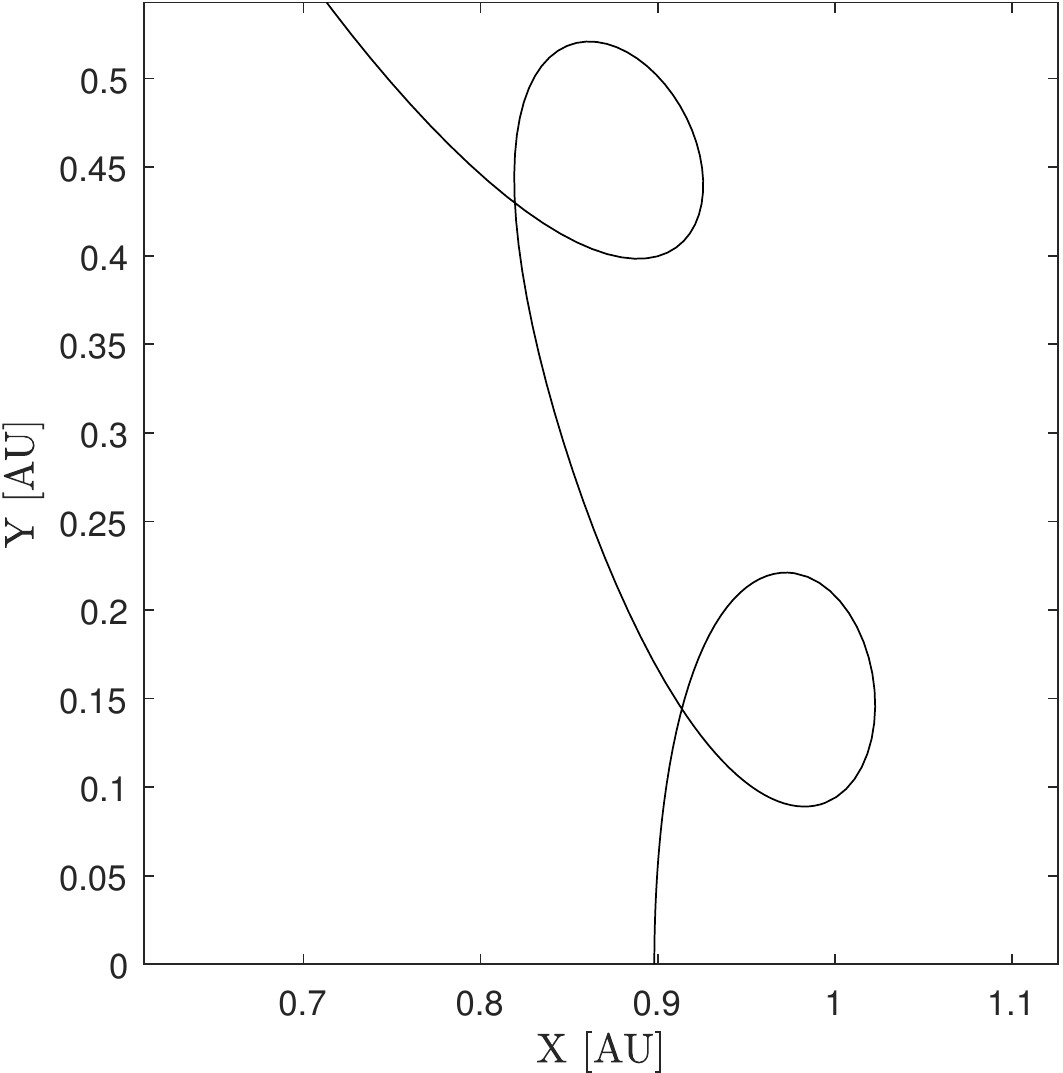}
		\caption*{A}
	\end{minipage}
	\begin{minipage}[b]{0.297\linewidth}
		\includegraphics[width=\textwidth]{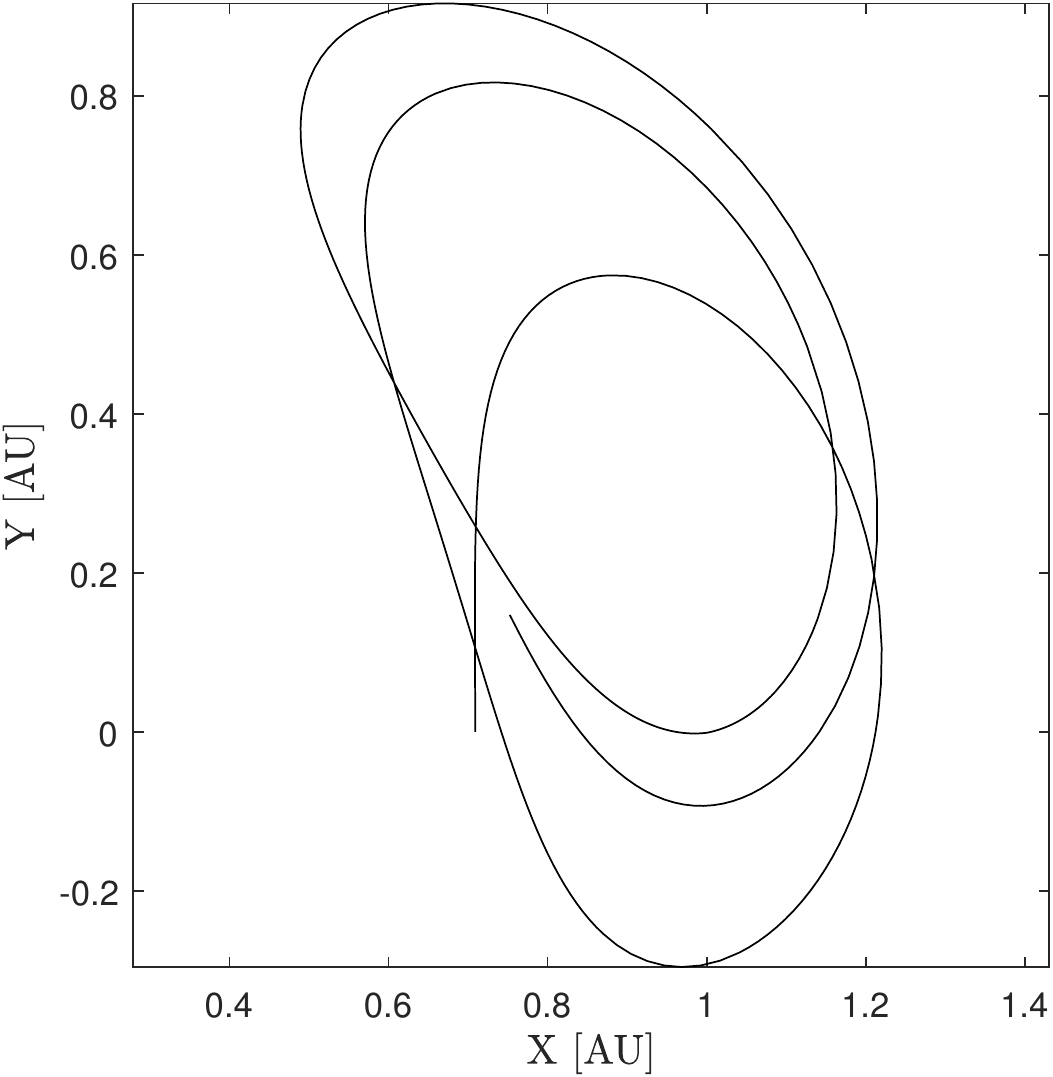}
		\caption*{B}
	\end{minipage}
	\begin{minipage}[b]{0.3\linewidth}
		\includegraphics[width=\textwidth]{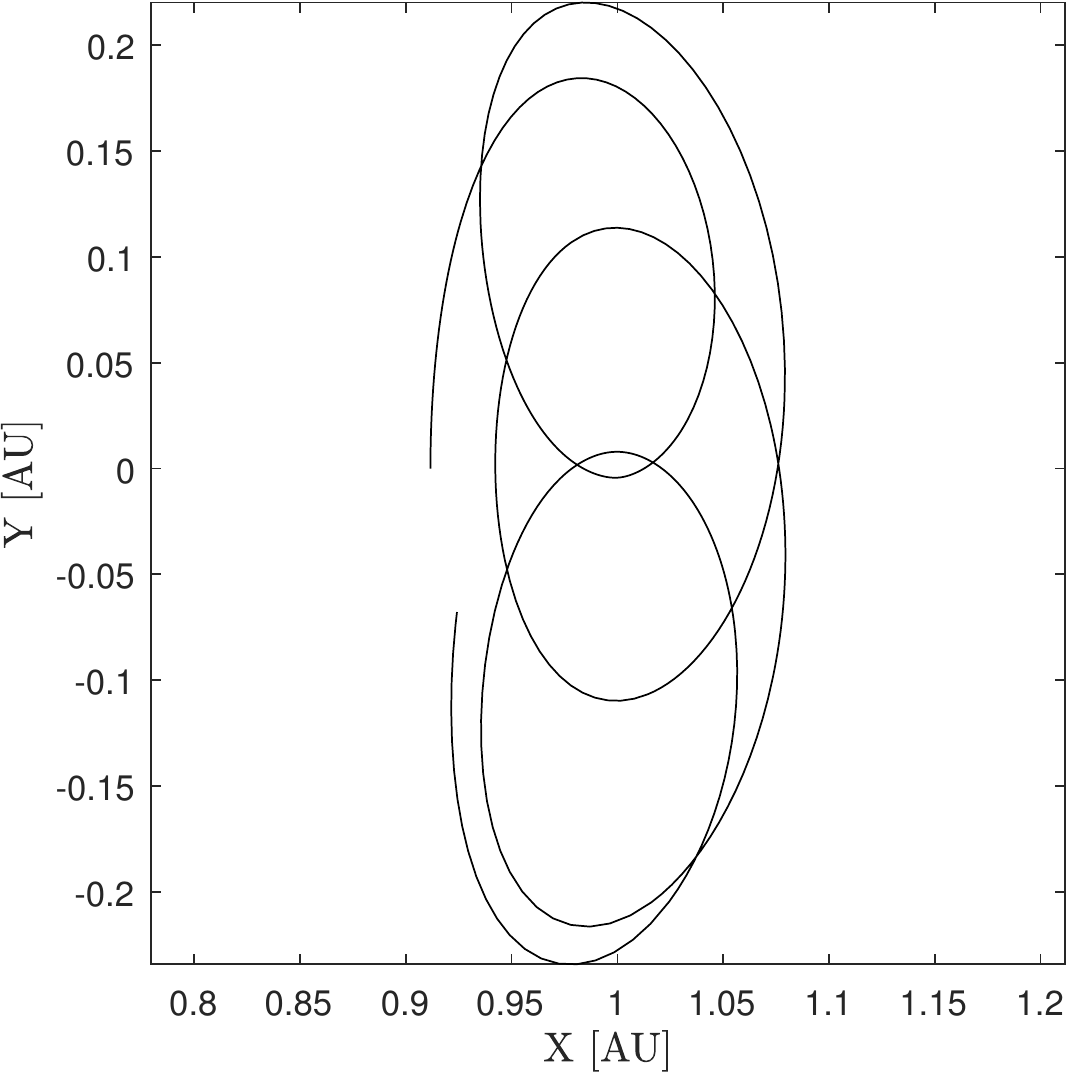}
		\caption*{C}
	\end{minipage}
	\begin{minipage}[b]{0.32\linewidth}
		\includegraphics[width=\textwidth]{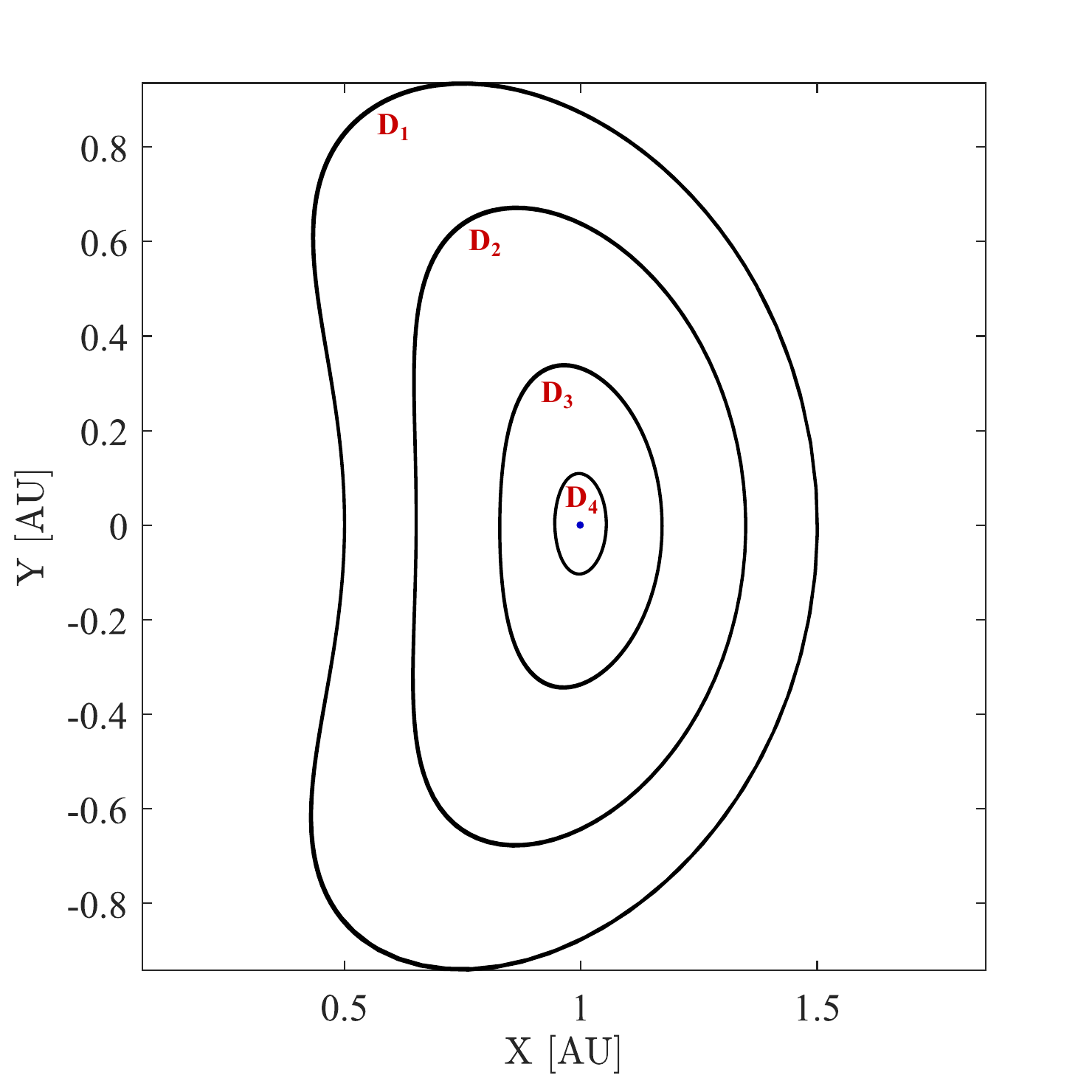}
		\caption*{D}
	\end{minipage}
	\begin{minipage}[b]{0.297\linewidth}
		\includegraphics[width=\textwidth]{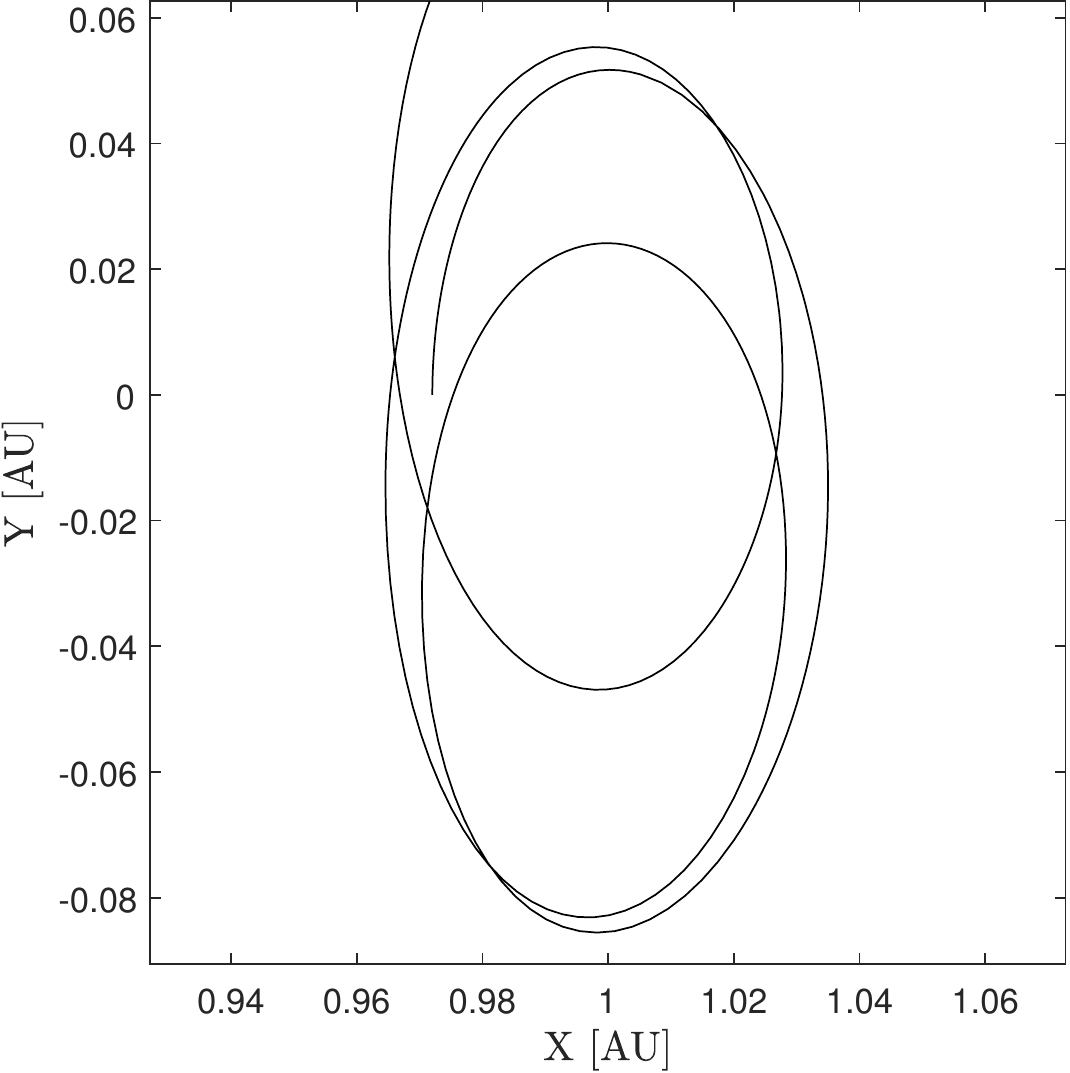}
		\caption*{E}
	\end{minipage}
	\begin{minipage}[b]{0.3\linewidth}
		\includegraphics[width=\textwidth]{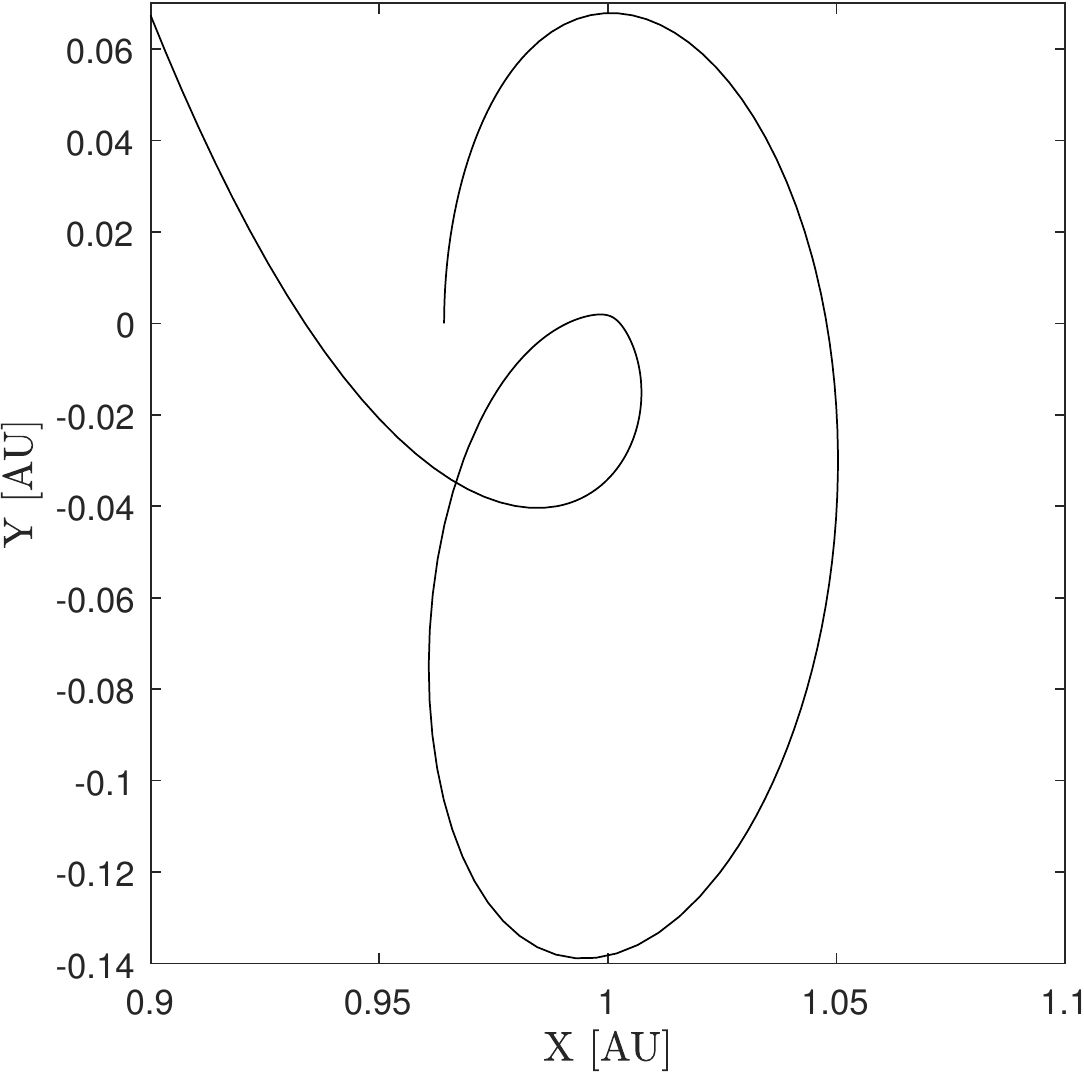}
		\caption*{F}
	\end{minipage}
	\begin{minipage}[b]{0.3\linewidth}
		\includegraphics[width=\textwidth]{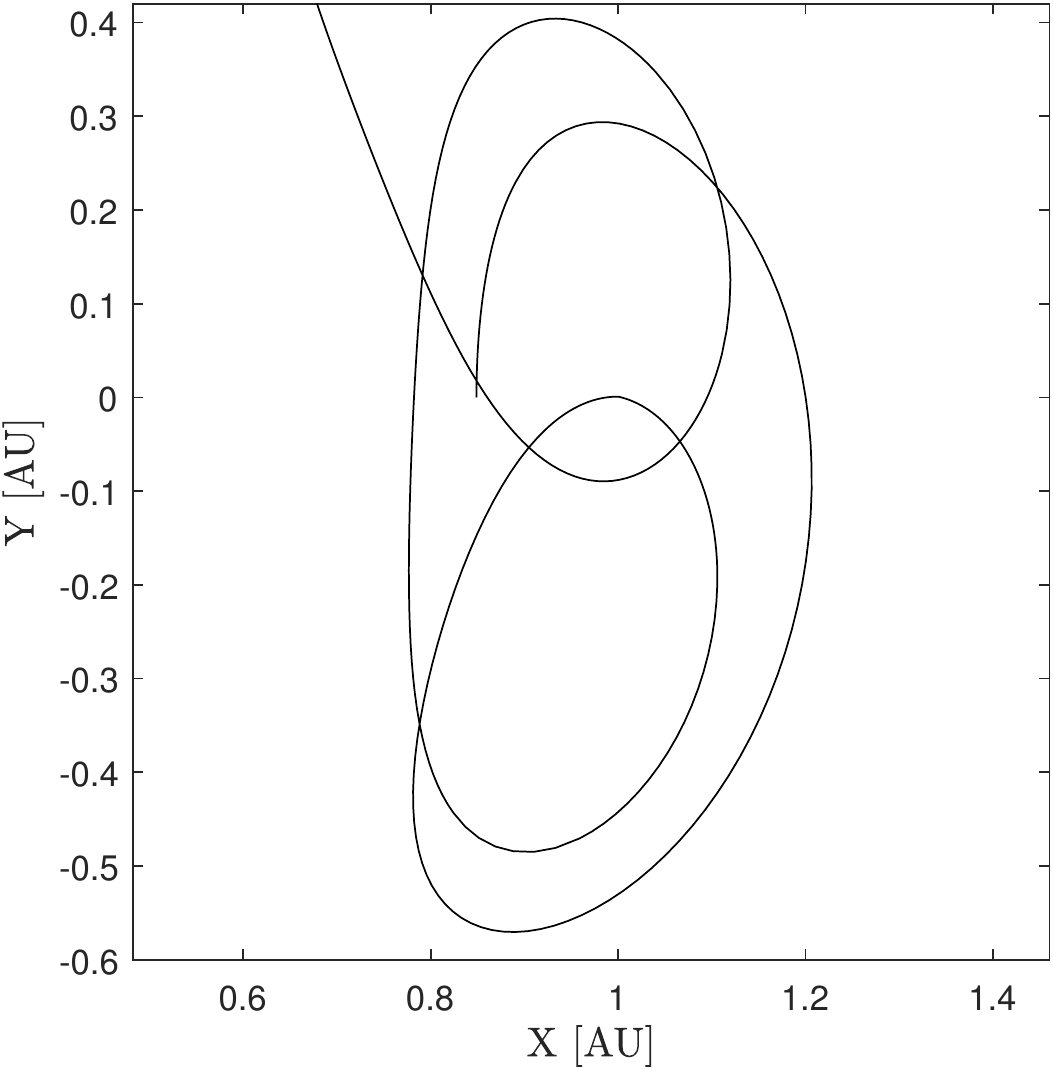}
		\caption*{G}
	\end{minipage}
	\begin{minipage}[b]{0.3\linewidth}
		\includegraphics[width=\textwidth]{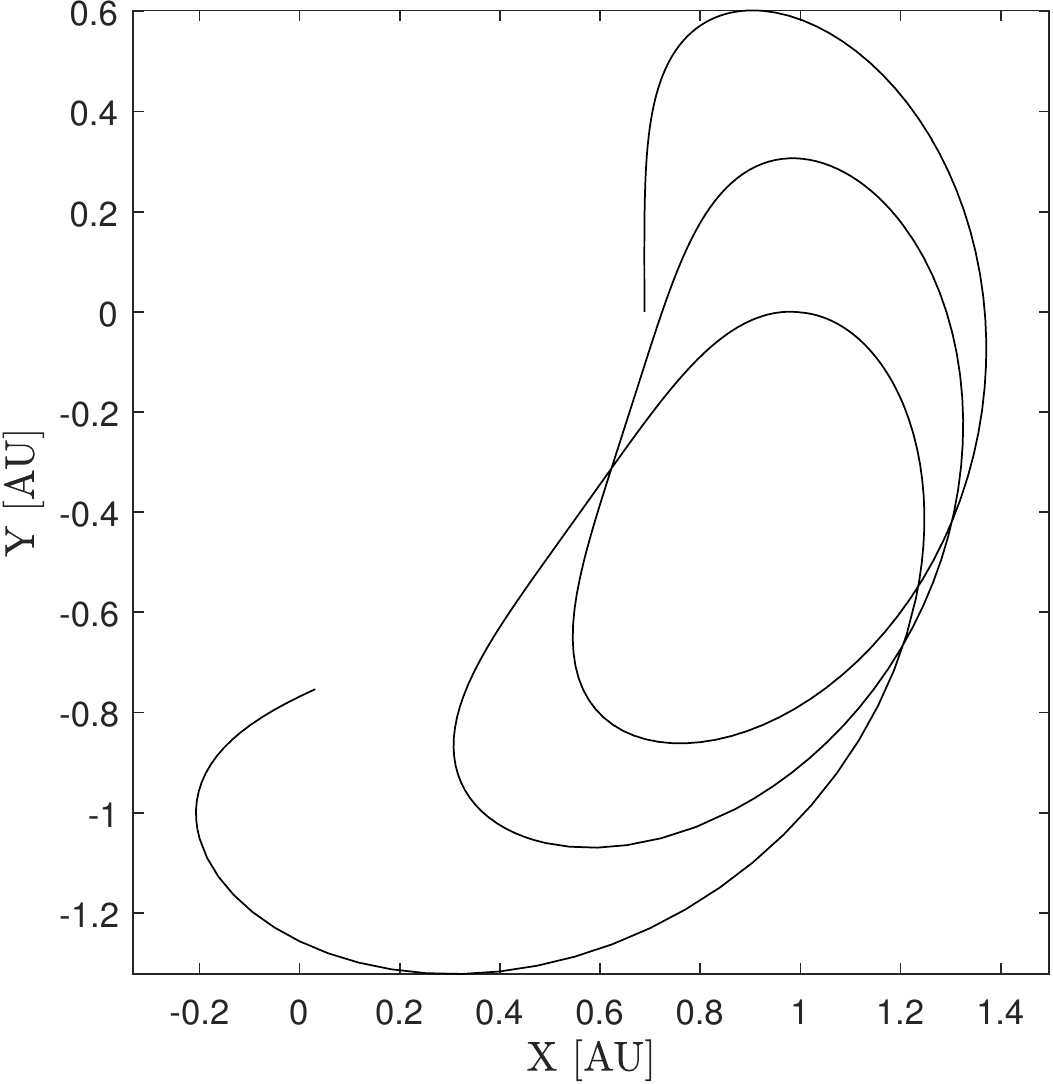}
		\caption*{H}
	\end{minipage}
	\begin{minipage}[b]{0.3\linewidth}
		\includegraphics[width=\textwidth]{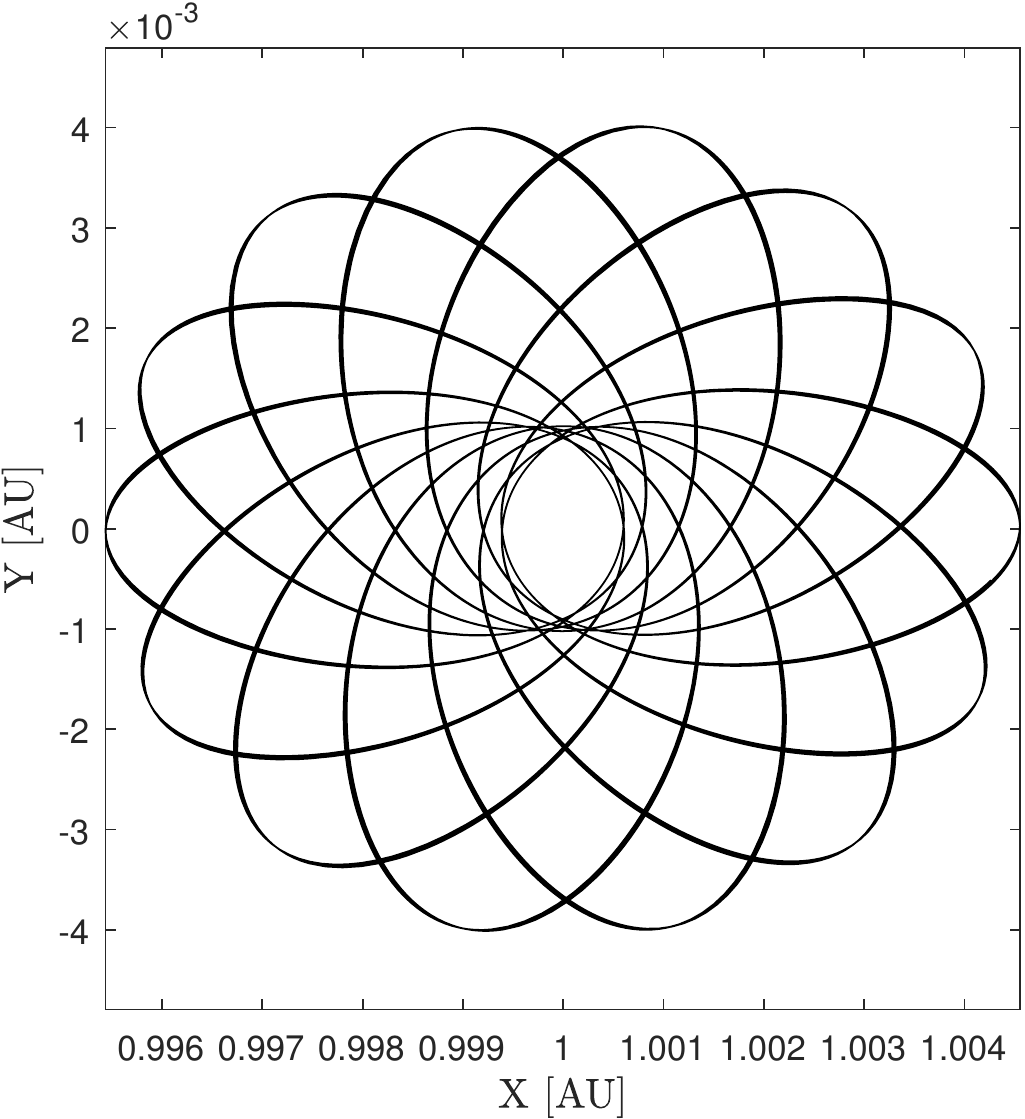}
		\caption*{I}
	\end{minipage}
	\caption{\label{fig:DROindivid}Individual depiction of the highlighted orbits of Figure \ref{fig:LCLMParticular}}
\end{figure}

Observing Figure \ref{fig:LCLMParticular}, \textit{regions of interest} can be identified: areas that indicate the biggest \textcolor{External}{likelihood} to find periodic orbits, shown in a lighter colour. Each of these regions will have examples depicted in Figure \ref{fig:DROindivid}, to highlight their individual characteristics.

The most obvious region of interest is the central light column, around a unitary semi-major axis. This area is analogous to family $f$; this can be seen on Orbits D of Figure \ref{fig:DROindivid}, depicting DROs of distinct sizes for each eccentricity marked in Figure \ref{fig:LCLMParticular}. 

After establishing this zone, the remaining regions are harder to classify. It is obvious that darker areas will not yield any kind of periodicity, as depicted by Orbit A of Figure \ref{fig:DROindivid}. However, the central column seems to have branches departing from it: two to the left, two to the right. Orbits that are very close to the Earth's sphere of influence, as is the case of Orbit I of Figure \ref{fig:DROindivid}, will mostly be found to be quasi-periodic motion. Since they are so close to the perturbing body, it is extremely difficult to fully define these complex and very sensitive orbits.

The bottom left and right-side branches have examples depicted on Orbits C, E, F and G of Figure \ref{fig:DROindivid}. These can be classified as possible P3DROs of varying energy levels, as they show a clear pattern that can possibly be differentially corrected. However, the upper branches of the Figure, depicting Orbits B and H, seem to have no periodicity at all: they are clear outliers. There can be two reasons for this occurrence: either conditions 1 and 2 are very closely met, or the same happens to condition 3. In the first case, this means that the orbit will distance itself from the Earth so fast throughout the propagation that the motion is no longer affected by Earth's perturbation, causing almost no change in the orbital elements. In the second case, the orbits may be quasi-periodic, which binds the motion to a small range of $\alpha$ values that do not shift considerably.

Using the LCLM with this number of points and initial conditions, no orbits of families $g$ and $g'$ appear to be found. The reasoning behind this fact is that these orbital types happen naturally much closer to the secondary than DROs and P3DROs, as it can be observed back in Figure \ref{fig:henon_dro}. This is a regime of motion that makes it extremely difficult for the PAP-KM to accurately convey these orbital families.

\section{Advantages and Limitations}

This chapter describes an initial study on a method to find periodic orbits around the secondary body. This framework uses the PAP-KM to map how a set of initial conditions changes after $P$ periods. The final result is the LCLM, which assigns the \textcolor{External}{likelihood} of a set of initial states to be a periodic orbit.

There is a clear benefit of using the PAP-KM to find periodic orbits: the mapping makes it so that the orbital element update is done only once for each period. In contrast, most cases found in literature require performing the full instantaneous propagation. Thus, this alternative setting is low-cost, follows a systematic method and has a higher fidelity than the Hill's problem, since it does not assume that $\mu$ tends to zero. 

There is one more consideration to be made regarding the complexity of the LCLM. The latter was defined as requiring three conditions for an orbit to be periodic. However, it is postulated that only two would be necessary: conditions 1 and 3, since the eccentricity does not change considerably when compared to the semi-major axis. Plus, the PAP-KM could be used to only update the semi-major axis and $\alpha$ values. Preliminary calculations under this assumptions have yielded LCLMs that are very similar to the ones shown throughout this Chapter, at a much lower computational cost. Nonetheless, the method is prone to errors, since the established conditions do not guarantee a closed orbit. Thus, the study of these requirements and the $FM$, namely the weights given to each of them, leaves many options to be explored.

Another difficulty is the fact that the PAP-KM only works for trajectories in the high energy spectrum, since the distance to the secondary has to be greater than its sphere of influence. Note that the smallest DRO computed in Orbits D of Figure \ref{fig:DROindivid} is the size of the biggest computed with the Hill's problem in Figure \ref{fig:henon_dro}. 

The search is done over orbital elements, contrasting with previous publications that characterize the orbits with Cartesian state vectors and energy. Naturally, the presented orbits are constrained by the values of $[a, e]$ utilised: more families could be computed by extending this search. It is likely to be very complicated to obtain examples of families $g$ or $g'$ with the LCLM, since they contain orbits that move much closer to the secondary. However, these are not as interesting in mission design, since they lack the stability of DROs.

The method can also be extended to asymmetrical orbits \cite{henon_assymmetrical}. However, for this scenario, the former short parametrisation of an orbit is no longer enough. In Cartesian elements, the horizontal velocity $\dot{\xi_0}$ must also be defined; in Keplerian elements, the starting condition could no longer be at periapsis. Thus, to get more diverse solutions, the dimensionality of the problem has to be increased accordingly: a systematic exploration is not as simple as in the case of symmetric orbits, since three initial values have to be simultaneously adjusted. \textcolor{External}{The same reasoning applies to the extension into three-dimensional orbits.}

\chapter{Concluding Remarks}
\section{Summary}

The goal of this investigation was two-fold: first, to develop and extend models of motion to study the third-body perturbation. Second, to verify how these can contribute to on-going research into low-energy trajectories, periodic orbits and their application in innovative concepts, such as asteroid capture missions.

The accurate modelling of the space environment is essential for every space mission. The utilisation of simplified models of motion, like the two-body problem, may be detrimental to particular mission designs. Additional disturbing accelerations can be greatly felt by the spacecraft in the form of third-body effects, which must be accounted for even when well outside the sphere of influence of a body. These effects can yet be exploited in order to obtain lower-cost trajectories.

However, the use of higher-order models to accurately convey the third-body effect and other disturbing accelerations can be quite expensive and time-consuming, especially when characterising the orbit in terms of rotating reference frames. For preliminary mission design, when a very large number of trajectories is computed and considered, lower-fidelity models can be used wisely, provided they are constrained by certain accuracy standards.

This document highlights existing low-fidelity models for the computation of the third-body effect and proposes four other novel ones. All of these are based on perturbation methods where the third-body effect is either described using accelerations or a disturbing function. The necessity of creating many methodologies comes from the possible application scenarios, which range from real-time computations to long-term propagations and mapping. The accuracy of these models is \textcolor{Internal}{generally} shown to be very similar to the CR3BP in the depicted mission design studies; in contrast, the computational cost was established to be much lower.

After developing the models of motion for the third-body perturbation, the related targetting methods were developed, together with tools to obtain low-energy and low-thrust trajectories. These allowed the exploration of the third-body effect for specific mission design cases. In particular, the perturbation of the Earth is used to reduce the capture trajectory costs in so-called Earth-resonant trajectories. This was achieved by concluding that the phasing of the spacecraft with the Earth is the main factor that determines the disturbing effect on the motion. Therefore, by moving the spacecraft accordingly, its relative phasing can be controlled and exploited. These findings were applied to asteroid capture missions, where several bodies were identified as adequate candidates for retrieval. Namely, six NEAs (targets 2016 RD34, 2012 TF79, 2011 MD, 2017 FJ3, 2017 BN93 and 2010 VQ98) yielded a very large increase in retrievable mass with respect to the state of the art. Thus, these trajectories may become valuable options for asteroid capture mission design: either for making larger targets accessible, or for lowering the fuel cost when retrieving smaller bodies.

\section{Recommendations}

By extending the analysis completed in this investigation, additional insights may be obtained into some of the covered topics. Recommendations for continuing this investigation include:

\begin{description}
	\item[On the Keplerian third-body potential and its applications:] this disturbing function and the related models of motion, both for conservative and non-conservative forces, have shown great promise in the depiction of the third-body perturbation. The computational \textcolor{External}{speed} of all the derived models can be improved by coding them in a lower-level language or improving the orbital propagation equations. Furthermore, since these models are based on the Lagrange planetary equations and Gauss' variational equations, the singularities related to the use of orbital elements can be resolved for an overall better model.
	
	\textcolor{Internal}{The Hamiltonian used to derive the K3BP is written as a Taylor expansion with $\mathpzc{O}(\mu^2)$. The equation can be developed such that higher-order terms are included. Some preliminary studies were made on this part, but no interesting increases in accuracy were achieved. However, this idea may be especially valuable for systems of greater gravitational parameter than the Sun-Earth one.}
	
	In terms of application scenarios, the provided computation of the third-body perturbation can be useful in several different mission designs. The case of Jovian and Saturnian moon tours, in which a spacecraft is subject to varied gravitational perturbations simultaneously, is one of them. Another would be the disposal of spacecraft at the end of life, by doing long-term propagation to guarantee that the object always remains at a minimum distance from the Earth.
	
	\item[On Earth-resonant capture missions:] the option to perform Earth encounters to decrease the capture $\Delta v$ opens a wide range of possibilities for mission design. One of them is to increase the number of Earth encounters: it is predicted that, when this value increases, the $\Delta v$ can be even further minimised. This would, however, increase the time taken by the trajectory immensely. A strategy to make the mission shorter could be to investigate the application of the $\Delta v_M$ manoeuvre when the spacecraft is closer to the asteroid, and not so many orbital periods before they meet. 
	
	Regarding the trajectory computation, the estimation using the filter can be further improved by using alternative ways to compute the manoeuvre, such as analytical Lambert arcs. This would possibly remove outliers, in order to obtain a more consistent list of target bodies. 
	
	The presented multi-fidelity framework for nearly-resonant encounters has similar potential applications to the previous point: from moon exploration missions to planetary protection analysis, amongst others. 
	
	\item[On the computation of periodic orbits:] given that this is a preliminary study on the subject, many topics are left to explore. The main one is the definition of \textit{Figure of Merit} and the conditions for a periodic orbit to exist in a mapping model. \textcolor{Both}{One metric that could better validate the usage of the LCLM and provide a nice tuning for the $FM$ would be the comparison with a list of known periodic orbits in the CR3BP. These orbits would be characterised by their semi-major axis and eccentricity, and could therefore be plotted on the LCLM to understand if their existence matches the high-likelihood regions of the latter.}
	
	The LCLM can be expanded so the initial states for the grid search include more orbital values and are extended to the spatial case ($i \neq 0$) or the asymmetrical case ($\nu \neq 0$). Furthermore, a differential correction process and a continuation method can both be implemented to further build the orbital families found in the PAP-KM. Although unlikely, other orbital types that are not represented in the Hill's problem may be found. Finally, the achieved corrected orbits can be used for mission design and exploration, or to compute low-energy trajectories connecting periodic orbits of the Solar System.	
\end{description}

\bibliographystyle{unsrtnat}
\bibliography{biblio}


\end{document}